\documentclass[aps,prb,onecolumn,superscriptaddress,amsmath,amsfonts]{revtex4}

\usepackage[dvipdfmx]{graphicx}
\usepackage{bm} 
\usepackage{braket}
\usepackage[usenames]{color}
\usepackage{comment}
\usepackage{amssymb}
\usepackage{multirow}
\usepackage{accents}

\makeatletter
  \def\widebar{\accentset{{\cc@style\underline{\mskip10mu}}}}
  \makeatother

\makeatletter
\renewcommand{\theequation}{%
\arabic{section}.\arabic{equation}}
\@addtoreset{equation}{section}
\makeatother

\begin{document}


\title{Topology of 
crystalline insulators and superconductors}

\author{Ken Shiozaki}
\affiliation{Department of Physics, Kyoto University, Kyoto 606-8502, Japan}
\author{Masatoshi Sato}
\affiliation{Department of Applied Physics, Nagoya University, Nagoya 464-8603, Japan}




\date{\today}

\begin{abstract}
We complete a classification of topological phases
 and their topological defects
in crystalline insulators and superconductors.
We consider topological phases and defects described by non-interacting
 Bloch and Bogoliubov de Gennes Hamiltonians that support additional
 order-two spatial symmetry, besides any of ten classes of
 symmetries defined by time-reversal symmetry and particle-hole symmetry.
The additional order-two spatial symmetry we consider is general and it includes
 ${\bm Z}_2$ global symmetry, mirror reflection,
 two-fold rotation, inversion, and their magnetic point group symmetries.  
We find that the topological periodic table shows a novel periodicity in
 the number of flipped coordinates under the order-two
 spatial symmetry, in addition to the Bott-periodicity in the space
 dimensions. 
Various symmetry protected topological phases and gapless modes 
will be identified and discussed in a unified framework.
We also present topological classification of symmetry protected Fermi points.
The bulk classification and the surface Fermi point
 classification provide a novel realization of 
the bulk-boundary correspondence in terms of the K-theory.
\end{abstract}



\maketitle

\tableofcontents

\section{Introduction}
\label{Intro}
Symmetry and topology have been two important principles in physics,   
both of which result in quantum numbers and the conservation laws. 
In many-body systems, symmetry can be broken spontaneously as a
collective phenomenon. The spontaneous symmetry breaking, which is
characterized by local order parameters, describes many quantum phases such as
ferromagnetism and superconductivity.

Topology also describes quantum phases that are
not captured by spontaneously symmetry breaking. 
Instead of local order parameters, those quantum phases are characterized by topological numbers of wave functions.
Such quantum phases are called as topological phase. \cite{volovik2003universe}
Integer and fractional quantum Hall systems are two representative examples of topological phases.
\footnote{
More specifically, 
these two states are classified into two different categories of
topological phase:
Integer quantum Hall states belong to a short range
entangled topological phase, but fractional quantum Hall states belong
to a long range entangled one.
Whereas short range entangled topological phases does not have topological
degeneracy, i.e. they have a unique ground state
on a closed real space manifold,
long range entangled ones show topological degeneracy.
The presence of symmetry crucially enriches possible short range
entangled topological phases, which referred to as symmetry protected
topological phase. \cite{PhysRevB.87.155114}
In particular, those in 
non-interacting fermionic system are called as topological
insulator and superconductor, which we will discuss in this paper.
}
The ground state wave functions of these quantum Hall states host non-zero Chern numbers, 
which directly explain the quantization of the Hall conductivity. \cite{thouless1982quantized,
 kohmoto1985topological, niu1985quantized}
In general, a topologically nontrivial phase can not adiabatically deform into
a topologically trivial one, and it is robust under perturbations
and/or disorders unless the bulk gap closes.


It has been recently discovered that topological phases are
 enriched by general symmetries of time-reversal and charge conjugation
 \cite{hasan2010colloquium, qi2011topological, volovik2011topology, tanaka2012symmetry, ando2013topological, budich2013adiabatic, fruchart2013introduction}.
Those non-spatial symmetries can persist even in the presence of
 disorders and/or perturbations. 
For instance, non-magnetic disorders retain time-reversal symmetry (TRS), and
 thus a non-trivial topological phase accompanied by TRS is robust against non-magnetic disorders.
Quantum spin Hall states \cite{kane2005quantum, kane2005z_, bernevig2006quantum} and
 topological insulators \cite{moore2007topological, fu2007topological,
 fu2007topological2, roy2009z_, qi2008topological} support such topological phases
 protected by TRS.
In a similar manner, charge conjugation symmetry specific to
superconductivity makes it possible to realize a novel topological
state of matters, topological superconductor. \cite{read2000paired,
 ivanov2001non, kitaev2001unpaired, sato2003non, fu2008superconducting,
 linder2009unconventional, qi2009time,
 schnyder2008classification, sato2009topologicalp,
 sato2010topological, tanaka2009theory, sato2009topological, sato2009non, PhysRevLett.105.097002, sau2010generic,
 alicea2010majorana, sato2010non, sato2010existence, lutchyn2010, oreg2010, PhysRevLett.103.107002, PhysRevLett.108.147003, 
PhysRevB.87.134519, PhysRevB.90.045130, PhysRevB.90.041301, PhysRevB.89.155140, wan2014turning}
Topological phases enriched by
 those general non-spatial symmetries are classified for non-interacting fermionic systems,
 \cite{schnyder2008classification, kitaev2009periodic,
 teo2010topological, stone2011symmetries, abramovici2012clifford, wen2012symmetry}
in terms of the Altland-Zirnbauer (AZ) tenfold symmetry
 classes.~\cite{altland1997nonstandard}

Whereas the classification based on the non-spatial symmetries
successfully captures topological nature of general systems,  
real materials often have other symmetries specific to their structures
such as translational and point group symmetries.
Those additional symmetries also give rise to a non-trivial
topology of bulk wave functions  
and gapless states on boundaries. \cite{volovik1987, ryu2002topological, sato2006nodal, teo2008surface, 
beri2010topologically, yada2011, sato2011topology, 
schnyder2011topological, wan2011topological, yang2011quantum,
burkov2011weyl, fu2007topological2, hatsugai2010symmetry, 
hughes2011inversion, turner2010entanglement, 
sato2009topologicalp, sato2010topological, fu2010odd}
It had been naively anticipated that the gapless boundary modes
are fragile against disorders because these specific symmetries are
microscopically sensitive to small perturbations, but recent studies of
topological crystalline insulators have shown that if the symmetries
are preserved on average, then the existence of some gapless boundary states
is rather robust.~\cite{mong2012quantum, ringel2012strong,
fu2012topology, fulga2012statistical} 
Moreover, surface gapless states protected by the mirror reflection
crystal symmetry have been observed
experimentally.~\cite{fu2011topological, hsieh2012topological, tanaka2012experimental, dziawa2012topological,
xu2012observation}
Motivated by those progresses, various symmetries and corresponding 
topological phases have been elucidated in insulators 
\cite{fang2012bulk, slager2012space, jadaun2013topological, fang2013entanglement, chiu2013classification, liu2013topological, alexandradinata2014spinless}
and superconductors.~\cite{teo2013existence, ueno2013symmetry, zhang2013topological, liu2013majorana, benalcazar2013classification}
In particular, various symmetry protected Majorana fermions have been
predicted in spinful unconventional superconductors or superfluids
\cite{ueno2013symmetry, zhang2013topological, tsutsumi2013upt3, sato2014mirror}.

In this paper, we complete a topological classification
of crystalline insulators and superconductors that support additional
order-two spatial
symmetry besides ten classes of discrete AZ symmetries.
Our classification reproduces previous
results for additional
reflection symmetry~\cite{chiu2013classification,morimoto2013topological}, but 
the symmetry we consider is general, and it also includes
global ${\bm Z}_2$ symmetry, two-fold rotation, and inversion.
Furthermore, the additional symmetry can be anti-unitary. 
Although ordinary point group symmetries are given by unitary operators, 
systems in a magnetic field or with a magnetic order often support an
anti-unitary symmetry as a magnetic point group symmetry.
The magnetic symmetry also has been known to provide non-trivial
topological phases in various systems \cite{sato2009topological,mong2010antiferromagnetic, 
mizushima2012symmetry, mizushima2013topological, fang2013topological, liu2013antiferromagnetic, 
kotetes2013classification, fang2013new, zhang2014topological}.


Our approach here provides a unified classification of topological
phases and defects in crystalline insulators and superconductors with 
additional order-two symmetry.
The topological classification we obtain indicates that topological
defects can be considered as boundary states in lower dimensional systems.
The resultant topological periodic table shows a novel periodicity in
 the number of the flipped coordinates under the order-two additional
 spatial symmetry, in addition to the Bott-periodicity in the space
 dimensions. 
Using the new topological periodic table, various symmetry protected
topological gapless modes at topological defects are identified in a
unified manner.

In addition, we also present a topological classification of Fermi points in the
crystalline insulators and superconductors.
The bulk topological classification and the Fermi
point classification show the bulk-boundary
correspondence in terms of the K-theory.  

  
The organization of this paper is as follows. 
In Sec. \ref{Set}, we explain the formalism we adapt in this paper. 
In this paper, we use the approach based on the
K-theory. \cite{atiyah1964clifford, atiyah1966x, dupont1969symplectic, karoubi2008k}
Our main results are summarized in Sec. \ref{Main}.
We show relations
between K-groups with different order-two additional spatial
symmetries and dimensions. 
The derivation and proof are given in Sec. \ref{TCI}.  
In Sec. \ref{sec:property}, we discuss properties of the obtained K-groups in the
presence of additional
symmetry. A novel periodicity in the number of flipped coordinates under
the additional symmetry is pointed out. 
We also find that the K-groups naturally implement topological defects as
boundaries of lower dimensional crystalline insulators/superconductors.
Crystalline weak topological indices are argued in Sec. \ref{sec:weak}. 
In Sec. \ref{Tab}, we present topological classification tables of crystalline
insulators/superconductors and their defect zero modes with order-two
additional spatial symmetry. 
The topological periodic tables are classified into four families.
Various symmetry protected topological phases and their gapless defect
modes are identified in a unified framework.
We also apply our formalism to a classification of Fermi point
protected by additional order-two symmetry in Sec. \ref{sec:Fermi_point}.
By combing the results in Secs. \ref{Main} and \ref{sec:Fermi_point}, the bulk-boundary
correspondence of K-groups are presented. 
In Sec. \ref{sec:Majorana_Ising}, we demonstrate that the Ising character of Majorana
fermions is a result of symmetry protected topological phases.
In Sec. \ref{sec:pump}, we apply our theory to anomalous topological pumps 
in Josephson junctions, in which crystalline symmetry is not essential
to lead to a new topological classification.
We conclude the present paper with some discussions in Sec. \ref{Conc}.  

Some technical details are presented in Appendices.
In Appendix \ref{appendix:map}, following
Ref. \onlinecite{teo2010topological}, we introduce useful maps between
Hamiltonians in different dimensions.
The isomorphic maps introduced here are used in
Sec. \ref{TCI}.
We review the dimensional hierarchy of AZ classes 
in the absence of additional symmetry in Appendix \ref{AZ}.
The classifying spaces of AZ classes with additional order-two symmetry
are summarized in Appendix \ref{Sec_Clif}.
The definition and the basic properties of Chern numbers, winding
numbers, and $\mathbb{Z}_2$ topological numbers 
which are used in this paper,  
are given in Appendix \ref{appendix:topological_invariant}.
Throughout this paper, we use the notation $s_{\mu}$, $\tau_{\mu}$ and
$\sigma_{\mu}$ $(\mu=0,1,2,3)$ to represent the Pauli matrices in the
spin, Nambu and orbital spaces, respectively.     

\section{Formalism}
\label{Set}
In this section, we briefly give our set up of the classification problem. 
The reader who only concerns the classification table with an additional symmetry, please see Sec. \ref{Tab}. 

\subsection{Spatially Modulated Hamiltonian}

In this paper, we consider band-insulators and superconductors which are
described by Bloch and Bogoliubov de Gennes (BdG) Hamiltonians, respectively.
In addition to uniform ground states, we also consider
topological defects of these systems. 
Away from the topological defects, 
the systems are gapped, and they are
described by
spatially modulated Bloch and
BdG Hamiltonians,
\cite{volovik2003universe,teo2010topological}
\begin{eqnarray}
{\cal H}({\bm k}, {\bm r}).
\label{eq:hamiltonian}
\end{eqnarray}
Here the base space of the Hamiltonian is composed of momentum
${\bm k}$, defined in the $d$-dimensional Brillouin zone $T^d$, and
real-space coordinates ${\bm r}$ of a $D$-dimensional sphere
$S^D$ surrounding a defect. 
For instance,
the Hamiltonian of a point defect in three-dimensions is given by $\mathcal{H}(k_x,k_y,k_z,r_1,r_2)$, where $(r_1,
r_2)$ are the coordinates of a two-dimensional sphere $S^2$ surrounding
the point defect.
Another example is a line defect in three-dimensions, in which the
Hamiltonian is $\mathcal{H}(k_x,k_y,k_z,r_1)$ where $r_1$ is
a parameter of a circle $S^1$ enclosing the line defect.
The case of $D=0$ corresponds to a uniform system.

As mentioned above, the exact base space is $T^d\times S^D$, but instead
we consider a simpler space $S^{d+D}$ in the following.
This simplification does not affect on ``strong'' topological nature of
the system.    
Although the difference of the base space may result in ``weak''
topological indices of the system, 
they can be obtained as``strong'' topological indices in lower
dimensions, as will be argued in Sec.\ref{sec:weak}.
Therefore, generality is not lost by the simplification.

Below, we treat ${\bm k}$ and ${\bm r}$ in the Hamiltonian as
classical variables,
i.e. momentum operators
$\hat{\bm{k}}$ and coordinate operators $\hat{\bm{r}}$ are commute with
each other. 
This semiclassical approach is justified if the
characteristic length of the spatial inhomogeneity is 
sufficiently longer than that of the quantum coherence.
A realistic Hamiltonian would not satisfy this
semiclassical condition,  but if there is no bulk gapless mode,
then the Hamiltonian can be adiabatically
deformed so as to satisfy the condition.
Because the adiabatic deformation does not close the bulk energy gap,
it retains the topological nature of the system. \cite{niemi1986fermion, teo2010topological,
sato2011topology, shiozaki2012index}

\subsection{Symmetries}

\subsubsection{Altland-Zirnbauer Symmetry Classes}

In the present paper, we classify the topological phases that have
an additional symmetry, beside any of the ten AZ symmetry
classes.
Here we briefly review the AZ symmetry classes. 

The AZ symmetry classes are defined by the presence or absence of
TRS, particle-hole symmetry (PHS) and/or chiral
symmetry (CS).
The AZ symmetries, TRS, PHS, and CS, imply 
\begin{eqnarray}
&&T{\cal H}({\bm k}, {\bm r})T^{-1}={\cal H}(-{\bm k}, {\bm r}),
\nonumber\\ 
&&C{\cal H}({\bm k}, {\bm r})C^{-1}=-{\cal H}(-{\bm k}, {\bm r}),
\nonumber\\ 
&&\Gamma{\cal H}({\bm k}, {\bm r})\Gamma^{-1}=-{\cal H}({\bm k}, {\bm r}),
\label{eq:AZsymmetry}
\end{eqnarray}  
respectively, where $T$ and $C$ are anti-unitary operators and $\Gamma$
is a unitary operator.
For spin-1/2 fermions, time-reversal operator $T$ is given by
$T=is_y{\cal K}$
with the Pauli matrix $s_i$ in the spin space and the complex conjugation
operator ${\cal K}$, which obeys $T^2=-1$.
In the absence of the spin-orbit interaction, spin rotation symmetry
allow a different time-reversal symmetry $T={\cal K}$ with $T^2=1$.
PHS is naturally realized in superconductors as $C=\tau_x {\cal K}$ with the
Pauli matrix $\tau_i$ acting on the Nambu space of the BdG Hamiltonian,
where $C^2=1$, but again
spin-rotation symmetry can introduce another PHS with $C^2=-1$.
Finally, CS can be 
obtained by combination of TRS and PHS, $\Gamma=e^{i\alpha}TC$. With a
suitable choice of the phase $\alpha$, one can always place the relation
$\Gamma^2=1$.

In terms of the sign of $T^2$ and $C^2$, the Hamiltonians are classified
into ten symmetry classes listed in Table \ref{Classifying_space}.
The AZ symmetry classes are further divided into two complex classes  and
eight real classes:
In the absence of time-reversal invariance
and particle-hole symmetry, the Hamiltonian belongs to one of two complex
classes, A or AIII. 
The presence of the anti unitary symmetries  $T$ and $C$
introduces a real structure of the Hamiltonian, and thus the
remaining eight classes are called as real AZ classes. 

Below, we choose a convention that $T$ and $C$ commutes with each other,
i.e. $[T,C]=0$:
Because Eq.(\ref{eq:AZsymmetry}) yields $[TCT^{-1}C^{-1},{\cal H}({\bm
k},{\bm r})]=0$ for any Hamiltonians with TRS and PHS, the unitary operator
$TCT^{-1}C^{-1}$ should be proportional to the identity,
$TCT^{-1}C^{-1}=e^{i\beta}1$. The phase $\beta$ can be removed by a
re-definition of the relative phase between $T$ and $C$ without changing
the sign of $T^2$ and $C^2$, which leads to $[T,C]=0$. 

\begin{table*}[!]
\begin{center}
\caption{AZ symmetry classes and their classifying spaces. 
The top two rows ($s=0,1$ (mod 2)) are complex AZ classes, and the bottom eight
 rows ($s=0,1,\dots,7$ (mod 8)) are
 real AZ classes. The second column represents the names of the AZ classes. 
The third to fifth columns indicate the absence (0) or the presence
 $(\pm 1)$ of TRS, PHS and CS, respectively, where $\pm 1$ means the
 sign of $T^2 = \epsilon_T$ and $C^2 = \epsilon_C$.
The sixth column shows the symbols of the classifying space.
}
\begin{tabular}[t]{ccccccccccc}
\hline \hline
s & AZ class & TRS & PHS & CS & $\mathcal{C}_s$ or $\mathcal{R}_s$ & classifying space & $\pi_0(\mathcal{C}_s)$ or $\pi_0(\mathcal{R}_s)$ \\
\hline
0 & A & $0$ & $0$ & $0$ & $\mathcal{C}_0$ & $(U(n+m)/U(n)\times U(m)) \times \mathbb{Z}$ & $\mathbb{Z}$ \\
1 & AIII & $0$ & $0$ & $1$ & $\mathcal{C}_1$ & $U(n)$ & $0$ \\
\hline
0 & AI & $+1$ & $0$ & $0$ & $\mathcal{R}_0$ & $(O(n+m)/O(n)\times O(m)) \times \mathbb{Z}$ & $\mathbb{Z}$ \\
1 & BDI & $+1$ & $+1$ & $1$ & $\mathcal{R}_1$ & $O(n)$ & $\mathbb{Z}_2$ \\
2 & D & $0$ & $+1$ & $0$ & $\mathcal{R}_2$ & $O(2 n)/U(n)$ & $\mathbb{Z}_2$ \\
3 & DIII & $-1$ & $+1$ & $1$ & $\mathcal{R}_3$ & $U(2 n)/Sp(n)$ & $0$ \\
4 & AII & $-1$ & $0$ & $0$ & $\mathcal{R}_4$ & $(Sp(n+m)/Sp(n)\times Sp(m)) \times \mathbb{Z}$ & $2\mathbb{Z}$ \\
5 & CII & $-1$ & $-1$ & $1$ & $\mathcal{R}_5$ & $Sp(n)$ & $0$ \\
6 & C & $0$ & $-1$ & $0$ & $\mathcal{R}_6$ & $Sp(n)/U(n)$ & $0$ \\
7 & CI & $+1$ & $-1$ & $1$ & $\mathcal{R}_7$ & $U(n)/O(n)$ & $0$ \\
\hline \hline
\end{tabular}
\label{Classifying_space}
\end{center}
\end{table*}

\subsubsection{Order-Two Spatial Symmetry}
\label{sec:OTS}

In addition to the AZ symmetries, we assume an
additional symmetry of Hamiltonians.
As an additional symmetry, we consider general order-two spatial symmetry.
Order-two symmetry $S$ implies that the symmetry operation in twice
trivially acts on the Hamiltonian,
\begin{eqnarray}
[S^2, {\cal H}({\bm k}, {\bm r})]=0, 
\quad S=U, A, 
\label{eq:order-two}
\end{eqnarray}
where $S$ can be either unitary $U$ or anti-unitary $A$. 
The order-two unitary symmetry includes reflection, two-fold spatial
rotation and inversion.
It also permits global ${\bm Z}_2$ symmetry such as a two-fold spin
rotation.
The anti-unitary case admits order-two magnetic point group symmetries.

Under an order-two spatial symmetry, the momentum ${\bm k}$ in the base
space of the Hamiltonian transforms as
\begin{eqnarray}
{\bm k}\rightarrow
\left\{
\begin{array}{rl}
O{\bm k},   & \mbox{for $S=U$} \\
-O{\bm k}, & \mbox{for $S=A$}
\end{array}
\right. ,
\end{eqnarray}
with an orthogonal matrix $O$ satisfying $O^2=1$.
Note that like time-reversal operator, the anti-linearity of $A$
results in the minus sign of the transformation law of ${\bm k}$.   
In a diagonal basis of $O$, this transformation reduces to
\begin{eqnarray}
{\bm k}\rightarrow 
\left\{
\begin{array}{rl}
(-{\bm k}_{\parallel}, {\bm k}_{\perp}),
& \mbox{for $S=U$} \\
({\bm k}_{\parallel}, -{\bm k}_{\perp}),
& \mbox{for $S=A$}
\end{array}
\right. ,
\end{eqnarray}
with ${\bm k}_{\parallel}=(k_1, k_2, \cdots,k_{d_{\parallel}})$ and
${\bm k}_{\perp}=(k_{d_{\parallel}+1},k_{d_{\parallel}+2}\cdots, k_{d})$.

In contrast to non-spatial AZ symmetries,  the spatial coordinate
${\bm r}$ of the $D$-dimensional sphere surrounding a topological defect also
transforms non-trivially under order-two spatial symmetry.
To determine the transformation law, we specify the coordinate ${\bm r}$
of the $D$-dimensional sphere.
First, to keep the additional symmetry, the topological defect should
be invariant under $S$.
Therefore, the additional symmetry $S$ maps the $D$-dimensional sphere
(with a radius $a>0$) given by 
\begin{eqnarray}
{\bm n}^2=a^2,
\quad
{\bm n}=(n_1,n_2,\cdots, n_{D+1}),
\end{eqnarray}
into itself, inducing the transformation
\begin{eqnarray}
{\bm n}\rightarrow O'{\bm n},
\end{eqnarray}
where $O'$ is an orthogonal matrix with $O^{'2}=1$.
The transformation of ${\bm n}$ can be rewritten as
\begin{eqnarray}
{\bm n}\rightarrow (-{\bm n}_{\parallel}, {\bm n}_{\perp}), 
\end{eqnarray}
with ${\bm n}_{\parallel}=(n_1, n_2, \cdots,n_{D_{\parallel}})$ and
${\bm n}_{\perp}=(n_{D_{\parallel}+1},n_{D_{\parallel}+2}\cdots, n_{D+1})$
in a diagonal basis $O'$. 
When  $D_{\parallel}\le  D$,
we can introduce the coordinate ${\bm r}$ of
the $D$-dimensional sphere by the stereographic projection of ${\bm n}$
\begin{eqnarray}
r_i=\frac{n_i}{a-n_{D+1}}, 
\quad 
(i=1,\cdots D),
\end{eqnarray}
which gives a simple transformation law of ${\bm r}$ as 
\begin{eqnarray}
{\bm r}\rightarrow (-{\bm r}_{\parallel}, {\bm r}_{\perp}), 
\end{eqnarray}
with ${\bm r}_{\parallel}=(r_1, r_2, \cdots,r_{D_{\parallel}})$ and
${\bm r}_{\perp}=(r_{D_{\parallel}+1},r_{D_{\parallel}+2}\cdots, r_{D})$.
Below, we assume $D_{\parallel}\le  D$, since the bulk-boundary
correspondence for topological defects works only in this case.

Now the order-two unitary symmetry is expressed as
\begin{eqnarray}
U{\cal H}({\bm k}, {\bm r})U^{-1}={\cal H}(-{\bm k}_{\parallel}, {\bm
 k}_{\perp}, -{\bm r}_{\parallel}, {\bm r}_{\perp}),
\end{eqnarray}
and the order-two anti-unitary symmetry is
\begin{eqnarray}
A{\cal H}({\bm k}, {\bm r})A^{-1}={\cal H}({\bm k}_{\parallel}, -{\bm
 k}_{\perp}, -{\bm r}_{\parallel}, {\bm r}_{\perp}). 
\end{eqnarray}
We suppose that 
\begin{eqnarray}
S^2=\epsilon_S=\pm 1,
\end{eqnarray}
and $S$
commutes or anticommutes with coexisting AZ symmetries, 
\begin{eqnarray}
ST=\eta_T TS, 
\quad
SC=\eta_C CS,  
\quad
S\Gamma=\eta_\Gamma \Gamma S.
\end{eqnarray}
where $\eta_T=\pm 1$, 
$\eta_C=\pm 1$, and $\eta_\Gamma=\pm 1$.
For a faithful representation of order-two symmetry, the sign
$\epsilon_S$ of $S^2$ must
be $1$, but a spinor representation of rotation makes it
possible to obtain $\epsilon_S=-1$.
For instance, two-fold spin rotation $S=e^{i\pi s_i/2}$ $(i=1,2,3)$
obeys $S^2=-1$.
Note that  when $S=U$, we can set $\epsilon_S=1$ by
multiplying $S$ by the imaginary unit $i$, but this changes the
(anti-)commutation relations with $T$ and/or $C$ at the same time.

Our classification framework also works even for
order-two {\it anti-symmetry} $\widebar{S}$ defined by
\begin{eqnarray}
&&\widebar{U}{\cal H}({\bm k}, {\bm r})\widebar{U}^{-1}
=-{\cal H}(-{\bm k}_{\parallel}, {\bm
 k}_{\perp}, -{\bm r}_{\parallel}, {\bm r}_{\perp}),
\\
&&\widebar{A}{\cal H}({\bm k}, {\bm r})\widebar{A}^{-1}
=-{\cal H}({\bm k}_{\parallel}, -{\bm
 k}_{\perp}, -{\bm r}_{\parallel}, {\bm r}_{\perp}),
\end{eqnarray}
where $\widebar{S}$ can be either unitary $\widebar{U}$ or anti-unitary $\widebar{A}$.
Such an anti-symmetry can be realized by combining  any of order-two
symmetries with CS or PHS.
In a similar manner as $S$, we define
$\epsilon_{\widebar{S}}$, $\widebar{\eta}_{T}$, $\widebar{\eta}_{C}$ and $\widebar{\eta}_{\Gamma}$ by
\begin{eqnarray}
(\widebar{S})^2=\epsilon_{\widebar{S}},
\quad 
\widebar{S}T=\widebar{\eta}_T T\widebar{S}, 
\quad
\widebar{S}C=\widebar{\eta}_C C\widebar{S},  
\quad
\widebar{S}\Gamma=\widebar{\eta}_\Gamma \Gamma \widebar{S}.
\end{eqnarray}

\subsection{Stable equivalence and K-group}
In principle, the classification of topological insulators and
superconductors are provided by a homotopy classification of maps  
from the base space $({\bm k}, {\bm r})\in S^{d+D}$
to the classifying space of Hamiltonians $\mathcal{H}({\bm k}, {\bm r})$,  
subject to a given set of symmetries:  
If the maps are smoothly
connected to each other, they belong to the same topological phase, but
if not, they are in topologically different phases.

Hamiltonians we consider here support an energy gap
separating positive and negative energy bands, relative to the Fermi level.
Such Hamiltonians $\mathcal{H}(\bm{k},{\bm r})$ are
adiabatically deformed so that the all empty (occupied) bands have
the same energy +1 (-1). 
If there are no
symmetries, the flattened Hamiltonians are  
characterized by unitary matrices $U(n+m)$ that diagonalize the
Hamiltonians,  divided by unitary rotations $U(n) \times U(m)$ of
the conduction bands and valence bands.   
The classifying space is therefore $U(n+m)/U(n)\times U(m)$.
Symmetries impose some constraints on the classifying space.  

Following the idea of stable equivalence, we extend the classifying space by
adding extra trivial bands \cite{kitaev2009periodic}:
Two sets of Hamiltonians $\mathcal{H}_1$, $\mathcal{H}_2$ are stable
equivalent $\mathcal{H}_1\sim \mathcal{H}_2$, if they can be
continuously deformed into each other by adding extra trivial bands.
One can then identify  
a family of Hamiltonians that are stable equivalent to each other.
We use a notation $[{\cal H}]$ to represent a set of Hamiltonians that
are stable equivalent to ${\cal H}$.  
The stable equivalence classes 
make it possible to supply {\it addition} in the
classifying space of Hamiltonians: 
$[\mathcal{H}_1] \oplus [\mathcal{H}_2] := [\mathcal{H}_1
\oplus \mathcal{H}_2]$,  
where $\oplus$ implies the direct sum of matrices. 
The identity $[0]$ expresses the trivial insulating Hamiltonian,
 and  $[\mathcal{H} \oplus (-\mathcal{H})]$ is ensured to be $[0]$. 
The last relation yields that the inverse of $[{\cal H}]$ is $-[\mathcal{H}]
=[-\mathcal{H}]$. 
As a result, the stable equivalent classes form an Abelian group, which
is called the K-group.
From the definition, it is evident that the stable equivalence retains 
topological natures.
The extended classifying spaces subject to AZ symmetries are
listed in Table \ref{Classifying_space}.

For topological insulators and superconductors in ten AZ symmetry
classes, the following relations
summarize their classification \cite{teo2010topological}
\begin{eqnarray}
&&K_{\mathbb{C}}(s;d,D) = K_{\mathbb{C}}(s-d+D;
 0,0)=\pi_0({\cal C}_{s-d+D}), 
\quad (s=0,1\, (\mbox{mod 2}))
\label{KCAZ}
\\
&&K_{\mathbb{R}}(s;d,D) = K_{\mathbb{R}}(s-d+D; 0,0)=\pi_0({\cal R}_{s-d+D}),
\quad (s=0,1,\cdots, 7\, (\mbox{mod 8}))
\label{KRAZ}
\end{eqnarray}
where $K_{\mathbb{C}}(s;d,D)$ ($K_{\mathbb{R}}(s;d,D)$) denotes the
K-group of maps from $({\bm k}, {\bm r})\in S^{d+D}$ to the
extended classifying space ${\cal C}_s$ $({\cal R}_s)$ of $s$ complex
(real) AZ class in Table \ref{Classifying_space}.
The case of $D=0$ corresponds to the bulk topological classification,
and the presence of topological defects shifts the dimension of the system.

The existence of an order-two spatial symmetry $S$ gives additional
constraints on the classifying space. 
In the subsequent sections, we provide the resulting K-group of the
homotopy classification.

\section{K-group in the presence of additional symmetry}
\label{Main}

In this section, 
we present the K-groups for topological crystalline
insulators/superconductors and their topological defects  protected by
an additional order-two symmetry. 
The derivation and proof are given in Sec. \ref{TCI}.

\subsection{Complex AZ classes (A and AIII) with additional order-two
  unitary symmetry}
The complex AZ classes, A and AIII,  are
characterized by the absence of TRS and PHS.
Whereas no AZ symmetry is imposed on Hamiltonians in class A,  
Hamiltonians in class AIII is invariant under CS,  
\begin{equation}\begin{split}
\Gamma \mathcal{H}(\bm{k},\bm{r}) \Gamma^{-1} = - \mathcal{H} (\bm{k},\bm{r}). 
\end{split}\end{equation}
Now we impose an additional order-two symmetry $U$,
\begin{eqnarray}
U{\cal H}({\bm k}, {\bm r})U^{-1}={\cal H}(-{\bm k}_{\parallel}, {\bm
 k}_{\perp}, -{\bm r}_{\parallel}, {\bm r}_{\perp}),
\end{eqnarray}
or order-two antisymmetry $\widebar{U}$
\begin{eqnarray}
\widebar{U}{\cal H}({\bm k}, {\bm r})\widebar{U}^{-1}
=-{\cal H}(-{\bm k}_{\parallel}, {\bm
 k}_{\perp}, -{\bm r}_{\parallel}, {\bm r}_{\perp}).
\end{eqnarray}
on complex AZ classes.
Since there is no anti-unitary symmetry, a phase factor of $U$ and
$\widebar{U}$ do not change
the topological classification, and thus the sign of $U^2$ and $(\widebar{U})^2$
can be fixed to be $1$.  
For class AIII, we specify the commutation/anti-commutation relation
between $U$ and $\Gamma$ ($\widebar{U}$ and $\Gamma$) by $U_{\eta_\Gamma}$
($\widebar{U}_{\eta_\Gamma}$).
Note that $\widebar{U}_{\eta_\Gamma}$ in class AIII is
essentially the same as $U_{\eta_{\Gamma}}$ because they can be
converted to each other by the relation $\widebar{U}_{\eta_\Gamma}=\Gamma U_{\eta_\Gamma}$.

We denote the obtained K-group by 
\begin{eqnarray}
K^U_{\mathbb{C}}(s, t;d, d_{\parallel},D, D_{\parallel}).
\end{eqnarray}
Here  $d$ ($D$) is the total space dimension (defect co-dimension), and
$d_{\parallel}$ ($D_{\parallel}$) is the number of the flipping momenta (defect
surrounding parameters) under the additional symmetry transformation, as
was introduced in Sec.\ref{sec:OTS}.    
The label $s=0,1$ (mod 2) indicates the AZ class ($s=0$ for class A
and $s=1$ for class AIII) to which the Hamiltonian belongs,  
and $t=0,1$ (mod 2) specifies the coexisting additional
unitary symmetry as in Table. \ref{Symmetry_type_UC}.

\begin{table*}[!]
\begin{center}
\caption{
Possible types ($t=0,1$ (mod 2)) of order-two additional unitary
 symmetries in complex AZ class ($s=0,1$ (mod 2)).  
$U$ and $\bar{U}$ represent symmetry and antisymmetry, respectively. 
The subscript of $U_{\eta_{\Gamma}}$ and $\bar{U}_{\eta_{\Gamma}}$
 specifies the relation $\Gamma U = \eta_{\Gamma} U \Gamma$. 
Symmetries in the same parenthesis are equivalent.}
\begin{tabular}[t]{ccccccccccc}
\hline \hline
$s$ & AZ class & $t=0$ & $t=1$ \\
\hline
0 & A & $U$ & $\widebar{U}$ \\
1 & AIII & ($U_+$, $\widebar{U}_+$) & ($U_-$, $\widebar{U}_-$) \\
\hline \hline
\end{tabular}
\label{Symmetry_type_UC}
\end{center}
\end{table*}

In Sec. \ref{TCI}, we prove the following relation: 
\begin{eqnarray}
K^U_{\mathbb{C}}(s,t;d, d_{\parallel},D, D_{\parallel})
= K^U_{\mathbb{C}}(s-d+D, t-d_{\parallel} + D_{\parallel};0,0,0,0). 
\label{KCU} 
\end{eqnarray}
This relation implies that topological natures of the system can be
deduced from those in $0$-dimension.
As we show in Appendix \ref{Sec_Clif}, 
the classifying spaces of the 0-dimensional K-group reduce to
complex Clifford algebra, and we can obtain
\begin{eqnarray}
&&K^U_{\mathbb{C}}(s,t=0; 0,0,0,0) = \pi_0 (\mathcal{C}_s \times
 \mathcal{C}_s) = \pi_0(\mathcal{C}_s) \oplus \pi_0(\mathcal{C}_s) , 
\nonumber\\
&&K^U_{\mathbb{C}}(s,t=1; 0,0,0,0) = \pi_0 (\mathcal{C}_{s+1}),
\label{KCU_classifying_space}
\end{eqnarray}
where $\mathcal{C}_s (s=0,1)$ represents the classifying space of complex AZ
classes. (See Table \ref{Classifying_space}.)

\subsection{Complex AZ classes (A and AIII) with additional order-two
  antiunitary symmetry}
\label{sec:CAZAS}

Next, we consider order-two {\it anti-unitary} symmetry $A$ or $\widebar{A}$
as an additional symmetry:
\begin{eqnarray}
A \mathcal{H}({\bm k}, {\bm r}) A^{-1} 
= \mathcal{H} (\bm{k}_{\parallel},-{\bm k}_{\perp},-\bm{r}_{\parallel}, 
\bm{r}_{\perp}), 
\\
\widebar{A} \mathcal{H}({\bm k}, {\bm r}) \widebar{A}^{-1} 
= - \mathcal{H} (\bm{k}_{\parallel},-{\bm k}_{\perp},-\bm{r}_{\parallel}, 
\bm{r}_{\perp}).
\end{eqnarray}
As listed in Table \ref{Symmetry_type_AC}, 
two different order-two anti-unitary symmetries $A^{\pm}$ and their
corresponding  anti-symmetries $\widebar{A}^{\pm}$
are possible in class A, depending on the sign of $A^2$ or $(\widebar{A})^2$,
i.e. $(A^{\epsilon_A})^2=\epsilon_A$, $(\widebar{A}^{\epsilon_A})^2=\epsilon_A$.
In a similar manner,  class AIII have two different types of additional
anti-unitary symmetries, 
$A^{\epsilon_A}_{\eta_\Gamma}$ $(\epsilon_A=\pm 1, \eta_A=\pm 1)$, and
their corresponding anti-symmetries, 
$\widebar{A}^{\epsilon_A}_{\eta_\Gamma}$ $(\epsilon_A=\pm 1, \eta_A=\pm
1)$,  
where $\epsilon_A$ represents the sign of $A^2$ or $(\widebar{A})^2$ and
$\eta_{\Gamma}$ indicates the
commutation ($\eta_\Gamma=1$) or the anti-commutation ($\eta_{\Gamma}=-1$)
relation between $A$ and $\Gamma$ or those between $A$ and $\Gamma$.
Note that $A_{\eta_\Gamma}^{\epsilon_A}$ and
$\widebar{A}_{\eta_\Gamma}^{\epsilon_A \eta_\Gamma}$ are equivalent in class
AIII since they can be related to each other as 
$A^{\epsilon_A}_{\eta_{\Gamma}} = \Gamma \widebar A^{\epsilon_A
\eta_{\Gamma}}_{\eta_{\Gamma}}$.

The existence of the anti-unitary symmetry introduces {\it real}
structures in complex AZ classes.  
Actually, by regarding $({\bm k}_{\perp}, {\bm r}_{\parallel} )$ as
``momenta'', and $({\bm k}_{\parallel}, {\bm r}_{\perp})$ as ``spatial
coordinates'', $A$ and $\widebar{A}$ can be considered as TRS and PHS,
respectively.
From this identification, a system in complex AZ class with an
additional anti-unitary symmetry can be mapped into a real AZ class, as
summarized in Table \ref{Symmetry_type_AC}.
As a result, the K-group of Hamiltonians with the symmetry $s$
($s=0,1,2,\dots,7$ (mod 8)) of Table
\ref{Symmetry_type_AC} 
\begin{eqnarray}
K^A_{\mathbb{C}}(s;d,d_{\parallel},D, D_{\parallel}),   
\end{eqnarray}
reduces to the K-group of real AZ classes in Eq. (\ref{KRAZ}),
\begin{eqnarray}
K^A_{\mathbb{C}}(s;d, d_{\parallel},D,
 D_{\parallel})=
K_{\mathbb{R}}(s;d-d_{\parallel}+D_{\parallel},
 D-D_{\parallel}+d_{\parallel}).    
\end{eqnarray}
where $d$ ($D$) is the total space dimension (defect co-dimension), and
$d_{\parallel}$ ($D_{\parallel}$) is the number of the flipping momentum (defect
surrounding parameter) under the additional symmetry transformation. 
From Eq. (\ref{KRAZ}), we have 
\begin{eqnarray}
K^A_{\mathbb{C}}(s;d,d_{\parallel},D,D_{\parallel})
= K^A_{\mathbb{C}}(s-d+D+2(d_{\parallel}-D_{\parallel});0,0,0,0),
\label{eq:KCA}
\end{eqnarray}
with 
\begin{eqnarray}
K^A_{\mathbb{C}}(s;0,0,0,0)=\pi_0({\cal R}_s).  
\end{eqnarray}

\begin{table*}[!]
\begin{center}
\caption{
Possible types ($s=0,1,\dots, 7$ (mod 8)) of order-two additional
 anti-unitary symmetries in complex AZ class.
$A$ and $\bar A$ represent symmetry and
 anti-symmetry, respectively.
The superscript of $A^{\epsilon_A}$ and $A^{\epsilon_A}_{\eta_{\Gamma}}$
represent the sign of the square $A^2 = \epsilon_A$, and the subscript
 of $A^{\epsilon_A}_{\eta_{\Gamma}}$ specifies the (anti-)commutation relation 
 $\Gamma A = \eta_{\Gamma} A \Gamma$.
Symmetries in the same parenthesis are equivalent.}
\begin{tabular}[t]{cccc}
\hline \hline
s & AZ class & Coexisting symmetry & Mapped AZ class  \\
\hline
0 & A & $A^{+}$ & AI
\\
1 & AIII & ($A_+^+$, $\widebar{A}_+^+$) & BDI 
\\
2& A & $\widebar{A}^+$ & D
\\
3& AIII&($A^-_-$,$\widebar{A}^+_-$) & DIII
\\
4 & A & $A^{-}$ & AII 
\\
5 & AIII &($A^-_+$, $\widebar{A}^-_+$) & CII
\\
6 & A & $\widebar A^{-}$ & C
\\
7 & AIII &($A_-^+$, $\widebar{A}_-^-$) & CI
\\
\hline \hline
\end{tabular}
\label{Symmetry_type_AC}
\end{center}
\end{table*}

\subsection{Real AZ classes with additional order-two symmetry}

Hamiltonians in eight real AZ classes are invariant under TRS, 
\begin{equation}\begin{split}
T \mathcal{H}(\bm{k},\bm{r}) T^{-1} = \mathcal{H} (-\bm{k},\bm{r})
\end{split}\end{equation}
and/or PHS, 
\begin{equation}\begin{split}
C \mathcal{H}(\bm{k},\bm{r}) C^{-1} = - \mathcal{H} (-\bm{k},\bm{r}). 
\end{split}\end{equation}
In addition to TRS and/or PHS, we enforce one of order-two
unitary/antiunitary spatial symmetries, $U$, $\widebar{U}$, $A$, and $\widebar{A}$ 
on the Hamiltonians,
\begin{eqnarray}
U \mathcal{H}(\bm{k},\bm{r}) U^{-1} = \mathcal{H} (-\bm{k}_{\parallel},
 \bm{k}_{\perp}, -\bm{r}_{\parallel},\bm{r}_{\perp}), 
\\
\widebar{U} \mathcal{H}(\bm{k},\bm{r})\widebar{U}^{-1} 
= -\mathcal{H} (-\bm{k}_{\parallel},
 \bm{k}_{\perp}, -\bm{r}_{\parallel},\bm{r}_{\perp}), 
\\
A \mathcal{H}(\bm{k},\bm{r}) A^{-1} = \mathcal{H} (\bm{k}_{\parallel},
-\bm{k}_{\perp}, -\bm{r}_{\parallel},\bm{r}_{\perp}), 
\\
\widebar{A} \mathcal{H}(\bm{k},\bm{r})\widebar{A}^{-1} 
= -\mathcal{H} (\bm{k}_{\parallel},
-\bm{k}_{\perp}, -\bm{r}_{\parallel},\bm{r}_{\perp}).
\end{eqnarray}

In class AI and AII, which support TRS, we have the following
equivalence relations between the additional symmetries,
\begin{eqnarray}
U^{\epsilon_U}_{\eta_T} = i U^{-\epsilon_U}_{-\eta_T} = T A^{\eta_T
\epsilon_T\epsilon_U}_{\eta_T} = 
iT A^{\eta_T \epsilon_T \epsilon_U}_{-\eta_T},
\\
\widebar U^{\epsilon_U}_{\eta_T} = i \widebar U^{-\epsilon_U}_{-\eta_T} 
= T \widebar A^{\eta_T \epsilon_T \epsilon_U}_{\eta_T}
= iT \widebar A^{\eta_T \epsilon_T \epsilon_U}_{-\eta_T}, 
\end{eqnarray}
where the superscript $\epsilon_S=\pm $ of $S$ $(S=U, \widebar{U}, A, \widebar{A})$
denotes the sign of $S^2$, and 
the subscript $\eta_T$ of $S$ specifies the commutation ($\eta_T=+$) or
anti-commutation $(\eta_T=-)$ relation between $S$ and $T$.
In a similar manner, in class D and C, PHS leads to the following
equivalence relations
\begin{eqnarray}
U^{\epsilon_U}_{\eta_C} = i U^{-\epsilon_U}_{-\eta_C} 
= C \widebar A^{\eta_C \epsilon_C \epsilon_U}_{\eta_C}
= iC \widebar A^{\eta_C \epsilon_C \epsilon_U}_{-\eta_C}, 
\\
\widebar U^{\epsilon_U}_{\eta_C} = i \widebar U^{-\epsilon_U}_{-\eta_C} 
= C A^{\eta_C \epsilon_C \epsilon_U}_{\eta_C}
= iC A^{\eta_C \epsilon_C \epsilon_U}_{-\eta_C},
\end{eqnarray}
where the superscript $\epsilon_S=\pm$  denotes the sign of $S^2$ and the
subscript $\eta_C=\pm$ denotes the commutation $(\eta_C=+)$ or
anti-commutation $(\eta_C=-)$ relation between $S$ and $C$.  
Finally, in class BDI, DIII, CII and CI, we obtain
\begin{eqnarray}
U^{\epsilon_U}_{\eta_T,\eta_C} = i U^{-\epsilon_U}_{-\eta_T,-\eta_C} 
= T A^{\eta_T\epsilon_T\epsilon_U}_{\eta_T,\eta_C} = 
iT A^{\eta_T \epsilon_T \epsilon_U}_{-\eta_T, -\eta_C}
= C \widebar A^{\eta_C \epsilon_C \epsilon_U}_{\eta_T, \eta_C}
= iC \widebar A^{\eta_C \epsilon_C \epsilon_U}_{-\eta_T, -\eta_C}, 
\\
\widebar U^{\epsilon_U}_{\eta_T, \eta_C} 
= i \widebar U^{-\epsilon_U}_{-\eta_T, -\eta_C} 
= T \widebar A^{\eta_T \epsilon_T \epsilon_U}_{\eta_T, \eta_C}
= iT \widebar A^{\eta_T \epsilon_T \epsilon_U}_{-\eta_T,-\eta_C}, 
= C A^{\eta_C \epsilon_C \epsilon_U}_{\eta_T, \eta_C}
= iC A^{\eta_C \epsilon_C \epsilon_U}_{-\eta_T,-\eta_C}.
\end{eqnarray}
These equivalence relations classify order-two symmetries into four
families ($t=0,1,2,3$), as summarized in Table \ref{Symmetry_type}.
Here one should note that unitary symmetries can be converted
into to anti-unitary ones by multiplying TRS or PHS.
Therefore, the presence of a unitary symmetry for real AZ classes gives
the same K-groups as those with an additional anti-unitary symmetry.

\begin{table*}[!]
\begin{center}
\caption{
Possible types ($t=0,1,2,3$ (mod 4)) of order-two additional symmetries
 in real AZ class ($s=0,1,\dots,7$ (mod 8)).
$U$ and $\bar U$ represent unitary symmetry and anti-symmetry,
 respectively, and $A$ and $\bar A$ represent anti-unitary symmetry and
 anti-symmetry, respectively. 
The superscript of $S$ ($S=U,\bar{U},A, \bar{A}$) indicates the sign of
$S^2$, and the subscript of $S$ specifies the commutation(+)/anti-commutation(-)
 relation between $S$ and TRS and/or PHS. 
For BDI, DIII, CII and CI, where both TRS and PHS exist, $S$ has two
 subscripts, in which the first one specifies the
 (anti-)commutation relation between $S$ and $T$ and the second one
 specifies that between $S$ and $C$.    
Symmetries in the same parenthesis are  equivalent.
}
\begin{tabular}[t]{ccccccccccc}
\hline \hline
$s$ & AZ class & $t=0$ & $t=1$ & $t=2$ & $t=3$ \\
\hline
\multirow{2}{*}{0} 
& \multirow{2}{*}{AI} & ($U^+_+$, $U^-_-$) 
& ($\widebar U^+_-$, $\widebar U^-_+$)
& ($U^+_-$, $U^-_+$)
& ($\widebar U^+_+$, $\widebar U^-_-$)
\\
& & ($A^+_+$, $A^+_-$) 
& ($\widebar A^-_+$, $\widebar A^-_-$) 
& ($A^-_-$, $A^-_+$) 
& ($\widebar A^+_+$, $\widebar A^+_-$)  
\\[3pt] 
\hline
\multirow{2}{*}{1} 
& \multirow{2}{*}{BDI} 
& ($U_{++}^+$, $U_{--}^-$, $\widebar{U}_{++}^+$, $\widebar{U}_{--}^-$)
& ($U^+_{+-}$, $U^-_{-+}$, $\widebar U^+_{-+}$, $\widebar U^-_{+-}$)
& ($U^+_{--}$, $U^-_{++}$, $\widebar U^+_{--}$, $\widebar U^-_{++}$)
& ($U^+_{-+}$, $U^-_{+-}$, $\widebar U^+_{+-}$, $\widebar U^-_{-+}$)
\\
& & ($A^+_{++}$, $A^+_{--}$, $\widebar A^+_{++}$, $\widebar A^+_{--}$) 
& ($A^+_{+-}$, $A^+_{-+}$, $\widebar A^-_{-+}$, $\widebar A^-_{+-}$) 
& ($A^-_{--}$, $A^-_{++}$, $\widebar A^-_{--}$, $\widebar A^-_{++}$) 
& ($A^-_{-+}$, $A^-_{+-}$, $\widebar A^+_{+-}$, $\widebar A^+_{-+}$) 
\\[3pt]
\hline
\multirow{2}{*}{2}
& \multirow{2}{*}{D}
& ($U^+_+$, $U^-_-$)
& ($\widebar U^+_+$, $\widebar U^-_-$) 
& ($U^+_-$, $U^-_+$) 
& ($\widebar U^+_-$, $\widebar U^-_+$) 
\\
& & ($\widebar A^+_+$, $\widebar A^+_-$) 
& ($A^+_+$, $A^+_-$) 
& ($\widebar A^-_-$, $\widebar A^-_+$) 
& ($A^-_+$, $A^-_-$) 
\\[3pt]
\hline
\multirow{2}{*}{3} 
& \multirow{2}{*}{DIII} 
& ($U^+_{++}$, $U^-_{--}$, $\widebar U^-_{++}$, $\widebar U^+_{--}$)
& ($U^+_{-+}$, $U^-_{+-}$, $\widebar U^+_{-+}$, $\widebar U^-_{+-}$)
& ($U^+_{--}$, $U^-_{++}$, $\widebar U^-_{--}$, $\widebar U^+_{++}$)
& ($U^+_{+-}$, $U^-_{-+}$, $\widebar U^+_{+-}$, $\widebar U^-_{-+}$)
\\
& & ($A^-_{++}$, $A^-_{--}$, $\widebar A^+_{++}$, $\widebar A^+_{--}$)
& ($A^+_{-+}$, $A^+_{+-}$, $\widebar A^+_{-+}$, $\widebar A^+_{+-}$)
& ($A^+_{--}$, $A^+_{++}$, $\widebar A^-_{--}$, $\widebar A^-_{++}$)
& ($A^-_{+-}$, $A^-_{-+}$, $\widebar A^-_{+-}$, $\widebar A^-_{-+}$)
\\[3pt] 
\hline
\multirow{2}{*}{4}
& \multirow{2}{*}{AII}
& ($U^+_+$, $U^-_-$)
& ($\widebar U^+_-$, $\widebar U^-_+$)
& ($U^+_-$, $U^-_+$)
& ($\widebar U^+_+$, $\widebar U^-_-$) 
\\
& &($A^-_+$, $A^-_-$) 
&($\widebar A^+_-$, $\widebar A^+_+$) 
&($A^+_-$, $A^+_+$) 
&($\widebar A^-_+$, $\widebar A^-_-$) 
\\[3pt]
\hline
\multirow{2}{*}{5} 
& \multirow{2}{*}{CII} 
&($U^+_{++}$, $U^-_{--}$, $\widebar U^+_{++}$, $\widebar U^-_{--}$)
&($U^+_{+-}$, $U^-_{-+}$, $\widebar U^+_{-+}$, $\widebar U^-_{+-}$)
&($U^+_{--}$, $U^-_{++}$, $\widebar U^+_{--}$, $\widebar U^-_{++}$)
&($U^+_{-+}$, $U^-_{+-}$, $\widebar U^+_{+-}$, $\widebar U^-_{-+}$)
\\
& &($A^-_{++}$, $A^-_{--}$, $\widebar A^-_{++}$, $\widebar A^-_{--}$)
& ($A^-_{+-}$, $A^-_{-+}$, $\widebar A^+_{-+}$, $\widebar A^+_{+-}$)
& ($A^+_{--}$, $A^+_{++}$, $\widebar A^+_{--}$, $\widebar A^+_{++}$)
& ($A^+_{-+}$, $A^+_{+-}$, $\widebar A^-_{+-}$, $\widebar A^-_{-+}$)
\\[3pt]
\hline
\multirow{2}{*}{6}
& \multirow{2}{*}{C}
& ($U^+_+$, $U^-_-$) 
& ($\widebar U^+_+$, $\widebar U^-_-$) 
& ($U^+_-$, $U^-_+$)
& ($\widebar U^+_-$, $\widebar U^-_+$)
\\ 
& &($\widebar A^-_+$, $\widebar A^-_-$) 
&($A^-_+$, $A^-_-$) 
&($\widebar A^+_-$, $\widebar A^+_+$) 
&($A^+_-$, $A^+_+$)
\\[3pt] 
\hline
\multirow{2}{*}{7} 
&\multirow{2}{*}{CI} 
&($U^+_{++}$, $U^-_{--}$, $\widebar U^-_{++}$, $\widebar U^+_{--}$)
&($U^+_{-+}$, $U^-_{+-}$, $\widebar U^+_{-+}$, $\widebar U^-_{+-}$)
&($U^+_{--}$, $U^-_{++}$, $\widebar U^-_{--}$, $\widebar U^+_{++}$) 
&($U^+_{+-}$, $U^-_{-+}$, $\widebar U^+_{+-}$, $\widebar U^-_{-+}$)
\\ 
& &($A^+_{++}$, $A^+_{--}$, $\widebar A^-_{++}$, $\widebar A^-_{--}$)
& ($A^-_{-+}$, $A^-_{+-}$, $\widebar A^-_{-+}$, $\widebar A^-_{+-}$)
& ($A^-_{--}$, $A^-_{++}$, $\widebar A^+_{--}$, $\widebar A^+_{++}$)
& ($A^+_{+-}$, $A^+_{-+}$, $\widebar A^+_{+-}$, $\widebar A^+_{-+}$)
\\[3pt]
\hline \hline
\end{tabular}
\label{Symmetry_type}
\end{center}
\end{table*}

We denote the K-group for real AZ class ($s=0,1,\dots, 7$ (mod 8)) with
additional order-two unitary (anti-unitary) symmetry ($t=0,1,2,3$ (mod 4)) as
\begin{equation}\begin{split}
K^{U}_{\mathbb{R}}(s, t ;d,d_{\parallel},D,D_{\parallel}),
\quad
(K^{A}_{\mathbb{R}}(s, t ;d,d_{\parallel},D,D_{\parallel})),
\end{split}\end{equation}
where $d$ ($D$) is the total space dimension (defect co-dimension), and
$d_{||}$ ($D_{||}$) is the number of the flipping momentum (defect
surrounding parameter) against the additional symmetry transformation.  
The equivalence between unitary and anti-unitary symmetries for real AZ
classes implies
\begin{eqnarray}
K^{U}_{\mathbb{R}}(s, t ;d,d_{\parallel},D,D_{\parallel})=
K^{A}_{\mathbb{R}}(s, t ;d,d_{\parallel},D,D_{\parallel}). 
\end{eqnarray}
In Sec. \ref{TCI}, we prove the following relation: 
\begin{eqnarray}
K^{U/A}_{\mathbb{R}}(s, t;d,d_{\parallel},D,D_{\parallel})
= K^{U/A}_{\mathbb{R}}(s-d+D,t- d_{\parallel} + D_{\parallel};0,0,0,0). 
\label{KR}
\end{eqnarray}
In Appendix \ref{Sec_Clif}, we show 
\begin{eqnarray}
&&K^{U/A}_{\mathbb{R}}(s,t=0; 0,0,0,0) = \pi_0 (\mathcal{R}_s \times \mathcal{R}_s) = \pi_0 (\mathcal{R}_s) \oplus \pi_0 (\mathcal{R}_s), \\
&&K^{U/A}_{\mathbb{R}}(s,t=1; 0,0,0,0) = \pi_0 (\mathcal{R}_{s-1}), \\
\label{Eq::Additional_Symmetry_t=2}&&K^{U/A}_{\mathbb{R}}(s,t=2; 0,0,0,0) = \pi_0 (\mathcal{C}_s), \\
&&K^{U/A}_{\mathbb{R}}(s,t=3; 0,0,0,0) = \pi_0 (\mathcal{R}_{s+1}),
\end{eqnarray}
where $\mathcal{R}_s$ ($s=0,1,\dots,7$ (mod 8))  and $\mathcal{C}_s$
($s=0,1$ (mod 2))
represent the classifying spaces of the real and complex AZ classes. 


\section{Properties of topological table and K-group}
\label{sec:property}

\subsection{New periodicity in flipped dimensions}

The K-groups (\ref{KCU}), (\ref{eq:KCA}), and (\ref{KR}) have common
general properties. 
First, the K-groups do not depend on $d$, $D$, $d_{\parallel}$ and
$D_{\parallel}$ separately, but they depend on their differences
$\delta=d-D$ and $\delta_{\parallel}=d_{\parallel} - D_{\parallel}$.
Second, in addition to the mod 2 or mod 8 Bott-periodicity in space dimension
$\delta$, there exists a novel periodic structure in flipped dimensions
$\delta_{\parallel}$, 
due to two-fold or fourfold periodicity in type $t$ of 
additional symmetries.
Consequently, the presence of order-two additional symmetry
provides four different families of periodic tables for topological
crystalline insulators and superconductors and their topological defects: 
(i) $\delta_{\parallel}=0$ family: The additional symmetry in this
family includes non-spatial unitary symmetry such as two-fold spin
rotation, where no spatial parameter is flipped in the bulk.
(ii) $\delta_{\parallel}=1$ family: This family includes bulk topological
phases protected by reflection symmetry, where one direction of the momenta is
flipped.
(iii) $\delta_{\parallel}=2$ family:  
Bulk topological phases protected by two-fold spatial rotation are
categorized into this family.
(iv) $\delta_{\parallel}=3$ family:
Inversion symmetric bulk topological phases are classified into this family.
Note that the correspondence between these additional symmetries
and the families is shifted by $D_{\parallel}$ in the presence of
topological defects.

\subsection{Defect gapless states as boundary states}

The differences
$\delta=d-D$ and $\delta_{\parallel} =d_{\parallel} - D_{\parallel}$
have simple graphical meanings:
First, we notice that 
a topological defect surrounded by $S^D$
in $d$-dimensions defines a $(\delta-1)$-dimensional submanifold,
since $D$ is the defect codimension.
For instance, a line defect in three dimensions has $\delta=2$ ($d=3$,
$D=1$), and thus it defines one-dimensional submanifold.
Then, we also find that  $\delta_{\parallel}$ indicates the number of flipped
coordinates of the submanifold under the additional symmetry.  
For instance, see topological defects in $\delta_{\parallel} = 0$ family,
illustrated in Fig. \ref{local_sym}.  
Although the surrounding parameters of the topological defects
transform nontrivially under the additional reflection or two-fold rotation, 
we find that the defects themselves are
unaffected by the additional symmetries. 
In a similar manner, for topological defects of $\delta_{\parallel} = 1$
($\delta_{\parallel}=2$) family in
Fig. \ref{reflection_sym} (Fig. \ref{pi_rotation_sym}),  one-direction
(two-directions) in the defect submanifold is (are)
flipped under additional symmetries.
In other words, 
additional symmetries in $\delta_{\parallel}=1$ ($\delta_{\parallel}=2$)
family act on defect submanifolds in the same manner as reflection (two-fold
rotation) whatever the original transformations are.

These graphical meanings provide a natural explanation why the K-groups
depend solely on $\delta$ and $\delta_{\parallel}$:   
As first suggested by Read and Green,\cite{read2000paired} 
the $(\delta-1)$-dimensional defect submanifold can be considered as a
boundary of a $\delta$-dimensional insulator/superconductor. 
Then, the above geometrical observation implies that additional
symmetries induce an effective symmetry with $\delta_{\parallel}$
flipped directions in the $(\delta-1)$-dimensional defect submanifold, and
thus also induce the same effective symmetry in the $\delta$-dimensional
insulator/superconductor. 
Consequently, the K-group of the topological defect reduces to that of the
$\delta$-dimensional crystalline insulator/superconductor with the
$\delta_{\parallel}$ flipped additional symmetry.



\section{Topological periodic table in the presence of additional
 order-two symmetry}
\label{Tab}

In the previous section, we have presented the K-groups for topological
crystalline insulators
and superconductors and their topological defects protected by order-two
additional symmetry. 
The K-groups give exhaustive topological periodic tables for the
symmetry protected topological phases.
We clarify the Abelian group structures such as $\mathbb{Z}$ or
$\mathbb{Z}_2$.
Whereas we do not give all of the explicit expressions of the
corresponding topological invariants, we
illustrate how the topological tables work by using concrete examples. 
In the following subsections, we focus on additional unitary and
antiunitary symmetries. 
We omit here classification tables for additional
antisymmetries because most of antisymmetries reduce to unitary or antiunitary
symmetries by the symmetry equivalence relation.\footnote{
An unitary antisymmetry in class A, AI, and AII does not reduce to a conventional
symmetry, but the realization of such antisymmetry is difficult in
the condensed matter systems.
}
Also, we omit antiunitary symmetry in the time-reversal symmetric AZ classes 
because antiunitary symmetry naturally realizes as a combination of  time-reversal and point group symmetries,  
i.e. magnetic point group symmetry, in TRS broken systems.

\subsection{$\delta_{\parallel}= 0$ family}
\begin{figure}[!]
 \begin{center}
  \includegraphics[width=\linewidth, trim=0cm 0cm 0cm 0cm]{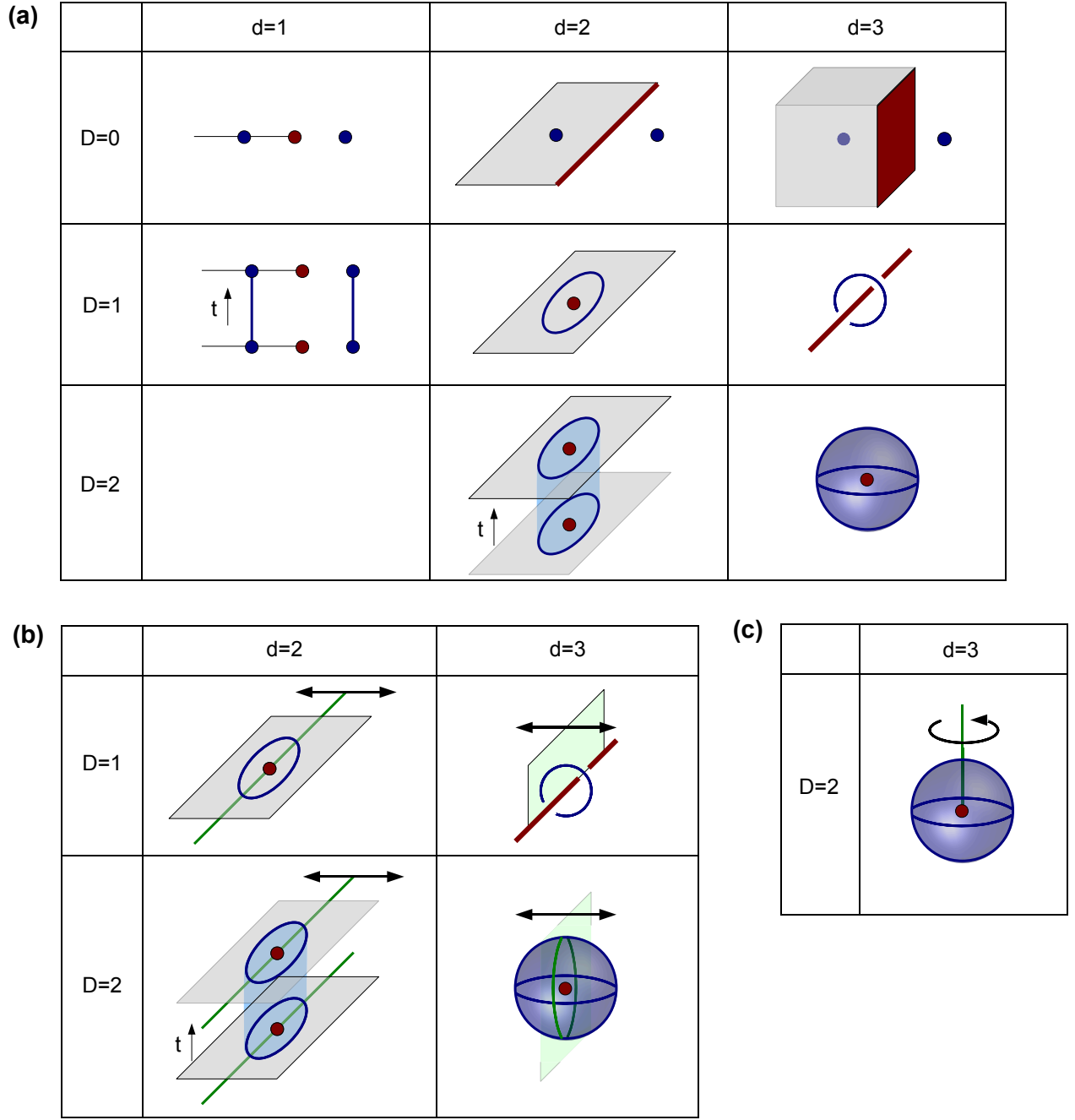}
 \end{center}
 \caption{Topological defects and adiabatic pump protected by order-two
 additional symmetries with
 $\delta_{\parallel}=d_{\parallel}-D_{\parallel}=0$. The additional
 symmetries are (a) global ${\bm Z}_2$ symmetry, (b) reflection
 symmetry and (c) $\pi$-rotation symmetry, respectively. 
 The spatial position of topological defects is unchanged under the symmetry
 transformation of $\delta_{\parallel}=0$ family. }
 \label{local_sym}
\end{figure}

In this subsection, we consider additional symmetries with
$\delta_{\parallel}= 0$ (mod 4).
In condensed matter contexts, relevant symmetries include order-two global
symmetry such two-fold spin rotation ($d_{\parallel} = D_{\parallel} =0$), 
reflection with a line and point defect in the mirror plane  ($d_{\parallel}= D_{\parallel} = 1$), 
two-fold spatial rotation with a point defect on the rotation axis
($d_{\parallel} = D_{\parallel} = 2$), 
as illustrated in Fig. \ref{local_sym}. 
We summarize the classification table for $\delta_{\parallel}=0$ (mod 4)
with order-two unitary symmetries in Table \ref{TabLU} and that with antiunitary
symmetries in Table \ref{TabLA}, respectively. 
\begin{table*}[!]
\begin{center}
\caption{
Classification table for topological crystalline insulators and
superconductors and their topological defects in the presence of
 order-two additional unitary symmetry with flipped parameters
 $\delta_{\parallel}\equiv d_{\parallel} - D_{\parallel} = 0$ (mod 4).
Here $\delta = d-D$.
}
\begin{tabular}[t]{ccccccccccc}
\hline \hline
Symmetry & Class 
& $\mathcal{C}_q$ or $\mathcal{R}_q$ 
& $\delta=0$ & $\delta=1$ & $\delta=2$ & $\delta=3$ 
& $\delta=4$ & $\delta=5$ & $\delta=6$ & $\delta=7$ \\
\hline
$U$  & A &$\mathcal{C}_0 \times \mathcal{C}_0$
& $\mathbb{Z}\oplus\mathbb{Z}$ 
& $0$ & $\mathbb{Z} \oplus \mathbb{Z}$ & $0$ &
$\mathbb{Z} \oplus \mathbb{Z}$ & $0$ &
$\mathbb{Z} \oplus \mathbb{Z}$ & $0$ \\ 
$U_+$& AIII   &$\mathcal{C}_1 \times \mathcal{C}_1$& $0$ 
& $\mathbb{Z} \oplus \mathbb{Z}$ & $0$ 
& $\mathbb{Z} \oplus \mathbb{Z}$ & $0$ 
& $\mathbb{Z} \oplus \mathbb{Z}$ & $0$ & $\mathbb{Z} \oplus \mathbb{Z}$ \\
$U_-$& AIII   &$\mathcal{C}_0$& $\mathbb{Z}$ & $0$ 
& $\mathbb{Z}$ & $0$ & $\mathbb{Z}$ & $0$ & $\mathbb{Z}$ & $0$ \\
\hline 
\multirow{8}{*}{$U^+_+,U^-_-,U^+_{++},U^-_{--}$}
 & AI   &$\mathcal{R}_0 \times \mathcal{R}_0$
& $\mathbb{Z}\oplus \mathbb{Z}$ & $0$ & $0$ 
& $0$ & $2\mathbb{Z}\oplus 2\mathbb{Z}$ & $0$ 
& $\mathbb{Z}_2\oplus\mathbb{Z}_2$ & $\mathbb{Z}_2\oplus\mathbb{Z}_2$ \\
 & BDI  &$\mathcal{R}_1 \times \mathcal{R}_1$
& $\mathbb{Z}_2\oplus\mathbb{Z}_2$ 
& $\mathbb{Z}\oplus\mathbb{Z}$ & $0$ & $0$ & $0$ 
& $2\mathbb{Z}\oplus 2\mathbb{Z}$ & $0$ & $\mathbb{Z}_2\oplus\mathbb{Z}_2$ \\
 & D    &$\mathcal{R}_2 \times \mathcal{R}_2$
& $\mathbb{Z}_2\oplus\mathbb{Z}_2$ & $\mathbb{Z}_2\oplus\mathbb{Z}_2$ 
& $\mathbb{Z}\oplus\mathbb{Z}$ & $0$ & $0$ & $0$ 
& $2\mathbb{Z}\oplus 2\mathbb{Z}$ & $0$ \\
& DIII &$\mathcal{R}_3 \times \mathcal{R}_3$& $0$ 
& $\mathbb{Z}_2\oplus\mathbb{Z}_2$ & $\mathbb{Z}_2\oplus\mathbb{Z}_2$ 
& $\mathbb{Z}\oplus\mathbb{Z}$ & $0$ & $0$ & $0$ 
& $2 \mathbb{Z}\oplus 2\mathbb{Z}$ \\
& AII  &$\mathcal{R}_4 \times \mathcal{R}_4$
& $2 \mathbb{Z}\oplus 2\mathbb{Z}$ & $0$ 
& $\mathbb{Z}_2\oplus\mathbb{Z}_2$ 
& $\mathbb{Z}_2\oplus\mathbb{Z}_2$ & $\mathbb{Z}\oplus\mathbb{Z}$ 
& $0$ & $0$ & $0$ \\
& CII  &$\mathcal{R}_5 \times \mathcal{R}_5$& $0$ 
& $2 \mathbb{Z}\oplus 2\mathbb{Z}$ & $0$ & $\mathbb{Z}_2\oplus\mathbb{Z}_2$ 
& $\mathbb{Z}_2\oplus\mathbb{Z}_2$ & $\mathbb{Z}\oplus\mathbb{Z}$ 
& $0$ & $0$ \\
& C    &$\mathcal{R}_6 \times \mathcal{R}_6$
& $0$ & $0$ & $2 \mathbb{Z}\oplus 2\mathbb{Z}$ 
& $0$ & $\mathbb{Z}_2\oplus\mathbb{Z}_2$ 
& $\mathbb{Z}_2\oplus\mathbb{Z}_2$ & $\mathbb{Z}\oplus\mathbb{Z}$ 
& $0$ \\
& CI   &$\mathcal{R}_7 \times \mathcal{R}_7$& $0$ & $0$ 
& $0$ & $2 \mathbb{Z}\oplus 2\mathbb{Z}$ & $0$ 
& $\mathbb{Z}_2\oplus\mathbb{Z}_2$ & $\mathbb{Z}_2\oplus\mathbb{Z}_2$ 
& $\mathbb{Z}\oplus\mathbb{Z}$ \\
\hline
$U^+_{+-},U^-_{-+}$ & BDI  &$\mathcal{R}_0$& $\mathbb{Z}$ 
& $0$ & $0$ & $0$ & $2 \mathbb{Z}$ & $0$ & $\mathbb{Z}_2$ & $\mathbb{Z}_2$ \\
$U^+_{-+},U^-_{+-}$ & DIII &$\mathcal{R}_2$& $\mathbb{Z}_2$ 
& $\mathbb{Z}_2$ & $\mathbb{Z}$ & $0$ & $0$ & $0$ & $2\mathbb{Z}$ & $0$ \\
$U^+_{+-},U^-_{-+}$ & CII  &$\mathcal{R}_4$& $2 \mathbb{Z}$ & $0$ 
& $\mathbb{Z}_2$ & $\mathbb{Z}_2$ & $\mathbb{Z}$ & $0$ & $0$ & $0$ \\
$U^+_{-+},U^-_{+-}$ & CI   &$\mathcal{R}_6$& $0$ & $0$ & $2 \mathbb{Z}$ 
& $0$ & $\mathbb{Z}_2$ & $\mathbb{Z}_2$ & $\mathbb{Z}$ & $0$ \\
\hline
$U^+_-,U^-_+$       & AI, D, AII, C      &$\mathcal{C}_0$
& $\mathbb{Z}$ & $0$  & $\mathbb{Z}$ & $0$ & $\mathbb{Z}$ 
& $0$ & $\mathbb{Z}$ & $0$ \\
$U^+_{--},U^-_{++}$ & BDI, DIII, CII, CI &$\mathcal{C}_1$& $0$ 
& $\mathbb{Z}$ & $0$  & $\mathbb{Z}$ & $0$ & $\mathbb{Z}$ & $0$ 
& $\mathbb{Z}$ \\
\hline
$U^+_{-+},U^-_{+-}$ & BDI  &$\mathcal{R}_2$& $\mathbb{Z}_2$ 
&$\mathbb{Z}_2$ 
& $\mathbb{Z}$ & $0$ & $0$ & $0$ & $2 \mathbb{Z}$ & $0$ \\
$U^+_{+-},U^-_{-+}$ & DIII &$\mathcal{R}_4$& $2 \mathbb{Z}$ & $0$ & $\mathbb{Z}_2$ & $\mathbb{Z}_2$ & $\mathbb{Z}$ & $0$ & $0$ & $0$ \\
$U^+_{-+},U^-_{+-}$ & CII  &$\mathcal{R}_6$& $0$ & $0$ & $2 \mathbb{Z}$ & $0$ & $\mathbb{Z}_2$ & $\mathbb{Z}_2$ & $\mathbb{Z}$ & $0$ \\
$U^+_{+-},U^-_{-+}$ & CI   &$\mathcal{R}_0$& $\mathbb{Z}$ & $0$ & $0$ & $0$ & $2 \mathbb{Z}$ & $0$ & $\mathbb{Z}_2$ & $\mathbb{Z}_2$ \\
\hline \hline
\end{tabular}
\label{TabLU}
\end{center}
\end{table*}

\subsubsection{Spin Chern insulator ($U^-_+$ in class AII)}
The simplest example of the symmetry protected topological phases is 
a quantum spin Hall insulator preserving the
$z$-component of spin.
The system has TRS, and it is
also invariant under the two-fold spin rotation along the $z$-axis,
which is generated by $U=is_z$.
Since the additional symmetry $U=is_z$  commutes with $T$, the system is
categorized into class AII with $U_+^-$ in two-dimensions. Thus the
topological nature is characterized by $\mathbb{Z}$, as is seen in Table
\ref{TabLU}.

The corresponding topological number is known as the spin Chern number:
In the presence of non-spatial unitary symmetry $U=is_z$, the
Hamiltonian ${\cal H}(k_x, k_y)$ of the system can be block-diagonal in
the eigen basis of $U$ with the eigenvalue $U=\pm i$. 
The anti-unitarity of $T$ implies that TRS do not close in the each eigen
sector, so each block of the Hamiltonian loses
a real structure caused by TRS.
In other words, $U = i s_z$ plays a role of the imaginary
unit $i$.
Such an effect
is called as {\it complexification},
which induces a complex structure in real AZ class.
As a result,  change of symmetry class, AII $\to$ A, occurs. 
The class A Hamiltonian is obtained by {\it forgetting} the real
structure,
\begin{eqnarray}
\tilde{\mathcal{H}}(k_x,k_y) 
:= \frac{1}{2} \mathrm{Tr}_{s} \left[s_z \mathcal{H}(k_x,k_y)\right],
\end{eqnarray}
and the topological invariant is given by the 1st Chern character,  
\begin{eqnarray}
Ch_{1} = \frac{i}{2 \pi} \int \mathrm{tr} \mathcal{\tilde F}, 
\label{SpinChern} 
\end{eqnarray}
where $\mathcal{\tilde F}$ is the Berry curvature of the complexified
Hamiltonian $\tilde{\mathcal{H}}(k_x,k_y)$.
The topological invariant Eq. (\ref{SpinChern}) is the spin Chern number.

To illustrate the complexification and the spin Chern number, 
consider the  model Hamiltonian given by 
\begin{eqnarray}
&&\mathcal{H}(k_x,k_y) 
= m(k_x,k_y)\sigma_z+ v k_x \sigma_x s_z + v k_y \sigma_y, 
\nonumber\\
&&m(k_x, k_y)=m_0-m_2 (k_x^2+k_y^2),
\end{eqnarray}
where  
$m_0$ is the a mass, and $v$ is a velocity. 
Here we have also introduced a cut-off $m_2$.
In terms of $U=is_z$, the Hamiltonian is rewritten as ${\cal H}({\bm
k})=m({\bm k})\sigma_z-ivk_x\sigma_x U+vk_y\sigma_y$, and thus 
the complexified Hamiltonian, $\tilde{\cal {H}}({\bm k})=m({\bm
k})\sigma_z+v k_x \sigma_x+vk_y \sigma_y$, is given by
replacing $U$ with $i$. The spin Chern number $Ch_1$ of this model is 
${\rm sgn}(m_0 m_2)$.

\subsubsection{Mirror-odd two-dimensional topological superconductor ($U^-_-$ in class D)}
\label{sec:mirrorodd2D}

Consider a time-reversal broken (class D) superconductor in
two dimensions:
\begin{eqnarray}
{\cal H}_{\rm BdG}(k_x, k_y)=
\left(
\begin{array}{cc}
\epsilon(k_x, k_y) & \Delta(k_x, k_y)\\
\Delta^{\dagger}(k_x, k_y) & -\epsilon^{T}(-k_x, -k_y)
\end{array}
\right) 
\end{eqnarray}
As an additional symmetry, we assume here the mirror reflection symmetry
with respect to the $xy$-plane.
The reflection symmetry implies $M \epsilon({\bm
k})M^{\dagger}=\epsilon({\bm k})$ with $M=is_z$, but the gap function
$\Delta({\bm k})$ can be mirror-even, $M\Delta({\bm k})M^T=\Delta({\bm
k})$, or mirror-odd, $M\Delta({\bm k})M^T=-\Delta({\bm k})$. 
Even in the latter case, the BdG Hamiltonian can be invariant under the
mirror reflection by performing simultaneously the $U(1)$ gauge symmetry
$\Delta({\bm k})\rightarrow \Delta({\bm k})e^{i\theta}$ with $\theta=\pi$.

First, examine the mirror odd case. In this case, the BdG Hamiltonian
${\cal H}_{\rm BdG}({\bm k})$ commutes with $\tilde{M}=is_z \tau_0$.
Since $\tilde{M}$ anticommutes with PHS, $C=\tau_x {\cal K}$, 
the additional symmetry $\tilde{M}$ is identified with $U_{-}^{-}$ in 
class D.
From Table \ref{TabLU}, the topological index is $\mathbb{Z}\oplus\mathbb{Z}$.
 
The ${\mathbb{Z}}\oplus{\mathbb{Z}}$ structure can be understood as a
pair of spinless class D superconductors:
From the commutation relation $[{\cal H}({\bm k}), \tilde{M}]=0$, the
BdG Hamiltonian can be block-diagonal into a pair of spinless systems
with different eigen values of $\tilde{M}=\pm i$.
The anti-unitarity of $C$ and the anti-commutation relation $\{C,
\tilde{M}\}=0$ imply that each spinless system retains PHS, and thus
it can be considered as a spinless class D superconductor.
Since each two-dimensional class D superconductor is
characterized by the 1st Chern number, we obtain the
${\mathbb{Z}}\oplus{\mathbb{Z}}$ structure.

The model Hamiltonian is given by 
\begin{eqnarray}
\mathcal{H}_{\mathrm{BdG}}(k_x,k_y) 
&=&\begin{pmatrix}
\frac{k^2}{2m} -\mu-h_z s_z & 
\frac{\Delta_p}{k_{\rm F}}(k_x s_x + k_y s_y) i s_y \\
-i s_y \frac{\Delta_p}{k_{\rm F}} (k_x s_x + k_y s_y) 
& -\frac{k^2}{2 m} +\mu+h_zs_z
\end{pmatrix}
\nonumber\\
&=&\left(\frac{k^2}{2m}-\mu\right)\tau_z-h_z
 s_z\tau_z-\frac{\Delta_p}{k_{\rm F}}k_xs_z\tau_x-\frac{\Delta_p}{k_{\rm
 F}}k_y\tau_y,
\end{eqnarray}
where we have introduced the Zeeman term $h_z s_z$ in order to break TRS.
In the diagonal basis with $\tilde M=\pm i$, we have
\begin{eqnarray}
{\cal H}_{\rm BdG}^{\pm}(k_x, k_y)=
 \left(\frac{k^2}{2m}-\mu\mp h_z\right)\tau_z
\mp \frac{\Delta_p}{k_{\rm F}}k_x \tau_x-\frac{\Delta_p}{k_{\rm
 F}}k_y\tau_y, 
\end{eqnarray}
where each of ${\cal H}_{\rm BdG}^{\pm}(k_x, k_y)$ supports PHS, i.e. $C{\cal
H}_{\rm BdG}^{\pm}({\bm k})C^{-1}=-{\cal H}^{\pm}_{\rm BdG}(-{\bm k})$. 
The topological invariant for each sector is
\begin{equation}\begin{split}
Ch^{\pm}_{1} = \frac{i}{2 \pi} \int \mathrm{tr} \mathcal{F}^{\pm}, 
\end{split}\end{equation}
where $\mathcal{F}^{\pm}$ is the Berry curvature of 
$\mathcal{H}_{\rm BdG}^{\pm}(k_x,k_y)$. 
The Abelian group $\mathbb{Z} \oplus \mathbb{Z}$ is characterized by the
two integers $(Ch_1^{+},Ch_1^{-})$.  
Note that $Ch_1^{+}$ and $Ch_1^{-}$ can be different from each other by
adjusting $h_z$, which also confirms the direct sum structure of
$\mathbb{Z}\oplus\mathbb{Z}$.   

The presence of a vortex shifts $\delta$ as $\delta=1$. 
From Table \ref{TabLU}, the topological index of the vortex is given by
$\mathbb{Z}_2 \oplus \mathbb{Z}_2$.
In a thin film of $^3$He-A under perpendicular Zeeman fields, one
can create
an integer quantum vortex, in which a pair of Majorana zero mode exist
due to the mirror symmetry.\cite{sato2014mirror}
The mirror protected Majorana zero mode gives rise to non-Abelian
statistics of integer quantum vortices.\cite{sato2014mirror}


\subsubsection{Mirror-even two-dimensional topological superconductor ($U^-_+$ in class D)}

Now consider the mirror even case, where the mirror reflection operator
for the BdG Hamiltonian is given by $\tilde M=is_z\tau_z$. From the
commutation relation between $\tilde M$ and $C$, $\tilde M$ is
identified as $U_+^-$ in class D. 
The topological index is $\mathbb{Z}$.  

Again the BdG Hamiltonian ${\cal H}_{\rm BdG}({\bm k})$ can be
block-diagonal in the eigen basis of $\tilde M$. However, in
contrast to the mirror odd case, each spinless sector does not support
PHS, and thus it belongs to class A. 
Moreover, because the spinless sectors are exchanged by $C$ to keep
PHS in the whole system, they
can not be independent, and thus they should have the essentially same
structure.
Hence, the topological index is not a
direct sum, 
${\mathbb{Z}}\oplus{\mathbb{Z}}$, but a single ${\mathbb{Z}}$.  

The model Hamiltonian is given by
\begin{eqnarray}
\mathcal{H}_{\mathrm{BdG}}(k_x,k_y) 
&=&\begin{pmatrix}
\frac{k^2}{2m} -\mu-h_z s_z & 
i\frac{\Delta_p}{k_{\rm F}}(k_x+ik_y) s_z s_y \\
-i s_y \frac{\Delta_p}{k_{\rm F}} (k_x-ik_y) s_z
& -\frac{k^2}{2 m} +\mu+h_zs_z
\end{pmatrix}
\nonumber\\
&=&\left(\frac{k^2}{2m}-\mu\right)\tau_z-h_z
 s_z\tau_z+\frac{\Delta_p}{k_{\rm F}}k_xs_x\tau_x-\frac{\Delta_p}{k_{\rm
 F}}k_ys_x \tau_y.
\end{eqnarray}
In the diagonal basis of $M_{\rm BdG}=\pm i$,  we obtain
\begin{eqnarray}
{\cal H}^{\pm}_{\rm BdG}(k_x, k_y)=
 \left(\frac{k^2}{2m}-\mu\right)\tau_z \pm h_z \tau_0
+\frac{\Delta_p}{k_{\rm F}}k_x \tau_x-\frac{\Delta_p}{k_{\rm
 F}}k_y \tau_y.
\end{eqnarray}
Contrary to the mirror odd case, the Zeeman field $h_z$ merely shifts
the origin of energy, so the first Chern numbers $Ch_1^{\pm}$ of the two
sectors coincide, i.e. $Ch_1^+=Ch_1^-$.

\subsubsection{Superconducting nanowire with Rashba SO interaction and
   Zeeman fields ($A^+_+$,
   $A^+_-$ in class D)} 
Consider a time-reversal broken (class D) superconductor with the
spin-orbit interaction in one-dimension,\cite{lutchyn2010, oreg2010} 
\begin{eqnarray}
\mathcal{H}_{\mathrm{BdG}}(k_x) 
&=&\begin{pmatrix}
\frac{k_x^2}{2 m} -\mu + \lambda k_x s_y 
+ \bm{h} \cdot \bm{s}& \Delta i s_y \\
-i s_y \Delta & -\frac{k_x^2}{2 m} +\mu - \lambda k_x s_y 
- \bm{h} \cdot \bm{s}^T
\end{pmatrix} \\
&=& \left( \frac{k_x^2}{2 m} -\mu \right) \tau_z + \lambda k_x s_y
\tau_z - \Delta s_y \tau_y + h_x s_x \tau_z + h_y s_y + h_z s_z \tau_z, 
\label{1DNanowire}
\end{eqnarray}
where $\lambda k_x s_y \tau_z$ is the Rashba spin-orbit interaction
term, $\Delta$ is an $s$-wave pairing, and $\bm{h}$ is the Zeeman
field.  
Equation (\ref{1DNanowire}) is the low-energy effective Hamiltonian
describing a one-dimensional nanowire with the Rashba spin-orbit interaction and
a proximity induced $s$-wave superconductivity.  
In the absence of the Zeeman field, TRS, $T= i s_y \mathcal{K}$, and
mirror reflection symmetry with respect to $zx$-plane, $\mathcal{M}_{zx}
= is_y$,  are preserved.  
The Zeeman field breaks both TRS and the mirror reflection symmetry, 
however, if $h_y = 0$ it retains an antiunitary
symmetry which is obtained as their combination $A = \mathcal{M}_{zx} T
= \mathcal{K}$:\cite{mizushima2013topological} $A \mathcal{H}(-k_x) A^{-1} = \mathcal{H}(k_x)$.
This system hosts topological superconductivity when $|{\bm h}|>\sqrt{\Delta^2+\mu^2}$.~\cite{sato2009non,sato2010non}

As the symmetry operator $A$ commutes with the particle-hole
transformation $C=\tau_x \mathcal{K}$, it is labeled as $A^+_+$ in
class D of Table \ref{TabLA}. 
%
The anti-symmetry $A = \mathcal{K}$ defines an emergent
spinless TRS \cite{sato2009topological, tewari2012topological, mizushima2013topological} because of
$A^2 = 1$, which changes the AZ symmetry class effectively as D $\to$ BDI.  
The topological number $\mathbb{Z}$ in Table \ref{TabLA} (
$A^{+}_{+}$ in class D with $\delta=1$) is the
winding number of the emergent class BDI, 
\begin{equation}\begin{split}
N_1 = \frac{1}{4 \pi i} \int \mathrm{tr} \left[ \tau_x \mathcal{H}^{-1}
		 d \mathcal{H} \right],  
\end{split}\end{equation}
with the chiral operator $\tau_x = C A$. 
Note that since the emergent class BDI is not accidental but it is
originated from the symmetry of the configuration, 
the same topological characterization works even for multi-band
nanowires as far as the wire
configuration respects the symmetry \cite{mizushima2013topological}.

In the above, we have assumed an $s$-wave pairing, but 
even for other unconventional pairings,\cite{wong2012, nakosai2013, zhang2013topological} one can
obtain a similar topological characterization if the gap function has a
definite parity under the mirror reflection with respect to the $zx$ plane:  
If the pairing is even under the mirror reflection $\mathcal{M}_{zx}$,
the same antiunitary symmetry $A^{+}_{+}$ characterizes the system, but 
even if the pairing is mirror-odd, 
a similar emergent TRS is obtained as $A = \tau_z \mathcal{K}$ by
combining TRS and the mirror operator of this case
$\mathcal{M}_{zx} = i s_y \tau_z$ \cite{ueno2013symmetry}. 
Because the particle-hole transformation $C = \tau_x \mathcal{K}$
anti-commutes with the latter $A$, it is labeled as $A^+_-$ in class D
of Table \ref{TabLA}.   
The corresponding topological number is $\mathbb{Z}$ again in one-dimension.

\begin{table*}[!]
\begin{center}
\caption{
Classification table for topological crystalline insulators and
 superconductors and their topological defects in the presence of
 order-two additional antiunitary symmetry with flipped parameters
 $\delta_{\parallel}=d_{\parallel} - D_{\parallel} = 0$ (mod 4).  
Here $\delta=d-D$.
}
\begin{tabular}[t]{ccccccccccc}
\hline \hline
Symmetry & Class & $\mathcal{C}_q$ or $\mathcal{R}_q$ & $\delta=0$ & $\delta=1$ & $\delta=2$ & $\delta=3$ & $\delta=4$ & $\delta=5$ & $\delta=6$ & $\delta=7$ \\
\hline
$A^+$  & A      &$\mathcal{R}_0$& $\mathbb{Z}$ & $0$ & $0$ & $0$ & $2\mathbb{Z}$ & $0$ & $\mathbb{Z}_2$ & $\mathbb{Z}_2$ \\
$A^-$  & A      &$\mathcal{R}_4$& $2\mathbb{Z}$ & $0$ & $\mathbb{Z}_2$ & $\mathbb{Z}_2$ & $\mathbb{Z}$ & $0$ & $0$ & $0$ \\
$A^+_+$& AIII &$\mathcal{R}_1$& $\mathbb{Z}_2$ & $\mathbb{Z}$ & $0$ & $0$ & $0$ & $2\mathbb{Z}$ & $0$ & $\mathbb{Z}_2$ \\
$A^-_-$& AIII &$\mathcal{R}_3$& $0$ & $\mathbb{Z}_2$ & $\mathbb{Z}_2$ & $\mathbb{Z}$ & $0$ & $0$ & $0$ & $2\mathbb{Z}$ \\
$A^-_+$& AIII &$\mathcal{R}_5$& $0$ & $2\mathbb{Z}$ & $0$ & $\mathbb{Z}_2$ & $\mathbb{Z}_2$ & $\mathbb{Z}$ & $0$ & $0$ \\
$A^+_-$& AIII &$\mathcal{R}_7$& $0$ & $0$ & $0$ & $2\mathbb{Z}$ & $0$ & $\mathbb{Z}_2$ & $\mathbb{Z}_2$ & $\mathbb{Z}$ \\
\hline 
$A^+_+,A^+_-$ & D &$\mathcal{R}_1$& $\mathbb{Z}_2$ & $\mathbb{Z}$ & $0$ & $0$ & $0$ & $2\mathbb{Z}$ & $0$ & $\mathbb{Z}_2$ \\
$A^-_+,A^-_-$ & C &$\mathcal{R}_5$& $0$ & $2\mathbb{Z}$ & $0$ & $\mathbb{Z}_2$ & $\mathbb{Z}_2$ & $\mathbb{Z}$ & $0$ & $0$  \\
\hline
$A^-_+,A^-_-$ & D &$\mathcal{R}_3$& $0$ & $\mathbb{Z}_2$ & $\mathbb{Z}_2$ & $\mathbb{Z}$ & $0$ & $0$ & $0$ & $2\mathbb{Z}$ \\
$A^+_+,A^+_-$ & C &$\mathcal{R}_7$& $0$ & $0$ & $0$ & $2\mathbb{Z}$ & $0$ & $\mathbb{Z}_2$ & $\mathbb{Z}_2$ & $\mathbb{Z}$ \\
\hline \hline
\end{tabular}
\label{TabLA}
\end{center}
\end{table*}

\subsubsection{Vortex in two-dimensional superconductors with
magnetic in-plane reflection symmetry ($A^+_+$, $A^+_-$ in class D)}
\label{sec:vortex2Dmagneticmirror}

Consider a two-dimensional time-reversal invariant superconductor,  
\begin{eqnarray}
\mathcal{H}_{\mathrm{BdG}}(k_x,k_y) 
=\begin{pmatrix}
\epsilon(k_x,k_y) & \Delta (k_x,k_y) \\
\Delta^{\dag}(k_x,k_y) & -\epsilon^T(-k_x,-k_y)
\end{pmatrix}, 
\end{eqnarray}
with in-plane mirror reflection symmetry that flips the $x$-direction.
The mirror symmetry implies
\begin{eqnarray}
M_{x} \epsilon(k_x, k_y) M^{\dagger}_{x} = \epsilon(-k_x,k_y),  
\quad
M_{x}=is_x,
\end{eqnarray}
in the normal part, but in a manner similar to Sec. \ref{sec:mirrorodd2D}, two
different realizations (mirror even and mirror odd) are possible in the
gap function
\begin{eqnarray}
M_{x} \Delta(k_x,k_y) M^T_{x} = \pm \Delta(-k_x,k_y), 
\end{eqnarray}
due to the U(1) gauge symmetry.
The mirror symmetry is summarized as
\begin{eqnarray}
\tilde{M}_{x} \mathcal{H}_{\mathrm{BdG}}(k_x,k_y)
		 \tilde{M}^{\dagger}_{x} 
= \mathcal{H}_{\mathrm{BdG}}(-k_x,k_y),
\end{eqnarray}
with $\tilde{M}_{x} = i s_x \tau_z$ $(\tilde{M}_{x}=i s_x\tau_0)$ for
the mirror even (odd) gap function.  


Now explore topological properties of a vortex in this system.
Applying a magnetic field normal to the system, one can create a vortex.
The adiabatic (semiclassical) BdG Hamiltonian with a vortex is given by
\begin{equation}\begin{split}
\mathcal{H}_{\mathrm{BdG}}(k_x,k_y,\phi)
= \begin{pmatrix}
\epsilon(k_x,k_y) & \Delta(k_x,k_y,\phi) \\
\Delta^{\dag}(k_x,k_y,\phi) & -\epsilon^{T}(-k_x,-k_y)
\end{pmatrix}, 
\end{split}\end{equation}
where 
$\phi$ denotes the angle around the vortex measured from the
$y$-axis.
Since $\phi$ transforms as $\phi\rightarrow -\phi$ under the mirror
reflection, the vortex configuration   
$\Delta(k_x,k_y, \phi)\sim \Delta(k_x,k_y)e^{i\phi}$ breaks the mirror
reflection symmetry as well as TRS, but the combination of
these two symmetries remains,   
\begin{eqnarray}
A_{x} \mathcal{H}_{\mathrm{BdG}}(k_x, k_y, \phi)A^{-1}_{x}=
 \mathcal{H}_{\mathrm{BdG}}(k_x, -k_y,-\phi) 
\end{eqnarray}
with $A_x= T\tilde{M}_x $. 
The magnetic in-plane reflection symmetry $A_x$ is labeled as $A^+_-$ or
$A^{+}_+$ in class D of Table \ref{TabLA}, and thus the topological
index of the vortex ($\delta=1$, $\delta_{\parallel}=0$) is given by
$\mathbb{Z}$.

A vortex in two-dimensional chiral $p_x+ip_y$ superconductors also has
the same magnetic in-plane reflection symmetry.
Although chiral $p_x+ip_y$ gap functions explicitly break TRS as well as the
in-plane reflection symmetry, they preserve the magnetic in-plane reflection
symmetry up to the U(1) gauge symmetry. Consequently, a vortex also
preserves the magnetic in-plane reflection symmetry, and thus 
the topological
index of the vortex is also given by
$\mathbb{Z}$.

In the mirror-symmetric subspace defined by $k_x=0$, $\phi=0$ or
$k_x=0$, $\phi=\pi$, the magnetic in-plane reflection symmetry in class
D implies the presence of
CS, 
\begin{eqnarray}
\Gamma_x \mathcal{H}(0, k_y, \phi)\Gamma_x^{-1}=
\mathcal{H}(0, k_y, \phi),
\quad (\phi=0, \pi)
\end{eqnarray}
where $\Gamma_x=CA_x$ with the particle-hole operator $C$. 
Using CS, one can define two one-dimensional winding
numbers as \begin{equation}\begin{split}
N_1^{\phi=0,\pi} 
= 
\frac{1}{4 \pi i}\int \left.
\mathrm{tr} [ \Gamma_{x} \mathcal{H}_{\mathrm{BdG}}^{-1}(0, k_y, \phi) d_{k_y}
		 \mathcal{H}_{\mathrm{BdG}}(0, k_y, \phi)]\right|_{\phi=0, \pi}. 
	    \end{split}\end{equation}
Among these two $\mathbb{Z}$ indices, only the
difference is relevant to topologically stable zero modes in the vortex.
Indeed if they are the same, i.e. $N_1^{0}=N_1^{\pi}$, 
the vortex can be smoothly deformed into the bulk without a topological
obstruction, and thus vortex zero modes, even if they exist, 
disappear.
This means that the $\mathbb{Z}$ index of 
the vortex, which ensures the topological stability of vortex zero
modes, is proportional to $N_1^{0} - N_1^{\pi}$.



To determine the proportional constant, consider
a representative Hamiltonian with the same magnetic in-plane reflection
symmetry,
\begin{eqnarray}
&&\mathcal{H}_1 = 
\begin{pmatrix}
\frac{k^2}{2m}-\mu & i\Delta e^{i \phi} (k_x+i k_y) \\
-i\Delta e^{-i \phi} (k_x-i k_y) & -\frac{k^2}{2m}+\mu
\end{pmatrix}, 
\end{eqnarray}
where the particle-hole transformation and the magnetic reflection are
given by $C = \tau_x K$ and $A = \tau_z K$, respectively.
This model supports a single zero mode localized at the
vortex,\cite{kopnin1991} and its topological index is
\begin{eqnarray}
\left. (N_1^{0}, N_1^{\pi}) \right|_{\mathcal{H}_1} = (1,-1).
\end{eqnarray}
Therefore, in order for the $\mathbb{Z}$ index of the vortex, $N_1^{\rm
vortex}$, to be equal to the number
of vortex zero modes,   
the proportional constant should be 1/2, 
\begin{eqnarray} 
N_1^{\rm vortex} =\frac{N_1^0-N_1^\pi}{2}.
\label{eq:vortexN1}
\end{eqnarray}

\subsubsection{Zero mode in a magnetic in-plane mirror reflection
   symmetric heterostructure ($A^+_+$ in class D)}

In the previous subsection,  we considered a vortex in a two-dimensional
superconductors, but
a similar zero mode protected by the magnetic in-plane mirror 
can be realized in a heterostructure of a topological insulator,
an $s$-wave superconductor and a ferromagnet. 
Consider a $\pi$-junction of an $s$-wave superconductor 
on the top of a topological
insulator. 
At the $\pi$-junction, there is a one-dimensional helical Majorana
gapless mode,\cite{fu2008superconducting}   
which becomes a domain wall Majorana zero energy
bound state in the simultaneous presence of a ferromagnetic
kink.\cite{shiozaki2012index}
The low-energy effective Hamiltonian of this model is 
\begin{equation}\begin{split}
\mathcal{H}(x,y)
&= \begin{pmatrix}
- i s_y \partial_x + i s_x \partial_y + h_x(x)s_x & i \Delta(y) s_y \\
-is_y \Delta(y) & i s_y \partial_x + i s_x \partial_y - h_x(x) s_x
\end{pmatrix} \\
&= - i s_y \tau_z \partial_x + i s_x \tau_0 \partial_y 
- \Delta(y) s_y \tau_y + h_x(x) s_x \tau_z, 
\end{split}\label{Domain_Wall_Majorana}\end{equation}
where $\Delta(y)$ is a proximity induced $s$-wave superconducting order
of surface Dirac fermions on a topological insulator,
and $h_x(x)$ is a ferromagnet induced exchange field that satisfies
$h_x(-x) = - h_x(x)$.
The system is invariant under the magnetic in-plane mirror reflection
\begin{eqnarray}
A_x \mathcal{H}(x,y) A_x^{-1} =\mathcal{H}(-x,y),
\end{eqnarray}
with $A_x=\-is_z \tau_zK$.
Assuming that $\Delta(y>0)=-\Delta(y<0)=\Delta_0>0$, 
we have a zero energy state
\begin{equation}\begin{split}
\xi(x,y) 
= 
\begin{pmatrix}
i \\
0 \\
1 \\
0 \\
\end{pmatrix} e^{-\int^x h_x(x') dx'} e^{-\int^y \Delta(y') dy' } , 
\end{split}\end{equation}
if $h_x(x>0) > 0$, 
and 
\begin{equation}\begin{split}
\xi(x,y) 
= 
\begin{pmatrix}
0 \\
1 \\
0 \\
i \\
\end{pmatrix} e^{\int^x h_x(x') dx'} e^{-\int^y \Delta(y') dy' } , 
\end{split}\end{equation}
if $h_x(x>0 ) < 0$, respectively. 
The existence of the zero mode is ensured by the $\mathbb{Z}$ index 
that is defined in a manner similar to  Eq.(\ref{eq:vortexN1}):
In the semiclassical limit, the Hamiltonian
Eq.(\ref{Domain_Wall_Majorana}) reads
\begin{eqnarray}
\mathcal{H}(k_x,k_y, x,y)= k_x s_y\tau_z
-k_ys_x\tau_0-\Delta(y)s_y\tau_y+h_x(x)s_x\tau_z,
\end{eqnarray}
which is chiral symmetric at $k_x=x=0$,
\begin{eqnarray}
\{\Gamma_x, \mathcal{H}(0,k_y,0,y)\}=0 
\end{eqnarray}
with $\Gamma_x=s_z\tau_y$.
The $\mathbb{Z}$ index is given by
Eq.(\ref{eq:vortexN1}) with the identification of $y=\cos\phi$.
By adding a regularization term in the gap function,
$\Delta(y)\rightarrow\Delta(y)-\delta (k_x^2+k_y^2)$ ($\delta>0$), 
one can evaluate the $\mathbb{Z}$ index of this model as $1$.

\subsubsection{$\mathbb{Z}$ Majorana point defect zero mode protected by
   magnetic  $\pi$-rotation symmetry ($A^+_+$, $A^+_-$ in class D)}

We argue here Majorana zero modes which are localized at a point
defect in three dimensions and protected by magnetic $\pi$-rotation
symmetry around the $z$-axis. 
The BdG Hamiltonian is given in the form of 
$\mathcal{H}_{\mathrm{BdG}}(k_x,k_y,k_z,\phi,\theta)$ where $\phi$ and
$\theta$ are the azimuthal and polar angles of a sphere surrounding the
point defect.
The magnetic $\pi$-rotation symmetry around the $z$-axis is expressed by 
\begin{equation}\begin{split}
A \mathcal{H}_{\mathrm{BdG}}(k_x,k_y,k_z,\phi,\theta) 
A^{-1} = \mathcal{H}_{\mathrm{BdG}}(k_x,k_y,-k_z,\phi+\pi,\theta), 
\end{split}\end{equation}
where $A$ is either $A = s_x \tau_0
\mathcal{K}$ or $s_x \tau_z \mathcal{K}$,
depending on the parity of the gap function under the magnetic 
$\pi$-rotation.
As $d=3$, $d_{\parallel}=2$, $D=2$, and $D_{\parallel}=2$ in this
transformation,  it is labeled as $A^+_+$ or $A^+_-$ in class D with
$\delta_{\parallel}=0$ and $\delta=1$. 
From Table \ref{TabLA}, the zero modes are topologically characterized
by $\mathbb{Z}$. 

The $\mathbb{Z}$ index is defined as follows.
Because the system also has the PHS 
$C \mathcal{H}_{\mathrm{BdG}}({\bm k},\phi,\theta) C^{-1} =
\mathcal{H}_{\mathrm{BdG}}(-{\bm k},\phi,\theta)$ with $C = \tau_x
\mathcal{K}$, 
we can obtain 
\begin{eqnarray}
\Gamma \mathcal{H}_{\mathrm{BdG}}(k_x,k_y,k_z,\phi, \theta) \Gamma^{-1} 
= -\mathcal{H}_{\mathrm{BdG}}(-k_x,-k_y,k_z,\phi+\pi, \theta), 
\end{eqnarray}
with $\Gamma=AC$, by combining the magnetic $\pi$-rotation and the PHS.
Therefore, the BdG Hamiltonian has a ``$\pi$-rotation CS''
\begin{eqnarray}
\Gamma \mathcal{H}_{\mathrm{BdG}}(0,0,k_z,\theta=0,\pi) \Gamma^{-1} 
= -\mathcal{H}_{\mathrm{BdG}}(0,0,k_z,\theta=0,\pi), 
\end{eqnarray}
on the $\pi$-rotation symmetric subspace defined by $\theta=0,\pi$ and
$k_x=k_y=0$. 
Here the BdG Hamiltonian does not depend on $\phi$ 
at $\theta=0, \pi$,
which are the north and south poles of the sphere surrounding the point defect.
The $\pi$-rotation CS enables us to define two one-dimensional winding numbers
\begin{eqnarray}
N_1^{\theta=0, \pi} 
= \frac{1}{4 \pi i} \int \left.  
\mathrm{tr}\left[
\Gamma \mathcal{H}^{-1}_{\mathrm{BdG}}(0,0,k_z,\theta) d_{k_z}
\mathcal{H}_{\mathrm{BdG}}(0,0,k_z,\theta)
\right] 
\right|_{\theta=0,\pi}.
\end{eqnarray}
From an argument similar to that in Sec. \ref{sec:vortex2Dmagneticmirror},
we can show that only the
difference between $N_1^0$ and $N_1^{\pi}$ is relevant to the zero
modes. 
The $\mathbb{Z}$ topological invariant of Majorana
zero modes is given by
\begin{equation}\begin{split}
N_1^{\rm defect} = \frac{N_1^{0} - N_1^{\pi}}{2}. 
\end{split}\end{equation}



\subsection{$\delta_{\parallel}=1$ family}

\begin{figure}[!]
 \begin{center}
  \includegraphics[width=\linewidth, trim=0cm 0cm 0cm 0cm]{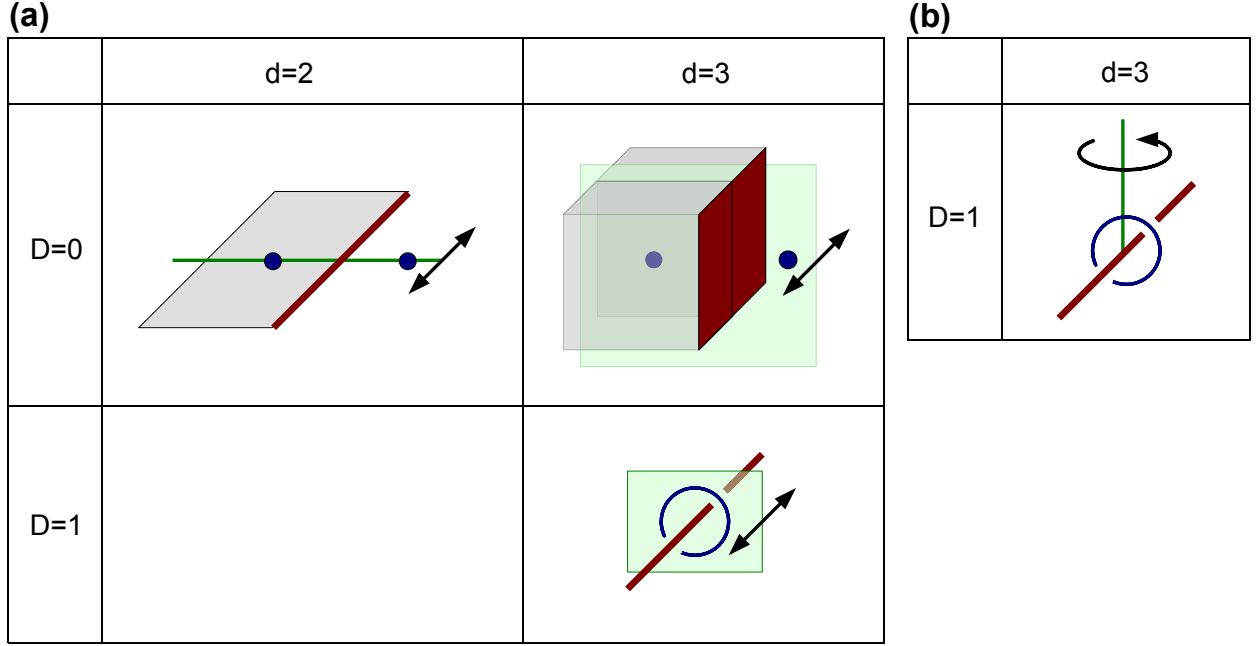}
 \end{center}
 \caption{Topological defects protected by order-two additional symmetries with
 $\delta_{\parallel}=d_{\parallel}-D_{\parallel}=1$. The additional
 symmetries are (a) reflection symmetry and (b) $\pi$-rotation symmetry,
 respectively.  
 The spatial position of topological defects is transformed as
 reflection under the symmetry transformation of $\delta_{\parallel}=1$ family.}
 \label{reflection_sym}
\end{figure}

In this subsection, we consider additional symmetries with
$\delta_{\parallel}= 1$ (mod 4).
In condensed matter contexts, relevant symmetries include 
reflection symmetry ($d_{\parallel} =1, D_{\parallel} = 0$) and 
$\pi$-rotation symmetry with one flipping defect surrounding parameter ($d_{\parallel} =2, D_{\parallel} = 1$) as shown in Fig. \ref{reflection_sym}. 
A common nature of the $\delta_{\parallel}=1$ family is that the additional
symmetries act on defect submanifolds as reflection. 
We summarize the classification table for $\delta_{\parallel}=1$ (mod 4)
with additional unitary symmetry in Table \ref{TabUR} and that with
antiunitary symmetry in Table \ref{TabAR}, respectively. 
A complete classification of the bulk topological phase with reflection
symmetry was given by Chiu et al. \cite{chiu2013classification}, and
Morimoto-Furusaki \cite{morimoto2013topological}. 
New results are the classification of topological defects, and that with
antiunitary symmetry.
In the following subsections, we illustrate some examples. 
\begin{table*}[!]
\begin{center}
\caption{
Classification table for topological crystalline insulators and
 superconductors and their topological defects in the presence of
 order-two additional unitary symmetry with flipped parameters
 $\delta_{\parallel}=d_{\parallel} - D_{\parallel} = 1$ (mod 4).  
Here $\delta=d-D$.
}
\begin{tabular}[t]{ccccccccccc}
\hline \hline
Symmetry & Class & $\mathcal{C}_q$ or $\mathcal{R}_q$ & $\delta=0$ & $\delta=1$ & $\delta=2$ & $\delta=3$ & $\delta=4$ & $\delta=5$ & $\delta=6$ & $\delta=7$ \\
\hline
$U$  & A      &$\mathcal{C}_1$& $0$ & $\mathbb{Z}$ & $0$ & $\mathbb{Z}$ & $0$ & $\mathbb{Z}$ & $0$ & $\mathbb{Z}$ \\
$U_+$& AIII   &$\mathcal{C}_0$& $\mathbb{Z}$ & $0$ & $\mathbb{Z}$ & $0$ & $\mathbb{Z}$ & $0$ & $\mathbb{Z}$ & $0$ \\
$U_-$& AIII   &$\mathcal{C}_1 \times \mathcal{C}_1$& $0$ & $\mathbb{Z} \oplus \mathbb{Z}$ & $0$ & $\mathbb{Z} \oplus \mathbb{Z}$ & $0$ & $\mathbb{Z} \oplus \mathbb{Z}$ & $0$ & $\mathbb{Z} \oplus \mathbb{Z}$ \\
\hline 
\multirow{8}{*}{$U^+_+,U^-_-,U^+_{++},U^-_{--}$}
 & AI   &$\mathcal{R}_1$& $\mathbb{Z}_2$ & $\mathbb{Z}$ & $0$ & $0$ & $0$ & $2\mathbb{Z}$ & $0$ & $\mathbb{Z}_2$ \\
 & BDI  &$\mathcal{R}_2$& $\mathbb{Z}_2$ & $\mathbb{Z}_2$ & $\mathbb{Z}$ & $0$ & $0$ & $0$ & $2\mathbb{Z}$ & $0$ \\
 & D    &$\mathcal{R}_3$& $0$ & $\mathbb{Z}_2$ & $\mathbb{Z}_2$ & $\mathbb{Z}$ & $0$ & $0$ & $0$ & $2\mathbb{Z}$ \\
 & DIII &$\mathcal{R}_4$& $2\mathbb{Z}$ & $0$ & $\mathbb{Z}_2$ & $\mathbb{Z}_2$ & $\mathbb{Z}$ & $0$ & $0$ & $0$ \\
 & AII  &$\mathcal{R}_5$& $0$ & $2\mathbb{Z}$ & $0$ & $\mathbb{Z}_2$ & $\mathbb{Z}_2$ & $\mathbb{Z}$ & $0$ & $0$  \\
 & CII  &$\mathcal{R}_6$& $0$ & $0$ & $2\mathbb{Z}$ & $0$ & $\mathbb{Z}_2$ & $\mathbb{Z}_2$ & $\mathbb{Z}$ & $0$ \\
 & C    &$\mathcal{R}_7$& $0$ & $0$  & $0$ & $2\mathbb{Z}$ & $0$ & $\mathbb{Z}_2$ & $\mathbb{Z}_2$ & $\mathbb{Z}$ \\
 & CI   &$\mathcal{R}_0$& $\mathbb{Z}$ & $0$ & $0$  & $0$ & $2\mathbb{Z}$ & $0$ & $\mathbb{Z}_2$ & $\mathbb{Z}_2$ \\
\hline
$U^+_{+-},U^-_{-+}$ & BDI  &$\mathcal{R}_1 \times \mathcal{R}_1$& $\mathbb{Z}_2\oplus\mathbb{Z}_2$ & $\mathbb{Z}\oplus\mathbb{Z}$ & $0$ & $0$ & $0$ & $2\mathbb{Z}\oplus 2\mathbb{Z}$ & $0$ & $\mathbb{Z}_2\oplus\mathbb{Z}_2$ \\
$U^+_{-+},U^-_{+-}$ & DIII &$\mathcal{R}_3 \times \mathcal{R}_3$& $0$ & $\mathbb{Z}_2\oplus\mathbb{Z}_2$ & $\mathbb{Z}_2\oplus\mathbb{Z}_2$ & $\mathbb{Z}\oplus\mathbb{Z}$ & $0$ & $0$ & $0$ & $2\mathbb{Z}\oplus 2\mathbb{Z}$ \\
$U^+_{+-},U^-_{-+}$ & CII  &$\mathcal{R}_5 \times \mathcal{R}_5$& $0$ & $2\mathbb{Z}\oplus 2\mathbb{Z}$ & $0$ & $\mathbb{Z}_2\oplus\mathbb{Z}_2$ & $\mathbb{Z}_2\oplus\mathbb{Z}_2$ & $\mathbb{Z}\oplus\mathbb{Z}$ & $0$ & $0$ \\
$U^+_{-+},U^-_{+-}$ & CI   &$\mathcal{R}_7 \times \mathcal{R}_7$& $0$ & $0$ & $0$ & $2\mathbb{Z}\oplus 2\mathbb{Z}$ & $0$ & $\mathbb{Z}_2\oplus\mathbb{Z}_2$ & $\mathbb{Z}_2\oplus\mathbb{Z}_2$ & $\mathbb{Z}\oplus\mathbb{Z}$ \\
\hline
\multirow{8}{*}{$U^+_-,U^-_+,U^+_{--},U^-_{++}$}
 & AI   &$\mathcal{R}_7$& $0$ & $0$  & $0$ & $2\mathbb{Z}$ & $0$ & $\mathbb{Z}_2$ & $\mathbb{Z}_2$ & $\mathbb{Z}$ \\
 & BDI  &$\mathcal{R}_0$& $\mathbb{Z}$ & $0$ & $0$  & $0$ & $2\mathbb{Z}$ & $0$ & $\mathbb{Z}_2$ & $\mathbb{Z}_2$ \\
 & D    &$\mathcal{R}_1$& $\mathbb{Z}_2$ & $\mathbb{Z}$ & $0$ & $0$ & $0$ & $2\mathbb{Z}$ & $0$ & $\mathbb{Z}_2$ \\
 & DIII &$\mathcal{R}_2$& $\mathbb{Z}_2$ & $\mathbb{Z}_2$ & $\mathbb{Z}$ & $0$ & $0$ & $0$ & $2\mathbb{Z}$ & $0$ \\
 & AII  &$\mathcal{R}_3$& $0$ & $\mathbb{Z}_2$ & $\mathbb{Z}_2$ & $\mathbb{Z}$ & $0$ & $0$ & $0$ & $2\mathbb{Z}$ \\
 & CII  &$\mathcal{R}_4$& $2\mathbb{Z}$ & $0$ & $\mathbb{Z}_2$ & $\mathbb{Z}_2$ & $\mathbb{Z}$ & $0$ & $0$ & $0$ \\
 & C    &$\mathcal{R}_5$& $0$ & $2\mathbb{Z}$ & $0$ & $\mathbb{Z}_2$ & $\mathbb{Z}_2$ & $\mathbb{Z}$ & $0$ & $0$  \\
 & CI   &$\mathcal{R}_6$& $0$ & $0$ & $2\mathbb{Z}$ & $0$ & $\mathbb{Z}_2$ & $\mathbb{Z}_2$ & $\mathbb{Z}$ & $0$ \\
\hline
$U^+_{-+},U^-_{+-}$ & BDI, CII  &$\mathcal{C}_1$& $0$ & $\mathbb{Z}$ & $0$ & $\mathbb{Z}$ & $0$ & $\mathbb{Z}$ & $0$ & $\mathbb{Z}$ \\
$U^+_{+-},U^-_{-+}$ & DIII, CI  &$\mathcal{C}_1$& $0$ & $\mathbb{Z}$ & $0$ & $\mathbb{Z}$ & $0$ & $\mathbb{Z}$ & $0$ & $\mathbb{Z}$ \\
\hline \hline
\end{tabular}
\label{TabUR}
\end{center}
\end{table*}

\subsubsection{Topological number $\mathbb{Z}\oplus\mathbb{Z}$}
First, we give a concrete expression of the topological number
$\mathbb{Z}\oplus\mathbb{Z}$ in Table \ref{TabUR}.
This number is denoted by ``$\mathbb{Z}^1$'' in the classification
table by Chiu, et. al. \cite{chiu2013classification} 
The topological number consists of two topological invariants. 
For odd (even) spatial dimensions $d$, 
one is the winding number $N_{2n+1}$ (the Chern character), and the other 
is the mirror Chern number (the mirror winding number).
While the former topological invariant can be defined without the
additional symmetry, the latter cannot.

For example, we consider class AIII system with a $U_-$ additional
symmetry in three dimensions. 
The Hamiltonian has the following symmetry: 
\begin{eqnarray}
\Gamma \mathcal{H}(k_x,k_y,k_z) \Gamma^{-1} = - \mathcal{H}(k_x,k_y,k_z),  
\end{eqnarray}
\begin{eqnarray}
U \mathcal{H}(k_x,k_y,k_z) U^{-1} = \mathcal{H}(-k_x,k_y,k_z), 
\quad \{U,\Gamma\} = 0.
\end{eqnarray}
The winding number is defined as
\begin{eqnarray}
N_3 
= \frac{1}{48 \pi^2} \int_{S^3} \mathrm{tr} \Gamma \left[ \mathcal{H}^{-1} d \mathcal{H} \right]^3. 
\end{eqnarray}
Note that reflection symmetry $U_-$ does not eliminate the
winding number because $U \Gamma U^{-1} = - \Gamma$ and  $U \left[
\mathcal{H}^{-1} d \mathcal{H} \right]^3 U^{-1} = - \left[
\mathcal{H}^{-1} d \mathcal{H} \right]^3$.  

In addition to $N_3$, we can introduce the first Chern number on the
mirror invariant plane with $k_x=0$: 
On the mirror invariant plane, the Hamiltonian $\mathcal{H}(0,k_y,k_z)$
can be block diagonal in the basis of eigenstates of $U=\pm$ since it
commutes with $U$, i.e. $[U,
\mathcal{H}(0,k_y,k_z)]=0$. 
Then the first Chern number is defined as 
\begin{eqnarray}
Ch_1^{\pm} 
= \frac{i}{2 \pi} \int_{S^2} \mathrm{tr} \mathcal{F}^{\pm}, 
\end{eqnarray}
where $\mathcal{F}^{\pm}$ is the Berry curvature of the Hamiltonian
$\mathcal{H}(0,k_y,k_z)$ in the $U= \pm$ sector. 
Here note that the two Chern numbers $Ch_1^{+}$ and $Ch_1^{-}$ are not
independent. 
In fact, the total first Chern number should be trivial in the sense of
strong topological index in three
dimensions,
$Ch_1^{+}+Ch_1^{-} = 0$ \cite{avron1983homotopy}.
Hence the meaningful topological invariant is only the difference
between $Ch_1^{+}$ and $Ch_2^{-}$,
\begin{eqnarray}
N_{M\mathbb{Z}} = \frac{Ch_1^{+}-Ch_1^{-}}{2}.  
\end{eqnarray}
Consequently, the K-group is characterized by $N_3$ and $N_{M\mathbb{Z}}$: 
\begin{eqnarray}
( N_3, N_{M\mathbb{Z}} ) \in \mathbb{Z}\oplus\mathbb{Z}. 
\end{eqnarray}

\subsubsection{Second descendant $\mathbb{Z}_2$ index in real AZ classes
   with $U^+_{-}$,
   $U^-_+$, $U^+_{--}$, $U^-_{++}$}

Consider a system in $s$ real AZ class with $U^+_{-}$, $U^-_+$,
$U^+_{--}$,
$U^-_{++}$ ($t=2$ of Table \ref{Symmetry_type}) in $d$-dimensions.
The system is time-reversal invariant 
\begin{eqnarray}
T \mathcal{H}(\bm{k}) T^{-1} = \mathcal{H}(-\bm{k}),
\quad T^2 = \epsilon_T,  
\end{eqnarray}
and/or particle-hole symmetric
\begin{eqnarray}
C \mathcal{H}(\bm{k}) C^{-1} = -\mathcal{H}(-\bm{k}), 
\quad C^2 = \epsilon_C,
\end{eqnarray}
with $\epsilon_T=\pm$ and $\epsilon_C=\pm$.
It is also invariant under the additional reflection symmetry
(because $\delta_{\parallel}=1$) 
\begin{eqnarray}
U \mathcal{H}(k_x, \bm{k}_{\perp }) U^{-1} 
= \mathcal{H}(-k_x, \bm{k}_{\perp }). 
\end{eqnarray}
When $d=s-3$ (mod.8), the topological index of this system is given by
the second descendant $\mathbb{Z}_2$, as is seen in Table \ref{TabUR}.
Now we would like to discuss how to define this $\mathbb{Z}_2$ number.

As was discussed in Ref. \onlinecite{chiu2013classification}, the second descendant $\mathbb{Z}_2$
cannot be defined on the reflection symmetric plane with $k_x=0$:
On the reflection invariant plane, 
the Hamiltonian is decomposed into two eigensectors of $U=\pm$, 
$\mathcal{H}(0, \bm{k}_{\perp}) =
\mathcal{H}_{U=+}(\bm{k}_{\perp}) \oplus
\mathcal{H}_{U=-}(\bm{k}_{\perp})$, since $\mathcal{H}(0,{\bm
k}_{\perp})$ commutes with $U$.
However, because $T$ and/or $C$ interchange two eigensectors,
$\mathcal{H}_{U=+}({\bm k}_{\perp})$ and $\mathcal{H}_{U=-}({\bm k}_{\perp})$,  
each sector is neither time-reversal symmetric or particle-hole
symmetric. 
Therefore, they belong to a complex AZ class, and thus no
$\mathbb{Z}_2$ number can be defined.
Furthermore, the original $s$ real AZ class 
is topologically trivial in $d=s-3$ (mod. 8) dimensions.  
From these observations, Ref. \onlinecite{chiu2013classification} had
concluded that the $\mathbb{Z}_2$ index cannot be properly defined. 

To the contrary, however, we find that 
the $\mathbb{Z}_2$
topological invariant can be defined by generalizing the Moore-Balents argument
\cite{moore2007topological} if $d\geq 2$. 
For this purpose, it is convenient to reparameterize the base momentum
space $(k_x, {\bm k}_{\perp})\in S^{d}$ as
$(k_x, \theta_{\perp}, \bm{k}'_{\perp})$ 
where $\theta_{\perp} \in [0,\pi]$ denotes the polar angle 
of $S^d$ that is invariant under the reflection $U$.
In this parameterization, the Hamiltonian obeys
\begin{eqnarray}
\label{Eq::Symmetry_B2_TRS}
T \mathcal{H}(k_x, \theta_{\perp}, \bm{k}'_{\perp}) T^{-1} 
= \mathcal{H}(-k_x, \pi-\theta_{\perp}, -\bm{k}'_{\perp}),
\quad T^2 = \epsilon_T, 
\end{eqnarray}
and/or
\begin{eqnarray}
\label{Eq::Symmetry_B2_PHS}
C \mathcal{H}(k_x, \theta_{\perp}, \bm{k}'_{\perp}) C^{-1}
=-\mathcal{H}(-k_x, \pi-\theta_{\perp}, -\bm{k}'_{\perp}),
\quad C^2 = \epsilon_C, 
\end{eqnarray}
and 
\begin{eqnarray}
\label{Eq::Symmetry_B2_Reflection}
U \mathcal{H}(k_x, \theta_{\perp}, \bm{k}'_{\perp}) U^{-1} 
= \mathcal{H}(-k_x, \theta_{\perp}, \bm{k}'_{\perp}).
\end{eqnarray}
The Hamiltonian
$\mathcal{H}(k_x, \pi/2, \bm{k}'_{\perp})$ at the equator
$\theta_{\perp}=\pi/2$ belongs to the same $s$ real AZ class with the
same $U$ $(t=2)$ but in $d-1$ dimensions, so its K-group is 
\begin{equation}\begin{split}
K^U_{\mathbb{R}}(s,t=2; d-1,d_{\parallel}=1,0,0) = \pi_{0}(\mathcal{R}_3)=0. 
\end{split}\end{equation}
This means that the equator $\theta_{\perp}=\pi/2$ can be smoothly
deformed into a reference Hamiltonian $\mathcal{H}_0$ 
{\it with keeping} the $(s, t=2)$ symmetries in $d-1$ dimensions. 
We denote this deformation as $\mathcal{H}_1(k_x, \bm{k}'_{\perp},
\theta)$, $\theta \in [\pi/2,\pi]$, with
$\mathcal{H}_1(k_x, \bm{k}'_{\perp}, \pi/2)=
\mathcal{H}(k_x,\pi/2, \bm{k}'_{\perp})$ 
and 
$\mathcal{H}_1(k_x,\bm{k}'_{\perp}, \pi)=\mathcal{H}_0$. 
Combining with the north hemisphere of the original Hamiltonian, 
we obtain an Hamiltonian
$\tilde{\mathcal{H}}_1(k_x, \bm{k}'_{\perp}, \theta)$ on $S^d$ as  
\begin{equation}\begin{split}
\tilde{\mathcal{H}}_1(k_x, \bm{k}'_{\perp}, \theta) 
= \left\{ \begin{array}{ll}
\mathcal{H}(k_x, \theta_{\perp}=\theta, \bm{k}'_{\perp}) & (0\leq \theta\leq \pi/2),  \\
\mathcal{H}_1(k_x, \bm{k}'_{\perp}, \theta) & (\pi/2\leq \theta\leq \pi). \\
\end{array} \right. 
\end{split}\end{equation}
The combined Hamiltonian $\tilde{\mathcal{H}}_1(k_x, \bm{k}'_{\perp},
\theta)$ breaks TRS and PHS, but it keeps the reflection symmetry 
\begin{equation}\begin{split}
U \tilde{\mathcal{H}}_1(k_x,\bm{k}'_{\perp}, \theta) U^{-1} 
= \tilde{\mathcal{H}}_1(-k_x, \bm{k}'_{\perp}, \theta).
\end{split}\end{equation}
Also it has CS 
\begin{eqnarray}
(TC) \tilde{\mathcal{H}}_1(k_x, \bm{k}'_{\perp}, \theta) (TC)^{-1} 
= -\tilde{\mathcal{H}}_1(k_x, \bm{k}'_{\perp}, \theta) 
\end{eqnarray}
when $s$ is odd. So it belongs to a complex $s$ AZ class with $U$
($t=0$ $\mbox{mod.2}$), which K-group is
\begin{equation}\begin{split}
K^U_{\mathbb{C}}
(s,t=0; d,1,0, 0) = \pi_0(\mathcal{C}_{s+d+1}) =\mathbb{Z},
\end{split}\end{equation}
for $d=s-3$ (mod.8).
Hence, $\tilde{\mathcal{H}}_1(k_x, \bm{k}'_{\perp}, \theta)$ defines
an integer topological invariant $N$. 

In general, this integer $N$ depends on how we extend
$\tilde{\mathcal{H}}_1(k_x, \bm{k}'_{\perp}, \theta)$.
If we choose another extension 
$\tilde{\mathcal{H}}_2(\theta,k_x,\tilde{\bm{k}})$, 
the resulting integer $N'$ differs from $N$. 
We can show, however, the difference between $N$ and $N'$ is even,
and thus its parity $(-1)^N$ is defined uniquely:   
The difference $N-N'$ is evaluated by calculating the same topological
number for the following Hamiltonian
$\mathcal{H}_{12}(k_x,\bm{k}'_{\perp}, \theta)$, 
\begin{equation}\begin{split}
\tilde{\mathcal{H}}_{12}(k_x,\bm{k}'_{\perp}, \theta) 
= \left\{ \begin{array}{ll}
\mathcal{H}_2(k_x, {\bm k}'_{\perp}, \pi-\theta) & (0\leq \theta\leq \pi/2),  \\
\mathcal{H}_1(k_x, {\bm k}'_{\perp}, \theta) & (\pi/2\leq \theta\leq \pi). \\
\end{array} \right. 
\end{split}\label{Eq::Def_H12}\end{equation}
Then, since $\mathcal{H}_1$ and $\mathcal{H}_2$ keep the original $(s, t=2)$
symmetries in $d-1$ dimensions with a coordinate parameter $\theta$, 
the obtained topological number of $\mathcal{H}_{12}(k_x,\bm{k}'_{\perp},
\theta)$ is restricted by the K-group
\begin{equation}\begin{split}
K^U_{\mathbb{R}}(s,t=2,d-1,1,1,0) = \pi_{0}(\mathcal{R}_4)= 2 \mathbb{Z}, 
\end{split}\end{equation}
which implies that $N-N'$ must be even. 
Therefore, the parity of $N$, i.e. $(-1)^{N}$ provides a well-defined
$\mathbb{Z}_2$ topological invariant.



To confirm the validity of the above definition, 
we calculate the $\mathbb{Z}_2$ number of a two-dimensional model. 
In two dimensions, the relevant real AZ class is CII ($s=5$) and the
model Hamiltonian reads 
\begin{equation}\begin{split}
\mathcal{H}(k_x,k_y) = k_x s_x \sigma_x \tau_x + k_y s_z \sigma_x \tau_0+
		 (1-k^2) s_z \sigma_y \tau_z,
\end{split}\label{Eq::B2_model}\end{equation}
with $T = i s_y \mathcal{K}$, $C = i s_y \sigma_z
\mathcal{K}$ and $U=s_z$. 
The equator $\theta_{\perp}=\pi/2$ and the north
(south) hemisphere in the above correspond to the $k_y=0$ line and the
upper (down) plane with $k_y>0$ ($k_y<0$), respectively. 
On the equator $k_y=0$, the Hamiltonian, $\mathcal{H}(k_x,0) 
= k_y s_z \sigma_x\tau_x + (1-k^2) s_z \sigma_y \tau_z$, 
has an extra symmetry preserving mass term $M =s_0 \sigma_x \tau_z$,
which enables us to deform the Hamiltonian on the south hemisphere as 
\begin{equation}\begin{split}
\mathcal{H}_1(k_x,k_y) = k_x s_x \sigma_x \tau_x 
+ k_y s_0\sigma_x \tau_z + (1-k^2) s_z \sigma_y \tau_z, \ \ (k_y<0). 
\end{split}\end{equation}
Then, $\tilde{\mathcal{H}}_1(k_x,k_y)$ is 
\begin{equation}\begin{split}
\tilde{\mathcal{H}}_1(k_x,k_y) 
= 
\left\{ \begin{array}{ll}
k_x s_x \sigma_x \tau_x + k_y s_z \sigma_x \tau_0+ (1-k^2) s_z \sigma_y \tau_z & (k_y>0), \\
k_x s_x \sigma_x \tau_x + k_y s_0\sigma_x \tau_z + (1-k^2) s_z \sigma_y \tau_z & (k_y<0), \\
\end{array} \right.
\end{split}\end{equation}
which has CS with $\Gamma=TC=\sigma_z$ as well as
the reflection symmetry with $U=s_z$.
The $\mathbb{Z}$ topological invariant of
$\tilde{\mathcal{H}}_1(k_x,k_y)$ is obtained as the
mirror winding number: 
On the mirror symmetric line $k_x=0$, $\tilde{\mathcal{H}}_1(0,k_y)$ is
decomposed into two mirror eigensector with $U=\pm$, i.e. 
$\tilde{\mathcal{H}}_1(0,k_y) = \tilde{\mathcal{H}}^+_1(0,k_y) \oplus
\tilde{\mathcal{H}}^-_1(0,k_y)$. 
Due to $[\Gamma, U]=0$, the decomposed Hamiltonians also have CS.
Then the mirror winding number is defined by $N^M_1 = (N^+_1-N^-_1)/2$ with 
$N_1^{\pm} = 1/(4 \pi i) \int \mathrm{tr} \left[  \Gamma
(\tilde{\mathcal{H}}^{\pm}_1(0,k_y))^{-1} d_{k_y}
\tilde{\mathcal{H}}^{\pm}_1(0,k_y) \right]$, which is found to be $1$.   
%
We can also find that 
if we take another deformation $\mathcal{H}_2(k_x,k_y)$ as 
\begin{equation}\begin{split}
\mathcal{H}_2(k_x,k_y) = k_x s_x \sigma_x \tau_x - k_y s_0\sigma_x \tau_z + (1-k^2) s_z \sigma_y \tau_z, \ \ (k_y<0), 
\end{split}\end{equation}
then the corresponding mirror winding number is $-1$.
Therefore, the parity of the mirror winding number is uniquely
determined to be odd, although the mirror winding number itself is not
determined uniquely.
From this calculation, we can conclude that the original model
(\ref{Eq::B2_model}) has a nontrivial $\mathbb{Z}_2$ topological invariant.  


Before closing this subsection, we would like to mention a subtle
instability of the present symmetry protected phase.
It has been shown that the present topological phase can be deformed
into a topologically trivial state if one admits a mass term breaking
the translation symmetry.\cite{chiu2013classification}
However, at the same time, it has 
been argued that surface gapless
states of this phase remain critical when the mass term is random
and spatially uniform on average.\cite{morimoto2013topological}
Our results here also indicate the existence of a proper topological
number, which also supports the validity of the
topological phase discussed here.

\begin{table*}[!]
\begin{center}
\caption{
Classification table for topological crystalline insulators and
 superconductors and their topological defects in the presence of
 order-two additional antiunitary symmetry with flipped parameters
 $\delta_{\parallel}=d_{\parallel} - D_{\parallel} = 1$ (mod 4). Here
 $\delta=d-D$.
}
\begin{tabular}[t]{ccccccccccc}
\hline \hline
Symmetry & Class & $\mathcal{C}_q$ or $\mathcal{R}_q$ & $\delta=0$ & $\delta=1$ & $\delta=2$ & $\delta=3$ & $\delta=4$ & $\delta=5$ & $\delta=6$ & $\delta=7$ \\
\hline
$A^+$  & A      &$\mathcal{R}_2$& $\mathbb{Z}_2$ & $\mathbb{Z}_2$ & $\mathbb{Z}$ & $0$ & $0$ & $0$ & $2\mathbb{Z}$ & $0$ \\
$A^-$  & A      &$\mathcal{R}_6$& $0$ & $0$ & $2\mathbb{Z}$ & $0$ & $\mathbb{Z}_2$ & $\mathbb{Z}_2$ & $\mathbb{Z}$ & $0$ \\
$A^+_+$& AIII &$\mathcal{R}_3$& $0$ & $\mathbb{Z}_2$ & $\mathbb{Z}_2$ & $\mathbb{Z}$ & $0$ & $0$ & $0$ & $2\mathbb{Z}$ \\
$A^-_-$& AIII &$\mathcal{R}_5$& $0$ & $2\mathbb{Z}$ & $0$ & $\mathbb{Z}_2$ & $\mathbb{Z}_2$ & $\mathbb{Z}$ & $0$ & $0$  \\
$A^-_+$& AIII &$\mathcal{R}_7$& $0$ & $0$  & $0$ & $2\mathbb{Z}$ & $0$ & $\mathbb{Z}_2$ & $\mathbb{Z}_2$ & $\mathbb{Z}$ \\
$A^+_-$& AIII &$\mathcal{R}_1$& $\mathbb{Z}_2$ & $\mathbb{Z}$ & $0$ & $0$ & $0$ & $2\mathbb{Z}$ & $0$ & $\mathbb{Z}_2$ \\
\hline 
$A^+_+,A^+_-$ & D  &$\mathcal{R}_2 \times \mathcal{R}_2$& $\mathbb{Z}_2\oplus\mathbb{Z}_2$ & $\mathbb{Z}_2\oplus\mathbb{Z}_2$ & $\mathbb{Z}\oplus\mathbb{Z}$ & $0$ & $0$ & $0$ & $2\mathbb{Z}\oplus 2\mathbb{Z}$ & $0$ \\
$A^-_+,A^-_-$ & C  &$\mathcal{R}_6 \times \mathcal{R}_6$& $0$ & $0$ & $2\mathbb{Z}\oplus 2\mathbb{Z}$ & $0$ & $\mathbb{Z}_2\oplus\mathbb{Z}_2$ & $\mathbb{Z}_2\oplus\mathbb{Z}_2$ & $\mathbb{Z}\oplus\mathbb{Z}$ & $0$ \\
\hline
$A^-_+,A^-_-$ & D  &$\mathcal{C}_0$& $\mathbb{Z}$ & $0$ & $\mathbb{Z}$ & $0$ & $\mathbb{Z}$ & $0$ & $\mathbb{Z}$ & $0$ \\
$A^+_+,A^+_-$ & C  &$\mathcal{C}_0$& $\mathbb{Z}$ & $0$ & $\mathbb{Z}$ & $0$ & $\mathbb{Z}$ & $0$ & $\mathbb{Z}$ & $0$ \\
\hline \hline
\end{tabular}
\label{TabAR}
\end{center}
\end{table*}

\subsubsection{Mirror reflection symmetric vortex in three-dimensional
   superconductors ($U^-_-$ in class D)}

Mirror reflection symmetry may protect Majorana gapless modes propagating
a vortex in three dimensions. 
Consider a superconductor in three dimensions,
\begin{eqnarray}
\mathcal{H}_{\rm BdG}({\bm k})=
\begin{pmatrix}
\epsilon({\bm k})& \Delta({\bm k})\\
\Delta^{\dagger}({\bm k}) & -\epsilon^T(-{\bm k})
\end{pmatrix}. 
\end{eqnarray}
As was mentioned in
Sec. \ref{sec:mirrorodd2D}, 
mirror reflection symmetry with respect to the $xy$-plane implies that
the normal part is invariant under the mirror reflection
\begin{eqnarray}
M_{xy}\epsilon(k_x,k_y,k_z)M_{xy}^{\dagger}=\epsilon(k_x, k_y, -k_z),
\quad M_{xy}=is_z 
\end{eqnarray}
but the gap function can be either mirror even or mirror odd
\begin{eqnarray}
M_{xy}\Delta(k_x, k_y, k_z)M^T_{xy}=\pm \Delta(k_x,k_y, -k_z). 
\end{eqnarray}
When the gap function is mirror even (mirror odd), $\mathcal{H}_{\rm
BdG}({\bm k})$ obeys 
\begin{eqnarray}
\tilde{M}_{xy}\mathcal{H}_{\rm BdG}(k_x, k_y, k_z)\tilde{M}_{xy}^{\dagger}= 
\mathcal{H}_{\rm BdG}(k_x, k_y, -k_z),
\end{eqnarray}
with $\tilde{M}_{xy}={\rm diag}(M_{xy}, M^*_{xy})=is_z\tau_z$ 
($\tilde{M}_{xy}={\rm diag}(M_{xy}, -M^*_{xy})=is_z\tau_0$).

A straight vortex extended in the $z$-direction does not break
the mirror reflection symmetry.
For the adiabatic BdG Hamiltonian with the vortex, the mirror symmetry is
expressed as
\begin{eqnarray}
\tilde{M}_{xy}\mathcal{H}_{\rm BdG}(k_x, k_y, k_z, \phi)
\tilde{M}_{xy}^{\dagger}= 
\mathcal{H}_{\rm BdG}(k_x, k_y, -k_z, \phi),
\end{eqnarray}
where $\phi$ is the angle around the vortex.
For mirror even gap superconductors, 
$\tilde{M}_{xy}=is_z\tau_z$ is labeled as $U^{-}_+$ in class D, 
while for mirror odd superconductors, $\tilde{M}_{xy}=is_z\tau_0$ is labeled
as $U^{-}_-$ in class D. 
Since $\delta=2$ and $\delta_{\parallel}=1$, the topological index of
the vortex is 0 for mirror even gap functions and $\mathbb{Z}_2$ for
mirror odd gap functions. See Table \ref{TabUR}.
The $\mathbb{Z}_2$ index in the mirror odd case is given in the
following manner.
On the mirror symmetric subspace with $k_z=0$, 
the BdG Hamiltonian commutes with $\tilde{M}_{xy}$, and thus it is
decomposed into two mirror eigensectors with $M_{xy} =\pm i$,  
\begin{equation}\begin{split}
\mathcal{H}_{\mathrm{BdG}}(k_x,k_y,0,\phi) 
= \begin{pmatrix}
\mathcal{H}^{i}_{\mathrm{BdG}}(k_x,k_y,0,\phi) & \\
 & \mathcal{H}^{-i}_{\mathrm{BdG}}(k_x,k_y,0,\phi)
\end{pmatrix}. 
\end{split}\end{equation}
Each mirror subsector
is mapped to itself by the particle-hole transformation due to the
anticommutation relation $\{C,
\tilde{M}_{xy}\}=0$ in the mirror odd case.
Therefore it supports its own PHS, 
which enables us to define the mirror $\mathbb{Z}_2$ numbers by
\begin{equation}\begin{split}
\nu_{\pm i} 
= 
2\cdot \frac{1}{2} \left( \frac{i}{2 \pi} \right)^2 
\int Q^{\pm i}_3 \ \ (\mathrm{mod}\ 2). 
\end{split}\end{equation}
Here $Q^{\pm}_3$ is the Chern-Simons 3-form, $Q_3 = \mathrm{tr} \left[
\mathcal{A} d \mathcal{A} + \frac{2}{3} \mathcal{A}^3 \right]$, of the
$M_{xy}=\pm i$ sector, and the integral is performed on the
three-dimensional sphere of $(k_x, k_y, \phi)$.
We can also show that 
the sum of the  mirror $\mathbb{Z}_2$ numbers is trivial,
i.e. $\nu_{i}+\nu_{-i}=0$ (mod 2):
First of all, the sum of the mirror $\mathbb{Z}_2$ numbers coincides with the
integral of the Chern-Simon 3-form of {\it the total Hamiltonian}, which
can be defined 
on any three dimensional sphere of $(k_x, k_y, \phi)$ even with a nonzero $k_z$.
Moreover, the latter integral is also quantized to be 0 or 1 (mod.2) and
is independent of $k_z$
because of the combined symmetry of PHS and the mirror reflection symmetry. 
Its value, however, should be zero, since the Hamiltonian is smoothly
connected into a 
topologically 
trivial one by taking $k_z\rightarrow \infty$. 
As a result, the sum of the mirror $\mathbb{Z}_2$ numbers is also zero.
This means that we have only a single independent $\mathbb{Z}_2$ number.


We can also show that 
when the $\mathbb{Z}_2$ number is nontrivial, there are a pair of Majorana
gapless modes propagating the vortex.
For instance, consider a vortex (o-vortex) in
$^3$He-B phase.\cite{salomaa1987quantized, silaev2009spectrum}
The adiabatic Hamiltonian describing the o-vortex is 
\begin{equation}\begin{split}
\mathcal{H}_{\mathrm{BdG}}(k_x,k_y,k_z,\phi) 
&= 
\begin{pmatrix}
\frac{k^2}{2m}-\mu & \frac{\Delta e^{i \phi}}{k_F} \bm{k} \cdot
 \bm{s} i s_y \\ 
-i s_y \frac{\Delta e^{-i \phi}}{k_F} \bm{k} \cdot \bm{s} &
 -\frac{k^2}{2m}+\mu  
\end{pmatrix}, 
\end{split}\end{equation}
which reduces to
\begin{equation}\begin{split}
\mathcal{H}^{\pm i}_{\mathrm{BdG}}(k_x,k_y,0,\phi) 
&= 
\begin{pmatrix}
\frac{k^2}{2m}-\mu & \frac{\Delta e^{i \phi}}{k_F} (\mp k_x + i k_y ) \\
\frac{\Delta e^{-i \phi}}{k_F} (\mp k_x - i k_y ) & -\frac{k^2}{2m}+\mu 
\end{pmatrix}. 
\end{split}\end{equation}
when $k_z=0$.
Since each mirror subsector is
nothing but a spinless chiral $p$-wave superfluid with a vortex, 
it supports a zero mode, which gives a pair of propagating modes totally. 
The topological invariant is $\nu_{i} = 1$ (mod 2). 
We also find that our $\mathbb{Z}_2$ number ensures the existence of
similar vortex gapless modes \cite{tsutsumi2013upt3, nagai2012} in an odd-parity
superconducting states of UPt$_3$\cite{stewart1984, machida2012, tsutsumi2012} and Cu$_x$Bi$_2$Se$_3$.\cite{hor2010,
fu2010odd, sasaki2011topological, hao2011, hsieh2012,yamakage2012}


\subsubsection{$2 \mathbb{Z}$ chiral doublet edge modes protected by the antiunitary reflection symmetry ($A^-$ in class A)}
From Table \ref{TabAR}, two-dimensional class A insulators with an
antiunitary reflection symmetry $A^-$ are topologically characterized
by an even integer $2\mathbb{Z}$, which implies that topologically
protected edge modes appear in a pair.  
This can be understood by quasi Kramers degeneracy originated from
the antiunitary reflection symmetry.

To illustrate this, consider an antiunitary reflection symmetry
\begin{eqnarray}
A\mathcal{H}(k_x, k_y)A^{-1}=\mathcal{H}(k_x, -k_y).
\end{eqnarray}
Note here that it corresponds to reflection of $x$, $x\rightarrow
-x$ since anti-unitarity changes the sign of momentum ${\bm k}$. 
An edge parallel to the $x$-direction preserves the reflection
symmetry, and thus if the first Chern number of the system is non-zero, 
there exists a chiral edge state $a_{k_x}$ described by the
effective Hamiltonian,  
\begin{eqnarray}
H = \sum_{k_x} v k_x a^{\dag}_{k_x} a_{k_x}.  
\end{eqnarray}
In a manner similar to the Kramers theorem, one can prove that the
antiunitary reflection symmetry with $A^2=-1$ results in degeneracy
of the edge state, but in contrast to TRS,
the resultant degenerate states $b_{k_x}$ have the same energy dispersion,
since the antiunitary reflection $A$ acts as
\begin{eqnarray}
a_{k_x} \rightarrow b_{k_x}, 
\quad
b_{k_x} \rightarrow -a_{k_x}. 
\end{eqnarray}
Indeed, the antiunitary invariance of $H$ leads to double chiral edge
modes with the same energy dispersion: 
\begin{eqnarray}
H = \sum_{k_x} v k_x \left( a^{\dag}_{k_x} a_{k_x} + b^{\dag}_{k_x}
		 b_{k_x} \right). 
\end{eqnarray}
Correspondingly, the first Chern number of the system should be an even integer.

\subsubsection{$\mathbb{Z}\oplus\mathbb{Z}$ superconductor protected by emergent spinless reflection TRS ($A^+_+, A^+_-$ in class D)}
Two-dimensional class D superconductors with an antiunitary reflection
symmetry with $A^2=1$ are characterized by a set of topological numbers
$\mathbb{Z}\oplus\mathbb{Z}$. 
(See $A^+_-$, $A^+_-$ in class D with $\delta=2$ of Table \ref{TabAR}.)
The PHS and the antiunitary symmetry are expressed as
\begin{eqnarray}
&&C \mathcal{H}(-k_x, -k_y) C^{-1} = - \mathcal{H}(k_x,k_y), 
\nonumber\\
&&A \mathcal{H}(k_x, -k_y) A^{-1} = \mathcal{H}(k_x,k_y), \quad A^2 = 1,  
\end{eqnarray}
where the reflection in the $x$-direction has been assumed.
One of the topological numbers is the 1st Chern character $Ch_1$, which
can be nonzero even in the presence of the antiunitary reflection.
The other is the winding number $N_1$ defined on the
high-symmetric line $k_x = 0$, where the Hamiltonian
$\mathcal{H}(0,k_y)$ effectively supports the class BDI symmetry if one
identifies $A$ with TRS.
The K-group $\mathbb{Z} \oplus \mathbb{Z}$ is spanned by the basis $e_1
= (Ch_1 = 1, N_1= 1)$ and $e_2=(Ch_1 = -1, N_1= 1)$
where the representative Hamiltonians $\mathcal{H}^{(Ch_1, N_1)}$ are given by  
\begin{eqnarray}
\mathcal{H}^{(\pm 1, 1)}(k_x,k_y) = \pm k_x \tau_y + k_y \tau_x 
+ [m-\epsilon (k_x^2+k_y^2)]\tau_z, 
\end{eqnarray}
with $C = \tau_x \mathcal{K}$, $A = \tau_z \mathcal{K}$, and $m, \epsilon>0$. 

Combining the representative Hamiltonians in the above, 
\begin{eqnarray}
\left(
\begin{array}{cc}
\mathcal{H}^{(1,1)}(k_x,k_y) & 0\\
0 & \mathcal{H}^{(-1,1)}(k_x,k_y) 
\end{array}
\right),
\end{eqnarray}
one obtains the system with $(Ch_1 = 0, N_1 = 2)$. This system hosts a
helical gapless Majorana state protected by the reflection symmetry $A$.

\subsubsection{Vortex in three-dimensional superconductors with magnetic
$\pi$-rotation symmetry ($A^+_+, A^+_-$ in class D)}

Consider a three-dimensional time-reversal invariant superconductor (or
superfluid) with
an additional $\pi$-rotation symmetry.
If one creates a vortex in this system, it breaks TRS, but if the vortex is
straight and perpendicular to the rotation axis of the $\pi$-rotation,
as illustrated in Fig.\ref{reflection_sym} (b),    
the system can be invariant under the combination of time-reversal and
the $\pi$-rotation. 
%
Supposing a vortex extended in the $z$-direction and the magnetic $\pi$-rotation around the $x$-axis,  
the magnetic $\pi$-rotation symmetry $A$ is expressed as 
\begin{equation}\begin{split}
A \mathcal{H}_{\mathrm{BdG}}(k_x,k_y,k_z,\phi) A^{-1} = \mathcal{H}_{\mathrm{BdG}}(-k_x,k_y,k_z,-\phi), \ \ A = \tau_z s_z \mathcal{K}, 
\end{split}\end{equation}
where $\mathcal{H}_{\mathrm{BdG}}(k_x, k_y, k_z, \phi)$ is the BdG
Hamiltonian with a vortex, and $\phi$ is the angle around the vortex
measured from the $x$-axis.
Since  $A$ anticommutes with $C = \tau_x \mathcal{K}$, 
it is labeled as $A^+_-$ with $\delta=2$ and $\delta_{\parallel}=1$ ($d=3$, $D=1$, $d_{\parallel}=2$ and
$D_{\parallel}=1$) in class D.
From Table \ref{TabAR}, the topological index is $\mathbb{Z}\oplus\mathbb{Z}$. 
One of the $\mathbb{Z}$ indices is the second Chern number 
\begin{equation}\begin{split}
Ch_2 = \frac{1}{2} \left( \frac{i}{2 \pi} \right)^2 \int \mathrm{tr} \mathcal{F}^2
\end{split}\end{equation}
where $\mathcal{F} = \mathcal{F}(\bm{k},\phi)$ is the Berry curvature of
$\mathcal{H}_{\rm
BdG}(k_x, k_y, k_z, \phi)$, and the trace is taken for all negative
energy states.
The other $\mathbb{Z}$ index is 
defined on the $k_z=0$ plane.
On the $k_z=0$ plane, the magnetic $\pi$-rotation coincides with the
magnetic in-plane reflection, and thus the BdG Hamiltonian is topologically the
same as that in Sec. \ref{sec:vortex2Dmagneticmirror}.
Consequently, the BdG Hamiltonian is chiral symmetric at $\phi=0, \pi$,
and zero modes with $k_z=0$ localized at the vortex are
characterized by
\begin{equation}\begin{split}
N_1^{\mathrm{strong}} = \frac{N_1^{0}-N_1^{\pi}}{2}, 
\end{split}\end{equation}
where $N_1^{0/\pi}$ is given by 
\begin{equation}\begin{split}
N^{0/\pi}_1 = \frac{1}{4 \pi i} \int \mathrm{tr} \Gamma \mathcal{H}_{\mathrm{BdG}}^{-1}(k_x,0,0,0/\pi) d \mathcal{H}_{\mathrm{BdG}}(k_x,0,0,0/\pi), 
\end{split}\end{equation}
with the chiral operator $\Gamma =s_z \tau_y$.  


When these $\mathbb{Z}$ indices are nonzero, the bulk-boundary
correspondence implies that there exist one-dimensional
gapless Majorana modes propagating the vortex.  
These gapless modes propagate upward or downward, and we call the
former mode as right mover and the latter as left mover. 
%
%
Also, thanks to the CS above, 
each gapless state has a definite chirality of $\Gamma$ at $k_z=0$.
Hence, 
%
a gapless state localized at the vortex has 
two characters $(\alpha, \Gamma)$, where $\alpha(={\rm R}, {\rm L})$
denotes the direction of the movement and $\Gamma$ denotes the
chirality of $\Gamma$ at $k_z=0$.
If we express the number of vortex gapless states with
$(\alpha,\Gamma)$ by $N(\alpha,\Gamma)$,   
then, $Ch_2$ and $N_1^{\mathrm{strong}}$ are related to $N(\alpha,
\Gamma)$ as
\begin{eqnarray}
&&Ch_2 = N({\rm R},+)+N({\rm R},-)-N({\rm L},+)-N({\rm L},-), 
\nonumber\\
&&N_1^{\mathrm{strong}} =
 N({\rm R},+)-N({\rm R},-)+N({\rm L},+)-N({\rm L},-).  
\end{eqnarray}
Such a magnetic $\pi$-rotation symmetric vortex can be realized in
$^3$He-B phase\cite{salomaa1987quantized} or
Cu$_x$Bi$_2$Se$_3$.\cite{hor2010, fu2010odd,
sasaki2011topological,hao2011, hsieh2012, yamakage2012}

\subsection{$\delta_{\parallel}=2$ family}
\begin{figure}[!]
 \begin{center}
  \includegraphics[width=0.5\linewidth, trim=0cm 0cm 0cm 0cm]{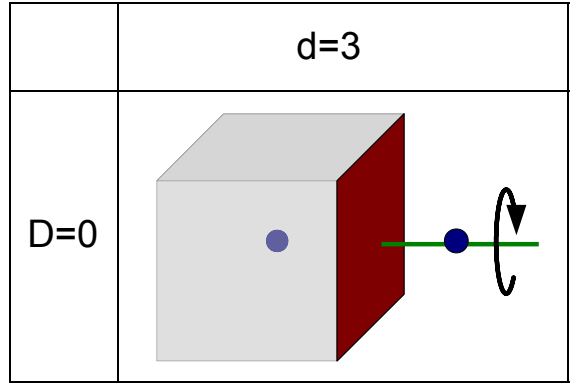}
 \end{center}
 \caption{Topological defects protected by order-two additional symmetry with
 $\delta_{\parallel}=d_{\parallel}-D_{\parallel}=2$. The additional
 symmetry is (a) $\pi$-rotation symmetry. 
 The spatial position of topological defects is transformed as
 $\pi$-rotation under the symmetry transformation of
 $\delta_{\parallel}=2$ family.}
 \label{pi_rotation_sym}
\end{figure}

In this subsection, we discuss topological phases protected by
additional symmetries with $\delta_{\parallel}= 2$ (mod 4).
Relevant systems are $\pi$-rotation symmetric
insulators and their surface defects ($d_{\parallel} =2$,
$D_{\parallel} = 0$) illustrated in Fig. \ref{pi_rotation_sym}.
We summarize the classification table
for
$d_{\parallel}=2$ (mod 4) 
with additional unitary symmetry in Table \ref{TabUP}  
and that with additional antiunitary symmetry in
Table \ref{TabAP}, respectively.
\begin{table*}[!]
\begin{center}
\caption{
Classification table for topological crystalline insulators and
 superconductors and their topological defects in the presence of 
 order-two additional unitary symmetry with flipped parameters
 $\delta_{\parallel}=d_{\parallel} - D_{\parallel} = 2$ (mod 4). 
Here $\delta=d-D$.
}
\begin{tabular}[t]{ccccccccccc}
\hline \hline
Symmetry & Class & $\mathcal{C}_q$ or $\mathcal{R}_q$ & $\delta=0$ & $\delta=1$ & $\delta=2$ & $\delta=3$ & $\delta=4$ & $\delta=5$ & $\delta=6$ & $\delta=7$ \\
\hline
$U$  & A      &$\mathcal{C}_0 \times \mathcal{C}_0$& $\mathbb{Z} \oplus \mathbb{Z}$ & $0$ & $\mathbb{Z} \oplus \mathbb{Z}$ & $0$ & $\mathbb{Z} \oplus \mathbb{Z}$ & $0$ & $\mathbb{Z} \oplus \mathbb{Z}$ & $0$ \\
$U_+$& AIII   &$\mathcal{C}_1 \times \mathcal{C}_1$& $0$ & $\mathbb{Z} \oplus \mathbb{Z}$ & $0$ & $\mathbb{Z} \oplus \mathbb{Z}$ & $0$ & $\mathbb{Z} \oplus \mathbb{Z}$ & $0$ & $\mathbb{Z} \oplus \mathbb{Z}$ \\
$U_-$& AIII   &$\mathcal{C}_0$& $\mathbb{Z}$ & $0$ & $\mathbb{Z}$ & $0$ & $\mathbb{Z}$ & $0$ & $\mathbb{Z}$ & $0$ \\
\hline
$U^+_+,U^-_-$       & AI, D, AII, C      &$\mathcal{C}_0$& $\mathbb{Z}$ & $0$  & $\mathbb{Z}$ & $0$ & $\mathbb{Z}$ & $0$ & $\mathbb{Z}$ & $0$ \\
$U^+_{++},U^-_{--}$ & BDI, DIII, CII, CI &$\mathcal{C}_1$& $0$ & $\mathbb{Z}$ & $0$  & $\mathbb{Z}$ & $0$ & $\mathbb{Z}$ & $0$ & $\mathbb{Z}$ \\
\hline
$U^+_{+-},U^-_{-+}$ & BDI  &$\mathcal{R}_2$& $\mathbb{Z}_2$ & $\mathbb{Z}_2$ & $\mathbb{Z}$ & $0$ & $0$ & $0$ & $2\mathbb{Z}$ & $0$ \\
$U^+_{-+},U^-_{+-}$ & DIII &$\mathcal{R}_4$& $2\mathbb{Z}$ & $0$ & $\mathbb{Z}_2$ & $\mathbb{Z}_2$ & $\mathbb{Z}$ & $0$ & $0$ & $0$ \\
$U^+_{+-},U^-_{-+}$ & CII  &$\mathcal{R}_6$& $0$ & $0$ & $2\mathbb{Z}$ & $0$ & $\mathbb{Z}_2$ & $\mathbb{Z}_2$ & $\mathbb{Z}$ & $0$ \\
$U^+_{-+},U^-_{+-}$ & CI   &$\mathcal{R}_0$& $\mathbb{Z}$ & $0$ & $0$ & $0$ & $2\mathbb{Z}$ & $0$ & $\mathbb{Z}_2$ & $\mathbb{Z}_2$ \\
\hline 
\multirow{8}{*}{$U^+_-,U^-_+,U^+_{--},U^-_{++}$}
 & AI   &$\mathcal{R}_0 \times \mathcal{R}_0$& $\mathbb{Z}\oplus\mathbb{Z}$ & $0$ & $0$ & $0$ & $2\mathbb{Z}\oplus 2\mathbb{Z}$ & $0$ & $\mathbb{Z}_2\oplus\mathbb{Z}_2$ & $\mathbb{Z}_2\oplus\mathbb{Z}_2$ \\
 & BDI  &$\mathcal{R}_1 \times \mathcal{R}_1$& $\mathbb{Z}_2\oplus\mathbb{Z}_2$ & $\mathbb{Z}\oplus\mathbb{Z}$ & $0$ & $0$ & $0$ & $2\mathbb{Z}\oplus 2\mathbb{Z}$ & $0$ & $\mathbb{Z}_2\oplus\mathbb{Z}_2$ \\
 & D    &$\mathcal{R}_2 \times \mathcal{R}_2$& $\mathbb{Z}_2\oplus\mathbb{Z}_2$ & $\mathbb{Z}_2\oplus\mathbb{Z}_2$ & $\mathbb{Z}\oplus\mathbb{Z}$ & $0$ & $0$ & $0$ & $2\mathbb{Z}\oplus 2\mathbb{Z}$ & $0$ \\
 & DIII &$\mathcal{R}_3 \times \mathcal{R}_3$& $0$ & $\mathbb{Z}_2\oplus\mathbb{Z}_2$ & $\mathbb{Z}_2\oplus\mathbb{Z}_2$ & $\mathbb{Z}\oplus\mathbb{Z}$ & $0$ & $0$ & $0$ & $2\mathbb{Z}\oplus 2\mathbb{Z}$ \\
 & AII  &$\mathcal{R}_4 \times \mathcal{R}_4$& $2\mathbb{Z}\oplus 2\mathbb{Z}$ & $0$ & $\mathbb{Z}_2\oplus\mathbb{Z}_2$ & $\mathbb{Z}_2\oplus\mathbb{Z}_2$ & $\mathbb{Z}\oplus\mathbb{Z}$ & $0$ & $0$ & $0$ \\
 & CII  &$\mathcal{R}_5 \times \mathcal{R}_5$& $0$ & $2\mathbb{Z}\oplus 2\mathbb{Z}$ & $0$ & $\mathbb{Z}_2\oplus\mathbb{Z}_2$ & $\mathbb{Z}_2\oplus\mathbb{Z}_2$ & $\mathbb{Z}\oplus\mathbb{Z}$ & $0$ & $0$ \\
 & C    &$\mathcal{R}_6 \times \mathcal{R}_6$& $0$ & $0$ & $2\mathbb{Z}\oplus 2\mathbb{Z}$ & $0$ & $\mathbb{Z}_2\oplus\mathbb{Z}_2$ & $\mathbb{Z}_2\oplus\mathbb{Z}_2$ & $\mathbb{Z}\oplus\mathbb{Z}$ & $0$ \\
 & CI   &$\mathcal{R}_7 \times \mathcal{R}_7$& $0$ & $0$ & $0$ & $2\mathbb{Z}\oplus 2\mathbb{Z}$ & $0$ & $\mathbb{Z}_2\oplus\mathbb{Z}_2$ & $\mathbb{Z}_2\oplus\mathbb{Z}_2$ & $\mathbb{Z}\oplus\mathbb{Z}$ \\
\hline
$U^+_{-+},U^-_{+-}$ & BDI  &$\mathcal{R}_0$& $\mathbb{Z}$ & $0$ & $0$ & $0$ & $2\mathbb{Z}$ & $0$ & $\mathbb{Z}_2$ & $\mathbb{Z}_2$ \\
$U^+_{+-},U^-_{-+}$ & DIII &$\mathcal{R}_2$& $\mathbb{Z}_2$ & $\mathbb{Z}_2$ & $\mathbb{Z}$ & $0$ & $0$ & $0$ & $2\mathbb{Z}$ & $0$ \\
$U^+_{-+},U^-_{+-}$ & CII  &$\mathcal{R}_4$& $2\mathbb{Z}$ & $0$ & $\mathbb{Z}_2$ & $\mathbb{Z}_2$ & $\mathbb{Z}$ & $0$ & $0$ & $0$ \\
$U^+_{+-},U^-_{-+}$ & CI   &$\mathcal{R}_6$& $0$ & $0$ & $2\mathbb{Z}$ & $0$ & $\mathbb{Z}_2$ & $\mathbb{Z}_2$ & $\mathbb{Z}$ & $0$ \\
\hline \hline
\end{tabular}
\label{TabUP}
\end{center}
\end{table*}

\subsubsection{$\pi$-rotation Chern number and $\pi$-rotation winding number}
In a manner similar to the mirror Chern number and the mirror winding number,  
we can define the {\it $\pi$-rotation Chern number} and the {\it $\pi$-rotation
winding number} in the presence of two-fold ($\pi$) rotation symmetry.

To define these topological numbers, we first introduce $\pi$-rotation
subsectors.
The presence of $\pi$-rotation symmetry implies
\begin{equation}\begin{split}
U \mathcal{H}(k_x,k_y,\bm{k}_{\perp }) U^{-1} 
= \mathcal{H}(-k_x,-k_y,\bm{k}_{\perp }). 
\end{split}\end{equation}
On the symmetric subspace $k_x=k_y=0$ of $\pi$-rotation, 
the Hamiltonian is decomposed into two $\pi$-rotation subsectors which
are eigenstates of $U$,  
\begin{eqnarray}
\mathcal{H}(0,0,\bm{k}_{\perp }) = \mathcal{H}_+(0,0,\bm{k}_{\perp })
		 \oplus \mathcal{H}_-(0,0,\bm{k}_{\perp }),
\end{eqnarray}
since the Hamiltonian commutes with $U$ on the $\pi$-rotation invariant
subspace.

In even $2n$-dimensions, we can define the $\pi$-rotation Chern number
by
\begin{equation}\begin{split}
Ch^{\Pi}_{n-1} := \frac{Ch^+_{n-1}-Ch^-_{n-1}}{2}, 
\end{split}\end{equation}
where $Ch^{\pm}_{n-1}$ is the $(n-1)$-th Chern number of
$\mathcal{H}_{\pm}(0,0,{\bm k}_{\perp})$. 
Since the original Chern number is identically zero in the presence of TRS or CS
in $(4p-2)$-dimensions, or in the presence of PHS in $4p$-dimensions, 
the meaningful $\pi$-rotation Chern number can be obtained only when
$\mathcal{H}_{\pm}(0,0,{\bm k}_{\perp})$ does not have such symmetries.   
For example, consider a $\pi$-rotation symmetric class DIII
system in four dimensions.  
There are four types of $\pi$-rotation with $U^2=1$: $U^+_{++}$,
$U^+_{--}$, $U^+_{+-}$, $U^+_{-+}$.  
In the former two cases, $U^+_{++}$, $U^+_{--}$, the $\pi$-rotation Chern
number is identically zero because the $\pi$-rotation subsectors support
CS in two dimensions, i.e. $[\Gamma, \mathcal{H}_{\pm}(0,0,{\bm
k}_{\perp})]=0$ with $\Gamma=CT$.
$U^+_{+-}$ also forbids a non-zero $\pi$-rotation Chern number
because TRS in two dimensions remains in the $\pi$-rotation subsectors,
because of $[U,T]=0$. 
A nonzero $\pi$-rotation Chern number is possible only in
the last case $U^+_{-+}$ since no AZ symmetry remains
in $\pi$-rotation subsectors.

On the other hand, the $\pi$-rotation winding number can be defined
in odd ($2n+1$)-dimensions.
In order to define the $\pi$-rotation winding number, we need CS that
commutes with $\pi$-rotation, $[U, \Gamma]=0$ .
In this case, CS remains even in the $\pi$-rotation subsectors.
Then, the winding number $N^{\pm}_{2n+1}$ for each $\pi$-rotation
subsector is given by
\begin{equation}\begin{split}
N^{\pm}_{2n-1} = \frac{n!}{(2 \pi i)^n(2n)!} \int \mathrm{tr} \Gamma \left[ \mathcal{H}_{\pm}^{-1} d \mathcal{H}_{\pm} \right]^{2n-1}, 
\end{split}\end{equation}
with $\mathcal{H}_{\pm} = \mathcal{H}_{\pm}(0,0,\bm{k}_{\perp })$. 
The $\pi$-rotation winding number $N^{\Pi}_{2n-1}$ is defined as the
difference between $N^+_{2n-1}$ and $N^-_{2n-1}$: 
\begin{equation}\begin{split}
N^{\Pi}_{2n-1} := \frac{N^+_{2n-1}-N^-_{2n-1}}{2}. 
\end{split}\end{equation}

\begin{table*}[!]
\begin{center}
\caption{
Classification table for topological crystalline insulators and
 superconductors and their topological defects in the presence of 
order-two additional antiunitary symmetry with flipped parameters
 $\delta_{\parallel}=d_{\parallel} - D_{\parallel} = 2$ (mod 4). 
Here $\delta=d-D$.
}
\begin{tabular}[t]{ccccccccccc}
\hline \hline
Symmetry & Class & $\mathcal{C}_q$ or $\mathcal{R}_q$ & $\delta=0$ & $\delta=1$ & $\delta=2$ & $\delta=3$ & $\delta=4$ & $\delta=5$ & $\delta=6$ & $\delta=7$ \\
\hline
$A^+$  & A      &$\mathcal{R}_4$& $2\mathbb{Z}$ & $0$ & $\mathbb{Z}_2$ & $\mathbb{Z}_2$ & $\mathbb{Z}$ & $0$ & $0$ & $0$ \\
$A^-$  & A      &$\mathcal{R}_0$& $\mathbb{Z}$ & $0$ & $0$ & $0$ & $2\mathbb{Z}$ & $0$ & $\mathbb{Z}_2$ & $\mathbb{Z}_2$ \\
$A^+_+$& AIII &$\mathcal{R}_5$& $0$ & $2\mathbb{Z}$ & $0$ & $\mathbb{Z}_2$ & $\mathbb{Z}_2$ & $\mathbb{Z}$ & $0$ & $0$  \\
$A^-_-$& AIII &$\mathcal{R}_7$& $0$ & $0$ & $0$ & $2\mathbb{Z}$ & $0$ & $\mathbb{Z}_2$ & $\mathbb{Z}_2$ & $\mathbb{Z}$ \\
$A^-_+$& AIII &$\mathcal{R}_1$& $\mathbb{Z}_2$ & $\mathbb{Z}$ & $0$ & $0$ & $0$ & $2\mathbb{Z}$ & $0$ & $\mathbb{Z}_2$ \\
$A^+_-$& AIII &$\mathcal{R}_3$& $0$ & $\mathbb{Z}_2$ & $\mathbb{Z}_2$ & $\mathbb{Z}$ & $0$ & $0$ & $0$ & $2\mathbb{Z}$ \\
\hline 
$A^+_+,A^+_-$ & D &$\mathcal{R}_3$& $0$ & $\mathbb{Z}_2$ & $\mathbb{Z}_2$ & $\mathbb{Z}$ & $0$ & $0$ & $0$ & $2\mathbb{Z}$ \\
$A^-_+,A^-_-$ & C &$\mathcal{R}_7$& $0$ & $0$ & $0$ & $2\mathbb{Z}$ & $0$ & $\mathbb{Z}_2$ & $\mathbb{Z}_2$ & $\mathbb{Z}$ \\
\hline
$A^-_+,A^-_-$ & D &$\mathcal{R}_1$& $\mathbb{Z}_2$ & $\mathbb{Z}$ & $0$ & $0$ & $0$ & $2\mathbb{Z}$ & $0$ & $\mathbb{Z}_2$ \\
$A^+_+,A^+_-$ & C &$\mathcal{R}_5$& $0$ & $2\mathbb{Z}$ & $0$ & $\mathbb{Z}_2$ & $\mathbb{Z}_2$ & $\mathbb{Z}$ & $0$ & $0$  \\
\hline \hline
\end{tabular}
\label{TabAP}
\end{center}
\end{table*}

\subsubsection{$\mathbb{Z}_2$ topological insulator protected by the magnetic $\pi$-rotation symmetry ($A^+$ in class A)}
Here we demonstrate a topologically nontrivial phase which is protected by
the combined symmetry of time-reversal and a $\pi$-rotation.  
The combined antiunitary symmetry we consider is $A = -iU T = s_x
\mathcal{K}$ where $U = is_z$ is the $\pi$-rotation around the
$z$-axis and $T = i s_y \mathcal{K}$ is time-reversal.
In three dimensions, the antiunitary symmetry implies
\begin{equation}\begin{split}
A \mathcal{H}(k_x, k_y, k_z)A^{-1} = \mathcal{H}(k_x, k_y, -k_z). 
\end{split}\end{equation}
The antiunitary symmetry $A$ is categorized as $A^+$ because of $A^2 =
1$, and thus the topological index is $\mathbb{Z}_2$ in  three dimensions,
as is shown in Table \ref{TabAP} of class A with $A^+$. 
The $\mathbb{Z}_2$ topological invariant is defined by the Chern-Simons 3-form, 
\begin{equation}\begin{split}
\nu = 2 \cdot \frac{1}{2} \left( \frac{i}{2 \pi} \right)^2 \int
		 \mathrm{tr} \left( \mathcal{A} d \mathcal{A} +
		 \frac{2}{3} \mathcal{A}^3 \right) \ \ (\mbox{mod 2}). 
\end{split}\end{equation}


The model Hamiltonian of this topological phase is given by
\begin{eqnarray}
{\cal H}(k_x, k_y, k_z)= s_x \sigma_0 (k_x-h)+ s_y \sigma_0 k_y+s_z\sigma_z
 k_z+s_z\sigma_y m(k),
\quad
m(k)=m_0-((k_x-h)^2+k_y^2+k_z^2),
\end{eqnarray}
where we have introduced orbital degrees of freedom $\sigma_\mu$, and
the antiunitary operator $A$ acts on the orbital space trivially as
$A=s_x\sigma_0\mathcal{K}$. 
The sign of $m_0$ provides the $\mathbb{Z}_2$ number of the above model:
When $m_0$ is positive (negative), the system is topologically
non-trivial (trivial). 
Indeed, the non-trivial $\mathbb{Z}_2$ number implies the existence of
a gapless
Dirac fermion on a surface parallel to the $z$-axis, which preserves the
$\pi$-rotation symmetry above. 
The wave function of the surface Dirac fermion localized at $z=0$ is solved as
\begin{eqnarray}
{\bm \varphi}(z)=(e^{\kappa_+ z}-e^{\kappa_- z})
\left(
\begin{array}{c}
1 \\
-1
\end{array}
\right)_{\sigma}
\otimes u_s(k_x,k_y) 
\end{eqnarray}
with the boundary condition ${\bm \varphi}(0)={\bm
\varphi}(-\infty)=0$, 
where $\kappa_{\pm}=1/2\pm \sqrt{-m_0+(k_x-h)^2+k_y^2+1/4}$ and
$(1,1)_{\sigma}^{T}$
is the spinor in the orbital space. 
The spinor $u_{s=\pm}$ in the spin space satisfies 
\begin{eqnarray}
{\cal H}^{\rm D}(k_x, k_y)u_{\pm}=\pm \sqrt{(k_x-h)^2+k_y^2}u_{\pm} 
\label{eq:u_pm}
\end{eqnarray}
where ${\cal H}^{\rm D}(k_x, k_y)$ is the low energy effective
Hamiltonian of the Dirac fermion.
\begin{eqnarray}
{\cal H}^{\rm D}(k_x,k_y)=s_x (k_x-h)+s_y k_y, 
\end{eqnarray}
Here note that $m_0$ needs to be positive in order to satisfy the
boundary condition ${\bm \varphi}(0)={\bm \varphi(-\infty)}=0$.
Otherwise, $\kappa_-$ becomes negative even for small $k_x-h$ and $k_y$, 
and thus ${\bm \varphi(z)}$ never converges when $z\rightarrow -\infty$.   

The $\mathbb{Z}_2$ character of this phase is also evident in the
surface state.
From the additional symmetry of the low energy effective surface Hamiltonian
\begin{eqnarray}
A{\cal H}^{\rm D}(k_x,k_y)A^{-1}={\cal H}^{\rm D}(k_x, k_y), 
\quad A=s_x{\cal K},
\end{eqnarray}
the Berry phase $\gamma(C)$, which is defined as a line integral along
the path $C$ enclosing a degenerate point of a surface state  
\begin{eqnarray}
\gamma(C)=i\oint_C \mathrm{tr}{\cal A},
\end{eqnarray}
is quantized as $e^{i\gamma(C)}=\pm 1$. 
The Berry phase defines a $\mathbb{Z}_2$ number of the surface state.
Using the Berry curvature ${\cal A}$ of the Dirac fermion, 
\begin{eqnarray}
{\cal A}=u_-^{\dagger}du_-, 
\end{eqnarray}
with $u_-$ in Eq.(\ref{eq:u_pm}), 
we find that the
surface Dirac fermion supports a non-trivial $\mathbb{Z}_2$ number,
i.e. $e^{i\gamma(C)}=-1$, 
and thus it cannot be gapped into a topologically trivial state as far
as the additional symmetry is preserved.

\subsubsection{$^3$He-B slab with parallel magnetic fields ($A_+^+$ in
   class D)}

In superfluid $^3$He-B, the gap function 
$\hat \Delta = i(\Delta/k_{\rm F}) \bm{k} \cdot \bm{s} s_{y}$ 
preserves the $SO_{L+S}(3)$ rotation symmetry as well as TRS.\cite{vollhardt1990}
The presence of a surface partially breaks the
$SO_{L+S}(3)$ rotation symmetry, but it still preserves the
$SO_{L+S}(2)$ rotation normal to the surface, say the spin-orbit rotation 
around $z$-axis.

If we apply magnetic field parallel to the surface (say in the
$y$-direction), both TRS and the $SO_{L+S}(2)$ symmetry are broken.
However, the magnetic $\pi$-rotation symmetry, which operator acts as
combination of time-reversal and the $\pi$-rotation of $SO(2)_{L+S}$,  remains.
It defines the antiunitary symmetry 
\begin{equation}\begin{split}
A \mathcal{H}_{\mathrm{BdG}}(k_x,k_y,-k_z) A^{-1} = \mathcal{H}_{\mathrm{BdG}}(k_x,k_y,k_z), 
\end{split}\end{equation}
for the BdG Hamiltonian
\begin{equation}\begin{split}
\mathcal{H}_{\mathrm{BdG}}(k_x,k_y,k_z)
&= \begin{pmatrix}
\frac{k^2}{2 m}-\mu + h_x s_x + h_y s_y & \frac{\Delta}{k_F} \bm{k} \cdot \bm{s} i s_y \\
- \frac{\Delta}{k_F} i s_y \bm{k} \cdot \bm{s} & -\frac{k^2}{2 m}+\mu -
		  h_x s_x + h_y s_y 
\end{pmatrix},
\end{split}\label{He-B_pi-rot}\end{equation}
where $A = T U(\pi)=is_x\tau_z \mathcal{K}$ with 
$T = i s_y \tau_0 \mathcal{K}$ and $U(\pi)=is_z \tau_z \in SO_{L+S}(2)$. 
Considering the sign of $A^2$ and the commutation relation between $A$ and PHS, 
we find that the additional symmetry is labeled as $A^+_+$ in the class
D of Table \ref{TabAP}.  
In three dimensions, the topological index of this system is  
$\mathbb{Z}$.

The $\mathbb{Z}$ index is the one-dimensional $\pi$-rotation
winding number.
\begin{equation}\begin{split}
N_1 = \frac{1}{4 \pi i} \int \mathrm{tr} \left[
\Gamma \mathcal{H}^{-1}_{\mathrm{BdG}}(0,0,k_z)d \mathcal{H}_{\mathrm{BdG}}(0,0,k_z)\right], 
\end{split}\end{equation}
with  $\Gamma = - A C = s_x \tau_y$. 
For the BdG Hamiltonian (\ref{He-B_pi-rot}) with small $h_y$, 
$N_1 = -2
\mathrm{sgn} (\Delta)$.
The non-trivial value of $N_1$ explains the reason why 
helical surface Majorana fermions in $^3$He-B can stay gapless 
under magnetic fields parallel to the surface:\cite{mizushima2012symmetry} 
Although the class DIII topological superconductivity of
$^3$He-B is lost by magnetic fields breaking TRS,
the additional magnetic $\pi$-rotation symmetry gives an extra
topological superconductivity to $^3$He-B.





\subsubsection{Inversion symmetric quantum (spin) Hall states  ($U$ in
   class A, $U^+_+$ in class AII)}
\label{Inversion_QSH}


Here we consider inversion symmetric quantum Hall states which satisfy   
\begin{eqnarray}
P \mathcal{H}(k_x,k_y) P^{-1} = \mathcal{H}(-k_x,-k_y),
\quad P^2 = 1.
\end{eqnarray}
Since the inversion $P$ is labeled as $U$ in class A of Table \ref{TabUP},
its topological index is $\mathbb{Z}\oplus\mathbb{Z}$.
One of the $\mathbb{Z}$ indices is the first Chern number
$Ch_1=i/(2\pi)\int {\rm tr}\mathcal{F}$, which is directly related to
the Hall conductance $\sigma_{xy}$ of the system as
$\sigma_{xy}=(e^2/h)Ch_1$.\cite{thouless1982quantized, kohmoto1985topological}
The other $\mathbb{Z}$ index is defined at symmetric points of inversion,
i.e. ${\bm k}=(0,0)\equiv\Gamma_0$ and ${\bm k}=\infty\equiv \Gamma_{\infty}$:
As the Hamiltonian $\mathcal{H}(\Gamma_i)$ at $\Gamma_i$ ($i=0,\infty$)
commutes with $P$, it can be
block-diagonal into two subsectors with different parity $P=\pm$
\begin{eqnarray}
\mathcal{H}(\Gamma_i) = \mathcal{H}_{P=+}(\Gamma_i) \oplus
\mathcal{H}_{P=-}(\Gamma_i)
\end{eqnarray}
Now let us denote $\# \Gamma_i^{\pm}$ to be the number of occupied states of
$\mathcal{H}_{P=\pm}(\Gamma_i)$.
Although a set of numbers
$\{\#\Gamma_i^{\pm}\}$
characterizes the
Hamiltonian, there are some constraints. 
First, for a full gapped system, 
the total number of occupied states is momentum-independent, 
so we have 
\begin{eqnarray}
\#\Gamma_0^+ +\#\Gamma_0^-=\#\Gamma_\infty^+ + \#\Gamma_\infty^- 
\end{eqnarray}
Furthermore, by adding $p^{\pm}$ trivial bands with $P=\pm$, 
we find that two sets of numbers, $\{\#\Gamma_i^{\pm}\}$ and
$\{\#\Gamma_i^{\pm}+p^\pm\}$, specify the same stable-equivalent Hamiltonian.
%
Consequently, the topological index, which should be unchanged under the
stable equivalence, is
given by
\begin{eqnarray}
[\Gamma_{0,\infty}]=\#\Gamma_0^+-\#\Gamma_\infty^+
=-(\#\Gamma_0^--\#\Gamma_\infty^-). 
\label{eq:Zinversion2d}
\end{eqnarray} 

It has been known that the following formula holds between $Ch_1$ and
$[\Gamma_{0,\infty}]$\cite{sato2010topological,hughes2011inversion, turner2012quantized}, 
\begin{eqnarray}
(-1)^{Ch_1}=(-1)^{[\Gamma_{0,\infty}]}.
\label{eq:parityChern}
\end{eqnarray}
We find that the K-theory simplifies the derivation of this formula:
Let us consider two
representative Hamiltonians of quantum Hall states
\begin{equation}\begin{split}
\mathcal{H}_1 &= k_x \sigma_x + k_y \sigma_x + (1-k_x^2 -k_y^2) \sigma_z, \\
\mathcal{H}_2 &= k_x \sigma_x -k_y \sigma_x + (1-k_x^2 -k_y^2) \sigma_z, \\
\end{split}\end{equation}
with $P=\sigma_z$, which
topological indices are 
$
(Ch_1, [\Gamma_{0,\infty}])|_{\mathcal{H}_1} = (1,-1)
$ and 
$
(Ch_1, [\Gamma_{0,\infty}])|_{\mathcal{H}_2} = (-1,-1),
$
respectively.
Then, because any inversion
symmetric quantum Hall state ${\cal H}$ is stable-equivalent to a direct sum of
$\mathcal{H}_1$ and $\mathcal{H}_2$,
\begin{equation}\begin{split}
[\mathcal{H}] =l_1 [\mathcal{H}_1] \oplus l_2 [\mathcal{H}_2],
\end{split}\end{equation}
its topological numbers, $(Ch_1, [\Gamma_{0,\infty}])|_{\mathcal{H}}=
(l_1-l_2, -l_1-l_2)$, obey
Eq.(\ref{eq:parityChern}) as  
$(-1)^{Ch_1}=(-1)^{l_1-l_2}=(-1)^{-l_1-l_2}=(-1)^{[\Gamma_{0,\infty}]}$.


If we consider an inversion symmetric quantum spin Hall state, instead
of a quantum Hall state, the system also supports TRS.
In this case, $P$ is labeled as $U^+_+$ in class AII of Table \ref{TabUP}.
Now the topological index reduces to $\mathbb{Z}$, since TRS makes  
$Ch_1$ to be zero. 
In a manner similar to Eq.(\ref{eq:Zinversion2d}), the remaining
topological index is given by 
\begin{eqnarray}
[\Gamma_{0,\infty}]=\frac{\#\Gamma_0^+-\#\Gamma_\infty^+}{2}
=-\frac{\#\Gamma_0^--\#\Gamma_\infty^-}{2}. 
%
\label{eq:Zinversion2d20}
\end{eqnarray} 
Here, in comparison with Eq.(\ref{eq:Zinversion2d}), the
$\mathbb{Z}$ index in Eq.(\ref{eq:Zinversion2d20}) is divided by 2, in
order to remove the trivial factor 2 caused by the Kramers degeneracy.

Like ordinary quantum spin Hall
states, 
we can also introduce
the Kane-Mele $\mathbb{Z}_2$ invariant $(-1)^\nu$\cite{kane2005z_}, 
but it is written by $[\Gamma_{0,\infty}]$, 
\begin{eqnarray}
(-1)^{\nu}=(-1)^{[\Gamma_{0,\infty}]}.
\label{eq:parityKaneMele2D}
\end{eqnarray}
as was shown by Fu and
Kane\cite{fu2007topological}.
Again, the K-theory provides a simple derivation of this formula:
Using a representative Hamiltonian of inversion symmetric quantum
spin Hall states
\begin{eqnarray}
\mathcal{H}_1 
= k_x s_x \sigma_x + k_y s_y \sigma_x + (1-k_x^2 -k_y^2)\sigma_z, 
\quad T = i s_y \mathcal{K}, \quad P=\sigma_z,
\label{Eq::representative model of QSHE}
\end{eqnarray}
with the topological indices
$
((-1)^\nu, [\Gamma_{0,\infty}])|_{\mathcal{H}_1}=(-1,-1)
$,
the K-theory implies that any inversion symmetric quantum spin Hall
state ${\cal H}$
is stable-equivalent to a direct sum of 
$\mathcal{H}_1$, 
\begin{equation}\begin{split}
[\mathcal{H}] =l_1 [\mathcal{H}_1]. 
\end{split}\end{equation}
Therefore, the topological indices of $\mathcal{H}$ is given by $((-1)^\nu,
[\Gamma_{0,\infty}])|_{\mathcal{H}} = ((-1)^{l_1},-l_1)$, and thus
Eq.(\ref{eq:parityKaneMele2D}) holds.
This parity formula is useful to evaluate the $\mathbb{Z}_2$
invariant of real materials \cite{fu2007topological}.

\subsubsection{Odd parity superconductors in two dimensions ($U^+_-$ in
   class D) }
\label{2DoddSC}

Now consider topological properties of odd parity superconductors in two
dimensions,\cite{sato2009topological, sato2010topological} 
where the normal dispersions are inversion symmetric, 
$P \epsilon(-\bm{k}) P^{\dag} = \epsilon(\bm{k})$, 
and the pairing functions are odd under inversion, $P
\hat{\Delta}(-\bm{k}) P^{T}= - \hat{\Delta}(\bm{k})$,  
with a unitary matrix  $P$.
Combining with $U(1)$ gauge symmetry, 
the inversion symmetry of the BdG Hamiltonian 
\begin{equation}\begin{split}
\mathcal{H}_{\mathrm{BdG}}(\bm{k})
= 
\begin{pmatrix}
\epsilon(\bm{k}) & \hat{\Delta}(\bm{k}) \\
\hat{\Delta}^{\dag}(\bm{k}) & -\epsilon^{T}(-\bm{k})
\end{pmatrix}
\end{split}\end{equation}
is expressed as
\begin{eqnarray}
\tilde P \mathcal{H}_{\mathrm{BdG}}(-\bm{k}) \tilde P^{\dagger}
= \mathcal{H}_{\mathrm{BdG}}(\bm{k}),
\quad
\tilde P = {\rm diag}
(P, -P^*).
\end{eqnarray}
Because $\tilde{P}^2=1$ and $\{\tilde{P},C\}=0$, 
$\tilde{P}$ is labeled as  $U^+_-$ in class D.
From Table \ref{TabUP}, its topological index is
$\mathbb{Z}\oplus\mathbb{Z}$ in two dimensions.
Like an inversion symmetric quantum Hall state, 
one of the $\mathbb{Z}\oplus\mathbb{Z}$ index is the first Chern number 
$Ch_1 = i/(2 \pi) \int \mathrm{tr} \mathcal{F}$, and  the other is
defined at the symmetric points ${\bm k}=(0,0)\equiv \Gamma_0$ and ${\bm
k}=\infty\equiv\Gamma_\infty$ of inversion.
Denoting the number of negative energy states with parity $\pm$ at $\Gamma_i$
as $\#\Gamma_i^{\pm}$, 
the latter topological index is given by
\begin{eqnarray}
[\Gamma_{0,\infty}]
=\#\Gamma_0^+-\#\Gamma_\infty^{+} 
=-(\#\Gamma_0^--\#\Gamma_\infty^{-}). 
\end{eqnarray}
Furthermore, we can also show 
\begin{eqnarray}
(-1)^{Ch_1}=(-1)^{[\Gamma_{0,\infty}]}, 
\label{eq:parityChern2}
\end{eqnarray}
in a manner similar to Eq.(\ref{eq:parityChern}).\cite{sato2010topological}

For ordinary odd-parity superconductors, the gap functions at
$\Gamma_i$ are often identically zero or they are much smaller than
the energy scale of the normal part.
The energy hierarchy between the normal and superconducting states
simplifies the evaluation of 
$[\Gamma_{0,\infty}]$:\cite{sato2009topological, sato2010topological}
Under these situations, without closing the bulk gap, $\Delta(\Gamma_i)$ can be
neglected in $\mathcal{H}(\Gamma_i)$,
\begin{eqnarray}
\mathcal{H}(\Gamma_i)=
\left(
\begin{array}{cc}
\epsilon(\Gamma_i) & 0 \\
0 & -\epsilon^T(\Gamma_i)
\end{array}
\right), 
\end{eqnarray}
and thus the normal dispersion $\epsilon(\Gamma_i)$ determines the BdG
spectrum at $\Gamma_i$: 
By using an eigenstate $|\varphi_i \rangle$ of
$\epsilon(\Gamma_i)$, a negative energy state of ${\cal H}(\Gamma_i)$ is
given by $(|\varphi_i \rangle, 0)^t$ [$(0, |\varphi^*_i\rangle)^t$]
if the state $|\varphi_i\rangle$ is below (above) the Fermi level.
Therefore, we obtain
\begin{eqnarray}
\#\Gamma_i^{\sigma}=\#\epsilon_{-}^{\sigma}(\Gamma_i)
+\#\epsilon_+^{-\sigma}(\Gamma_i)
\end{eqnarray}
where $\#\epsilon_-^{\sigma}(\Gamma_i)$
$[\#\epsilon^{\sigma}_+(\Gamma_i)]$ denotes the number 
of $P=\sigma$ bands in the normal
state below (above) the Fermi level.
Consequently, $[\Gamma_{0,\infty}]$ is recast into
\begin{eqnarray}
[\Gamma_{0,\infty}]&=&\#\epsilon_-^+(\Gamma_0)+\#\epsilon_+^-(\Gamma_0)
-\#\epsilon_-^+(\Gamma_\infty)-\#\epsilon_+^-(\Gamma_\infty)
\nonumber\\
&=&
\#\epsilon_-^+(\Gamma_0)-\#\epsilon_-^+(\Gamma_\infty)
-[\#\epsilon_-^-(\Gamma_0)-\#\epsilon_-^-(\Gamma_\infty)]
\nonumber\\
&=&
[\epsilon_-^+(\Gamma_{0,\infty})]-[\epsilon_-^-(\Gamma_{0,\infty})]
\label{eq:FermiN}
\end{eqnarray}
where
$[\epsilon_-^{\sigma}(\Gamma_{0,\infty})]\equiv \#\epsilon_-^\sigma(\Gamma_0)-\#\epsilon_-^\sigma(\Gamma_\infty)$,
and we have used the relation $\#\epsilon_-^\sigma(\Gamma_0)
+\#\epsilon_+^\sigma(\Gamma_0)=\#\epsilon_-^\sigma(\Gamma_\infty)
+\#\epsilon_+^\sigma(\Gamma_\infty)
$.
From Eqs. (\ref{eq:parityChern2}) and (\ref{eq:FermiN}), the parity of
the first Chern number is also evaluated as \cite{sato2009topological,
sato2010topological}
\begin{eqnarray}
(-1)^{Ch_1}=(-1)^{[\epsilon_-^+(\Gamma_{0,\infty})]+[\epsilon_-^-(\Gamma_{0,\infty})]}
=(-1)^{N_{\rm F}}, 
\label{eq:parityChern3}
\end{eqnarray}
where $N_{\rm F}$ is the number of the Fermi surfaces enclosing
$\Gamma_0$.

The Fermi surface formula (\ref{eq:parityChern3}) enables us to predict
topological superconductivity of odd parity superconductors without
detailed knowledge of the gap function.
In particular, remembering that when $(-1)^{Ch_1}=-1$ a vortex hosts a
single Majorana zero mode so to obey the non-Abelian anyon
statistics\cite{read2000paired, roy2010topological},  
Eq.(\ref{eq:parityChern3}) provide a simple criterion for non-Abelian
topological order:\cite{sato2010topological} If $N_F$ is odd, then the
odd-parity superconductor is in non-Abelian topological phase.

If odd parity superconductors support TRS,
$\tilde{P}$ is labeled as $U^+_{+-}$ in class DIII of Table \ref{TabUP}, 
and thus the topological index reduces to a single $\mathbb{Z}$ in two
dimensions. 
This is because
$Ch_1$ vanishes due to TRS. 
Removing the trivial factor $2$ caused by the Kramers degeneracy, 
the remaining topological index is given by 
\begin{eqnarray}
[\Gamma_{0,\infty}]
=\frac{\#\Gamma_0^+-\#\Gamma_\infty^{+}}{2} 
=-\frac{\#\Gamma_0^--\#\Gamma_\infty^{-}}{2}. 
\end{eqnarray} 
We can also introduce the Kane-Mele $\mathbb{Z}_2$ invariant
$(-1)^{\nu}$ as in the case of quantum spin Hall states. In a manner similar to
Eq.(\ref{eq:parityChern3}), for weak coupling odd-parity Cooper pairs, 
it is evaluated by the number $N_{\rm F}$ of the Fermi
surfaces enclosing $\Gamma_0$,\cite{sato2010topological} 
\begin{eqnarray}
(-1)^{\nu}=(-1)^{N_{\rm F}/2},
\end{eqnarray}
where the Kramers degeneracy in $N_{\rm F}$ is taken into account. 
This formula is also useful to clarify the topological
superconductivity of time-reversal invariant odd parity superconductors.

\subsection{$\delta_{\parallel}=3$ family}
Here we consider additional symmetries with $\delta_{\parallel}= 3$ (mod 4). 
In condensed matter systems, relevant symmetry is inversion. 
We summarize the classification table for $\delta_{\parallel}=3$ with
order-two unitary symmetry in Table \ref{TabUI} and 
that with order-two antiunitary symmetry in Table \ref{TabAI}, respectively. 
\begin{table*}[!]
\begin{center}
\caption{
Classification table for topological crystalline insulators and
 superconductors and their topological defects in the presence of
 order-two additional unitary symmetry with flipped parameters
 $\delta_{\parallel}=d_{\parallel} - D_{\parallel} = 3$ (mod 4). 
Here $\delta=d-D$.
}
\begin{tabular}[t]{ccccccccccc}
\hline \hline
Symmetry & Class & $\mathcal{C}_q$ or $\mathcal{R}_q$ & $\delta=0$ & $\delta=1$ & $\delta=2$ & $\delta=3$ & $\delta=4$ & $\delta=5$ & $\delta=6$ & $\delta=7$ \\
\hline
$U$  & A      &$\mathcal{C}_1$& $0$ & $\mathbb{Z}$ 
& $0$ & $\mathbb{Z}$ & $0$ & $\mathbb{Z}$ & $0$ & $\mathbb{Z}$ \\
$U_+$& AIII   &$\mathcal{C}_0$& $\mathbb{Z}$ & $0$ 
& $\mathbb{Z}$ & $0$ & $\mathbb{Z}$ & $0$ & $\mathbb{Z}$ & $0$ \\
$U_-$& AIII   &$\mathcal{C}_1 \times \mathcal{C}_1$
& $0$ & $\mathbb{Z} \oplus \mathbb{Z}$ & $0$ 
& $\mathbb{Z} \oplus \mathbb{Z}$ & $0$ 
& $\mathbb{Z} \oplus \mathbb{Z}$ & $0$ & $\mathbb{Z} \oplus \mathbb{Z}$ \\
\hline 
\multirow{8}{*}{$U^+_+,U^-_-,U^+_{++},U^-_{--}$}
 & AI   &$\mathcal{R}_7$& $0$ & $0$  & $0$ & $2\mathbb{Z}$ & $0$ & $\mathbb{Z}_2$ & $\mathbb{Z}_2$ & $\mathbb{Z}$ \\
 & BDI  &$\mathcal{R}_0$& $\mathbb{Z}$ & $0$ & $0$  & $0$ & $2\mathbb{Z}$ & $0$ & $\mathbb{Z}_2$ & $\mathbb{Z}_2$ \\
 & D    &$\mathcal{R}_1$& $\mathbb{Z}_2$ & $\mathbb{Z}$ & $0$ & $0$ & $0$ & $2\mathbb{Z}$ & $0$ & $\mathbb{Z}_2$ \\
 & DIII &$\mathcal{R}_2$& $\mathbb{Z}_2$ & $\mathbb{Z}_2$ & $\mathbb{Z}$ & $0$ & $0$ & $0$ & $2\mathbb{Z}$ & $0$ \\
 & AII  &$\mathcal{R}_3$& $0$ & $\mathbb{Z}_2$ & $\mathbb{Z}_2$ & $\mathbb{Z}$ & $0$ & $0$ & $0$ & $2\mathbb{Z}$ \\
 & CII  &$\mathcal{R}_4$& $2\mathbb{Z}$ & $0$ & $\mathbb{Z}_2$ & $\mathbb{Z}_2$ & $\mathbb{Z}$ & $0$ & $0$ & $0$ \\
 & C    &$\mathcal{R}_5$& $0$ & $2\mathbb{Z}$ & $0$ & $\mathbb{Z}_2$ & $\mathbb{Z}_2$ & $\mathbb{Z}$ & $0$ & $0$  \\
 & CI   &$\mathcal{R}_6$& $0$ & $0$ & $2\mathbb{Z}$ & $0$ & $\mathbb{Z}_2$ & $\mathbb{Z}_2$ & $\mathbb{Z}$ & $0$ \\
\hline
$U^+_{+-},U^-_{-+}$ & BDI, CII  &$\mathcal{C}_1$& $0$ & $\mathbb{Z}$ & $0$ & $\mathbb{Z}$ & $0$ & $\mathbb{Z}$ & $0$ & $\mathbb{Z}$ \\
$U^+_{-+},U^-_{+-}$ & DIII, CI  &$\mathcal{C}_1$& $0$ & $\mathbb{Z}$ & $0$ & $\mathbb{Z}$ & $0$ & $\mathbb{Z}$ & $0$ & $\mathbb{Z}$ \\
\hline
\multirow{8}{*}{$U^+_-,U^-_+,U^+_{--},U^-_{++}$}
 & AI   &$\mathcal{R}_1$& $\mathbb{Z}_2$ & $\mathbb{Z}$ & $0$ & $0$ & $0$ & $2\mathbb{Z}$ & $0$ & $\mathbb{Z}_2$ \\
 & BDI  &$\mathcal{R}_2$& $\mathbb{Z}_2$ & $\mathbb{Z}_2$ & $\mathbb{Z}$ & $0$ & $0$ & $0$ & $2\mathbb{Z}$ & $0$ \\
 & D    &$\mathcal{R}_3$& $0$ & $\mathbb{Z}_2$ & $\mathbb{Z}_2$ & $\mathbb{Z}$ & $0$ & $0$ & $0$ & $2\mathbb{Z}$ \\
 & DIII &$\mathcal{R}_4$& $2\mathbb{Z}$ & $0$ & $\mathbb{Z}_2$ & $\mathbb{Z}_2$ & $\mathbb{Z}$ & $0$ & $0$ & $0$ \\
 & AII  &$\mathcal{R}_5$& $0$ & $2\mathbb{Z}$ & $0$ & $\mathbb{Z}_2$ & $\mathbb{Z}_2$ & $\mathbb{Z}$ & $0$ & $0$  \\
 & CII  &$\mathcal{R}_6$& $0$ & $0$ & $2\mathbb{Z}$ & $0$ & $\mathbb{Z}_2$ & $\mathbb{Z}_2$ & $\mathbb{Z}$ & $0$ \\
 & C    &$\mathcal{R}_7$& $0$ & $0$  & $0$ & $2\mathbb{Z}$ & $0$ & $\mathbb{Z}_2$ & $\mathbb{Z}_2$ & $\mathbb{Z}$ \\
 & CI   &$\mathcal{R}_0$& $\mathbb{Z}$ & $0$ & $0$  & $0$ & $2\mathbb{Z}$ & $0$ & $\mathbb{Z}_2$ & $\mathbb{Z}_2$ \\
\hline
$U^+_{-+},U^-_{+-}$ & BDI  &$\mathcal{R}_1 \times \mathcal{R}_1$& $\mathbb{Z}_2\oplus\mathbb{Z}_2$ & $\mathbb{Z}\oplus\mathbb{Z}$ & $0$ & $0$ & $0$ & $2\mathbb{Z}\oplus 2\mathbb{Z}$ & $0$ & $\mathbb{Z}_2\oplus\mathbb{Z}_2$ \\
$U^+_{+-},U^-_{-+}$ & DIII &$\mathcal{R}_3 \times \mathcal{R}_3$& $0$ & $\mathbb{Z}_2\oplus\mathbb{Z}_2$ & $\mathbb{Z}_2\oplus\mathbb{Z}_2$ & $\mathbb{Z}\oplus\mathbb{Z}$ & $0$ & $0$ & $0$ & $2\mathbb{Z}\oplus 2\mathbb{Z}$ \\
$U^+_{-+},U^-_{+-}$ & CII  &$\mathcal{R}_5 \times \mathcal{R}_5$& $0$ & $2\mathbb{Z}\oplus 2\mathbb{Z}$ & $0$ & $\mathbb{Z}_2\oplus\mathbb{Z}_2$ & $\mathbb{Z}_2\oplus\mathbb{Z}_2$ & $\mathbb{Z}\oplus\mathbb{Z}$ & $0$ & $0$ \\
$U^+_{+-},U^-_{-+}$ & CI   &$\mathcal{R}_7 \times \mathcal{R}_7$& $0$ & $0$ & $0$ & $2\mathbb{Z}\oplus 2\mathbb{Z}$ & $0$ & $\mathbb{Z}_2\oplus\mathbb{Z}_2$ & $\mathbb{Z}_2\oplus\mathbb{Z}_2$ & $\mathbb{Z}\oplus\mathbb{Z}$ \\
\hline \hline
\end{tabular}
\label{TabUI}
\end{center}
\end{table*}

\subsubsection{Inversion symmetric topological insulators ($U$ in
   class A, $U^+_+$ in class AII)} 
\label{sec:inversion3D}

Class A systems in three dimensions cannot host a strong
topological phase in general. 
However, the presence of inversion
symmetry admits a strong crystalline
$\mathbb{Z}$ topological index. \cite{turner2010entanglement,
hughes2011inversion, turner2012quantized}
The inversion symmetry is expressed as 
\begin{eqnarray}
P \mathcal{H}(-\bm{k}) P^{-1} = \mathcal{H}(\bm{k}) 
\end{eqnarray}
with a unitary matrix $P$. Since $P$ is labeled as $U$ in class A of Table
\ref{TabUI}, its topological index is given by $\mathbb{Z}$ in
three dimensions. 
The $\mathbb{Z}$ index is defined at
symmetric points of inversion, ${\bm k}=(0,0,0)\equiv\Gamma_0$ and ${\bm
k}=\infty\equiv\Gamma_\infty$ in a manner similar to the two-dimensional
case discussed in Sec.\ref{Inversion_QSH}.
Since the Hamiltonian commutes with $P$ at $\bm{k} = \Gamma_i$
$(i=0,\infty)$, the Hamiltonian is decomposed into two eigensectors of
$P=\pm$ as
$
\mathcal{H}(\Gamma_i) = \mathcal{H}_{P=+}(\Gamma_i) \oplus \mathcal{H}_{P=-}(\Gamma_i).  
$
Then, the $\mathbb{Z}$ index is defined by the number $\#\Gamma_i^{\pm}$ of 
occupied states with parity $P=\pm$ at $\Gamma_i$. 
Using the same argument in Sec.\ref{Inversion_QSH}, we have
$\#\Gamma_0^++\#\Gamma_0^-=\#\Gamma_\infty^++\#\Gamma_\infty^-$, and the stable equivalence implies
that the index $\mathbb{Z}$ depends only on the difference
$(\#\Gamma_0^+-\#\Gamma_0^-)-(\#\Gamma_\infty^+-\#\Gamma_\infty^-)$. 
In three dimensions, however,
there exists
an extra global constraint: 
By regarding $k_z$ in $\mathcal{H}({\bm k})$ as a parameter, one can define   
the first Chern number $Ch_1(k_z)$, but for a full gapped system in
$S^3$, it must be zero since the two dimensional system, which is obtained
by fixing  $k_z$ of $S^3$ in the momentum
space, smoothly goes to a topologically trivial state
as $k_z\rightarrow \infty$.
This means that $Ch_1(k_z=0)=0$ on the inversion symmetric two dimensional
plane at $k_z=0$.
Therefore, from Eq.(\ref{eq:parityChern}),
$\#\Gamma_0^+-\#\Gamma_\infty^+$ must be even. 
Taking into account this constraint, the $\mathbb{Z}$ index is defined as 
\begin{eqnarray}
[\Gamma_{0,\infty}] 
=\frac{\#\Gamma_0^+-\#\Gamma_\infty^+}{2}. 
\label{eq:symmetricpointI}
\end{eqnarray}

For inversion symmetric insulators, the magnetoelectric polarizability, 
\begin{eqnarray}
\label{eq:Magnetoelectric_Polarizatioin}
P_3=\frac{-1}{8\pi^2}\int \mathrm{tr} \left( \mathcal{A}d \mathcal{A} +
\frac{2}{3} \mathcal{A}^3 \right)
\quad (\mathrm{mod}. \ 1)  
\end{eqnarray}
with the Berry connection $\mathcal{A}$ of $\mathcal{H}({\bm k})$, 
also defines a topological invariant: Because $P_3$ is defined modulo
integer and $P_3\rightarrow -P_3$ under inversion, the value of $P_3$ is
quantized to be $0$, $1/2$ for inversion symmetric insulators, which
means that $(-1)^{2P_3}$ defines a $\mathbb{Z}_2$ invariant.  
This $\mathbb{Z}_2$ invariant, however, is not independent of
$[\Gamma_{0,\infty}]$.
It holds that
\begin{eqnarray}
(-1)^{2P_3}=(-1)^{[\Gamma_{0,\infty}]}. 
\label{eq:P3N}
\end{eqnarray}
Therefore, the $\mathbb{Z}$ index $[\Gamma_{0,\infty}]$ fully characterizes the
topological phase of three dimensional inversion symmetric insulators,
as indicated by Table \ref{TabUI}.

If we impose TRS on inversion symmetric
 insulators, $P$ is labeled as $U_+^+$ in class AII of Table \ref{TabUI}.
The topological index in three dimensions is $\mathbb{Z}$, which is
 defined in a manner similar to Eq.(\ref{eq:symmetricpointI}).
\begin{eqnarray}
[\Gamma_{0,\infty}] 
=\frac{\#\Gamma_0^+-\#\Gamma_\infty^+}{2}. 
\end{eqnarray}
Note here that in contrast to the two dimensional case in
Sec.\ref{Inversion_QSH}, the Kramers degeneracy does not impose an extra
constraint because of the global constraint mentioned in the above.

%

Like an ordinary topological insulator, TRS also admits to
define the three-dimensional $\mathbb{Z}_2$ invariant.
\cite{moore2007topological, fu2007topological, roy2009topological}
However, it is not independent of $[\Gamma_{0,\infty}]$ again.
Indeed, the $\mathbb{Z}_2$ invariant can be expressed in terms of
the magnetoelectric polarizability,\cite{wang2010}
\begin{eqnarray}
(-1)^\nu=(-1)^{2P_3}, 
\end{eqnarray}
and thus Eq.(\ref{eq:P3N}) leads to
\begin{eqnarray}
(-1)^{\nu}=(-1)^{[\Gamma_{0,\infty}]}. 
\end{eqnarray}
By using the relation
$\#\Gamma_0^++\#\Gamma_0^-=\#\Gamma_\infty^++\#\Gamma_\infty^-$ that
holds for
full gapped systems, this equation is recast
into 
\begin{eqnarray}
(-1)^{\nu}=(-1)^{(\#\Gamma_0^-+\#\Gamma_\infty^-)/2}.
\end{eqnarray}
This is the Fu-Kane's parity formula for the $\mathbb{Z}_2$ invariant.\cite{fu2007topological}

\subsubsection{Odd parity superconductors in three dimensions ($U^+_-$ in
   class D, $U^+_{+-}$ in class DIII)}
\label{sec:3DoddSC}


We examine here topological phases in three-dimensional odd parity
superconductors.
As in two dimensions discussed in Sec.\ref{2DoddSC}, 
the inversion $\tilde{P}$ of the BdG Hamiltonian anticommutes with $C = \tau_x \mathcal{K}$,
and thus it is labeled as $U^+_{-}$ in class D of Table \ref{TabUI}.
In three dimensions, the topological
index is $\mathbb{Z}$.
The $\mathbb{Z}$ index is defined at symmetric points, ${\bm
k}=(0,0,0)\equiv \Gamma_0$ and ${\bm k}=\infty\equiv\Gamma_\infty$, in a
manner similar to that for three-dimensional inversion symmetric
topological insulators discussed in Sec.\ref{sec:inversion3D}, 
\begin{eqnarray}
[\Gamma_{0,\infty}] 
= \frac{\#\Gamma_0^+-\#\Gamma_\infty^+}{2}. 
\label{eq:symmetricoddSC}
\end{eqnarray}
where $\#\Gamma_i^\pm$ is the number of negative energy states with parity
$\tilde{P}=\pm$ at $\Gamma_i$.  
As well as inversion symmetric topological insulators, 
we can also introduce a $\mathbb{Z}_2$ index $(-1)^{2P_3}$ with the
gravitomagnetoelectric polarizability $P_3$ defined by
Eq.(\ref{eq:Magnetoelectric_Polarizatioin}) for the BdG Hamiltonian, but
it is not independent of $[\Gamma_{0,\infty}]$, again.
The relation 
\begin{eqnarray}
(-1)^{2P_3}=(-1)^{[\Gamma_{0,\infty}]}  
\label{eq:P3NSC}
\end{eqnarray}
holds, and thus the present topological phase is fully characterized by $[\Gamma_{0,\infty}]$. 
The formula Eq.(\ref{eq:P3NSC}) is useful to discuss the heat response
of odd-parity superconductors by using the axion-type low energy
effective Hamiltonian. \cite{ryu2012electromagnetic, wang2011topological,
nomura2012cross, shiozaki2013electromagnetic}
  
Using an argument given in Sec. \ref{2DoddSC},  
for weak pairing odd parity superconductors, 
one can evaluate $[\Gamma_{0,\infty}]$ by the electron spectrum in the
normal state,  
\begin{eqnarray}
[\Gamma_{0,\infty}]=\frac{[\#\epsilon_-^+(\Gamma_0)-\#\epsilon_-^-(\Gamma_0)]
-[\#\epsilon_-^+(\Gamma_\infty)-\#\epsilon_-^-(\Gamma_\infty)]}{2}
\label{eq:gammaepsilon3D}
\end{eqnarray}
where $\#\epsilon_-^{\pm}(\Gamma_i)$ denotes
the number 
of $P=\pm$ bands in the normal
state below the Fermi level at $\Gamma_i$.
($P$ is the inversion operator acting on electron in the normal
state. See Sec. \ref{2DoddSC}.)

If an odd parity superconductor has TRS as well, then
$\tilde{P}$ is labeled as $U^+_{+-}$ in class DIII. 
As is seen in Table \ref{TabUI}, its topological number is
enriched as $\mathbb{Z}\oplus\mathbb{Z}$ in three dimensions. 
One of the $\mathbb{Z}$ indices is $[\Gamma_{0,\infty}]$ in
Eq.(\ref{eq:symmetricoddSC}), and the additional $\mathbb{Z}$ index is
the three-dimensional winding number $N_3$ in class DIII.
Although the parity of $N_3$ is equal to the
parity of $[\Gamma_{0,\infty}]$ \cite{sato2010topological, fu2010odd}, 
\begin{eqnarray}
(-1)^{N_3}=(-1)^{[\Gamma_{0,\infty}]}, 
\label{eq:W3N}
\end{eqnarray}
a full description of the present topological phase needs both of
$[\Gamma_{0,\infty}]$ and $N_3$.
We can also relate the
parity of $N_3$ 
to the gravitomagnetoelectric polarization $P_3$ as
$(-1)^{2 P_3} = (-1)^{N_3}$.\cite{ryu2010topological, sasaki2011topological} 

In a weak pairing odd parity superconductor, from
Eq.(\ref{eq:gammaepsilon3D}), the formula Eq.(\ref{eq:W3N}) is recast into
\begin{eqnarray}
(-1)^{N_3}=(-1)^{[\sum_{\sigma=\pm}\#\epsilon_-^\sigma(\Gamma_0)
-\sum_{\sigma=\pm}\#\epsilon_-^\sigma(\Gamma_\infty)]/2
}
=(-1)^{N_{\rm  F}/2},  
\end{eqnarray}
where $N_{\rm F}$ is the number of the Fermi surfaces enclosing
$\Gamma_0$.\cite{sato2010topological, fu2010odd} 
Note here that $N_{\rm F}$ is even due to the Kramers degeneracy.
This formula means that an odd parity superconductor automatically realizes
topological superconductivity with non-zero $N_3$ if it has the Fermi surface
with odd $N_{\rm F}/2$.
Although a boundary breaks inversion symmetry,   
the Fermi surface criterion for topological odd-parity
superconductivity predicts the existence of surface helical
Majorana fermions 
since $N_3$ itself is well-defined even in the presence of boundary.

\subsubsection{$\mathbb{Z}_2$ topological phase protected by antiunitary inversion symmetry in three-dimensional class AIII system ($A^+_-$ in class AIII)}

Finally, we examine a three-dimensional class AIII system with an additional
antiunitary inversion symmetry $A^+_-$. 
As a class AIII system, the winding number $N_3$ can be
introduced by $N_3=1/(48 \pi^2)\int {\rm tr}[\Gamma ({\cal H}^{-1}d{\cal
H})^3]$, 
but the presence of $A^+_-$ makes $N_3$ identically zero
 because it imposes the constraint $A \Gamma(
\mathcal{H}^{-1}d \mathcal{H})^3 A^{-1} = -\Gamma( \mathcal{H}^{-1} d
\mathcal{H})^3$ on the integral.
Alternatively, one can introduce the following $\mathbb{Z}_2$
topological invariant:
Because the additional antiunitary inversion
\begin{eqnarray}
A \mathcal{H}(\bm{k}) A^{-1} = \mathcal{H}(\bm{k}), \ \ \{A,\Gamma\}=0, \ \ A^2 = 1, 
\end{eqnarray}
acts in the same way as the time-reversal in the {\it coordinate}
space, the system can be identified with those in class CI with three
coordinate parameters. 
Therefore, the alternative topological number can be introduced as the
third homotopy group of the classifying space of class CI, i.e. $\pi_3
(\mathcal{R}_7) = \mathbb{Z}_2$, which  
reproduces the topological index in Table \ref{TabAI}.

\begin{table*}[!]
\begin{center}
\caption{
Classification table for topological crystalline insulators and
 superconductors and their topological defects in the presence of order-two additional antiunitary symmetry with flipped parameters $\delta_{\parallel}=d_{\parallel} - D_{\parallel} = 3$ (mod 4). 
Here $\delta=d-D$.
}

\begin{tabular}[t]{ccccccccccc}
\hline \hline
Symmetry & Class & $\mathcal{C}_q$ or $\mathcal{R}_q$ & $\delta=0$ & $\delta=1$ & $\delta=2$ & $\delta=3$ & $\delta=4$ & $\delta=5$ & $\delta=6$ & $\delta=7$ \\
\hline
$A^+$  & A      &$\mathcal{R}_6$& $0$ & $0$ & $2\mathbb{Z}$ & $0$ & $\mathbb{Z}_2$ & $\mathbb{Z}_2$ & $\mathbb{Z}$ & $0$ \\
$A^-$  & A      &$\mathcal{R}_2$& $\mathbb{Z}_2$ & $\mathbb{Z}_2$ & $\mathbb{Z}$ & $0$ & $0$ & $0$ & $2\mathbb{Z}$ & $0$ \\
$A^+_+$& AIII &$\mathcal{R}_7$& $0$ & $0$  & $0$ & $2\mathbb{Z}$ & $0$ & $\mathbb{Z}_2$ & $\mathbb{Z}_2$ & $\mathbb{Z}$ \\
$A^-_-$& AIII &$\mathcal{R}_1$& $\mathbb{Z}_2$ & $\mathbb{Z}$ & $0$ & $0$ & $0$ & $2\mathbb{Z}$ & $0$ & $\mathbb{Z}_2$ \\
$A^-_+$& AIII &$\mathcal{R}_3$& $0$ & $\mathbb{Z}_2$ & $\mathbb{Z}_2$ & $\mathbb{Z}$ & $0$ & $0$ & $0$ & $2\mathbb{Z}$ \\
$A^+_-$& AIII &$\mathcal{R}_5$&  $0$ & $2\mathbb{Z}$ & $0$ & $\mathbb{Z}_2$ & $\mathbb{Z}_2$ & $\mathbb{Z}$ & $0$ & $0$  \\
\hline 
$A^+_+,A^+_-$ & D  &$\mathcal{C}_0$& $\mathbb{Z}$ & $0$ & $\mathbb{Z}$ & $0$ & $\mathbb{Z}$ & $0$ & $\mathbb{Z}$ & $0$ \\
$A^-_+,A^-_-$ & C  &$\mathcal{C}_0$& $\mathbb{Z}$ & $0$ & $\mathbb{Z}$ & $0$ & $\mathbb{Z}$ & $0$ & $\mathbb{Z}$ & $0$ \\
\hline
$A^-_+,A^-_-$ & D  &$\mathcal{R}_2 \times \mathcal{R}_2$& $\mathbb{Z}_2\oplus\mathbb{Z}_2$ & $\mathbb{Z}_2\oplus\mathbb{Z}_2$ & $\mathbb{Z}\oplus\mathbb{Z}$ & $0$ & $0$ & $0$ & $2\mathbb{Z}\oplus 2\mathbb{Z}$ & $0$ \\
$A^+_+,A^+_-$ & C  &$\mathcal{R}_6 \times \mathcal{R}_6$& $0$ & $0$ & $2\mathbb{Z}\oplus 2\mathbb{Z}$ & $0$ & $\mathbb{Z}_2\oplus\mathbb{Z}_2$ & $\mathbb{Z}_2\oplus\mathbb{Z}_2$ & $\mathbb{Z}\oplus\mathbb{Z}$ & $0$ \\
\hline \hline
\end{tabular}
\label{TabAI}
\end{center}
\end{table*}


\section{Weak crystalline topological indices}
\label{sec:weak}

Up to now, we have treated the base space of Hamiltonians as
a $d+D$-dimensional sphere $S^{d+D}$.
For band insulators, however, the actual base space is a direct
product of a $d$-dimensional torus $T^d$ and a $D$-dimensional sphere
$S^{D}$, i.e. $T^d\times S^D$, 
because of the periodic structure of the Brillouin zone.
The torus manifold gives rise to an extra topological structure.
For example, the K-group of $d$-dimensional topological band insulators
$(D=0)$ in the real AZ class
$s$ $(s=0,1,\cdots,7)$ is given as
\cite{kitaev2009periodic} 
\begin{eqnarray}
K_{\mathbb{R}}(s; T^d)
\cong \pi_0(\mathcal{R}_{s-d})
\bigoplus_{q=1}^{d-1} \begin{pmatrix}
d \\
q
\end{pmatrix} \pi_0(\mathcal{R}_{s-d+q}) 
, \quad (d \geq 1). 
\label{weak_index}
\end{eqnarray}
The first term reproduces the K-group of the Hamiltonians on $S^d$,
i.e. $K_{\mathbb{R}}(s;d,D=0)$, but there are extra terms that
define weak topological indices. 
Here Eq. (\ref{weak_index}) does not include zero-dimensional indices
$\pi_0(\mathcal{R}_s)$
since the base space d-dimensional torus $T^d$ does not have $Z_2$
distinct parts.\cite{ran2010weak} 

The extra terms in the presence of additional symmetry are more
complicated than those of the above case because there are two
different choices of lowering dimension,   
i.e. the parameters which is flipped by the additional symmetry
transformation or not. 
The complete K-group for topological crystalline band insulators
and superconductors in complex AZ classes
with additional unitary symmetry 
is given by 
\begin{equation}\begin{split}
K^U_{\mathbb{C}}(s,t;T^d) 
&\cong
\bigoplus_{0\leq q_{\perp} \leq d-d_{\parallel}, 0 \leq q_{\parallel} \leq d_{\parallel}, 0\leq q_{\perp}+q_{\parallel}\leq d-1}
\begin{pmatrix}
d-d_{\parallel} \\
q_{\perp} 
\end{pmatrix}
\begin{pmatrix}
d_{\parallel} \\
q_{\parallel} 
\end{pmatrix}
K^U_{\mathbb{C}}
(s,t;d-q_{\perp}-q_{\parallel}, d_{\parallel}-q_{\parallel},0,0) ,\ \ (d \geq 1)
\end{split}\end{equation}
where $s=0,1$ and $t=0,1$ denote the AZ class and the unitary symmetry
in Table \ref{Symmetry_type_UC}.
Similar results are obtained for those with additional antiunitary
symmetry and those in 
real AZ classes.

To illustrate weak crystalline topological indices, consider an odd-parity
superconductor in three dimensions.
The full K-group on the torus $T^3$ is given by
\begin{eqnarray}
&&K_{\mathbb{R}}^U(s=2,t=2; d=3, d_{\parallel}=3, 0,0;T^3) 
\nonumber\\
&&=K_{\mathbb{R}}^U(s=2,t=2; d=3, d_{\parallel}=3, 0,0;S^3)(=\mathbb{Z}) 
\nonumber\\
&&\hspace{3ex}
\bigoplus_{i=1}^3 K_{\mathbb{R}}^U(s=2,t=2; d=2, d_{\parallel}=2, 0,0;S_i^2)
\left(=\bigoplus^3(\mathbb{Z}\oplus\mathbb{Z})\right) 
\nonumber\\
&&\hspace{3ex}
\bigoplus^3_{i=1} K_{\mathbb{R}}^U(s=2,t=2; d=1, d_{\parallel}=1, 0,0;S_i^1) 
\left(=\bigoplus^3\mathbb{Z} \right)
\nonumber\\
&&=\bigoplus^{10}\mathbb{Z},
\end{eqnarray}
where $S^2_i$ and $S^1_i$ denote two-dimensional and
one-dimensional spheres that are obtained as high symmetric submanifolds
of the torus.
This equation implies that  there are ten $\mathbb{Z}$ crystalline
topological indices.
Among them, 
three indices are the weak first Chern numbers 
defined at fixed $k_i$ plane $T^2_i$ ($i=x,y,z$) in the Brillouin zone, 
\begin{eqnarray}
Ch^i_1 = \frac{i}{2 \pi} \frac{\epsilon_{ijk}}{2} \int_{T^2_i}\mathrm{tr}
 \mathcal{F}_{jk}(\bm{k})
\quad (i=x,y,z). 
\end{eqnarray}
The other seven $\mathbb{Z}$ indices are defined at the 
eight symmetric points $\Gamma_i$ ($i=1, \dots 8$) of inversion, which
satisfy $\bm{k}=-\bm{k}+{\bm G}$ with a reciprocal vector ${\bm G}$.
In the cubic lattice, these $\Gamma_i$ are $\Gamma_1=(0,0,0),
\Gamma_2=(\pi,0,0), \Gamma_3=(0,\pi,0), \Gamma_4=(0,0,\pi),
\Gamma_5=(\pi,\pi,0), \Gamma_6=(\pi,0,\pi), \Gamma_7=(0,\pi,\pi)$,
and $\Gamma_8=(\pi,\pi,\pi)$. 
The seven $\mathbb{Z}$ indices are
\begin{eqnarray}
[\Gamma_{i,8}]=\#\Gamma_i^+-\#\Gamma_8^+
=-(\#\Gamma_i^--\#\Gamma_8^-),
\quad
(i=1,\dots,7), 
\end{eqnarray}
where $\#\Gamma_i^\pm$ is the number of negative energy states with parity
$\tilde{P}=\pm$ at $\Gamma_i$.
Here we have used the relation $\#\Gamma_i^{+}+\#\Gamma_i^{-}
=\#\Gamma_j^++\#\Gamma_j^-$ for full
gapped systems.
In a manner similar to Eq.(\ref{eq:parityChern}), the parity of $Ch_1^i$ can be
expressed by $[\Gamma_{i,8}]$:
\begin{eqnarray}
&&(-1)^{Ch_1^1}
=(-1)^{[\Gamma_{1,8}]+[\Gamma_{3,8}]+[\Gamma_{4,8}]+[\Gamma_{7,8}]}
=(-1)^{[\Gamma_{2,8}]+[\Gamma_{5,8}]+[\Gamma_{6,8}]},
\nonumber\\
&&(-1)^{Ch_1^2}
=(-1)^{[\Gamma_{1,8}]+[\Gamma_{2,8}]+[\Gamma_{4,8}]+[\Gamma_{6,8}]}
=(-1)^{[\Gamma_{3,8}]+[\Gamma_{5,8}]+[\Gamma_{7,8}]},
\nonumber\\
&&(-1)^{Ch_1^3}
=(-1)^{[\Gamma_{1,8}]+[\Gamma_{2,8}]+[\Gamma_{3,8}]+[\Gamma_{5,8}]}
=(-1)^{[\Gamma_{4,8}]+[\Gamma_{6,8}]+[\Gamma_{7,8}]}. 
\label{eq:Chernlattice}
\end{eqnarray}
These relations provide a global constraint that the
summation of all seven $[\Gamma_{i,8}]$s must be even.
From the same argument in Secs.\ref{2DoddSC} and \ref{sec:3DoddSC}, in weak pairing odd-parity superconductors, we can also show
\begin{eqnarray}
[\Gamma_{i,8}]=[\#\epsilon_-^+(\Gamma_i)-\#\epsilon_-^-(\Gamma_i)]
-[\#\epsilon_-^+(\Gamma_8)-\#\epsilon_-^-(\Gamma_8)], 
\quad (i=1,\dots,7),
\label{eq:Nlattice}
\end{eqnarray}
where $\#\epsilon_-^{\pm}(\Gamma_i)$ denotes
the number 
of $P=\pm$ parity bands in the normal
state below the Fermi level at $\Gamma_i$.

In odd parity superconductors, we can
define other topological indices, but they can be expressed by the above
ten indices:
First, as was shown in Sec.\ref{sec:3DoddSC}, using the gravitomagnetoelectric
polarizability $P_3$,
one can define a $\mathbb{Z}_2$ index by $(-1)^{2P_3}$. This index is
written as
\begin{eqnarray}
(-1)^{2P_3}=(-1)^{\sum_{i=1}^8N_i/2}. 
\end{eqnarray}
Furthermore, PHS enables us to define $\mathbb{Z}_2$ 
indices $(-1)^{\nu_{ij}}$ with
\begin{eqnarray}
\nu_{ij}=\frac{i}{\pi}\oint_{C_{ij}} {\rm tr}\mathcal{A}, 
\end{eqnarray}
where $C_{ij}$ is a closed path that passes through $\Gamma_i$ and
$\Gamma_j$ in $T^3$ and is invariant under ${\bm k}\rightarrow -{\bm k}$.
These indices satisfies\cite{sato2010topological}
\begin{eqnarray}
(-1)^{\nu_{ij}}=(-1)^{[\Gamma_{i,8}]+[\Gamma_{j,8}]}. 
\label{eq:Z2lattice}
\end{eqnarray}

Note that the weak indices $Ch_1^i$ and $(-1)^{\nu_{ij}}$ are
well-defined even in the presence of boundaries which induce a parity
mixing of Cooper pairs in general.
Therefore, the bulk-boundary correspondence holds for $Ch_1^i$ and
$(-1)^{\nu_{ij}}$. 
Combing Eqs.(\ref{eq:Chernlattice}) and (\ref{eq:Z2lattice}) with
Eq.(\ref{eq:Nlattice}), 
details of surface gapless modes can be predicted by the knowledge of the Fermi
surface structure.\cite{sato2010topological}


\section{Dimensional hierarchy with order-two additional symmetry}
\label{TCI}

In this section, we establish the relations between the K-groups of
topological crystalline insulators and superconductors with order-two
additional symmetry in 
different dimensions.

\subsection{Additional order-two unitary symmetry in complex AZ classes}
\label{sec:HC}

In this case, due to the absence of antiunitary symmetry, the momentum
$\bm{k}$ and coordinate parameters $\bm{r}$ cannot be distinguished from
each other. Therefore, we have
\begin{eqnarray}
K^U_{\mathbb{C}}(s,t;d, d_{\parallel},D, D_{\parallel}) 
= K^U_{\mathbb{C}}(s,t;d+D,d_{\parallel}+D_{\parallel},0,0). 
\label{eq:d+D}
\end{eqnarray}
We can also derive the following relation, 
\begin{eqnarray}
K^U_{\mathbb{C}}(s,t;d+D,d_{\parallel}+D_{\parallel},0,0)
&=&K^U_{\mathbb{C}}(s+1,t;d+D+1,d_{\parallel}+D_{\parallel},0,0)
\nonumber\\
&=&K^U_{\mathbb{C}}(s+1,t+1;d+D+1,d_{\parallel}+D_{\parallel}+1,0,0), 
\label{KCU2} 
\end{eqnarray}
which leads to Eq.(\ref{KCU}).

To prove the dimensional hierarchy of the K-groups Eq.(\ref{KCU2}), we use
the dimension-raising maps, Eqs. (\ref{eq:DRC}) and (\ref{eq:DRNC}), and
their inverses, Eqs. (\ref{eq:saddlepoint}) and (\ref{eq:DLNC}).
These maps determine uniquely how order-two unitary symmetry of an original
Hamiltonian acts on the mapped Hamiltonian, and as a result, we can
obtain the relation between the K-groups. 
For instance, 
a Hamiltonian $\mathcal{H}(\bm{k},\bm
{r})$ in class A $(s=0)$ is mapped into a Hamiltonian
$\mathcal{H}(\bm{k},\bm {r},\theta)$ in class AIII $(s=1)$ with CS
$\Gamma={\bm 1}\otimes \tau_x$ by the dimension-raising map
\begin{eqnarray}
\mathcal{H}(\bm{k},\bm{r},\theta)
&= \sin \theta \mathcal{H}(\bm{k},\bm{r}) \otimes \tau_z 
+ \cos \theta {\bm 1}\otimes \tau_y. 
\label{eq:DR}
\end{eqnarray}
If the class A Hamiltonian $\mathcal{H}(\bm{k},\bm{r})$ has an
additional unitary symmetry $U$, which is labeled by $(s, t)=(0,0)$ in Table
\ref{Symmetry_type_UC},  
\begin{eqnarray}
U \mathcal{H}(\bm{k},\bm{r}) U^{-1} 
= \mathcal{H}(-\bm{k}_{\parallel},{\bm k}_{\perp}, -{\bm r}_{\parallel},
{\bm r}_{\perp}), 
\end{eqnarray}
then the mapped class AIII Hamiltonian
$\mathcal{H}(\bm{k},\bm{r},\theta)$ also has the
corresponding symmetries,  
\begin{eqnarray}
&&(U \otimes \tau_0) \mathcal{H}(\bm{k},{\bm r},\theta)
		 (U\otimes \tau_0)^{-1} =
\mathcal{H}(-\bm{k}_{\parallel},{\bm k}_{\perp}, -{\bm r}_{\parallel},
		 {\bm r}_{\perp} ,\theta),
\\ 
&&(U \otimes \tau_z)
 \mathcal{H}(\bm{k},{\bm r},\theta) (U\otimes
 \tau_z)^{-1} =
 \mathcal{H}(-\bm{k}_{\parallel},{\bm k}_{\perp},-{\bm r}_{\parallel},
 {\bm r}_{\perp},\pi-\theta). 
\end{eqnarray}
The former (latter) symmetry $U \otimes \tau_0$ $(U \otimes \tau_z)$
defines $U_+$ ($U_-$) in Table \ref{Symmetry_type_UC}, which belongs to
$(s,t)=(1,0)$ [$(s,t)=(1,1)$] in Table \ref{Symmetry_type_UC}, because it
(anti-)commutes with the chiral operator $\Gamma={\bm 1}\otimes \tau_x$.  
Also, in the former (latter) case, the trivial (non-trivial)
transformation of $\theta$ under the mapped symmetry implies that 
$\theta$ must be considered as a ${\bm k}_{\perp}$/${\bm r}_{\perp}$-type   
(${\bm k}_{\parallel}/{\bm r}_{\parallel}$-type) variable for the mapped
symmetry.
Therefore, Eq.(\ref{eq:DR}) provides the K-group homomorphism  
$
K^U_{\mathbb{C}}(0,0;d+D,d_{\parallel}+D_{\parallel},0,0)
\rightarrow
K^U_{\mathbb{C}}(1,0;d+D+1,d_{\parallel}+D_{\parallel},0,0)
$ and
$K^U_{\mathbb{C}}(0,0;d+D,d_{\parallel}+D_{\parallel},0,0)
\rightarrow
K^U_{\mathbb{C}}(1,1;d+D+1,d_{\parallel}+D_{\parallel}+1,0,0)
$.

In a similar manner, one can specify how other unitary symmetries in Table
\ref{Symmetry_type_UC} are mapped, and how $\theta$ transforms under the
mapped symmetries, as summarized in Table \ref{Tab_raise_complex_unitary}.
We also find that the dimension-lowering maps
Eqs.(\ref{eq:saddlepoint}) and (\ref{eq:DLNC}) provide the inverse of
these mappings.
Consequently, we have isomorphism between Hamiltonians with
different $(s,t)$s' of Table \ref{Symmetry_type_UC}, in the meaning of
stable equivalence, which establishes the K-group 
isomorphism of Eq.(\ref{KCU2}).

\begin{table*}[!]
\begin{center}
\caption{
Homomorphism from $K^U_{\mathbb{C}}(s,t,d,d_{\parallel},0,0)$ to $K^U_{\mathbb{C}}(s+1,t,d+1,d_{\parallel},0,0)$ 
and $K^U_{\mathbb{C}}(s+1,t+1,d+1,d_{\parallel}+1,0,0)$. 
}
\begin{tabular}[t]{cccccccccc}
\hline \hline
AZ Class & $t$ & Symmetry & Hamiltonian mapping & Type of $\theta$ & Mapped AZ class &
$\Gamma$
& Mapped $t$ & Mapped symmetry \\
\hline
\multirow{4}{*}{A} 
& 0 
& $U$ 
& \multirow{4}{*}
{$\sin \theta 
\mathcal{H}(\bm{k},{\bm r}) 
\otimes \tau_z + \cos \theta {\bm 1}\otimes \tau_y$} 
&$\bm{k}_{\perp}/{\bm r}_{\perp}$ 
& \multirow{4}{*}{AIII} 
& \multirow{4}{*}
{${\bm 1}\otimes \tau_x$} &0& $U_+ = U \otimes \tau_0$  \\
 & 1 & $\widebar U$ & &$\bm{k}_{\perp}/{\bm r}_{\perp}$
& & 
& 1 
& $U_- = \widebar U \otimes \tau_y$ \\ 
 & 0 & $U$ & & $\bm{k}_{\parallel}/{\bm r}_{\parallel}$
& & & 1 & $U_- = U \otimes \tau_z$ \\
 & 1 & $\widebar U$ & &$\bm{k}_{\parallel}/\bm{r}_{\parallel}$& & & 0 
& $U_+ = \widebar U \otimes \tau_x$ \\
\hline
\multirow{4}{*}{AIII} 
    & 0 & $U_+$ & \multirow{4}{*}
{$\sin \theta \mathcal{H}(\bm{k},\bm{r}) + \cos \theta \Gamma$} 
&$\bm{k}_{\perp}/{\bm r}_{\perp}$
& \multirow{4}{*}{A} & \multirow{4}{*}{} & 0 & $U = U_+$  \\
 & 1 & $U_-$ & &$\bm{k}_{\perp}/{\bm r}_{\perp}$& & & 1 
& $\widebar U = i \Gamma U_-$ \\
 & 0 & $U_+$ & & $\bm{k}_{\parallel}/{\bm r}_{\parallel}$
& & & 1 & $\widebar U = \Gamma U_+$ \\
 & 1 & $U_-$ & &$\bm{k}_{\parallel}/{\bm r}_{\parallel}$& & & 0 & $U=U_-$ \\
\hline \hline
\end{tabular}
\label{Tab_raise_complex_unitary}
\end{center}
\end{table*}

\subsection{Additional order-two antiunitary symmetry in complex AZ classes}
As was explained in Sec.\ref{sec:CAZAS}, the presence of additional order-two
antiunitary symmetry introduces real structures in complex AZ classes,
and consequently, with mapping of symmetries in
Table.\ref{Symmetry_type_AC}, the K-group of this case reduces to that
of real AZ classes, 
\begin{eqnarray}
K_{\mathbb{C}}^{A}(s;d,d_{\parallel}, D, D_{\parallel}) 
=K_{\mathbb{R}}(s;d-d_{\parallel}+D_{\parallel}, D-D_{\parallel}+d_{\parallel}).
\end{eqnarray}
From the dimensional hierarchy of real AZ classes, Eq.(\ref{eq:KisoR}), we have
\begin{eqnarray}
K_{\mathbb{C}}^{A}(s;d,d_{\parallel}, D, D_{\parallel}) 
&=&K_{\mathbb{C}}^{A}(s+1;d+1,d_{\parallel}, D, D_{\parallel})
\nonumber\\ 
&=&K_{\mathbb{C}}^{A}(s-1;d,d_{\parallel}, D+1, D_{\parallel}) 
\nonumber\\
&=&K_{\mathbb{C}}^{A}(s-1;d+1,d_{\parallel}+1, D, D_{\parallel})
\nonumber\\
&=&K_{\mathbb{C}}^{A}(s+1;d,d_{\parallel}, D+1, D_{\parallel}+1).
\label{KCA2}
\end{eqnarray}

\subsection{Additional order-two symmetry in real AZ classes}
We now outline the proof of the following isomorphism: 
\begin{eqnarray}
K^{U/A}_{\mathbb{R}}(s,t;d,d_{\parallel},D,D_{\parallel})
&=&K^{U/A}_{\mathbb{R}}(s+1,t;d+1,d_{\parallel},D,D_{\parallel})
\nonumber\\
&=&K^{U/A}_{\mathbb{R}}(s-1,t ;d, d_{\parallel},D+1, D_{\parallel}) 
\nonumber\\
&=&K^{U/A}_{\mathbb{R}}(s+1,t+1;d+1, d_{\parallel}+1,D, D_{\parallel})
\nonumber\\
&=&K^{U/A}_{\mathbb{R}}(s-1,t-1;d, d_{\parallel},D+1, D_{\parallel}+1), 
\label{KR2}
\end{eqnarray}
which leads Eq.(\ref{KR}). 

In order to prove Eq.(\ref{KR2}), 
we use the dimension-raising maps, Eqs.(\ref{eq:DRC}) and
(\ref{eq:DRNC}), and their inverse, Eqs.(\ref{eq:saddlepoint}) and
(\ref{eq:DLNC}), in a manner similar to Sec.\ref{sec:HC}.
From these maps, we can determine how additional symmetries in Table
\ref{Symmetry_type} are mapped.
We have directly determined the mapped symmetries for all 
128 = 8 $(s=0,\cdots,7)$ $\times$ 4 $(t=0,1,2,3)$
$\times$ 4 $(\mbox{type of $\theta$}, \bm{k}_{\parallel},
\bm{k}_{\perp}, \bm{r}_{\parallel}, \bm{r}_{\perp})$ possible
combinations. 
This procedure is straightforward but cumbersome, 
so we explain only the case of $(s,t)=(4,2)$ in details.
Other cases can be considered in the same manner.

A representative Hamiltonian of the K-group 
$K^{U/A}_{\mathbb{R}}(s=4,t=2;d, d_{\parallel},D,D_{\parallel})$ 
has the following symmetries, 
\begin{eqnarray}
&&T \mathcal{H}(\bm{k},\bm{r}) T^{-1} = \mathcal{H}(-\bm{k},\bm{r}),
\quad T^2 = -1, 
\\
&&U \mathcal{H}(\bm{k},\bm{r}) U^{-1} 
= \mathcal{H}(-\bm{k}_{\parallel},
\bm{k}_{\perp},-\bm{r}_{\parallel},\bm{r}_{\perp}),\quad
U^2 = 1,\quad \{T, U\} =0, 
\label{eq:42U}
\end{eqnarray}
where $U=U^{+}_-$ is one of the equivalent symmetries with 
$(s,t)=(4,2)$ in Table \ref{Symmetry_type}. 
(Note that the symmetries $U^+_-$,
$U^-_+$, $A^+_-$ and $A^+_+$ are equivalence to each other.) 
Equation (\ref{eq:DRNC}) provides two different mapped TRS
\begin{eqnarray}
&&(T \otimes \tau_0)
\mathcal{H}(\bm{k},{\bm r}, \theta) (T \otimes \tau_0)^{-1} =
\mathcal{H}(-\bm{k},\bm{r},\pi-\theta), \quad (T \otimes \tau_0)^2 = -1.
\label{eq:TRS31}
\\
&&(T \otimes \tau_z)
\mathcal{H}(\bm{k},{\bm r}, \theta) (T \otimes \tau_z)^{-1} =
\mathcal{H}(-\bm{k},\bm{r}, \theta), \quad (T \otimes \tau_z)^2 = -1,
\label{eq:TRS33}
\end{eqnarray}
where $\theta$ in each case transforms in a different manner.
By combining with the CS, $C={\bm 1}\otimes \tau_x$, of the
mapped Hamiltonian, we automatically obtain the corresponding PHS
\begin{eqnarray}
&&(T \otimes \tau_x)
\mathcal{H}(\bm{k},\bm{r}, \theta) (T \otimes \tau_x)^{-1} =
-\mathcal{H}(-\bm{k},\bm{r},\pi-\theta), \quad (T \otimes \tau_x)^2 = -1,
\label{eq:PHS31}
\\
&&(T \otimes \tau_y)
\mathcal{H}(\bm{k},\bm{r}, \theta) (T \otimes \tau_y)^{-1} =
-\mathcal{H}(-\bm{k},\bm{r},\theta), \quad (T \otimes \tau_y)^2 = 1.
\end{eqnarray}
The additional symmetry $U$ is also realized in two
different manners,
\begin{eqnarray}
&&(U \otimes \tau_z)
\mathcal{H}(\bm{k},\bm{r},\theta) 
(U \otimes \tau_z)^{-1} =
\mathcal{H}
(-\bm{k}_{\parallel},\bm{k}_{\perp},-\bm{r}_{\parallel},\bm{r}_{\perp},
\pi-\theta), 
\label{eq:U31}
\\
&&(U \otimes \tau_0)
\mathcal{H}(\bm{k},\bm{r},\theta) 
(U \otimes \tau_0)^{-1} =
\mathcal{H}
(-\bm{k}_{\parallel},\bm{k}_{\perp},-\bm{r}_{\parallel},\bm{r}_{\perp},
\theta).
\end{eqnarray}
Therefore, there are four possible combinations of the mapped TRS and
$U$, which correspond to four possible types of $\theta$. 
By fixing the type of $\theta$, we can select one of the combinations
and determine the type of $U^{\epsilon_U}_{\eta_T, \eta_C}$.
For instance, if one considers ${\bm k}_{\parallel}$-type $\theta$, then
$\theta$ transforms non-trivially under the mapped TRS and
$U$. Hence the
mapped TRS and U are given by Eqs.(\ref{eq:TRS31}) and (\ref{eq:U31}),
respectively, which specifies the mapped $U$ as $U^{+}_{-,+}$ labeled
by $(s,t)=(3,3)$ in Table \ref{Symmetry_type}.
This means that the dimension-raising map Eq.(\ref{eq:DRNC}) provides a
homomorphism,  
$K_{\mathbb{R}}(4,2;d, d_{\parallel},D,D_{\parallel})
\rightarrow
K_{\mathbb{R}}(3,1;d+1, d_{\parallel}+1,D,D_{\parallel}).
$

One can specify how other symmetries in Table
\ref{Symmetry_type} are mapped, and how $\theta$ transforms under the
mapped symmetries, as summarized in Tables \ref{Tab_raise_real_nonchiral}
and \ref{Tab_raise_real_chiral}.
We also find that the dimension-lowering maps
Eqs.(\ref{eq:saddlepoint}) and (\ref{eq:DLNC}) provide the inverse of
these mappings.
Consequently, we have isomorphism between Hamiltonians with
different $(s,t)$s' of Table \ref{Symmetry_type}, which establishes the K-group 
isomorphism of Eq.(\ref{KR2}).

\begin{table*}[!]
\begin{center}
\caption{
Homomorphism from $K^{U/A}_{\mathbb{R}}(s,t,d,d_{\parallel},D,D_{\parallel})$ to 
$K^{U/A}_{\mathbb{R}}(s+1,t,d+1,d_{\parallel},D,D_{\parallel})$, 
$K^{U/A}_{\mathbb{R}}(s+1,t+1,d+1,d_{\parallel}+1,D,D_{\parallel})$, 
$K^{U/A}_{\mathbb{R}}(s-1,t,d,d_{\parallel},D+1,D_{\parallel})$, and 
$K^{U/A}_{\mathbb{R}}(s-1,t-1,d,d_{\parallel},D+1,D_{\parallel}+1)$ 
for nonchiral classes. 
}
\begin{tabular}[t]{ccccccccccc}
\hline \hline
AZ class & $t$ & Symmetry & Type of $\theta$ &
 Mapped AZ class & TRS & PHS & $\Gamma$ & Mapped $t$ & Mapped symmetry \\
\hline
\multirow{4}{*}{AI/AII} & 0 & $U^+_+$ &$\bm{k}_{\perp}$& \multirow{4}{*}{BDI/CII} & \multirow{4}{*}{$T \otimes \tau_0$} & \multirow{4}{*}{$T \otimes \tau_x$} & \multirow{4}{*}{${\bm 1}\otimes \tau_x$} & 0 & $U^+_{++} = U^+_+ \otimes \tau_0$  \\
 & 1 & $\widebar U^+_-$ & $\bm{k}_{\perp}$& & & & & 1 & $\widebar U^+_{-+} = \widebar U^+_- \otimes \tau_z$ \\
 & 2 & $U^+_-$ & $\bm{k}_{\perp}$& & & & & 2 & $U^+_{--} = U^+_- \otimes \tau_0$ \\
 & 3 & $\widebar U^+_+$ & $\bm{k}_{\perp}$& && & & 3 & $\widebar U^+_{+-} = \widebar U^+_+ \otimes \tau_z$ \\
\hline
\multirow{4}{*}{AI/AII} & 0 & $U^+_+$ & $\bm{k}_{\parallel}$& \multirow{4}{*}{BDI/CII} & \multirow{4}{*}{$T \otimes \tau_0$} & \multirow{4}{*}{$T \otimes \tau_x$} & \multirow{4}{*}{${\bm 1}\otimes \tau_x$} &  1 & $U^+_{+-} = U^+_+ \otimes \tau_z$ \\
 & 1 & $\widebar U^+_-$ & $\bm{k}_{\parallel}$& & & & & 2 & $\widebar U^+_{--} = \widebar U^+_- \otimes \tau_0$ \\
 & 2 & $U^+_-$ & $\bm{k}_{\parallel}$& & & & & 3 & $U^+_{-+} = U^+_- \otimes \tau_z$ \\
 & 3 & $\widebar U^+_+$ & $\bm{k}_{\parallel}$& & & & & 0 & $\widebar U^+_{++} = \widebar U^+_+ \otimes \tau_0$ \\
\hline
\multirow{4}{*}{AI/AII} & 0 & $U^+_+$ &$\bm{r}_{\perp}$& \multirow{4}{*}{CI/DIII} & \multirow{4}{*}{$T \otimes \tau_z$} & \multirow{4}{*}{$T \otimes \tau_y$} & \multirow{4}{*}{${\bm 1}\otimes \tau_x$} & 0 & $U^+_{++} = U^+_+ \otimes \tau_0$  \\
 & 1 & $\widebar U^+_-$&$\bm{r}_{\perp}$& & & & & 1 & $\widebar U^+_{-+} = \widebar U^+_- \otimes \tau_z$ \\
 & 2 & $U^+_-$ & $\bm{r}_{\perp}$& & & & & 2 & $U^+_{--} = U^+_- \otimes \tau_0$ \\
 & 3 & $\widebar U^+_+$ & $\bm{r}_{\perp}$& & & & & 3 & $\widebar U^+_{+-} = \widebar U^+_+ \otimes \tau_z$ \\
\hline
\multirow{4}{*}{AI/AII} & 0 & $U^+_+$ & $\bm{r}_{\parallel}$& \multirow{4}{*}{CI/DIII} & \multirow{4}{*}{$T \otimes \tau_z$} & \multirow{4}{*}{$T \otimes \tau_y$} & \multirow{4}{*}{${\bm 1}\otimes \tau_x$} & 3 & $U^+_{+-} = U^+_+ \otimes \tau_z$ \\
 & 1 & $\widebar U^+_-$ & $\bm{r}_{\parallel}$& & & & & 0 & $\widebar U^+_{--} = \widebar U^+_- \otimes \tau_0$ \\
 & 2 & $U^+_-$ & $\bm{r}_{\parallel}$& & & & & 1 & $U^+_{-+} = U^+_- \otimes \tau_z$ \\
 & 3 & $\widebar U^+_+$ & $\bm{r}_{\parallel}$& & & & & 2 & $\widebar U^+_{++} = \widebar U^+_+ \otimes \tau_0$ \\
\hline\hline
\multirow{4}{*}{D/C} & 0 & $U^+_+$ &$\bm{k}_{\perp}$& \multirow{4}{*}{DIII/CI} & \multirow{4}{*}{$C \otimes \tau_y$} & \multirow{4}{*}{$C \otimes \tau_z$} & \multirow{4}{*}{${\bm 1}\otimes \tau_x$} & 0 & $U^+_{++} = U^+_+ \otimes \tau_0$  \\
 & 1 & $\widebar U^+_+$ & $\bm{k}_{\perp}$& & & & & 1 & $\widebar U^+_{-+} = \widebar U^+_+ \otimes \tau_z$ \\
 & 2 & $U^+_-$ & $\bm{k}_{\perp}$& & & & & 2 & $U^+_{--} = U^+_- \otimes \tau_0$ \\
 & 3 & $\widebar U^+_-$ & $\bm{k}_{\perp}$& && &  & 3 & $\widebar U^+_{+-} = \widebar U^+_- \otimes \tau_z$ \\
\hline
\multirow{4}{*}{D/C} & 0 & $U^+_+$ & $\bm{k}_{\parallel}$& \multirow{4}{*}{DIII/CI} & \multirow{4}{*}{$C \otimes \tau_y$} & \multirow{4}{*}{$C \otimes \tau_z$} & \multirow{4}{*}{${\bm 1}\otimes \tau_x$} &  1 & $U^+_{-+} = U^+_+ \otimes \tau_z$ \\
 & 1 & $\widebar U^+_+$ & $\bm{k}_{\parallel}$& & & & & 2 & $\widebar U^+_{++} = \widebar U^+_+ \otimes \tau_0$ \\
 & 2 & $U^+_-$ & $\bm{k}_{\parallel}$& & & & & 3 & $U^+_{+-} = U^+_- \otimes \tau_z$ \\
 & 3 & $\widebar U^+_-$ & $\bm{k}_{\parallel}$& & & & & 0 & $\widebar U^+_{--} = \widebar U^+_- \otimes \tau_0$ \\
\hline
\multirow{4}{*}{D/C} & 0 & $U^+_+$ &$\bm{r}_{\perp}$& \multirow{4}{*}{BDI/CII} & \multirow{4}{*}{$C \otimes \tau_x$} & \multirow{4}{*}{$C \otimes \tau_0$} & \multirow{4}{*}{${\bm 1}\otimes \tau_x$} & 0 & $U^+_{++} = U^+_+ \otimes \tau_0$  \\
 & 1 & $\widebar U^+_+$ &$\bm{r}_{\perp}$& & & & & 1 & $\widebar U^+_{-+} = \widebar U^+_+ \otimes \tau_z$ \\
 & 2 & $U^+_-$ & $\bm{r}_{\perp}$& & & & & 2 & $U^+_{--} = U^+_- \otimes \tau_0$ \\
 & 3 & $\widebar U^+_-$ & $\bm{r}_{\perp}$& & & & & 3 & $\widebar U^+_{+-} = \widebar U^+_- \otimes \tau_z$ \\
\hline
\multirow{4}{*}{D/C} & 0 & $U^+_+$ & $\bm{r}_{\parallel}$& \multirow{4}{*}{BDI/CII} & \multirow{4}{*}{$C \otimes \tau_x$} & \multirow{4}{*}{$C \otimes \tau_0$} & \multirow{4}{*}{${\bm 1}\otimes \tau_x$} & 3 & $U^+_{-+} = U^+_+ \otimes \tau_z$ \\
 & 1 & $\widebar U^+_+$ & $\bm{r}_{\parallel}$& & & & & 0 & $\widebar U^+_{++} = \widebar U^+_+ \otimes \tau_0$ \\
 & 2 & $U^+_-$ & $\bm{r}_{\parallel}$& & & & & 1 & $U^+_{+-} = U^+_- \otimes \tau_z$ \\
 & 3 & $\widebar U^+_-$ & $\bm{r}_{\parallel}$ & & & & & 2 & $\widebar U^+_{--} = \widebar U^+_- \otimes \tau_0$ \\
\hline \hline
\end{tabular}
\label{Tab_raise_real_nonchiral}
\end{center}
\end{table*}

\begin{table*}[!]
\begin{center}
\caption{
Homomorphism from $K^{U/A}_{\mathbb{R}}(s,t,d,d_{\parallel},D,D_{\parallel})$ to 
$K^{U/A}_{\mathbb{R}}(s+1,t,d+1,d_{\parallel},D,D_{\parallel})$, 
$K^{U/A}_{\mathbb{R}}(s+1,t+1,d+1,d_{\parallel}+1,D,D_{\parallel})$, 
$K^{U/A}_{\mathbb{R}}(s-1,t,d,d_{\parallel},D+1,D_{\parallel})$, and 
$K^{U/A}_{\mathbb{R}}(s-1,t-1,d,d_{\parallel},D+1,D_{\parallel}+1)$ 
for chiral classes. 
}
\begin{tabular}[t]{cccccccccc}
\hline \hline
AZ Class & $t$ & Symmetry & Type of $\theta$ & Mapped AZ class & TRS &
 PHS & $\Gamma$ & Mapped $t$ & Mapped symmetry \\
\hline
\multirow{4}{*}{BDI/CII}  & 0 & $U^+_{++}$ &$\bm{k}_{\perp}$ & \multirow{4}{*}{D/C} & & \multirow{4}{*}{$C$} & & 0 & $U^+_{+} = U^+_{++}$ \\
 & 1 & $\widebar U^+_{-+}$ &$\bm{k}_{\perp}$& & & & & 1 & $\widebar U^+_{+} = \widebar U^+_{-+}$ \\
 & 2 & $U^+_{--}$ & $\bm{k}_{\perp}$& & & & & 2 & $U^+_{-} = U^+_{--}$ \\
 & 3 & $\widebar U^+_{+-}$ & $\bm{k}_{\perp}$& && & & 3 & $\widebar U^+_{-} = \widebar U^+_{+-}$ \\
\hline
\multirow{4}{*}{BDI/CII} & 0 & $U^+_{++}$ & $\bm{k}_{\parallel}$& \multirow{4}{*}{D/C} & & \multirow{4}{*}{$C$} & & 1 & $\widebar U^+_{+} = \Gamma U^+_{++}$ \\
 & 1 & $\widebar U^+_{-+}$ & $\bm{k}_{\parallel}$& & & & & 2 & $U^+_{-} = i \Gamma \widebar U^+_{-+}$ \\
 & 2 & $U^+_{--}$ & $\bm{k}_{\parallel}$& & & & & 3 & $\widebar U^+_{-} = \Gamma U^+_{--}$ \\
 & 3 & $\widebar U^+_{+-}$ & $\bm{k}_{\parallel}$& & & & & 0 & $U^+_{+} = i \Gamma \widebar U^+_{+-}$ \\
\hline
\multirow{4}{*}{BDI/CII} & 0 & $U^+_{++}$ & $\bm{r}_{\perp}$ & \multirow{4}{*}{AI/AII} & \multirow{4}{*}{$T$} & & & 0 & $U^+_{+} = U^+_{++}$ \\
 & 1 & $\widebar U^+_{-+}$ &$\bm{r}_{\perp}$& & & & & 1 & $\widebar U^+_{-} = \widebar U^+_{-+}$ \\
 & 2 & $U^+_{--}$ & $\bm{r}_{\perp}$& & & & & 2 & $U^+_{-} = U^+_{--}$ \\
 & 3 & $\widebar U^+_{+-}$ & $\bm{r}_{\perp}$& && &  & 3 & $\widebar U^+_{+} = \widebar U^+_{+-}$ \\
\hline
\multirow{4}{*}{BDI/CII} & 0 & $U^+_{++}$ & $\bm{r}_{\parallel}$& \multirow{4}{*}{AI/AII} & \multirow{4}{*}{$T$} & & & 3 & $\widebar U^+_{+} = \Gamma U^+_{++}$ \\
 & 1 & $\widebar U^+_{-+}$ &$\bm{r}_{\parallel}$& & & & & 0 & $U^+_{+} = i \Gamma \widebar U^+_{-+}$ \\
 & 2 & $U^+_{--}$ & $\bm{r}_{\parallel}$& & & & & 1 & $\widebar U^+_{-} = \Gamma U^+_{--}$ \\
 & 3 & $\widebar U^+_{+-}$ & $\bm{r}_{\parallel}$& & & &  & 2 & $U^+_{-} = i \Gamma \widebar U^+_{+-}$ \\
\hline\hline
\multirow{4}{*}{DIII/CI} & 0 & $U^+_{++}$ &$\bm{k}_{\perp}$& \multirow{4}{*}{AII/AI} & \multirow{4}{*}{$T$} & & & 0 & $U^+_{+} = U^+_{++}$ \\
 & 1 & $\widebar U^+_{-+}$ & $\bm{k}_{\perp}$& & & & & 1 & $\widebar U^+_{-} = \widebar U^+_{-+}$ \\
 & 2 & $U^+_{--}$ & $\bm{k}_{\perp}$& & & & & 2 & $U^+_{-} = U^+_{--}$ \\
 & 3 & $\widebar U^+_{+-}$ &$\bm{k}_{\perp}$& && &  & 3 & $\widebar U^+_{+} = \widebar U^+_{+-}$ \\
\hline
\multirow{4}{*}{DIII/CI} & 0 & $U^+_{++}$ & $\bm{k}_{\parallel}$& \multirow{4}{*}{AII/AI} & \multirow{4}{*}{$T$} & & & 1 & $\widebar U^+_{-} = \Gamma U^+_{++}$ \\
 & 1 & $\widebar U^+_{-+}$ & $\bm{k}_{\parallel}$& & & & & 2 & $U^+_{-} = i \Gamma \widebar U^+_{-+}$ \\
 & 2 & $U^+_{--}$ & $\bm{k}_{\parallel}$& & & & & 3 & $\widebar U^+_{+} = \Gamma U^+_{--}$ \\
 & 3 & $\widebar U^+_{+-}$ & $\bm{k}_{\parallel}$& & & &  & 0 & $U^+_{+} = i \Gamma \widebar U^+_{+-}$ \\
\hline
\multirow{4}{*}{DIII/CI} & 0 & $U^+_{++}$ & $\bm{r}_{\perp}$& \multirow{4}{*}{D/C} & & \multirow{4}{*}{$C$} & & 0 & $U^+_{+} = U^+_{++}$ \\
 & 1 & $\widebar U^+_{-+}$ & $\bm{r}_{\perp}$& & & & & 1 & $\widebar U^+_{+} = \widebar U^+_{-+}$ \\
 & 2 & $U^+_{--}$ & $\bm{r}_{\perp}$& & & & & 2 & $U^+_{-} = U^+_{--}$ \\
 & 3 & $\widebar U^+_{+-}$ & $\bm{r}_{\perp}$& && &  & 3 & $\widebar U^+_{-} = \widebar U^+_{+-}$ \\
\hline
\multirow{4}{*}{DIII/CI} & 0 & $U^+_{++}$ & $\bm{r}_{\parallel}$& \multirow{4}{*}{D/C} & & \multirow{4}{*}{$C$} & & 3 & $\widebar U^+_{-} = \Gamma U^+_{++}$ \\
 & 1 & $\widebar U^+_{-+}$ & $\bm{r}_{\parallel}$& & & & & 0 & $U^+_{+} = i \Gamma \widebar U^+_{-+}$ \\
 & 2 & $U^+_{--}$ & $\bm{r}_{\parallel}$& & & & & 1 & $\widebar U^+_{+} = \Gamma U^+_{--}$ \\
 & 3 & $\widebar U^+_{+-}$ & $\bm{r}_{\parallel}$& & & &  & 2 & $U^+_{-} = i \Gamma \widebar U^+_{+-}$ \\
\hline \hline
\end{tabular}
\label{Tab_raise_real_chiral}
\end{center}
\end{table*}

\section{Topological classification of Fermi points with additional symmetry }
\label{sec:Fermi_point}

\subsection{K-group of Fermi points}

So far, we have argued topological classification of crystalline insulators and
 superconductors and 
 their topological defects in the presence of an additional order-two
 symmetry. 
In this section, we will show that a similar but a slightly different
 argument works for classification of topological stable Fermi points
 in the momentum space.

The topological classification of Fermi points is done by the
homotopy classification of Hamiltonians
$\mathcal{H}({\bm \kappa})$ where ${\bm \kappa}=(\kappa_1,\kappa_2,
\cdots, \kappa_d)$ is coordinates of a 
$d$-dimensional sphere $S^d$ surrounding Fermi points in the momentum
space.
Since the Hamiltonian ${\cal H}({\bm \kappa})$ defines a map from $S^d$
to a classifying space of topological
insulators, a similar K-group
argument applies to the classification of Fermi
points eventually.
However, as is shown below, the application is not straightforward but a careful
treatment of symmetry is needed. 

The obstruction encountered here is non-trivial transformation of ${\bm
\kappa} \in S^d$ under symmetry:
Consider a
Fermi point located at the origin in $(d+1)$-dimensions.
A $d$-dimensional sphere $S^{d}$, which is defined as ${\bm
k}^2=k_1^2+k_2^2+\cdots k_{d+1}^2=\epsilon^2$, 
encloses the Fermi point.
Although ${\bm k}$ transforms as ${\bm k}\rightarrow 
-{\bm k}$ under TRS and/or PHS, 
any $d$-dimensional coordinates ${\bm \kappa}=(\kappa_1,\dots,\kappa_d)$ of
$S^d$ does not transform such a simple way.
Therefore, one cannot directly apply our arguments so far to the Fermi points.

The key to resolve this difficulty is the dimension-raising map
introduced in Appendix \ref{sec:dimension-raising}:
Formally, one can raise the dimension of the surrounding $d$-dimensional
sphere, and map isomorphically a Hamiltonian
${\cal H}({\bm \kappa})$ on $S^d$ into ${\cal H}({\bm \kappa},
\kappa_{d+1})$ on $S^{d+1}$. 
Then, topological classification of the original Hamiltonian
${\cal H}({\bm \kappa})$ reduces to that of the mapped Hamiltonian
${\cal H}({\bm \kappa}, \kappa_{d+1})$, which is found to be done in
the framework developed so far.

The map from ${\cal H}({\bm \kappa})$ to ${\cal H}({\bm \kappa},
\kappa_{d+1})$ is constructed as follows.
If the original Hamiltonian supports CS $\Gamma$, then
the map is
\begin{eqnarray}
{\mathcal{H}}_{\rm nc}(\bm{\kappa}, \kappa_{d+1}=\theta)
= \sin\theta \mathcal{H}_{\rm c}(\bm{\kappa}) +
\cos\theta \Gamma, 
\quad \theta\in[0,\pi]
\label{eq:map1}
\end{eqnarray}
and if not, then
\begin{eqnarray}
{\mathcal{H}}_{\rm c}(\bm{\kappa}, \kappa_{d+1}=\theta)
= \sin \theta\mathcal{H}_{\rm nc}(\bm{\kappa})
\otimes \tau_z 
+ \cos\theta {\bm 1} \otimes \tau_y,
\quad \theta\in [0,\pi]
\label{eq:map2}
\end{eqnarray}
where the
subscripts ${\rm c}$ and ${\rm nc}$ of ${\cal H}$ denote the presence
and  absence of CS, respectively.
(The chiral operator of the latter Hamiltonian ${\cal H}_{\rm c}({\bm \kappa},
\kappa_{d+1})$ is given by ${\bm 1}\otimes \tau_x$.)  
Since ${\cal H}({\bm \kappa}, \kappa_{d+1})={\rm const.}$ at
$\kappa_{d+1}=0$ and $\kappa_{d+1}=\pi$, the $d$-dimensional
sphere ${\bm \kappa}\in S^d$ can be
contracted into a point either at $\kappa_{d+1}=0$ and $\kappa_{d+1}=\pi$.
The resultant space of $({\bm \kappa}, \kappa_{d+1})$ is identified as a
$d+1$-dimensional sphere $S^{d+1}$ where ${\bm \kappa}\in
S^{d}$ and $\kappa_{d+1}$ parametrize the ``circles of latitude'' and the
``meridian'' of the $d+1$-dimensional sphere, respectively, and
$\kappa_{d+1}=0$ and $\kappa_{d+1}=\pi$
point to the ``north and south poles''.
Because the
inverse map also can be constructed in the meaning of stable equivalence, 
as explained in Appendix \ref{sec:dimension-lowering},
the topological nature of the mapped Hamiltonian is the same
as that of the original one.

Let us now define TRS and/or PHS in the mapped Hamiltonian.
To define them, we need to determine the transformation law of the new variable
$\kappa_{d+1}$ under these symmetries, since $\kappa_{d+1}$ is
an artificial variable, and thus no a priori transformation law is given.
A convenient way is to treat the new variable 
$\kappa_{d+1}$ as ${\bm r}$-type, which means that $\kappa_{d+1}$ is
invariant under TRS and/or PHS.


From the construction, it is evident that the
TRS and/or PHS for the original Hamiltonian induce
a two fold rotation of the $d+1$-dimensional sphere:
If one represents the $d+1$-dimensional sphere $S^{d+1}$ as
$k_1^2+k_2^2\cdots+k_{d+1}^2+k_{d+2}^2=\epsilon^2$,
TRS and/or PHS act as $(k_1,k_2,\cdots, k_{d+1}, k_{d+2}) 
\rightarrow (-k_1,-k_2,\cdots, -k_{d+1}, k_{d+2})$ in a suitable basis.
Then, the following new reparametrization of $S^{d+1}$
\begin{eqnarray}
\kappa_i=\frac{k_i}{\epsilon+k_{d+2}},
\quad (i=1,\cdots, d+1),
\end{eqnarray}
simplifies the transformation law of $({\bm \kappa}, \kappa_{d+1})$ as
$({\bm \kappa}, \kappa_{d+1})\rightarrow (-{\bm \kappa}, -\kappa_{d+1})$.
Therefore, the mapped Hamiltonian is categorized as a
Hamiltonian discussed in Sec. \ref{Main}.

Here note that the mapped Hamiltonian ${\cal H}({\bm \kappa},
\kappa_{D+1})$ supports a different set of AZ symmetries than the
original one since
it loses or obtains CS.
With a careful analysis of the symmetry, we find that the
dimension-raising map shifts $s$ 
of the K-groups of the original Hamiltonians by $-1$.
Therefore, denoting the K-group of the Fermi point in AZ class $s$ as
$K^{\mathrm{FP}}_{\mathbb{F}}(s,d)$, 
we obtain 
\begin{eqnarray}
K^{\mathrm{FP}}_{\mathbb{F}}(s,d) = K_{\mathbb{F}}(s-1,d+1), 
\quad
(\mathbb{F}=\mathbb{C}, \mathbb{R}),
\label{eq:FermiK}
\end{eqnarray}
where the right hand side is {\it the K-group of topological
insulator and superconductors in AZ class $s$}.  
This relation reproduces the previous classification of
the Fermi points by Ho{\v{r}}ava \cite{hovrava2005stability} and
Zhao-Wang. \cite{zhao2013topological, zhao2014topological}

We can also classify Fermi points stabilized by an additional symmetry
besides AZ symmetries:
Under the assumption that the Fermi points are enclosed by a
$d$-dimensional sphere $S^d$ and they are invariant under the symmetries we
consider, the K-groups for the Fermi points can be related to {\it the
K-groups for
$d+1$ dimensional topological crystalline insulators and superconductors
in the presence of an
additional symmetry}: 
\begin{eqnarray}
&&K^{U;\mathrm{FP}}_{\mathbb{C}}(s,t;d, d_{\parallel}) 
= K^{U}_{\mathbb{C}}(s-1,t;d+1,d_{\parallel}, 0, 0), 
\nonumber\\
&&K^{A;\mathrm{FP}}_{\mathbb{C}}(s;d, d_{\parallel}) 
= K^A_{\mathbb{C}}(s-1;d+1,d_{\parallel},0,0), 
\nonumber\\
&&K^{U/A;\mathrm{FP}}_{\mathbb{R}}(s,t;d, d_{\parallel}) 
= K^{U/A}_{\mathbb{R}}(s-1,t;d+1,d_{\parallel},0,0), 
\label{eq:FermiK_additional}
\end{eqnarray}
where $d_{\parallel}$ is the number of flipped momenta under the
additional symmetry.

\subsection{Bulk-boundary correspondence of K-groups}

Equations (\ref{eq:FermiK}) and (\ref{eq:FermiK_additional}) provide
a novel realization of the bulk-boundary correspondence, 
\cite{hatsugai1993, volovik2003universe, essin2011bulk,
sato2011topology, zhao2014topological}
in terms of the
K-theory. 
First, consider Eq.(\ref{eq:FermiK}). 
From the dimensional hierarchy in Eqs. (\ref{eq:KisoC}) and (\ref{eq:KisoR}), 
Eq.(\ref{eq:FermiK}) is recast into 
\begin{eqnarray}
K^{\mathrm{FP}}_{\mathbb{F}}(s,d) = K_{\mathbb{F}}(s,d+2).
\label{K^FP_F}
\end{eqnarray}
The relation (\ref{K^FP_F}) is nothing but the bulk-boundary
correspondence:  
While the right hand side provides a bulk topological number
of a $(d+2)$-dimensional insulator or superconductor, the left hand
side ensures the existence of topologically stable surface Fermi points enclosed
by a sufficiently large $S^d$ in the $(d+1)$-dimensional surface momentum space.

In a similar manner, we can obtain the bulk-boundary correspondence of
the K-group in the presence of an additional symmetry. 
From the dimensional hierarchy Eqs. (\ref{KCU2}) and (\ref{KR2}) in the
presence of additional symmetry,
we obtain
\begin{eqnarray}
&&K^{U;\mathrm{FP}}_{\mathbb{C}}(s,t;d, d_{\parallel}) 
= K^{U}_{\mathbb{C}}(s,t;d+2,d_{\parallel}, 0, 0), 
\nonumber\\
&&K^{A;\mathrm{FP}}_{\mathbb{C}}(s;d, d_{\parallel}) 
= K^A_{\mathbb{C}}(s; d+2,d_{\parallel},0,0), 
\nonumber\\
&&K^{U/A;\mathrm{FP}}_{\mathbb{R}}(s,t;d, d_{\parallel}) 
= K^{U/A}_{\mathbb{R}}(s,t;d+2,d_{\parallel},0,0),
\label{eq:K^FP_additional}
\end{eqnarray}
where the right hand sides represent bulk $(d+2)$-dimensional topological
numbers of topological crystalline insulators and superconductors and
the left hand sides give $d$-dimensional topological numbers of
the corresponding $(d+1)$-dimensional surface states.
Both topological numbers ensure the stability of topological crystalline
phases.

Note that the number of the flipped momentum $d_{\parallel}$ is
the same in the both sides of Eq.(\ref{eq:K^FP_additional}). 
Otherwise, the boundary breaks the
additional symmetry in the bulk, and thus the bulk-boundary
correspondence does not hold anymore.


\subsection{Inversion symmetric Fermi points}

To obtain the bulk-boundary correspondence, at least one-direction in
the bulk should not be flipped under the additional symmetry.
Indeed, if this happens, surfaces normal to the non-flipped direction preserve
the additional symmetry.
This condition implies that Eq.(\ref{eq:K^FP_additional}) holds only when
$d_{\parallel}\le d+1$. 

Here note that the possible $d_{\parallel}$ can be larger than $d$,
i.e. $d_{\parallel}=d+1$.
In this case, the left hand side of Eq.(\ref{eq:K^FP_additional})
implies that  
the number of flipped coordinates of $S^d$ surrounding
Fermi points becomes larger than the total dimension $d$ of $S^d$. 
This can be understood as follows.
As was mentioned above, the bulk-boundary correspondence holds only for the
surface normal to a non-flipped direction of the additional symmetry.
Therefore, when $d_{\parallel}=d+1$, the additional symmetry flips
all directions parallel to the surface.
In other words, the additional symmetry induces inversion ${\bm
k}\rightarrow -{\bm k}$ on the surface.
In a manner similar to TRS and PHS, while the $d$-dimensional sphere
surrounding Fermi points,
${\bm k}^2\equiv k_1^2+k_2^2+\cdots+k_{d+1}^2=\epsilon^2$, preserves the
inversion symmetry, any $d$-dimensional coordinates ${\bm \kappa}$ of
$S^d$ transforms nontrivially under the inversion. This makes it
possible to  realize
$d_{\parallel}>d$.  
As well as TRS and PHS, 
the dimension-raising is needed to obtain a
simple transformation law of the surrounding coordinates. 

We notice that such an inversion symmetric Fermi point may support a
topological number in a unusual manner.
For example, consider an inversion symmetric Fermi point in class AIII
with $d=0$, $d_{\parallel}=1$ and $U_+$. 
From Eq.(\ref{eq:K^FP_additional}), the relevant K-group
$K_{\mathbb{C}}^{U;FP}(1,0;0,1)$
is evaluated as
$K_{\mathbb{C}}^U(1,0; 2,1,0,0)=\pi_0({\cal C}_0)=\mathbb{Z}$. 
Therefore, the Fermi point can be topologically stable.
Indeed, such a topologically stable Fermi point is realize in the
following model
\begin{eqnarray}
{\cal H}(k)=\sigma_x k,
\label{eq:model1_inversion}
\end{eqnarray}
with the chiral operator $\Gamma=\sigma_z$ and inversion operator
$U=\sigma_z$,
\begin{eqnarray}
\{
\Gamma, {\cal H}  
\}=0,
\quad
U{\cal H}(k)U^{\dagger}={\cal H}(-k). 
\end{eqnarray} 
The energy of this model is given by $E(k)=\pm k$, and thus 
there exists an inversion symmetric Fermi point at $k=0$. 
Although the Fermi point can be gapped by the mass terms $m\sigma_y$
and $m'\sigma_z$, these terms are not allowed by the chiral and
inversion symmetries. Therefore, the Fermi point is symmetry-protected.
The Hamiltonian of the Fermi point is given by
\begin{eqnarray}
{\cal H}(\kappa_0=\pm)=\pm \sigma_z,
\end{eqnarray}
where the ``sphere'' surrounding the Fermi point consists
of just two points $\kappa_0=\pm$ in the present case. 

To calculate the topological number of this class of model, we use the
Hamiltonian mapped by the dimension raising,
\begin{eqnarray}
{\cal H}(\kappa_0=\pm, \kappa_1=\theta)=\sin\theta {\cal
 H}(\kappa_0=\pm)+\cos\theta \Gamma,
\quad 
\theta\in[0,\pi]. 
\end{eqnarray}
Inversion of the original Hamiltonian induces the following
additional symmetry,
\begin{eqnarray}
U{\cal H}(+, \theta)U^{\dagger}={\cal H}(-, \theta). 
\end{eqnarray}
Since the mapped Hamiltonian commutes with $U$ at the high-symmetric points
of this symmetry, i.e. at $\theta=0, \pi$, the energy eigenstates at
these points are
decomposed into two subsets with different eigenvalues of $U$.
Then, we can introduce a topological number by
\begin{eqnarray}
N=\frac{(N_+(0)-N_-(0))-(N_+(\pi)-N_-(\pi))}{2} = N_+(0)-N_+(\pi), 
\end{eqnarray}
where $N_{\pm}$ is the number of negative energy states with the
eigenvalue $U=\pm$.     
We find that $N=1$ in the above model (\ref{eq:model1_inversion}), 
which ensures topological stability of the Fermi point at $k=0$.

Here note that at the high-symmetric points, the mapped Hamiltonian
reduces to the chiral operator $\pm \Gamma$ of the original Hamiltonian.
Therefore, in contrast to ordinary topological numbers, the topological
number of the inversion symmetric Fermi point is not directly 
evaluated from the original Hamiltonian ${\cal H}(\kappa_0=\pm)$, but it
is implicitly encoded in the chiral operator $\Gamma$.  

Now let see how the topological number of the chiral operator stabilizes
the Fermi points.
In general, a Fermi point of this class is described by the following
Dirac Hamiltonian
\begin{eqnarray}
{\cal H}=\gamma k, 
\label{eq:model2_inversion}
\end{eqnarray}
with the chiral operator $\Gamma$ and the inversion $U$
\begin{eqnarray}
\{
\Gamma, {\cal H}  
\}=0,
\quad
U{\cal H}(k)U^{\dagger}={\cal H}(-k),
\quad 
[U, \Gamma]=0. 
\label{eq:Fermi_inversion_sym}
\end{eqnarray} 
If the Fermi point at $k=0$ is topologically unstable, then there exists a mass
term $M$ consistent with Eq.(\ref{eq:Fermi_inversion_sym}).
As the mass term satisfies
\begin{eqnarray}
\{\gamma, M\}=0,
\quad
\{\Gamma, M\}=0,
\quad
[U, M]=0, 
\end{eqnarray} 
it defines an extra CS $\Gamma'$ by
$\Gamma'=M/\sqrt{M^2}$.
The existence of the extra CS, however, implies that $N$ of
$\Gamma$ must be zero. Actually, 
using $\Gamma'$, 
one can interpolate $\Gamma(0)=\Gamma$ and $\Gamma(\pi)=-\Gamma$ smoothly by
$\Gamma(t)=\Gamma \cos t+ \Gamma' \sin t$,
which means $N=0$
since $\Gamma$ and $-\Gamma$ have an opposite topological number.
As a result, we can conclude that the topological number must be zero to
obtain a gap of the Fermi point.

\section{Majorana Ising spin character as a result of topological
 crystalline superconductivity}
\label{sec:Majorana_Ising}

In spinful superconductors or superfluids, zero energy (or gapless)
modes often show an anisotropic response to magnetic
fields.\cite{sato2009topological,chung2009detecting,nagato2009,shindou2010,
volovik2010, mizushima2011, silaev2011, mizushima2012symmetry, mizushima2013topological} 
Here we show that these anisotropic behaviors, which are called Majorana Ising
spin character,\cite{chung2009detecting} is a
result of symmetry protected topological phase.\cite{mizushima2012symmetry}
As is discussed below, the Ising spin character offers a new
mechanism for stability of zero energy modes
against disorders.

Let us consider $N$ zero modes $|u_0^{(a)}\rangle$ $(a=1,2,\dots,
N)$ localized on a defect or a boundary in
a superconductor or superfluid.  
Because of PHS, we can place the following condition,   
\begin{eqnarray}
C|u_0^{(a)}\rangle=|u_0^{(a)}\rangle,
\quad C=\tau_x {\cal K}. 
\label{eq:MajoranaM}
\end{eqnarray}
Then, in order to introduce quantum operators for the zero modes, we perform the
mode expansion of the quantized field $\hat{\bm \Psi}({\bm
x})=(\hat{\psi}_{\uparrow}({\bm x}), \hat{\psi}_{\downarrow}({\bm x}), 
\hat{\psi}^{\dagger}_{\uparrow}({\bm x}), 
\hat{\psi}^{\dagger}_{\downarrow}({\bm x}))^T$. 
Ignoring nonzero energy modes, $\hat{\bm \Psi}({\bm x})$ is expanded as 
\begin{eqnarray}
\hat{\bm \Psi}({\bm x})=\sum_a \gamma^{(a)}|u_0^{(a)}\rangle.
\label{eq:expansionM}
\end{eqnarray}
Since Eq. (\ref{eq:MajoranaM}) implies that the coefficients
$\gamma^{(a)}$ are real (self-conjugate), $\gamma^{(a)}$ represents a Majorana operator.
Furthermore, the anticommutation relation of 
$\hat{\bm
\Psi}$ leads to the Majorana relation
$\{\gamma^{(a)},
\gamma^{(b)}\}=2\delta_{ab}$, under a suitable normalization of
$|u_0^{(a)}\rangle$.

The local density operator and the spin density operators of the Majorana zero
modes are given by
\begin{eqnarray}
\rho({\bm x})\equiv \frac{1}{2}
\left[\hat{\psi}_{s}^{\dagger}({\bm x})\hat{\psi}_s({\bm x})-
\hat{\psi}_{s}({\bm x})\hat{\psi}_s^{\dagger}({\bm x})
\right],
\quad
S_i({\bm x})\equiv \frac{1}{4}
\left[\hat{\psi}^{\dagger}_s({\bm x})(\sigma_i)_{ss'}\hat{\psi}_{s'}({\bm x})
-\hat{\psi}_s({\bm x})(\sigma_i^t)_{ss'}\hat{\psi}^{\dagger}_{s'}({\bm x})
\right]. 
\end{eqnarray}
Substituting $\hat{\psi}_s$ and $\hat{\psi}_s^{\dagger}$ of Eq.(\ref{eq:expansionM})
into this, we obtain
\begin{eqnarray}
\rho({\bm
 x})=\frac{1}{2}\sum_{a,b}\left[
\gamma^{(a)},\gamma^{(b)}
\right]
u_{0,s}^{(a)*}({\bm x}) u_{0,s}^{(b)}({\bm x}),
\quad 
S_i({\bm x})=\frac{1}{4}\sum_{a,b}
\left[\gamma^{(a)},\gamma^{(b)}\right]
u_{0,s}^{(a)*}({\bm x})(\sigma_i)_{ss'}u_{0,s'}^{(b)}({\bm x}),
\label{eq:densityM}
\end{eqnarray} 
where $u_{0,s}^{(a)}({\bm x})$ represents the zero mode $|u_0^{(a)}\rangle$ in
the coordinate space.
The orthogonality condition of the zero modes implies that the
total density of the Majorana zero modes vanishes
\begin{eqnarray}
\int d{\bm x}\rho({\bm x})=0.
\end{eqnarray} 
When $N=1$, Eq.(\ref{eq:densityM}) also yields that $\rho({\bm x})$ and
$S_i({\bm x})$ are identically zero. 

When the system supports more than one Majorana zero mode (i.e. $N\ge
2$), the local operators $\rho({\bm x})$ and $S_i({\bm x})$
are not identically zero, in general.
Nevertheless, if the system hosts an additional antiunitary symmetry, $\rho({\bm
x})$ can vanish identically . 
For example, if the system has time-reversal symmetry with $T^2=-1$,  
Majorana zero modes form Kramer's pairs.
Thus if only a single pair of Majorana zero modes exist in the system,
the Kramers theorem implies that $\rho({\bm x})$ in Eq.(\ref{eq:densityM}) is
identically zero.

We here specify the antiunitary symmetry relevant to the
Majorana Ising character.
It is 
the magnetic point group
symmetry that is obtained by combining time-reversal $T$
and either of two-fold spin-rotation, mirror
reflection, or two-fold rotation.
Since the latter operations commonly rotate the spin by $\pi$,
the combined antiunitary symmetry acts on the spin space as
\begin{eqnarray}
A_{\rm spin}=s_a s_y {\cal K}, 
\label{eq:sasyK}
\end{eqnarray}
if the spin rotation axis is the $a$-direction.
For two-fold spin-rotation or two-fold rotation, the spin rotation axis
is obvious, and for the mirror reflection, it is normal to the mirror
reflection plane.  
To keep the antiunitary symmetry in the superconducting state, the gap
function must be even or odd under the antiunitary symmetry,
\begin{eqnarray}
A_{\rm spin}\Delta({\bm k})(A_{\rm spin}^*)^{-1}=\pm \Delta({\bm k}_{\parallel}, -{\bm
 k}_{\perp}),  
\end{eqnarray}
then in these cases, the BdG Hamiltonian 
\begin{eqnarray}
{\cal H}({\bm k})=\left(
\begin{array}{cc}
\epsilon({\bm k}) & \Delta({\bm k})\\
\Delta^{\dagger}({\bm k}) & -\epsilon^{T}(-{\bm k})
\end{array}
\right) 
\end{eqnarray}
retains the antiunitary
symmetry that acts on the BdG Hamiltonian in the following form,
\begin{eqnarray}
A=
\left(
\begin{array}{cc}
A_{\rm spin} & 0\\
0 & \pm A_{\rm spin}^*
\end{array}
\right),
\label{eq:magneticgroupM}
\end{eqnarray}
which can be rewritten as $A=s_as_y\tau_0{\cal K}$ or $A=s_as_y\tau_z
{\cal K}$. 
Since Eq. (\ref{eq:magneticgroupM})
satisfies $A^2=1$ and it commutes or anticommutes with PHS $C$, it is labeled as $A^{+}_{\pm}$.
%
%
%
%

When the system supports $A^+_{+}$ ($A^+_-$) symmetry, a CS $\Gamma$
can be introduced as combined symmetry of $A^{+}_+$ ($A^+_-$) and
PHS $C$, i.e.
$
\Gamma=C A_+^+ 
$
($\Gamma=iC A_-^+$). 
The CS defines an
integer topological number when $\delta=\delta_{\parallel}+1$.
If the topological number is $N$, the bulk-boundary
correspondence ensures the existence of the $N$ zero
energy states $|u_0^{(a)}\rangle$, which wave functions have a definite
eigenvalue of $\Gamma$, say, $\Gamma=1$,
\footnote{In general, 
the index theorem allows more than $N$ zero energy states if
some of them have the opposite eigenvalue,\cite{sato2011topology} i.e. $\Gamma=-1$ in the above
case, but extra zero energy states can be
gapped easily by small perturbation and only $N$ zero energy modes are
topologically stable.
}
\begin{eqnarray}
\Gamma|u_0^{(a)}\rangle=|u_0^{(a)}\rangle,
\quad
(a=1,2,\dots, N), 
\label{eq:chiralM}
\end{eqnarray} 
at the position of the defect or surface.
Note that one can place Eq.(\ref{eq:MajoranaM}) at the same time because
$\Gamma$ commutes with $C$.

%
%
%
Now we show that Eqs. (\ref{eq:MajoranaM}) and (\ref{eq:chiralM})
determine the spin structures of the zero modes.
%
%
These equations require that 
the components of $|u_0^{(a)}\rangle$ in the spin and
Nambu spaces are related to each other, implying
that the components of the quantized field $\hat{\bm \Psi}({\bm
 x})$ are dependent as well.
%
%
For instance, for $A=s_y^2\tau_0{\cal K}$, 
Eqs.(\ref{eq:MajoranaM}) and (\ref{eq:chiralM}) yield that
the zero modes
have a generic form as $|u_0^{(a)}\rangle=(\alpha^{(a)}_{\uparrow}, 
\alpha^{(a)}_{\downarrow}, \alpha^{(a)}_{\uparrow}, 
\alpha^{(a)}_{\downarrow})$ with real
functions $\alpha^{(a)}_{\uparrow, \downarrow}$.
Then substituting this into
Eq.(\ref{eq:expansionM}), we have
 $\hat{\psi}_{\uparrow}=\hat{\psi}_{\uparrow}^{\dagger}$ and 
$\hat{\psi}_{\downarrow}=\hat{\psi}_{\downarrow}^{\dagger}$, from which
one can show
that only $S_y$ is nonzero and the other density operators vanish.
In a similar manner, $\rho$ and $S_i$ for $A=s_as_y\tau_0{\cal K}$ or
$A=s_as_y\tau_z {\cal K}$ are evaluated as
\begin{eqnarray}
\rho=
S_{i\neq a}=0,
\quad 
S_a\neq 0
\label{eq:MajoranaIsing}
\end{eqnarray}
at the position of the zero modes.

Equation (\ref{eq:MajoranaIsing}) indicates that the Majorana zero modes
considered here couple to Zeeman magnetic fields only in
a particular direction, namely in the $a$-direction.
Our arguments presented here clarify that this Ising spin
character of Majorana zero modes is originated from the topological
phase protected by the magnetic point group symmetry in the above.

Although the magnetic point group symmetry is broken by
disorders,  
the Ising character implies that the Majorana zero modes can survive
even in the presence of non-magnetic disorders: Since the local density
of the Majorana zero modes $\rho$ vanishes at the position of the zero
modes, the coupling between the zero modes and non-magnetic disorders
are strongly suppressed. As a result, the Majorana zero modes remains to
be (nearly) zero modes even in the presence of non-magnetic disorders.







\section{Anomalous topological pump}
\label{sec:pump}

Recently, Zhang and Kane have discussed topological classification of
adiabatic pump cycles in Josephson junctions of time-reversal
invariant superconductors. 
They found that adiabatic parameters of the pump cycles, such as the phase
difference $\phi$ of the Josephson junction, may have a mixed behavior under
TRS and PHS, leading to new topological classes. \cite{zhang2013anomalous}
We argue here that our present framework is also applicable to such systems.

In the adiabatic pumps,  two different types of anomalous
parameters may appear. The first one ${\bm \phi}$ is odd (even) under TRS (PHS)
and the second one ${\bm \theta}$ is even (odd) under TRS (PHS).
In both types, unlike ${\bm k}$ and ${\bm r}$, the anomalous
parameters are odd under CS.
Since such an anomalous CS is not taken into account as the
original AZ symmetry, relevant topological phases are not
included in the original periodic table.

Our classification is naturally applicable to even such phases.
The anomalous CS  
\begin{eqnarray}
\Gamma \mathcal{H}({\bm k}, {\bm r}, {\bm \phi}, {\bm \theta})\Gamma^{-1}=
-\mathcal{H}({\bm k}, {\bm r}, -{\bm \phi}, -{\bm \theta})
\end{eqnarray}
with ${\bm k}=(k_1,\dots, k_{d_k})$, ${\bm r}=(r_1,\dots, d_{d_r})$,
${\bm \phi}=(\phi_1,\dots,\phi_{d_\phi})$ and
${\bm \theta}=(\phi_1,\dots,\phi_{d_\theta})$ 
is identified with an order-two antisymmetry $\bar{U}$ with
$d=d_r+d_k+d_{\phi}+d_{\theta}$,
$d_{\parallel}=d_{\phi}+d_{\theta}$, $D=0$ and $D_{\parallel}=0$.  
Therefore, its K-group is given by
$
K_{\mathbb{C}}^U(0,1;d_k+d_r+d_\phi+d_\theta,d_\phi+d_\theta, 0, 0).
$
Considering two-fold periodicity in $s$ and $t$ of
$K_{\mathbb{C}}^U(s,t;d, d_{\parallel}, D, D_{\parallel})$, 
we find that Eqs.(\ref{KCU}) and (\ref{KCU_classifying_space}) reproduce Table I of
Ref.\onlinecite{zhang2013anomalous}.   

In Josephson junctions, the anomalous CS is realized as the 
combination of the following TRS and PHS, 
\begin{eqnarray}
&&C\mathcal{H}({\bm k}, {\bm r}, {\bm \phi}, {\bm \theta})C^{-1}=
-\mathcal{H}(-{\bm k}, {\bm r}, {\bm \phi}, -{\bm \theta}),
\nonumber\\
&&T\mathcal{H}({\bm k}, {\bm r}, {\bm \phi}, {\bm \theta})T^{-1}=
\mathcal{H}(-{\bm k}, {\bm r}, -{\bm \phi}, {\bm \theta}).
\end{eqnarray}
These combinations are not allowed in the standard AZ classification
again, so either TRS or PHS is anomalous, but it can be handled in our
framework.
A possible identification of these symmetries in our framework is that
$T$ is the standard TRS but $C$ is the antiunitary antisymmetry $\bar{A}_{+}^+$ with
$d_k=d_{\perp}$,  $d_r=D_{\perp}$, $d_\phi=d_{\parallel}$ and
$d_\theta=D_\parallel$.
The K-group is given by $K_{\mathbb{R}}^{A}(4,1;d_k+d_\phi,
d_r+d_\theta, d_\phi, d_\theta)$, which reproduces the
$\mathbb{Z}_2\times\mathbb{Z}_2$ structure of Josephson
effects.\cite{zhang2013anomalous}

\section{Conclusion}
\label{Conc}
In this paper, we present a topological classification of crystalline
insulators and superconductors and their topological defects that support
order-two additional symmetry, besides AZ symmetries.  
The additional symmetry includes spin-rotation, reflection,
$\pi$-rotation, and inversion. 
Their
magnetic point group symmetries are also included.  
Using the dimensional hierarchy of K-groups,
we can reduce the topological classification of Hamiltonians into that of
simple matrices in zero-dimension.
The obtained results are summarized in Eqs. (\ref{KCU}), (\ref{eq:KCA})
and (\ref{KR}).
These K-groups suggest that defect zero modes can be considered as
boundary states of lower-dimensional crystalline insulators and superconductors.
We also classify Fermi points stabilized by the additional symmetry, and
derive the K-theory version of the bulk-boundary correspondence.
Various symmetry protected topological phases and gapless modes are
identified and discussed in a unified framework.


While we have completed a topological classification of crystalline
insulators and superconductors with order-two additional
symmetry, 
the full classification of topological crystalline
insulators and superconductors has not been yet done.
General crystalline symmetries admit higher-order symmetries such as
 $C_n$-rotation $(n=3,4,6)$, which are also responsible for non-trivial
 topological
 phases.\cite{fu2011topological,fang2012bulk,fang2013entanglement,teo2013existence,liu2013majorana, benalcazar2013classification, zhang2014topological,alexandradinata2014spinless} 
Even for these higher-order symmetries, the dimensional hierarchy of K-groups
may hold as Thom isomorphism, and thus a similar K-theory approach is
applicable,\cite{atiyah1969equivariant, freed2012twisted} but
we need a more sophisticated
representation theory beyond the Clifford algebra in order to clarify
these topological structures systematically.

\begin{acknowledgments}
We acknowledge useful discussions with S. Fujimoto, K. Gomi,
 T. Kawakami, S. Kobayashi, T. Mizushima, S. Ryu, Y. Tanaka, Y. Tsutsumi, and
 A. Yamakage.
Especially, K.S. thanks T. Morimoto for many fruitful discussions. 
This work was supported by the Grant-in-Aids for Scientific
Research from MEXT of Japan [Grants No.25287085 and No.22103005 (KAKENHI on Innovative Areas "Topological Quantum Phenomena")]. 
K.S. is supported by a JSPS Fellowship for Young Scientists. 

\end{acknowledgments}



\appendix

\makeatletter
\renewcommand{\theequation}{%
\thesection\arabic{equation}}
\@addtoreset{equation}{section}
\makeatother

\section{Dimensional shift of Hamiltonians}
\label{appendix:map}

To derive Eqs.(\ref{KCU2}), (\ref{KCA2}) and (\ref{KR2}), we need to shift
the dimension of Hamiltonians.
In this appendix, we present  
K-group isomorphic maps from
$d$-dimensional Hamiltonians to one-dimension higher $d+1$-dimensional
Hamiltonians,

\subsection{Dimension-raising map}
\label{sec:dimension-raising}

Here we introduce a map from a Hamiltonian on a $d+D$-dimensional
sphere  $({\bm k}, {\bm r})\in S^{d+D}$ to a Hamiltonian on $S^{d+D+1}$:
If the original Hamiltonian ${\cal H}({\bm k}, {\bm r})$ supports CS $\Gamma$, then the map is
\begin{eqnarray}
\mathcal{H}_{\rm nc}(\bm{k}, \bm{r},\theta) 
= \sin \theta \mathcal{H}_{\rm c}(\bm{k},{\bm r}) 
+ \cos \theta \Gamma, 
\quad \theta \in \left[0, \pi\right],  
\label{eq:DRC}
\end{eqnarray}
and if not, it is 
\begin{eqnarray}
\mathcal{H}_{\rm c}(\bm{k},\bm{r}, \theta) 
= \sin \theta \mathcal{H}_{\rm nc}(\bm{k},\bm{r})
 \otimes \tau_z + \cos \theta {\bm 1}\otimes \tau_y, 
\quad
\theta \in 
\left[0, \pi\right].   
\label{eq:DRNC}
\end{eqnarray}
Since the mapped Hamiltonian ${\cal H}({\bm k}, {\bm r}, \theta)$
is independent of $({\bm k}, {\bm r})\in S^{d+D}$ at $\theta=0$ and
$\theta=\pi$,
the base space $({\bm k}, {\bm r}, \theta)\in S^{d+D}\times [0,\pi]$ of the
mapped Hamiltonian 
can be identified as a $d+D+1$-dimensional sphere $S^{d+D+1}$ 
by shrinking $S^{d+D}$ to a point at $\theta=0$ and $\theta=\pi$,
respectively. 
The dimension-raising map interchanges a Hamiltonian with CS and a
Hamiltonian without CS.

\subsection{Dimension-lowering map}
\label{sec:dimension-lowering}

A dimension-lowering homomorphic map can be constructed as follows.
Consider a Hamiltonian ${\cal H}({\bm k}, {\bm r}, \theta)$ defined on
a $d+D+1$-dimensional sphere parametrized by $({\bm k}, {\bm r}, \theta)\in
S^{D+d+1}$. 
Here $\theta$ denotes the azimuthal angle of $S^{d+D+1}$, which points
the north pole (south pole) of $S^{d+D+1}$ at $\theta=0$ ($\theta=\pi$),
and $({\bm r}, {\bm k})$ parametrizes the $d+D$ dimensional circle of
latitude, so the Hamiltonian satisfies
\begin{eqnarray}
\mathcal{H}(\bm{k}, {\bm r},\theta = 0) = {\rm const.}, \ \ 
\mathcal{H}(\bm{k}, {\bm r}, \theta = \pi) ={\rm  const}', 
\label{boundary}
\end{eqnarray}
By using continuous deformation, we can also flatten the
Hamiltonian as
\begin{eqnarray}
\mathcal{H}^2(\bm{k}, {\bm r},\theta) = 1.
\end{eqnarray}
The above  parametrization $({\bm k}, {\bm r}, \theta)$ provides a
natural dimensional reduction $S^{d+D+1}\rightarrow S^{d+D}$ by
fixing $\theta$, say $\theta=\pi/2$. 
This procedure, however, does not ensure providing the inverse map of
Eqs.(\ref{eq:DRC}) and (\ref{eq:DRNC}), because  
the flattened Hamiltonian does not have the form of the right hand side
of Eqs.(\ref{eq:DRC}) or (\ref{eq:DRNC}) in general. 

To fix the form of the flattened Hamiltonian, following Teo and Kane, \cite{teo2010topological} we
introduce an artificial action $S[\mathcal{H}]$ of the Hamiltonian,
\begin{eqnarray}
S[\mathcal{H}] = \int d{\bm k} d{\bm r}d \theta \ 
\mathrm{Tr}[\partial_{\theta} \mathcal{H} \partial_{\theta} \mathcal{H}].
\end{eqnarray}
By continuous deformation of the Hamiltonian,
the value of action can reduce to reach its minimal value, where ${\cal
H}$ satisfies the saddle point equation with the constraint of ${\cal
H}^2=1$,  i.e. $\partial^2_{\theta}
\mathcal{H} + \mathcal{H} = 0$.  
Imposing the boundary condition (\ref{boundary}),  
we can fix the form of the Hamiltonian as the saddle point solution,
\begin{eqnarray}
\mathcal{H}(\bm{k},{\bm r},\theta) 
= \sin \theta \mathcal{H}_1(\bm{k},{\bm r}) 
+ \cos \theta \mathcal{H}_2, 
\label{eq:saddlepoint}
\end{eqnarray}
where the flatness condition ${\cal H}^2({\bm k}, {\bm r}, \theta)=1$
implies
\begin{eqnarray}
{\cal H}^2_1({\bm k}, {\bm r})=1,
\quad 
{\cal H}^2_2=1, 
\quad
\{\mathcal{H}_1(\bm{k}, {\bm r}), \mathcal{H}_2\} = 0.  
\label{eq:h1h2}
\end{eqnarray}
Then, by fixing $\theta=\pi/2$, we have a dimensional reduction from
${\cal H}({\bm k}, {\bm r}, \theta)$ to ${\cal H}_1({\bm k}, {\bm r})$

The last relation of Eq.(\ref{eq:h1h2}) means that ${\cal H}_2$ act as
CS on $\mathcal{H}_1(\bm{k}, {\bm r})$.
Therefore, if the original Hamiltonian ${\cal H}({\bm k}, {\bm r})$ does
not support CS, Eq.(\ref{eq:saddlepoint}) defines 
a dimensional reduction from non-chiral to chiral Hamiltonians. 
On the other hand, if the original Hamiltonian has CS
$\Gamma$, then ${\cal H}_1({\bm k}, {\bm r})$ hosts a couple of CSs, 
$\Gamma$ and ${\cal H}_2$, with $\{\Gamma$, ${\cal H}_2\}=0$.
Hence, ${\cal H}_1({\bm k}, {\bm r})$ has redundancy due to the
commutation relation $[{\cal H}_1({\bm k}, {\bm r}), \Gamma{\cal H}_2]=0$.
In the basis where $\Gamma={\bm 1}\otimes \tau_x$ and ${\cal
H}_2={\bm 1}\otimes \tau_y$, 
the redundancy of ${\cal H}_1({\bm k}, {\bm r})$ is resolved as
$
{\cal H}_1({\bm k}, {\bm r})={\cal H}_3({\bm k}, {\bm r})\otimes \tau_z,
$
and thus we obtain
\begin{eqnarray}
{\cal H}({\bm k}, {\bm r}, \theta)=\sin \theta {\cal H}_3({\bm k}, {\bm
 r})\otimes \tau_z+\cos\theta {\bm 1}\otimes \tau_y   
\label{eq:DLNC}
\end{eqnarray}
In this manner, a chiral Hamiltonian ${\cal H}({\bm k}, {\bm r},
\theta)$ is mapped to a non-chiral one ${\cal H}_3({\bm k}, {\bm r})$.

\section{Dimensional hierarchy of AZ classes}
\label{AZ}
In this section, we review the topological classification for AZ
symmetry classes.\cite{schnyder2008classification, kitaev2009periodic,teo2010topological}  
We provide the periodic table for the topological insulator and
superconductor by using of the K-group isomorphic map between different
dimensions and symmetries.  
Following Teo and Kane \cite{teo2010topological}, we argue the
dimensional hierarchy of the K-groups,   
\begin{eqnarray}
K_{\mathbb{C}}(s,d,D) = K_{\mathbb{C}}(s+1,d+1,D) = K_{\mathbb{C}}(s+1,d,D+1) 
\label{eq:KisoC}
\end{eqnarray}
for complex AZ classes and 
\begin{eqnarray}
K_{\mathbb{R}}(s,d,D) = K_{\mathbb{C}}(s+1,d+1,D) = K_{\mathbb{C}}(s-1,d,D+1) 
\label{eq:KisoR}
\end{eqnarray}
for real AZ classes.

\subsection{Complex AZ classes}
The complex AZ classes consist of two symmetry class, class A
for Hamiltonians with no symmetry and class AIII for those with the
presence of CS.  
The symmetry classes are labeled by $s = 0,1$ (mod $2$) as in Table \ref{Classifying_space}. 
For the complex AZ classes, because of the absence of antiunitary symmetry,
momentum $\bm{k}$ and coordinates $\bm{r}$ are not distinguished from
each other, and thus 
$K_{\mathbb{C}}(s,d,D) = K_{\mathbb{C}}(s,d+D,0)$.  

The dimensional raising maps, Eqs. (\ref{eq:DRC}) and (\ref{eq:DRNC}), 
interchange Hamiltonians with CS and those without CS,  and
thus they define a K-group homomorphism
$K_{\mathbb{C}}(s,d+D,0) \mapsto K_{\mathbb{C}}(s+1,d+D+1,0)$, where $s$
is also shifted by 1.
At the same time, the dimensional lowering maps,
Eqs. (\ref{eq:saddlepoint}) and (\ref{eq:DLNC}) define the inverse of
the K-group homomorphism, i.e. $K_{\mathbb{C}}(s+1;d+D+1,0) \mapsto
K_{\mathbb{C}}(s;d+D,0)$.
Consequently, we obtain the K-group isomorphism Eq.(\ref{eq:KisoC}).

\subsection{Real AZ classes}
The real AZ classes consist of eight symmetry classes which specified by
the presence of TRS and/or PHS. 
The eight symmetry classes are labeled by $s=0,\dots, 7$ (mod $8$) as shown in
Table \ref{Classifying_space}. 
In this paper, we take a convention that $T$ and $C$ commute with each
other: $[T,C] = 0$.
In this rule, the chiral operator $\Gamma$ (that is a Hermitian matrix) is
given by 
\begin{eqnarray}
\Gamma = 
\left\{ \begin{array}{ll}
TC & (s=1,5) \\
iTC & (s=3,7) \\
\end{array} \right. ,
\end{eqnarray}
where the following relation holds, 
\begin{eqnarray}
T \Gamma T^{-1} = C \Gamma C^{-1} = 
\left\{ \begin{array}{ll}
\Gamma & (s=1,5) \\
-\Gamma & (s=3,7) \\
\end{array} \right. . 
\label{GammaTC} 
\end{eqnarray}

For real AZ classes hosting CS
($s=1,3,5,7$), 
one can raise the dimension of the base space by using Eq.(\ref{eq:DRC}). 
The mapped Hamiltonian $\mathcal{H}(\bm{k}, {\bm r},\theta)$ supports
either TRS or PHS, but does not have both.  
The remaining symmetry depends on the type of $\theta$ one considers:
If one increases the dimension $d$ of the momentum space, 
the parameter $\theta$ should transform as $\theta \rightarrow \pi-\theta$
under TRS and PHS.
In contrast,  if one raises the dimension $D$ of the position space, 
$\theta$ does not transform under these symmetries. 
We call the former $\theta$ as ${\bm k}$-type, and the latter as ${\bm
r}$-type.
The difference in the transformation law of $\theta$ results in the
difference of 
the remaining symmetry.
For instance, consider the BDI class ($s=1$) and ${\bm k}$-type $\theta$.
In this case, because of Eq.(\ref{GammaTC}), one finds that the mapped Hamiltonian,
$
{\cal H}({\bm k}, {\bm r}, \theta)=\sin\theta {\cal H}({\bm k}, {\bm
r})+\cos \theta (TC), 
$
supports PHS.
For real AZ classes without CS
 ($s=0,2,4,6$), the dimensional raising map is provided by Eq.(\ref{eq:DRNC}). 
The mapped Hamiltonian $\mathcal{H}(\bm{k},\bm{r},\theta)$ has the
CS, $\{{\bm 1}\otimes \tau_x,
\mathcal{H}(\bm{k},\bm{r},\theta)\}=0$.
It also realizes TRS or PHS of the original Hamiltonian ${\cal H}({\bm
k}, {\bm r})$, in the form of 
$T\otimes \tau_a$ or $C\otimes
\tau_a$,
where the choice of
$\tau_a$ ($a=0, z$) depends on the type of $\theta$.
The mapped Hamiltonian also has the rest of AZ symmetries, which is
obtained by
combination of these symmetries.

We summarize the AZ symmetries of the mapped Hamiltonian for each real AZ
class in the lower part of Table \ref{Tab_raise}. 
From this table, one finds that the dimensional raising maps,
Eqs. (\ref{eq:DRC}) and (\ref{eq:DRNC}), shift the label $s$ of AZ
classes by $\pm 1$,  and thus they define K-group homomorphic maps,
$K_{\mathrm{R}}(s,d,D) 
\mapsto K_{\mathrm{R}}(s+1,d+1,D)$ and $K_{\mathrm{R}}(s,d,D)
\mapsto K_{\mathrm{R}}(s-1,d,D+1)$.  

In a manner similar to complex AZ classes, the dimensional lowering
maps, Eqs. (\ref{eq:saddlepoint})
and (\ref{eq:DLNC}), define the inverse of the K-group homomorphism, 
$K_{\mathrm{R}}(s+1,d+1,D) 
\mapsto K_{\mathrm{R}}(s,d,D)$ and $K_{\mathrm{R}}(s-1,d,D+1)
\mapsto K_{\mathrm{R}}(s,d,D)$:
Here note that
Eqs. (\ref{eq:saddlepoint}) and (\ref{eq:DLNC})
determine uniquely how TRS and/or PHS of higher dimensional Hamiltonians
act on the lower dimensional ones.
As a result, we have the K-group isomorphism, Eq.(\ref{eq:KisoR}).

\begin{table*}[!]
\begin{center}
\caption{
Homomorphism from $K_{\mathbb{R}}(s,d,D)$ to
 $K_{\mathbb{F}}(s+1,d+1,D)$ and $K_{\mathbb{F}}(s-1,d,D+1)$, $(\mathbb{F}=\mathbb{C}, \mathbb{R})$. 
}
\begin{tabular}[t]{ccccccc}
\hline \hline
AZ class & Hamiltonian mapping & Type of $\theta$ 
& Mapped AZ class & TRS & PHS & Chiral \\
\hline
A  & $\sin \theta \mathcal{H}(\bm{k},\bm{r}) \otimes\tau_z 
+ \cos \theta {\bm 1}\otimes\tau_y$ 
&$\bm{k}/\bm{r}$& AIII & & & ${\bm 1}\otimes\tau_x$ \\
AIII  & $\sin \theta \mathcal{H}(\bm{k},\bm{r}) + \cos \theta \Gamma$ 
&$\bm{k}/\bm{r}$& A & \\
\hline\hline 
\multirow{2}{*}{AI/AII} 
 & \multirow{2}{*}{
$\sin \theta \mathcal{H}(\bm{k},\bm{r}) \otimes\tau_z 
+ \cos \theta {\bm 1}\otimes\tau_y$} 
 &    $\bm{k}$ & BDI/CII & $T\otimes \tau_0$ & $T \otimes \tau_x$ 
& ${\bm 1}\otimes\tau_x$ \\
 & & $\bm{r}$ & CI/DIII & $T\otimes \tau_z$ & $T \otimes \tau_y$ 
& ${\bm 1}\otimes \tau_x$ \\
\hline
\multirow{2}{*}{BDI/CII} 
 & \multirow{2}{*}{$\sin \theta \mathcal{H}(\bm{k},\bm{r}) + \cos \theta (TC)$} 
 &    $\bm{k}$ & D/C & & $C$ & \\
 & & $\bm{r}$ & AI/AII & $T$ & & \\
\hline
\multirow{2}{*}{D/C} 
 & \multirow{2}{*}{$\sin \theta \mathcal{H}(\bm{k},\bm{r}) \otimes
     \tau_z + \cos \theta {\bm 1}\otimes\tau_y$} 
 &    $\bm{k}$ & DIII/CI & $C\otimes \tau_y$ & $C \otimes \tau_z$ 
& ${\bm 1}\otimes \tau_x$ \\
 & & $\bm{r}$ & BDI/CII & $C\otimes \tau_x$ & $C \otimes \tau_0$ 
& ${\bm 1}\otimes\tau_x$ \\
\hline
\multirow{2}{*}{DIII/CI} 
 & \multirow{2}{*}{$\sin \theta \mathcal{H}(\bm{k},\bm{r}) + \cos \theta (i TC)$} 
 &    $\bm{k}$ & AII/AI & $T$ & & \\
 & & $\bm{r}$ & D/C & & $C$ & \\
\hline \hline
\end{tabular}
\label{Tab_raise}
\end{center}
\end{table*}

\section{Classifying space of AZ classes with additional symmetry}
\label{Sec_Clif}

In this appendix, we show classifying spaces of real and complex
AZ classes in the presence of additional symmetry. 
The classifying spaces are identified by
counting distinct symmetry-allowed zero-dimensional Hamiltonians that cannot be
connected to each other by continuous deformation.
As flattened Hamiltonians and symmetry
operators form the Clifford algebra, the counting reduces to 
the extension problem of the Clifford algebra.\cite{kitaev2009periodic,
morimoto2013topological}  
Here we need to consider only additional unitary
symmetries: For complex AZ classes, the classifying spaces in the
presence of an antiunitary symmetry are obtained  as those of real AZ
classes without additional symmetry, as is shown in Sec.\ref{sec:CAZAS}.
For real AZ classes, antiunitary
symmetries are converted into unitary symmetries (See Table
\ref{Symmetry_type}).

\subsection{Complex AZ classes with additional order-two unitary symmetry}
The complex Clifford algebra $Cl_p$ is generated by a set of generators
$\{e_1, e_2, \dots, e_p\}$ with $\{e_i,e_j\} = 2 \delta_{ij}$, and  
the vector space is spanned by $2^p$s basis $\{ e^{n_1}_1 \otimes
e^{n_2}_2 \otimes \dots \otimes e^{n_p}_p\}_{n_i=0,1}$ with
$\mathbb{C}$-coefficients.

Symmetry operators in complex AZ classes, namely no operator in class A
and the chiral operator
$\Gamma$ in class AIII, are generators of the complex Clifford algebra,
$Cl_0$ and $Cl_1$, respectively. 
Since a flattened Hamiltonian ${\cal H}$ satisfies ${\cal H}^2=1$, and 
it also anticommutes with $\Gamma$ in class AIII, it extends the
Clifford algebra as 
\begin{equation}\begin{split}
Cl_p = \{e_1, e_2, \dots, e_p\} \rightarrow Cl_{p+1}=\{{\cal H},e_1,
		 e_2, \dots, e_p\}, 
\end{split}\end{equation}
where $p=0$ for class A and $p=1$ for class AIII.
($Cl_0=\{\emptyset\}$, $Cl_1=\{e_1=\Gamma\}$.)
The map from $Cl_p$ to $Cl_{p+1}$ defines the classifying space
$\mathcal{C}_p$, which obeys 
the Bott periodicity $\mathcal{C}_p \simeq \mathcal{C}_{p+2}$. 

The presence of an additional unitary symmetry affects on the extension
in two possible manners: 
(i) decoupling of the Clifford algebra, or (ii) adding another
generator of the Clifford algebra. 
We summarize the extensions and classifying spaces of complex AZ classes
with an additional unitary symmetry in Table \ref{Clifford}.

\subsection{Real AZ classes with additional order-two symmetry}
The real Clifford algebra $Cl_{p,q}$ is generated by a set of generators
$\{e_1, e_2, \dots, e_p, e_{p+1}, \dots e_{p+q}\}$
with $\{e_i,e_j\} = 2 \delta_{ij} (i \neq j)$ and $e_i^2 = -1 (i=1, \dots, p)$, $e^2_i=1 (i=p+1, \dots, p+q)$. 
The vector space is spanned by $2^{p+q}$s basis $\{ e^{n_1}_1 \otimes
e^{n_2}_2 \otimes \dots \otimes e^{n_{p+q}}_{p+q}\}_{n_i=0,1}$ with
$\mathbb{R}$-coefficient. 
Since symmetry operators of real AZ classes can be considered as
generators of real Clifford algebra, 
the classifying spaces of real AZ classes are derived by the counting the
distinct symmetry-allowed zero-dimensional Hamiltonians $H$, $\{H,e_i\}=0$,  
which define the extension problem of the Clifford algebra : 
\begin{equation}\begin{split}
Cl_{p,q} = \{e_1, \dots, e_{p+q}\} \rightarrow Cl_{p+1,q}=\{H,e_1,\dots, e_{p+q}\}, \ \ H^2=-1, 
\end{split}\end{equation}
or 
\begin{equation}\begin{split}
Cl_{p,q} = \{e_1, \dots, e_{p+q}\} \rightarrow Cl_{p,q+1}=\{H,e_1,\dots, e_{p+q}\}, \ \ H^2=1. 
\end{split}\end{equation}
The classifying space obtained in the former case is $\mathcal{R}_{p+2-q}$, and 
that obtained in the latter case is $\mathcal{R}_{q-p}$. 
The Bott periodicity implies $\mathcal{R}_p \simeq \mathcal{R}_{p+8}$. 
The presence of the additional unitary symmetry affect on the extension
in four possible manners : 
(i) decoupling the Clifford algebra, 
(ii) inducing a complex structure, 
(iii) adding another generator $e$ with $e^2=-1$, or
(iv)adding another generator $e$ with $e^2=1$. 
We summarize the classifying spaces of real AZ classes with an additional
unitary symmetry in Table \ref{Clifford}.

\begin{table*}[!]
\begin{center}
\caption{Classifying spaces of AZ classes with additional order-two symmetry. 
In the fourth column, $J$ is the pure imaginary constant.}
\begin{tabular}[t]{ccccccccc}
\hline \hline
AZ class & Symmetry & Extension & Generator & Classifying space \\
\hline
A & $U$ & $Cl_1 \rightarrow Cl_1 \otimes Cl_1$ 
& $U \rightarrow \{H\} \otimes U$ & $\mathcal{C}_0 \times \mathcal{C}_0$ \\
A & $\widebar U$ & $Cl_1 \rightarrow Cl_2$ 
& $\{\widebar U\} \rightarrow \{H,\widebar U\}$ & $\mathcal{C}_1$ \\
\hline
AIII & $U_+$ & $Cl_1\otimes Cl_1 \rightarrow Cl_2\otimes Cl_1$ 
& $\{\Gamma\}\otimes U_+ \rightarrow \{H,\Gamma\} \otimes U_+$ 
& $\mathcal{C}_1 \times \mathcal{C}_1$ \\
AIII & $\widebar U_-$ & $Cl_2 \rightarrow Cl_3$ 
& $\{\Gamma,\widebar U_-\} \rightarrow \{H,\Gamma,\widebar U_-\}$ 
& $\mathcal{C}_0$ \\
\hline \hline
AI & $U^+_+$ 
& $Cl_{0,2}\otimes Cl_{0,1} \rightarrow Cl_{1,2}\otimes Cl_{0,1}$ 
& $\{T,JT\}\otimes U^+_+ \rightarrow \{JH,T,JT\}\otimes U^+_+$ 
& $\mathcal{R}_0 \times \mathcal{R}_0$ \\
AI & $\widebar U^+_-$ & $Cl_{0,3} \rightarrow Cl_{1,3}$ 
& $\{T,JT,\widebar U^+_-\} \rightarrow \{JH,T,JT,\widebar U^+_-\}$ 
& $\mathcal{R}_7$ \\
AI & $U^-_+$ 
& $Cl_{0,2}\otimes Cl_{1,0} \rightarrow Cl_{1,2}\otimes Cl_{1,0}$ 
& $\{T,JT\}\otimes U^-_+ \rightarrow \{JH,T,JT\}\otimes U^-_+$ 
& $\mathcal{C}_0$ \\
AI & $\widebar U^-_-$ & $Cl_{1,2} \rightarrow Cl_{2,2}$ 
& $\{T,JT,\widebar U^-_-\} \rightarrow \{JH,T,JT,\widebar U^-_-\}$ 
& $\mathcal{R}_{1}$ \\
\hline
BDI & $U^+_{++}$ 
& $Cl_{1,2}\otimes Cl_{0,1} \rightarrow Cl_{1,3}\otimes Cl_{0,1}$ 
& $\{C,JC,JCT\}\otimes U^+_{++} \rightarrow \{H,C,JC,JCT\}\otimes
	     U^+_{++}$ 
& $\mathcal{R}_1 \times \mathcal{R}_1$ \\
BDI & $\widebar U^-_{+-}$ & $Cl_{2,2} \rightarrow Cl_{2,3}$ 
& $\{C,JC,JCT,\widebar U^-_{+-}\} \rightarrow \{H,C,JC,JCT,\widebar U^-_{+-}\}$ 
& $\mathcal{R}_{0}$ \\
BDI & $U^-_{++}$ 
& $Cl_{1,2}\otimes Cl_{1,0} \rightarrow Cl_{1,3}\otimes Cl_{1,0}$ 
& $\{C,JC,JCT\}\otimes U^-_{++} \rightarrow \{H,C,JC,JCT\}\otimes
	     U^-_{++}$ 
& $\mathcal{C}_1$ \\
BDI & $\widebar U^+_{+-}$ & $Cl_{1,3} \rightarrow Cl_{1,4}$ 
& $\{C,JC,JCT,\widebar U^+_{+-}\} \rightarrow \{H,C,JC,JCT,\widebar U^+_{+-}\}$ 
& $\mathcal{R}_{2}$ \\
\hline
D & $U^+_+$ & $Cl_{0,2}\otimes Cl_{0,1} \rightarrow Cl_{0,3}\otimes
	 Cl_{0,1}$ 
& $\{C,JC\}\otimes U^+_+ \rightarrow \{H,C,JC\}\otimes U^+_+$ 
& $\mathcal{R}_2 \times \mathcal{R}_2$ \\
D & $\widebar U^-_-$ & $Cl_{1,2} \rightarrow Cl_{1,3}$ 
& $\{C,JC,\widebar U^-_-\} \rightarrow \{H,C,JC,\widebar U^-_-\}$ 
& $\mathcal{R}_{1}$ \\
D & $U^-_+$ & $Cl_{0,2}\otimes Cl_{1,0} \rightarrow Cl_{0,3}\otimes
	 Cl_{1,0}$ 
& $\{C,JC\}\otimes U^-_+ \rightarrow \{H,C,JC\}\otimes U^-_+$ 
& $\mathcal{C}_0$ \\
D & $\widebar U^+_-$ & $Cl_{0,3} \rightarrow Cl_{0,4}$ 
& $\{C,JC,\widebar U^+_-\} \rightarrow \{H,C,JC,\widebar U^+_-\}$ 
& $\mathcal{R}_{3}$ \\
\hline
DIII & $U^+_{++}$ 
& $Cl_{0,3}\otimes Cl_{0,1} \rightarrow Cl_{0,4}\otimes Cl_{0,1}$ 
& $\{C,JC,JCT\}\otimes U^+_{++} \rightarrow \{H,C,JC,JCT\}\otimes
	     U^+_{++}$ 
& $\mathcal{R}_3 \times \mathcal{R}_3$ \\
DIII & $\widebar U^-_{+-}$ & $Cl_{1,3} \rightarrow Cl_{1,4}$ 
& $\{C,JC,JCT,\widebar U^-_{+-}\} \rightarrow \{H,C,JC,JCT,\widebar U^-_{+-}\}$ 
& $\mathcal{R}_{2}$ \\
DIII & $U^-_{++}$ 
& $Cl_{0,3}\otimes Cl_{1,0} \rightarrow Cl_{0,4}\otimes Cl_{1,0}$ 
& $\{C,JC,JCT\}\otimes U^-_{++} \rightarrow \{H,C,JC,JCT\}\otimes
	     U^-_{++}$ 
& $\mathcal{C}_1$ \\
DIII & $\widebar U^+_{+-}$ & $Cl_{0,4} \rightarrow Cl_{0,5}$ 
& $\{C,JC,JCT,\widebar U^+_{+-}\} \rightarrow \{H,C,JC,JCT,\widebar U^+_{+-}\}$ 
& $\mathcal{R}_{4}$ \\
\hline
AII & $U^+_+$ 
& $Cl_{2,0}\otimes Cl_{0,1} \rightarrow Cl_{3,0}\otimes Cl_{0,1}$ 
& $\{T,JT\}\otimes U^+_+ \rightarrow \{JH,T,JT\}\otimes U^+_+$ 
& $\mathcal{R}_4 \times \mathcal{R}_4$ \\
AII & $\widebar U^+_-$ & $Cl_{2,1} \rightarrow Cl_{3,1}$ 
& $\{T,JT,\widebar U^+_-\} \rightarrow \{JH,T,JT,\widebar U^+_-\}$ 
& $\mathcal{R}_{3}$ \\
AII & $U^-_+$ 
& $Cl_{2,0}\otimes Cl_{1,0} \rightarrow Cl_{3,0}\otimes Cl_{1,0}$ 
& $\{T,JT\}\otimes U^-_+ \rightarrow \{JH,T,JT\}\otimes U^-_+$ 
& $\mathcal{C}_0$ \\
AII & $\widebar U^-_-$ 
& $Cl_{3,0} \rightarrow Cl_{4,0}$ 
& $\{T,JT,\widebar U^-_-\} \rightarrow \{JH,T,JT,\widebar U^-_-\}$ 
& $\mathcal{R}_{5}$ \\
\hline
CII & $U^+_{++}$ 
& $Cl_{3,0}\otimes Cl_{0,1} \rightarrow Cl_{3,1}\otimes Cl_{0,1}$ 
& $\{C,JC,JCT\}\otimes U^+_{++} \rightarrow \{H,C,JC,JCT\}\otimes
	     U^+_{++}$ 
& $\mathcal{R}_5 \times \mathcal{R}_5$ \\
CII & $\widebar U^-_{+-}$ 
& $Cl_{4,0} \rightarrow Cl_{4,1}$ & $\{C,JC,JCT,\widebar U^-_{+-}\}
	     \rightarrow \{H,C,JC,JCT,\widebar U^-_{+-}\}$ 
& $\mathcal{R}_4$ \\
CII 
& $U^-_{++}$ & $Cl_{3,0}\otimes Cl_{1,0} \rightarrow Cl_{3,1}\otimes
	 Cl_{1,0}$ 
& $\{C,JC,JCT\}\otimes U^-_{++} \rightarrow \{H,C,JC,JCT\}\otimes
	     U^-_{++}$ 
& $\mathcal{C}_1$ \\
CII & $\widebar U^+_{+-}$ 
& $Cl_{3,1} \rightarrow Cl_{3,2}$ & $\{C,JC,JCT,\widebar U^+_{+-}\}
	     \rightarrow \{H,C,JC,JCT,\widebar U^+_{+-}\}$ 
& $\mathcal{R}_6$ \\
\hline
C & $U^+_+$ 
& $Cl_{2,0}\otimes Cl_{0,1} \rightarrow Cl_{2,1}\otimes Cl_{0,1}$ 
& $\{C,JC\}\otimes U^+_+ \rightarrow \{H,C,JC\}\otimes U^+_+$ 
& $\mathcal{R}_6 \times \mathcal{R}_6$ \\
C & $\widebar U^-_-$ 
& $Cl_{3,0} \rightarrow Cl_{3,1}$ 
& $\{C,JC,\widebar U^-_-\} \rightarrow \{H,C,JC,\widebar U^-_-\}$ 
& $\mathcal{R}_5$ \\
C & $U^-_+$ 
& $Cl_{2,0}\otimes Cl_{1,0} \rightarrow Cl_{2,1}\otimes Cl_{1,0}$ 
& $\{C,JC\}\otimes U^-_+ \rightarrow \{H,C,JC\}\otimes U^-_+$ 
& $\mathcal{C}_0$ \\
C & $\widebar U^+_-$ & $Cl_{2,1} \rightarrow Cl_{2,2}$ 
& $\{C,JC,\widebar U^+_-\} \rightarrow \{H,C,JC,\widebar U^+_-\}$ 
& $\mathcal{R}_7$ \\
\hline
CI & $U^+_{++}$ 
& $Cl_{2,1}\otimes Cl_{0,1} \rightarrow Cl_{2,2}\otimes Cl_{0,1}$ 
& $\{C,JC,JCT\}\otimes U^+_{++} \rightarrow \{H,C,JC,JCT\}\otimes
	     U^+_{++}$ 
& $\mathcal{R}_7 \times \mathcal{R}_7$ \\
CI & $\widebar U^-_{+-}$ 
& $Cl_{3,1} \rightarrow Cl_{3,2}$ 
& $\{C,JC,JCT,\widebar U^-_{+-}\} \rightarrow \{H,C,JC,JCT,\widebar U^-_{+-}\}$ 
& $\mathcal{R}_6$ \\
CI & $U^-_{++}$ 
& $Cl_{2,1}\otimes Cl_{1,0} \rightarrow Cl_{2,2}\otimes Cl_{1,0}$ 
& $\{C,JC,JCT\}\otimes U^-_{++} \rightarrow \{H,C,JC,JCT\}\otimes
	     U^-_{++}$ 
& $\mathcal{C}_1$ \\
CI & $\widebar U^+_{+-}$ 
& $Cl_{2,2} \rightarrow Cl_{2,3}$ & $\{C,JC,JCT,\widebar U^+_{+-}\} \rightarrow \{H,C,JC,JCT,\widebar U^+_{+-}\}$ & $\mathcal{R}_{0}$ \\
\hline \hline
\end{tabular}
\label{Clifford}
\end{center}
\end{table*}

\section{Topological invariants}
\label{appendix:topological_invariant}

In this appendix, we summarize the notation and the definition of
topological invariants used in this paper. 

\subsection{Topological invariants in zero-dimension}

Here, we introduce topological invariants in zero dimension.
A Hamiltonian $\mathcal{H}$ in zero dimension is merely a constant matrix, so
adding extra trivial bands to the Hamiltonian makes any change of the
Hamiltonian possible.
This means that no well-defined topological number of a single
Hamiltonian is possible in the meaning of the stable-equivalence. 
We need a couple of Hamiltonians $(\mathcal{H}_+, \mathcal{H}_-)$
to define a topological number.
We say that two coupled Hamiltonians, $(\mathcal{H}_{1+},
\mathcal{H}_{1-})$ and $(\mathcal{H}_{2+}, \mathcal{H}_{2-})$, are
stable equivalent if they are
continuously deformed into each other by adding the {\it same} extra bands 
to the coupled Hamiltonian.
In other words, the stable equivalence implies  
$(\mathcal{H}_+,\mathcal{H}_-)\sim
(\mathcal{H}_+\oplus\mathcal{H}_{\rm
ext},\mathcal{H}_-\oplus\mathcal{H}_{\rm ext})$ with an extra band
$\mathcal{H}_{\rm ext}$.




\subsubsection{$\mathbb{Z}$ ($2\mathbb{Z}$) invariant}
First, consider non particle-hole symmetric Hamiltonians. 
We assume here that $\mathcal{H}_+$ and $\mathcal{H}_-$ have the same
matrix dimension. 
Denoting the numbers of empty (occupied) states of $\mathcal{H}_{\pm
}$ by $n_{\pm}$ ($m_{\pm}$), the topological nature of the coupled Hamiltonians 
can be characterized by $n_+-m_+$ and $n_--m_-$ since there appears a
band crossing the Fermi level when these numbers are changed.  
Adding trivial $p$ empty bands and $q$ occupied ones to the
coupled Hamiltonian,  
we also have the stable equivalence between these numbers,
\begin{eqnarray}
(n_+-m_+, n_--m_-)\sim (n_+-m_++p-q, n_--m_-+p-q). 
\end{eqnarray}
Therefore, 
the topological number in zero dimension is defined as
\begin{equation}\begin{split}
Ch_0 := \frac{n_+ -m_+-n_- + m_-}{2},
\end{split}\end{equation}
because it should be invariant
under the stable equivalence.
Whereas $Ch_0$ can take any integer for class A and AI Hamiltonians, 
it takes only an even integer for class AII due to the Kramers
degeneracy of the spectrum.



\subsubsection{$\mathbb{Z}_2$ invariant}

For  Hamiltonians with PHS satisfying $C^2=1$, 
the following $\mathbb{Z}_2$ invariant can be introduced
\begin{equation}\begin{split}
\nu_0 = \mathrm{sgn} \left[\mathrm{Pf}(\mathcal{H}_+ \tau_x) \right] \mathrm{sgn} \left[\mathrm{Pf}(\mathcal{H}_- \tau_x) \right],
\end{split}\end{equation}
with $C=\tau_x\mathcal{K}$: 
First, PHS implies $
\mathcal{H}_{\pm}\tau_x=-(\mathcal{H}_{\pm}\tau_x)^T$, which enables us to
define the Pfaffian of $\mathcal{H}_{\pm}\tau_x$. 
Then, from the relation 
\begin{eqnarray}
\left[{\rm Pf}(\mathcal{H}_{\pm}\tau_x)\right]^*&=&{\rm Pf}(\mathcal{H}^*_{\pm}\tau_x)
\nonumber\\
&=&{\rm Pf}(\tau^T_x (\mathcal{H}_{\pm} \tau_x)^T \tau_x)
\nonumber\\
&=&{\rm Pf}(\mathcal{H}_{\pm} \tau_x)
\end{eqnarray}
the sign of ${\rm Pf}(\mathcal{H}_{\pm}\tau_x)$ is quantized
as $\pm 1$.
Taking into account the stable equivalence, 
we find that each of ${\rm Pf}(\mathcal{H}_{\pm}\tau_x)$ does not give
a $\mathbb{Z}_2$ invariant, but their product $\nu_0$ defines it.

In AZ classes, BDI and D in zero dimension support this $\mathbb{Z}_2$
invariant.
Class DIII also has PHS with $C^2=1$, but    
$\nu_0$ becomes trivial in this case because of the Kramers degeneracy.



\subsection{Chern number and winding number}
\label{appendix:Chern_winding}

Here we summarize the analytic expressions 
of integer $\mathbb{Z}$ topological invariants, i.e. 
the Chern numbers in even-dimensions, 
and the winding numbers in odd-dimensions.

\subsubsection{Q-function}
It is useful to introduce the so called ``Q-function''\cite{schnyder2008classification} defined by 
\begin{equation}\begin{split}
Q(\bm{k}, \bm {r}) 
= \sum_{E_{\alpha}(\bm{k}, \bm{r})>E_{\rm F}} 
\ket{u_{\alpha}(\bm{k}, \bm{r})} \bra{u_{\alpha}(\bm{k}, {\bm r})} 
- \sum_{E_{\alpha}(\bm{k},\bm{r})<E_{\rm F}} \ket{u_{\alpha}(\bm{k},\bm{r})} 
\bra{u_{\alpha}(\bm{k},\bm{r})}, 
\end{split}\end{equation}
where $|u_{\alpha}({\bm k},\bm{r})\rangle$ is an eigenstate of
$\mathcal{H}({\bm k},\bm{r})$ with an eigen energy $E_{\alpha}({\bm k}, {\bm r})$.
The Q-function is nothing but the flattened Hamiltonian of
$\mathcal{H}({\bm k})$, and it has the following properties: 
\begin{eqnarray}
Q^2(\bm{k},\bm{r}) = 1, 
\quad
Q(\bm{k},\bm{r}) \ket{u_{\alpha}(\bm{k},\bm{r})} = 
\left\{ \begin{array}{rl}
\ket{u_{\alpha}(\bm{k}, \bm{r})}, & (E_{\alpha}(\bm{k},\bm{r})>E_{\rm F})  \\
-\ket{u_{\alpha}(\bm{k},\bm{r})}, & (E_{\alpha}(\bm{k},\bm{r})<E_{\rm F})  \\
\end{array} \right. .
\end{eqnarray}
The symmetry of the Q-function is the same as the original Hamiltonian
$\mathcal{H}(\bm{k},\bm{r})$. 

\subsubsection{Chern number}
In the $2 n$-dimensional base space, 
the $n$-th Chern number $Ch_n$ is defined by
\begin{equation}\begin{split}
Ch_n = 
\frac{1}{n!} \left( \frac{i}{2 \pi} \right)^{n} \int \mathrm{tr} \mathcal{F}^n, 
\end{split}\label{Eq::Chern_Character}\end{equation}
with $\mathcal{F}= d \mathcal{A} + \mathcal{A} \wedge \mathcal{A}$.
Here $\mathcal{A}_{\alpha\beta} =
\Braket{u_{\alpha}|du_{\beta}}$ 
is the connection of occupied states $|u_{\alpha}({\bm k},{\bm r})\rangle$ of
$\mathcal{H}(\bm{k},{\bm r})$, and the trace is taken for all occupied states.
The Chern number is rewritten as 
\begin{equation}\begin{split}
Ch_n 
&= - \frac{1}{2^{2n+1}} \frac{1}{n!} \left( \frac{i}{2 \pi} \right)^{n} \int \mathrm{tr} Q (d Q)^{2 n}. \\
\end{split}\end{equation}
It is also useful to express the Chern number in terms of the Green
 function,\cite{ishikawa1986} $G(\omega,\bm{k},{\bm r}) = [i \omega -
 \mathcal{H}(\bm{k},\bm{r})]^{-1}$, when
 we discuss the electromagnetic and/or heat responses.  
The Chern number is recast into 
\begin{equation}\begin{split}
Ch_n = 
- \frac{n!}{\left( 2 \pi i \right)^{n+1} (2n+1)!}  \int \mathrm{tr} \left[ G d G^{-1} \right]^{2n+1}. 
\end{split}\end{equation}

Although the Chern number can be defined in any even dimensions, 
symmetry of the system sometimes prohibits a
non-zero Chern number.
For example, consider an antiunitary symmetry 
\begin{eqnarray}
A \mathcal{H}(\bm{k}_{\parallel},\bm{k}_{\perp })A^{-1} = \mathcal{H}(\bm{k}_{\parallel},-\bm{k}_{\perp }). 
\label{Symmetry-prohibition-A}
\end{eqnarray}
Since the Q-function has the same symmetry, $A
Q(\bm{k}_{\parallel},\bm{k}_{\perp }) A^{-1} =
Q(\bm{k}_{\parallel},-\bm{k}_{\perp })$, we find 
\begin{eqnarray}
Ch_n
&=&- \frac{1}{2^{2n+1}}\frac{1}{n!} 
\left( \frac{i}{2 \pi} \right)^{n} \int \mathrm{tr} 
\left[
A^{-1}Q(\bm{k}_{\parallel},-{\bm k}_{\perp}) 
dQ(\bm{k}_{\parallel}, -{\bm k}_{\perp})^{2 n}A 
\right]
\nonumber\\
&=&
- \frac{1}{2^{2n+1}}\frac{1}{n!} 
\left( \frac{i}{2 \pi} \right)^{n} \int \mathrm{tr} 
\left[
Q^*(\bm{k}_{\parallel},-{\bm k}_{\perp}) 
dQ^*(\bm{k}_{\parallel}, -{\bm k}_{\perp})^{2 n} 
\right]
\nonumber\\
&=&
- \frac{(-1)^{2n-d_{\parallel}}}{2^{2n+1}}\frac{1}{n!} 
\left( \frac{i}{2 \pi} \right)^{n} \int \mathrm{tr} 
\left[
Q^*(\bm{k}) 
dQ^*(\bm{k})^{2 n} 
\right]
\nonumber\\
&=& (-1)^{n-d_{\parallel}}Ch^*_n.
\nonumber\\
&=& (-1)^{n-d_{\parallel}}Ch_n,
\label{Eq::Sym_of_Chern_character}
\end{eqnarray}
where we have used the fact that $Ch_n$ is an integer in the last line.
The above equation yields $Ch_n = 0$ if $n=d_{\parallel}+1$ (mod 2). 



\subsubsection{Winding number}
In the $2 n+1$-dimensional base space,
the winding number is defined by, 
\begin{eqnarray}
N_{2n+1} = 
\frac{n!}{2 (2 \pi i)^{n+1} (2n+1)!} \int \mathrm{tr} \Gamma \left( \mathcal{H}^{-1} d \mathcal{H} \right)^{2n+1}, 
\label{Winding_Number}
\end{eqnarray}
if the Hamiltonian $\mathcal{H}({\bm k}, {\bm r})$ has CS, 
$\Gamma \mathcal{H}(\bm{k},{\bm r}) \Gamma^{-1} = - \mathcal{H}(\bm{k},
{\bm r})$. 
Equivalently, the winding number (\ref{Winding_Number}) is expressed by
the Q-function,
\begin{equation}\begin{split}
N_{2n+1} = 
\frac{(-1)^n n!}{2 (2 \pi i)^{n+1} (2n+1)!} \int \mathrm{tr} \Gamma Q \left( d Q \right)^{2n+1}. 
\end{split}\end{equation}
In the diagonal base of $\Gamma = {\rm diag}(1, -1)$, the Q-function is
off-diagonal, 
\begin{eqnarray}
Q(\bm{k},{\bm r}) = \begin{pmatrix}
0 & q(\bm{k}, {\bm r}) \\
q^{\dag}(\bm{k}, {\bm r}) & 0
\end{pmatrix},
\end{eqnarray}
so the winding number $N_{2n+1}$ is simplified as 
\begin{equation}\begin{split}
N_{2n+1} = 
\frac{n!}{(2 \pi i)^{n+1} (2n+1)!} \int \mathrm{tr} \left[ q d q^{\dag} \right]^{2n+1}.
\end{split}
\end{equation}

In a manner similar to the Chern numbers, 
symmetry of the system sometimes prohibits a
non-zero winding number.
For example, the antiunitary symmetry in
Eq.(\ref{Symmetry-prohibition-A}) leads
\begin{eqnarray}
N_{2n+1}
&=&\frac{(-1)^n n!}{2 (2 \pi i)^{n+1} (2n+1)!} 
\int \mathrm{tr}\left[
\Gamma A^{-1}Q({\bm k}_{\parallel}, -{\bm k}_{\perp}) 
d Q({\bm k}_{\parallel}, -{\bm k}_{\perp})^{2n+1} 
A\right]
\nonumber\\
&=&\frac{\eta_{\Gamma}(-1)^n n!}{2 (2 \pi i)^{n+1} (2n+1)!} 
\int \mathrm{tr}\left[
\Gamma Q^*({\bm k}_{\parallel}, -{\bm k}_{\perp}) 
d Q^*({\bm k}_{\parallel}, -{\bm k}_{\perp})^{2n+1} \right] 
\nonumber\\
&=&\frac{\eta_{\Gamma}(-1)^{2n+1-d_{\parallel}}(-1)^n n!}{2 (2 \pi
 i)^{n+1} (2n+1)!} \int \mathrm{tr}\left[
\Gamma Q^*({\bm k}) 
d Q^*({\bm k})^{2n+1} \right] 
\nonumber\\
&=& (-1)^{n-d_{\parallel}}\eta_{\Gamma}N_{2n+1}^*
\nonumber\\
&=& (-1)^{n-d_{\parallel}}\eta_{\Gamma}N_{2n+1},
\label{Eq::Sym_of_winding_number}
\end{eqnarray}
where $\eta_{\Gamma}=\pm$ specifies the commutation(+) or
anti-commutation(-) relation between $\Gamma$ and $A$. 
Hence, 
$N_{2n+1} = 0$ when 
$n=d_{\parallel}+(\eta_{\Gamma}-1)/2$ (mod 2).


\subsubsection{$2 \mathbb{Z}$ topological invariant}
In the real AZ classes, there are two integer K-groups, 
$K_{\mathbb{R}}(s;d,D)= \mathbb{Z}$ for $s=d-D$ (mod $8$) and 
$K_{\mathbb{R}}(s;d,D)= 2 \mathbb{Z}$ for $s=d-D+4$ (mod $8$),
where ``$2 \mathbb{Z}$'' means that the corresponding Chern number 
defined by Eq.(\ref{Eq::Chern_Character}) or the winding number defined
by Eq.(\ref{Winding_Number})) 
takes an even integer.
Here, we outline the proof why the topological number becomes even when
$s=d-D+4$ (mod $8$) and $d\geq 1$.

Consider a Hamiltonian $\mathcal{H}(\bm{k},\bm{r})$ in real AZ class
with $s=d-D+4$ (mod $8$).  
Choosing one of the momenta as the polar angle $\theta$ of the base
space $S^{d+D}$ and denoting the rest momenta by ${\bm k}'$,
AZ symmetries are expressed as
$T \mathcal{H}(\theta,\bm{k}',\bm{r}) T^{-1} =
 \mathcal{H}(\pi-\theta,-\bm{k}',\bm{r})$,
$C \mathcal{H}(\theta,\bm{k}',\bm{r}) C^{-1} =
-\mathcal{H}(\pi-\theta,-\bm{k}',\bm{r})$, and
$\Gamma \mathcal{H}(\theta,\bm{k}',\bm{r}) \Gamma^{-1} =
-\mathcal{H}(\theta,\bm{k}',\bm{r})$.
%
%
%
Thus, the Hamiltonian on the equator,
$\mathcal{H}(\theta=\pi/2,\bm{k}',\bm{r})$, retains all the AZ
symmetries that
the original Hamiltonian $\mathcal{H}({\bm k}, {\bm r})$ has. 
Furthermore, the equator Hamiltonian is found to be topologically trivial,
since its K-group is given as $K_{\mathbb{R}}(s;d-1,D) =
K_{\mathbb{R}}(s-d+1+D;0,0) =\pi_{0}(\mathcal{R}_{5}) = 0$ when
$s=d-D+4$ (mod 8).
This means that we can pinch the $(d+D)$-dimensional sphere $S^{d+D}$ on
the equator by deforming the equator Hamiltonian into a constant
Hamiltonian without breaking the symmetries.



After the pinching, the north and south hemispheres turn into a couple of
$(d+D)$-dimensional spheres $S^{d+D}$, and the original Hamiltonian
reduces to a couple of Hamiltonians in complex AZ class with the same
$s=d-D+4$ (mod 8). 
Since their K-groups obey 
$K_{\mathbb{C}}(s;d,D)=K_{\mathbb{C}}(s-d+D;0,0)=\pi_{0}(\mathcal{C}_4)=\mathbb{Z}$,   
the couple of Hamiltonian have definite integer topological numbers
$N_{\rm north}$ and $N_{\rm south}$, which are defined by
Eq. (\ref{Eq::Chern_Character}) or Eq. (\ref{Winding_Number}). 
These topological numbers, however, are not independent.
Because TRS and/or PHS in the original Hamiltonian exchange the north
and south hemispheres, $N_{\rm north}$ and $N_{\rm south}$ must be the same.  
Consequently, the topological number of the original Hamiltonian, which
is given by the sum of $N_{\rm north}$ and $N_{\rm south}$, must be
even,
%

\subsection{$\mathbb{Z}_2$ topological invariant}
\label{appendix:Z2_invariant}
In this appendix we summarize various arguments and formulas to define
$\mathbb{Z}_2$ invariants. i.e.
the dimensional reduction,\cite{qi2008topological, ryu2010topological} 
the Moore-Balents argument, \cite{moore2007topological} 
and the integral formulas. 

\begin{table*}[!]
\begin{center}
\caption{
Topological periodic table for topological insulators and
 superconductors.\cite{schnyder2008classification, kitaev2009periodic,teo2010topological} 
The superscripts on $\mathbb{Z}$ and $\mathbb{Z}_2$ specify the integral
 representation of the corresponding topological indices.
$\mathbb{Z}^{(\mathrm{Ch})}$ and $\mathbb{Z}^{(\mathrm{W})}$ are given by the Chern number
 Eq.(\ref{Eq::Chern_Character}) and the winding number
 Eq.(\ref{Winding_Number}), respectively.
$\mathbb{Z}_2^{(\mathrm{CS})}$ and $\mathbb{Z}_2^{(\mathrm{CS}_\mathrm{T})}$ represent the
 Chern-Simons integral Eq.(\ref{Chern-Simons_1st_descendant}) without and with 
the time-reversal constraint Eq.(\ref{Time-Reversal_constraint}), respectively.
$\mathbb{Z}_2^{(\mathrm{FK})}$ denotes the Fu-Kane invariant
Eq.(\ref{Fu-Kane_invariant}).
The $\mathbb{Z}_2$ invariants without any superscript are not expressed
 by these integrals, but they can be defined
 operationally  by the dimensional reduction or the
 Moore-Balents argument.
}
\begin{tabular}[t]{cccccccccccccc}
\hline \hline
$s$ &AZ class &TRS&PHS&chiral &$\mathcal{C}_s$ or $\mathcal{R}_s$ & $\delta=0$ & $\delta=1$ & $\delta=2$ & $\delta=3$ & $\delta=4$ & $\delta=5$ & $\delta=6$ & $\delta=7$ \\
\hline
$0$&A&$0$&$0$&$0$&$\mathcal{C}_0$& $\mathbb{Z}^{(\mathrm{Ch})}$ & $0$ & $\mathbb{Z}^{(\mathrm{Ch})}$ & $0$ & $\mathbb{Z}^{(\mathrm{Ch})}$ & $0$ & $\mathbb{Z}^{(\mathrm{Ch})}$ & $0$ \\
$1$&AIII&$0$&$0$&$1$&$\mathcal{C}_1$& $0$ & $\mathbb{Z}^{(\mathrm{W})}$ & $0$ & $\mathbb{Z}^{(\mathrm{W})}$ & $0$ & $\mathbb{Z}^{(\mathrm{W})}$ & $0$ & $\mathbb{Z}^{(\mathrm{W})}$ \\
\hline 
$0$&AI&$1$&$0$&$0$&$\mathcal{R}_0$& $\mathbb{Z}^{(\mathrm{Ch})}$ & $0$ & $0$  & $0$ & $2\mathbb{Z}^{(\mathrm{Ch})}$ & $0$ & $\mathbb{Z}_2^{(\mathrm{FK})}$ & $\mathbb{Z}_2^{(\mathrm{CS})}$ \\
$1$&BDI&$1$&$1$&$1$&$\mathcal{R}_1$& $\mathbb{Z}_2$ & $\mathbb{Z}^{(\mathrm{W})}$ & $0$ & $0$ & $0$ & $2\mathbb{Z}^{(\mathrm{W})}$ & $0$ & $\mathbb{Z}_2$ \\
$2$&D&$0$&$1$&$0$&$\mathcal{R}_2$& $\mathbb{Z}_2$ & $\mathbb{Z}_2^{(\mathrm{CS})}$ & $\mathbb{Z}^{(\mathrm{Ch})}$ & $0$ & $0$ & $0$ & $2\mathbb{Z}^{(\mathrm{Ch})}$ & $0$ \\
$3$&DIII&$-1$&$1$&$1$&$\mathcal{R}_3$& $0$ & $\mathbb{Z}_2^{(\mathrm{CS}_\mathrm{T})}$ & $\mathbb{Z}_2^{(\mathrm{FK})}$ & $\mathbb{Z}^{(\mathrm{W})}$ & $0$ & $0$ & $0$ & $2\mathbb{Z}^{(\mathrm{W})}$ \\
$4$&AII&$-1$&$0$&$0$&$\mathcal{R}_4$& $2\mathbb{Z}^{(\mathrm{Ch})}$ & $0$ & $\mathbb{Z}_2^{(\mathrm{FK})}$ & $\mathbb{Z}_2^{(\mathrm{CS})}$ & $\mathbb{Z}^{(\mathrm{Ch})}$ & $0$ & $0$ & $0$ \\
$5$&CII&$-1$&$-1$&$1$&$\mathcal{R}_5$& $0$ & $2\mathbb{Z}^{(\mathrm{W})}$ & $0$ & $\mathbb{Z}_2$ & $\mathbb{Z}_2$ & $\mathbb{Z}^{(\mathrm{W})}$ & $0$ & $0$  \\
$6$&C&$0$&$-1$&$0$&$\mathcal{R}_6$& $0$ & $0$ & $2\mathbb{Z}^{(\mathrm{Ch})}$ & $0$ & $\mathbb{Z}_2$ & $\mathbb{Z}_2^{(\mathrm{CS})}$ & $\mathbb{Z}^{(\mathrm{Ch})}$ & $0$ \\
$7$&CI&$1$&$-1$&$1$&$\mathcal{R}_7$& $0$ & $0$  & $0$ & $2\mathbb{Z}^{(\mathrm{W})}$ & $0$ & $\mathbb{Z}_2^{(\mathrm{CS}_\mathrm{T})}$ & $\mathbb{Z}_2^{(\mathrm{FK})}$ & $\mathbb{Z}^{(\mathrm{W})}$ \\
\hline \hline
\end{tabular}
\label{Periodic_Table}
\end{center}
\end{table*}

\subsubsection{Dimensional reduction}
\label{appendix:dimensionalreduction}

In our topological periodic tables,  a sequence of $\mathbb{Z}_2$ indices 
follows a $\mathbb{Z}$ index 
as the dimension of the system decreases.
This structure makes it possible to define the corresponding
$\mathbb{Z}_2$ invariants by dimensional reduction:\cite{qi2008topological,
ryu2010topological} 
Let us consider a $(d+2)$-dimensional Hamiltonian $\mathcal{H}({\bm k}, k_{d+1}, k_{d+2},
{\bm r})$
that is characterized by the $\mathbb{Z}$ index
mentioned in the above.
Then, we can construct maps from this Hamiltonian to one and two lower
dimensional Hamiltonians, by considering $\mathcal{H}({\bm k}, k_{d+1},
0, {\bm r})$ and $\mathcal{H}({\bm k}, 0, 0, {\bm r})$, respectively.
These maps define surjective homomorphic maps from $\mathbb{Z}$ to
$\mathbb{Z}_2$.  
%
%
%
As a result, the first and second descendant $\mathbb{Z}_2$ invariants
, $\nu_{\rm 1st}$ and $\nu_{\rm 2nd}$, of the lower dimensional
Hamiltonians are obtained as
\begin{eqnarray}
(-1)^{\nu_{\rm 1st}} =(-1)^{\nu_{\rm 2nd}}= (-1)^N,  
\end{eqnarray}
%
where $N$ is the integer topological invariant of $\mathcal{H}(\bm{k},
k_{d+1}, k_{d+2}, \bm{r})$. 
$N$ is the Chern number Eq.(\ref{Eq::Chern_Character}) for non-chiral class 
or the winding number Eq. (\ref{Winding_Number}) for chiral class. 
%

\subsubsection{Moore-Balents argument for second descendant $\mathbb{Z}_2$ index}

For the second descendant $\mathbb{Z}_2$ index ($d=s+D-2$) of Table
\ref{Periodic_Table} with $d\geq 1$,
there is another operational definition of the $\mathbb{Z}_2$ invariant,
which was first 
discussed by Moore and Balents.\cite{moore2007topological}
Consider a Hamiltonian $\mathcal{H}(\bm{k},\bm{r})$ in real AZ class
with $d=s+D-2$.
Choosing one of the momenta as the polar angle $\theta$ of the base sphere
$S^{d+D}$ and denoting the rest by ${\bm k}'$, 
the AZ symmetries are expressed as
$T \mathcal{H}(\theta,\bm{k}',\bm{r}) T^{-1} =
 \mathcal{H}(\pi-\theta,-\bm{k}',\bm{r})$,
$C \mathcal{H}(\theta,\bm{k}',\bm{r}) C^{-1} =
-\mathcal{H}(\pi-\theta,-\bm{k}',\bm{r})$, and
$\Gamma \mathcal{H}(\theta,\bm{k}',\bm{r}) \Gamma^{-1} =
-\mathcal{H}(\theta,\bm{k}',\bm{r})$.
Then, take only the north hemisphere ($0\le \theta \le \pi/2$) of the
system.
Although TRS and/or PHS cannot be retained only on the north
hemisphere, they are retained at its boundary, i.e. the equator.
Indeed the Hamiltonian on the equator $\mathcal{H}(\theta=\pi/2, {\bm k}',
{\bm r})$ has the same symmetry of the original Hamiltonian
$\mathcal{H}({\bm k}, {\bm r})$, and thus  its K-group is 
$K_{\mathbb{R}}(s;d-1,D)=\pi_0(\mathcal{R}_{3})=0$.

To define the topological number, we introduce another hemisphere in
the following manner. 
As I mentioned in the above, the K-group of the equator Hamiltonian is trivial.
Therefore the equator Hamiltonian can
smoothly shrink into a point $\mathcal{H}_0$ with keeping the
AZ symmetry of the  $(d-1)$-dimensional momentum space.
This deformation defines a Hamiltonian $\tilde{\mathcal{H}}(\theta, {\bm
k}', {\bm r})$ on a new hemisphere, say, a new south hemisphere,  where the new
Hamiltonian interpolates $\mathcal{H}(\theta=\pi/2, {\bm k}', {\bm r})$
at the equator $(\theta=\pi/2)$ to
$\mathcal{H}_0$ at the south pole ($\theta=\pi$).
Note here that $\theta$ of $\tilde{\mathcal{H}}(\theta,{\bm
k}', {\bm r})$ is just an interpolating parameter, and thus it transforms 
trivially under the AZ symmetries as
$T \tilde{\mathcal{H}}(\theta,\bm{k}',\bm{r}) T^{-1} =
\tilde{\mathcal{H}}(\theta,-\bm{k}',\bm{r})$,
$C \tilde{\mathcal{H}}(\theta,\bm{k}',\bm{r}) C^{-1} =
-\tilde{\mathcal{H}}(\theta,-\bm{k}',\bm{r})$, and
$\Gamma \tilde{\mathcal{H}}(\theta,\bm{k}',\bm{r}) \Gamma^{-1} =
-\tilde{\mathcal{H}}(\theta,\bm{k}',\bm{r})$.

Now define the topological number.
Sewing the new south and the original north hemispheres together, 
we have a Hamiltonian on a sphere. 
In contrast to the original Hamiltonian, the resultant Hamiltonian no longer has TRS and/or PHS since
$\theta$ transforms differently in the north hemisphere and the south
hemisphere.
It belongs to a complex AZ class (A or AIII), so it can
host a nonzero integer topological number $N$ given by the Chern number
$Ch_{(d+D)/2}$ or the winding
number $N_{d+D}$.
Its value, however, depends on the choice of the interpolating Hamiltonian 
$\tilde{\mathcal{H}}(\theta, {\bm k}',{\bm r})$ in general.
Therefore,
$N$ itself does not characterize the topological nature of
the original Hamiltonian. 
Nevertheless, its parity $(-1)^N$ is uniquely determined:
Take another interpolating Hamiltonian
$\tilde{\mathcal{H}}'(\theta,{\bm k}',{\bm r})$ which may give a different
integer $N'$. 
The difference between $N$ and $N'$ can be evaluated as
the topological number of
the Hamiltonian that is obtained by sewing the hemispheres of
$\tilde{\mathcal{H}}(\theta, {\bm k}',{\bm r})$ 
and
$\tilde{\mathcal{H}}'(\theta, {\bm k}',{\bm r})$ together. 
This time, the combined Hamiltonian keeps TRS and/or PHS which are
the
same as those of the original Hamiltonian except the ${\bm r}$-type
transformation of $\theta$.
Therefore, its K-group is $K_{\mathbb{R}}(s,d-1,D+1)=\pi_0(R_4)=2
\mathbb{Z}$, which implies that $N-N'$ must be even.
As a result, the parity $(-1)^N$ is unique, i.e. $(-1)^N=(-1)^{N'}$.
The parity defines the $\mathbb{Z}_2$ invariant of the original Hamiltonian. 

\subsubsection{Chern-Simons invariant for first descendant
   $\mathbb{Z}_2$ index in odd-dimensional non-chiral real class}

The integral representation of the first descendant $\mathbb{Z}_2$ invariant
in non-chiral real class is given by the Chern-Simons
form. \cite{qi2008topological}
Consider a Hamiltonian $\mathcal{H}(\bm{k},\bm{r})$ on the base space
$S^{d+D}$ with odd $d+D$.
The $\mathbb{Z}_2$ topological invariant is given by
\begin{equation}\begin{split}
\nu = \frac{2}{((d+D+1)/2)!} \left( \frac{i}{2 \pi} \right)^{(d+D+1)/2} \int_{S^{d+D}} CS_{d+D} \ \ (\mathrm{mod}\ 2). 
\end{split}\label{Chern-Simons_1st_descendant}\end{equation}
Here $CS_{d+D}$ is the Chern-Simons $(d+D)$-form given by \cite{nakahara2003geometry}
\begin{equation}\begin{split}
CS_{2n+1} = (n+1) \int_0^1 d t \mathrm{tr} \left( \mathcal{A} (t d \mathcal{A} + t^2 \mathcal{A}^2)^n \right) 
\end{split}\end{equation}
where $\mathcal{A}_{\alpha\beta}(\bm{k},\bm{r}) =
\Braket{u_{\alpha}(\bm{k},\bm{r})|d u_{\beta}(\bm{k},\bm{r})}$ is the connection defined by occupied states $\ket{u_{\alpha}(\bm{k},\bm{r})}$. 
Some of lower dimensional Chern-Simons forms are 
\begin{equation}\begin{split}
CS_1 &= \mathrm{tr} \mathcal{A} , \\
CS_3 &= \mathrm{tr} \left( \mathcal{A} d \mathcal{A} + \frac{2}{3} \mathcal{A}^3 \right) , \\
CS_5 &= \mathrm{tr} \left( \mathcal{A} (d \mathcal{A})^2 + \frac{3}{2} \mathcal{A}^3 d \mathcal{A} + \frac{3}{5} \mathcal{A}^5 \right) . \\
\end{split}\end{equation}
Here phases of the occupied states should be globally defined on the overall
parameter manifold $S^{d+D}$ so as the connection $\mathcal{A}$ is non-singular.
The $\mathbb{Z}_2$ nontriviality of this integral is ensured by the
dimensional reduction discussed in Sec.\ref{appendix:dimensionalreduction}.

The Chern-Simons invariant characterizes
the real AZ classes with $(s, \delta)=(2n, 2n-1)$ (mod 8) in
Table \ref{Periodic_Table}.
\subsubsection{Fu-Kane invariant for first and second descendant $\mathbb{Z}_2$
   indices in even-dimensional TRS class}

In the presence of TRS,  the $\mathbb{Z}_2$ invariant can be introduced
as
\cite{fu2006time}
\begin{equation}\begin{split}
\nu = \frac{1}{((d+D)/2)!} \left( \frac{i}{2 \pi} \right)^{(d+D)/2} \left[ \int_{S^{d+D}_{1/2}} \mathrm{tr} \mathcal{F}^{(d+D)/2} - \oint_{\partial S^{d+D}_{1/2}} CS_{d+D-1} \right] \ \ (\mathrm{mod}\ 2), 
\end{split}\label{Fu-Kane_invariant}\end{equation}
where $S^{d+D}_{1/2}$ is a (north) hemisphere of $S^{d+D}$ and $\partial S^{d+D}_{1/2} \cong
S^{d+D-1}$ is the 
equator.
We suppose here that the north hemisphere and the south one are
exchanged by TRS,  but the equator is invariant.
%
The valence band wave functions of the Chern-Simons form in
Eq.(\ref{Fu-Kane_invariant}) must be smoothly defined on the
equator $\partial S^{d+D}_{1/2}$ 
 (not on the hemisphere $S^{d+D}_{1/2}$). 
An appropriate gauge condition is
needed to obtain the $\mathbb{Z}_2$ nontriviality, and thus
we impose the time-reversal constraint for the valence band Bloch wave
functions $\{\Ket{u_n(\bm{k},\bm{r})}\}$ as \cite{fu2006time}
\begin{eqnarray}
w_{mn}(\bm{k},\bm{r}) = \Braket{u_m(-\bm{k},\bm{r}) | T
 u_n(\bm{k},\bm{r})} \equiv {\rm const}.  
\label{Time-Reversal_constraint}
\end{eqnarray}
on the equator $({\bm k}, {\bm r})\in \partial S^{d+D}_{1/2}$. 
The $\mathbb{Z}_2$ invariant  Eq.(\ref{Fu-Kane_invariant}) picks up an
obstruction to choosing the gauge satisfying
(\ref{Time-Reversal_constraint}) on overall Brillouin zone.

In Table \ref{Periodic_Table}, the Fu-Kane invariant characterizes 
the real AZ classes with $(s, \delta)=(4n+3, 4n+2)$ (mod 8) and those with  
$(s, \delta)=(4n+4, 4n+2)$ (mod 8).
Note that the Fu-Kane invariant is not applicable to class BDI and CII,
since the presence of CS that commutes with TRS makes the integral
Eq.(\ref{Fu-Kane_invariant}) trivial.


\subsubsection{Constrained Chern-Simons invariant for second descendant
   $\mathbb{Z}_2$ index in odd-dimensional chiral TRS class}

Consider a Hamiltonian $\mathcal{H}({\bm k}, {\bm r})$ on the base space
$S^{d+D}$ with odd $d+D$.
If the Hamiltonian has CS and TRS that anti-commute with each other, then 
the $\mathbb{Z}_2$ invariant of the Hamiltonian can be given in a form
of the Chern-Simons integral: 
To see this, first consider the
dimension raising map in Eq.(\ref{eq:DRC}),
\begin{eqnarray}
\mathcal{H}({\bm k}, {\bm r}, \theta)=\sin \theta \mathcal{H}({\bm k},
 {\bm r})+\cos\theta \Gamma. 
\end{eqnarray}
Since the mapped Hamiltonian on $S^{d+D+1}$ has TRS and is
even-dimensional, we can define the Fu-Kane $\mathbb{Z}_2$ invariant of
$\mathcal{H}(\theta, {\bm k}, {\bm r})$ as
\begin{equation}\begin{split}
\nu = \frac{1}{((d+D+1)/2)!} \left( \frac{i}{2 \pi} \right)^{(d+D+1)/2}
		 \left[ \int_{S^{d+D+1}_{1/2}} \mathrm{tr}
		 \mathcal{F}^{(d+D+1)/2} - \oint_{\partial S^{d+D+1}_{1/2}}
		 CS_{d+D} \right] \ \ (\mathrm{mod}\ 2).  
\end{split}
\label{Fu-Kane_invariant2}
\end{equation}
It is convenient to choose a $\bm k$-type $\theta$ and take the equator
as $\theta=\pi/2$.
Then, we can show  that the first term of Eq.(\ref{Fu-Kane_invariant2}) is
identically zero due to TRS and CS of $\mathcal{H}({\bm k}, {\bm
r})$.\cite{teo2010topological}
Also the equator is nothing but the original base space $S^{d+D}$,
Eq.(\ref{Fu-Kane_invariant2}) is recast into
\begin{equation}\begin{split}
\nu = \frac{1}{((d+D+1)/2)!} \left( \frac{i}{2 \pi} \right)^{(d+D+1)/2} \int_{S^{d+D}} CS_{d+D} \ \ (\mathrm{mod}\ 2), 
\end{split}\label{Chern-Simons_2nd_descendant}\end{equation}
with the time-reversal constraint Eq.(\ref{Time-Reversal_constraint})
on $({\bm k}, {\bm r})\in S^{d+D}$.\cite{teo2010topological}  
Note here that Eq.(\ref{Chern-Simons_2nd_descendant}) is a half of
Eq.(\ref{Chern-Simons_1st_descendant}) so the additional gauge constraint
Eq.(\ref{Time-Reversal_constraint}) is necessary to obtain the
$\mathbb{Z}_2$ nontriviality.

%

%

The constrained Chern-Simons invariant characterizes the real AZ classes
with $(s,\delta)=(4n+3,4n+1)$ (mod 8) in Table \ref{Periodic_Table}.


%


\begin{thebibliography}{150}
\expandafter\ifx\csname natexlab\endcsname\relax\def\natexlab#1{#1}\fi
\expandafter\ifx\csname bibnamefont\endcsname\relax
  \def\bibnamefont#1{#1}\fi
\expandafter\ifx\csname bibfnamefont\endcsname\relax
  \def\bibfnamefont#1{#1}\fi
\expandafter\ifx\csname citenamefont\endcsname\relax
  \def\citenamefont#1{#1}\fi
\expandafter\ifx\csname url\endcsname\relax
  \def\url#1{\texttt{#1}}\fi
\expandafter\ifx\csname urlprefix\endcsname\relax\def\urlprefix{URL }\fi
\providecommand{\bibinfo}[2]{#2}
\providecommand{\eprint}[2][]{\url{#2}}

\bibitem[{\citenamefont{Volovik}(2003)}]{volovik2003universe}
\bibinfo{author}{\bibfnamefont{G.~E.} \bibnamefont{Volovik}},
  \emph{\bibinfo{title}{The Universe in a helium droplet}}
  (\bibinfo{publisher}{Clarendon Press}, \bibinfo{year}{2003}).

\bibitem[{\citenamefont{Thouless et~al.}(1982)\citenamefont{Thouless, Kohmoto,
  Nightingale, and den Nijs}}]{thouless1982quantized}
\bibinfo{author}{\bibfnamefont{D.~J.} \bibnamefont{Thouless}},
  \bibinfo{author}{\bibfnamefont{M.}~\bibnamefont{Kohmoto}},
  \bibinfo{author}{\bibfnamefont{M.~P.} \bibnamefont{Nightingale}},
  \bibnamefont{and} \bibinfo{author}{\bibfnamefont{M.}~\bibnamefont{den Nijs}},
  \bibinfo{journal}{Phys. Rev. Lett.} \textbf{\bibinfo{volume}{49}},
  \bibinfo{pages}{405} (\bibinfo{year}{1982}).

\bibitem[{\citenamefont{Kohmoto}(1985)}]{kohmoto1985topological}
\bibinfo{author}{\bibfnamefont{M.}~\bibnamefont{Kohmoto}},
  \bibinfo{journal}{Ann. Phys.} \textbf{\bibinfo{volume}{160}},
  \bibinfo{pages}{343} (\bibinfo{year}{1985}).

\bibitem[{\citenamefont{Niu et~al.}(1985)\citenamefont{Niu, Thouless, and
  Wu}}]{niu1985quantized}
\bibinfo{author}{\bibfnamefont{Q.}~\bibnamefont{Niu}},
  \bibinfo{author}{\bibfnamefont{D.~J.} \bibnamefont{Thouless}},
  \bibnamefont{and} \bibinfo{author}{\bibfnamefont{Y.-S.} \bibnamefont{Wu}},
  \bibinfo{journal}{Phys. Rev. B} \textbf{\bibinfo{volume}{31}},
  \bibinfo{pages}{3372} (\bibinfo{year}{1985}).

\bibitem[{\citenamefont{Hasan and Kane}(2010)}]{hasan2010colloquium}
\bibinfo{author}{\bibfnamefont{M.~Z.} \bibnamefont{Hasan}} \bibnamefont{and}
  \bibinfo{author}{\bibfnamefont{C.~L.} \bibnamefont{Kane}},
  \bibinfo{journal}{Rev. Mod. Phys.} \textbf{\bibinfo{volume}{82}},
  \bibinfo{pages}{3045} (\bibinfo{year}{2010}).

\bibitem[{\citenamefont{Qi and Zhang}(2011)}]{qi2011topological}
\bibinfo{author}{\bibfnamefont{X.-L.} \bibnamefont{Qi}} \bibnamefont{and}
  \bibinfo{author}{\bibfnamefont{S.-C.} \bibnamefont{Zhang}},
  \bibinfo{journal}{Rev. Mod. Phys.} \textbf{\bibinfo{volume}{83}},
  \bibinfo{pages}{1057} (\bibinfo{year}{2011}).

\bibitem[{\citenamefont{Volovik}(2011)}]{volovik2011topology}
\bibinfo{author}{\bibfnamefont{G.~E.} \bibnamefont{Volovik}},
  \bibinfo{journal}{arXiv:1111.4627}  (\bibinfo{year}{2011}).

\bibitem[{\citenamefont{Tanaka et~al.}(2012{\natexlab{a}})\citenamefont{Tanaka,
  Sato, and Nagaosa}}]{tanaka2012symmetry}
\bibinfo{author}{\bibfnamefont{Y.}~\bibnamefont{Tanaka}},
  \bibinfo{author}{\bibfnamefont{M.}~\bibnamefont{Sato}}, \bibnamefont{and}
  \bibinfo{author}{\bibfnamefont{N.}~\bibnamefont{Nagaosa}},
  \bibinfo{journal}{J. Phys. Soc. Jpn.} \textbf{\bibinfo{volume}{81}},
  \bibinfo{pages}{011013} (\bibinfo{year}{2012}{\natexlab{a}}).

\bibitem[{\citenamefont{Ando}(2013)}]{ando2013topological}
\bibinfo{author}{\bibfnamefont{Y.}~\bibnamefont{Ando}}, \bibinfo{journal}{J.
  Phys. Soc. Jpn.} \textbf{\bibinfo{volume}{82}}, \bibinfo{pages}{102001}
  (\bibinfo{year}{2013}).

\bibitem[{\citenamefont{Budich and Trauzettel}(2013)}]{budich2013adiabatic}
\bibinfo{author}{\bibfnamefont{J.~C.} \bibnamefont{Budich}} \bibnamefont{and}
  \bibinfo{author}{\bibfnamefont{B.}~\bibnamefont{Trauzettel}},
  \bibinfo{journal}{Phys. Status Solidi RRL} \textbf{\bibinfo{volume}{7}},
  \bibinfo{pages}{109} (\bibinfo{year}{2013}).

\bibitem[{\citenamefont{Fruchart and
  Carpentier}(2013)}]{fruchart2013introduction}
\bibinfo{author}{\bibfnamefont{M.}~\bibnamefont{Fruchart}} \bibnamefont{and}
  \bibinfo{author}{\bibfnamefont{D.}~\bibnamefont{Carpentier}},
  \bibinfo{journal}{C.R. Phys.} \textbf{\bibinfo{volume}{14}},
  \bibinfo{pages}{779} (\bibinfo{year}{2013}).

\bibitem[{\citenamefont{Kane and Mele}(2005{\natexlab{a}})}]{kane2005quantum}
\bibinfo{author}{\bibfnamefont{C.~L.} \bibnamefont{Kane}} \bibnamefont{and}
  \bibinfo{author}{\bibfnamefont{E.~J.} \bibnamefont{Mele}},
  \bibinfo{journal}{Phys. Rev. Lett.} \textbf{\bibinfo{volume}{95}},
  \bibinfo{pages}{226801} (\bibinfo{year}{2005}{\natexlab{a}}).

\bibitem[{\citenamefont{Kane and Mele}(2005{\natexlab{b}})}]{kane2005z_}
\bibinfo{author}{\bibfnamefont{C.~L.} \bibnamefont{Kane}} \bibnamefont{and}
  \bibinfo{author}{\bibfnamefont{E.~J.} \bibnamefont{Mele}},
  \bibinfo{journal}{Phys. Rev. Lett.} \textbf{\bibinfo{volume}{95}},
  \bibinfo{pages}{146802} (\bibinfo{year}{2005}{\natexlab{b}}).

\bibitem[{\citenamefont{Bernevig and Zhang}(2006)}]{bernevig2006quantum}
\bibinfo{author}{\bibfnamefont{B.~A.} \bibnamefont{Bernevig}} \bibnamefont{and}
  \bibinfo{author}{\bibfnamefont{S.-C.} \bibnamefont{Zhang}},
  \bibinfo{journal}{Phys. Rev. Lett.} \textbf{\bibinfo{volume}{96}},
  \bibinfo{pages}{106802} (\bibinfo{year}{2006}).

\bibitem[{\citenamefont{Moore and Balents}(2007)}]{moore2007topological}
\bibinfo{author}{\bibfnamefont{J.~E.} \bibnamefont{Moore}} \bibnamefont{and}
  \bibinfo{author}{\bibfnamefont{L.}~\bibnamefont{Balents}},
  \bibinfo{journal}{Phys. Rev. B} \textbf{\bibinfo{volume}{75}},
  \bibinfo{pages}{121306} (\bibinfo{year}{2007}).

\bibitem[{\citenamefont{Fu et~al.}(2007)\citenamefont{Fu, Kane, and
  Mele}}]{fu2007topological}
\bibinfo{author}{\bibfnamefont{L.}~\bibnamefont{Fu}},
  \bibinfo{author}{\bibfnamefont{C.~L.} \bibnamefont{Kane}}, \bibnamefont{and}
  \bibinfo{author}{\bibfnamefont{E.~J.} \bibnamefont{Mele}},
  \bibinfo{journal}{Phys. Rev. Lett.} \textbf{\bibinfo{volume}{98}},
  \bibinfo{pages}{106803} (\bibinfo{year}{2007}).

\bibitem[{\citenamefont{Fu and Kane}(2007)}]{fu2007topological2}
\bibinfo{author}{\bibfnamefont{L.}~\bibnamefont{Fu}} \bibnamefont{and}
  \bibinfo{author}{\bibfnamefont{C.~L.} \bibnamefont{Kane}},
  \bibinfo{journal}{Phys. Rev. B} \textbf{\bibinfo{volume}{76}},
  \bibinfo{pages}{045302} (\bibinfo{year}{2007}).

\bibitem[{\citenamefont{Roy}(2009{\natexlab{a}})}]{roy2009z_}
\bibinfo{author}{\bibfnamefont{R.}~\bibnamefont{Roy}}, \bibinfo{journal}{Phys.
  Rev. B} \textbf{\bibinfo{volume}{79}}, \bibinfo{pages}{195321}
  (\bibinfo{year}{2009}{\natexlab{a}}).

\bibitem[{\citenamefont{Qi et~al.}(2008)\citenamefont{Qi, Hughes, and
  Zhang}}]{qi2008topological}
\bibinfo{author}{\bibfnamefont{X.-L.} \bibnamefont{Qi}},
  \bibinfo{author}{\bibfnamefont{T.~L.} \bibnamefont{Hughes}},
  \bibnamefont{and} \bibinfo{author}{\bibfnamefont{S.-C.} \bibnamefont{Zhang}},
  \bibinfo{journal}{Phys. Rev. B} \textbf{\bibinfo{volume}{78}},
  \bibinfo{pages}{195424} (\bibinfo{year}{2008}).

\bibitem[{\citenamefont{Read and Green}(2000)}]{read2000paired}
\bibinfo{author}{\bibfnamefont{N.}~\bibnamefont{Read}} \bibnamefont{and}
  \bibinfo{author}{\bibfnamefont{D.}~\bibnamefont{Green}},
  \bibinfo{journal}{Phys. Rev. B} \textbf{\bibinfo{volume}{61}},
  \bibinfo{pages}{10267} (\bibinfo{year}{2000}).

\bibitem[{\citenamefont{Ivanov}(2001)}]{ivanov2001non}
\bibinfo{author}{\bibfnamefont{D.~A.} \bibnamefont{Ivanov}},
  \bibinfo{journal}{Phys. Rev. Lett.} \textbf{\bibinfo{volume}{86}},
  \bibinfo{pages}{268} (\bibinfo{year}{2001}).

\bibitem[{\citenamefont{Kitaev}(2001)}]{kitaev2001unpaired}
\bibinfo{author}{\bibfnamefont{A.~Y.} \bibnamefont{Kitaev}},
  \bibinfo{journal}{Phys. Usp.} \textbf{\bibinfo{volume}{44}},
  \bibinfo{pages}{131} (\bibinfo{year}{2001}).

\bibitem[{\citenamefont{Sato}(2003)}]{sato2003non}
\bibinfo{author}{\bibfnamefont{M.}~\bibnamefont{Sato}}, \bibinfo{journal}{Phys.
  Lett. B} \textbf{\bibinfo{volume}{575}}, \bibinfo{pages}{126}
  (\bibinfo{year}{2003}).

\bibitem[{\citenamefont{Fu and Kane}(2008)}]{fu2008superconducting}
\bibinfo{author}{\bibfnamefont{L.}~\bibnamefont{Fu}} \bibnamefont{and}
  \bibinfo{author}{\bibfnamefont{C.~L.} \bibnamefont{Kane}},
  \bibinfo{journal}{Phys. Rev. Lett.} \textbf{\bibinfo{volume}{100}},
  \bibinfo{pages}{096407} (\bibinfo{year}{2008}).

\bibitem[{\citenamefont{Linder et~al.}(2010)\citenamefont{Linder, Tanaka,
  Yokoyama, Sudb{\o}, and Nagaosa}}]{linder2009unconventional}
\bibinfo{author}{\bibfnamefont{J.}~\bibnamefont{Linder}},
  \bibinfo{author}{\bibfnamefont{Y.}~\bibnamefont{Tanaka}},
  \bibinfo{author}{\bibfnamefont{T.}~\bibnamefont{Yokoyama}},
  \bibinfo{author}{\bibfnamefont{A.}~\bibnamefont{Sudb{\o}}}, \bibnamefont{and}
  \bibinfo{author}{\bibfnamefont{N.}~\bibnamefont{Nagaosa}},
  \bibinfo{journal}{Phys. Rev. Lett.} \textbf{\bibinfo{volume}{104}},
  \bibinfo{pages}{067001} (\bibinfo{year}{2010}).

\bibitem[{\citenamefont{Qi et~al.}(2009)\citenamefont{Qi, Hughes, Raghu, and
  Zhang}}]{qi2009time}
\bibinfo{author}{\bibfnamefont{X.-L.} \bibnamefont{Qi}},
  \bibinfo{author}{\bibfnamefont{T.~L.} \bibnamefont{Hughes}},
  \bibinfo{author}{\bibfnamefont{S.}~\bibnamefont{Raghu}}, \bibnamefont{and}
  \bibinfo{author}{\bibfnamefont{S.-C.} \bibnamefont{Zhang}},
  \bibinfo{journal}{Phys. Rev. Lett.} \textbf{\bibinfo{volume}{102}},
  \bibinfo{pages}{187001} (\bibinfo{year}{2009}).

\bibitem[{\citenamefont{Schnyder et~al.}(2008)\citenamefont{Schnyder, Ryu,
  Furusaki, and Ludwig}}]{schnyder2008classification}
\bibinfo{author}{\bibfnamefont{A.~P.} \bibnamefont{Schnyder}},
  \bibinfo{author}{\bibfnamefont{S.}~\bibnamefont{Ryu}},
  \bibinfo{author}{\bibfnamefont{A.}~\bibnamefont{Furusaki}}, \bibnamefont{and}
  \bibinfo{author}{\bibfnamefont{A.~W.~W.} \bibnamefont{Ludwig}},
  \bibinfo{journal}{Phys. Rev. B} \textbf{\bibinfo{volume}{78}},
  \bibinfo{pages}{195125} (\bibinfo{year}{2008}).

\bibitem[{\citenamefont{Sato}(2009)}]{sato2009topologicalp}
\bibinfo{author}{\bibfnamefont{M.}~\bibnamefont{Sato}}, \bibinfo{journal}{Phys.
  Rev. B} \textbf{\bibinfo{volume}{79}}, \bibinfo{pages}{214526}
  (\bibinfo{year}{2009}).

\bibitem[{\citenamefont{Sato}(2010)}]{sato2010topological}
\bibinfo{author}{\bibfnamefont{M.}~\bibnamefont{Sato}}, \bibinfo{journal}{Phys.
  Rev. B} \textbf{\bibinfo{volume}{81}}, \bibinfo{pages}{220504(R)}
  (\bibinfo{year}{2010}).

\bibitem[{\citenamefont{Tanaka et~al.}(2009{\natexlab{a}})\citenamefont{Tanaka,
  Yokoyama, Balatsky, and Nagaosa}}]{tanaka2009theory}
\bibinfo{author}{\bibfnamefont{Y.}~\bibnamefont{Tanaka}},
  \bibinfo{author}{\bibfnamefont{T.}~\bibnamefont{Yokoyama}},
  \bibinfo{author}{\bibfnamefont{A.~V.} \bibnamefont{Balatsky}},
  \bibnamefont{and} \bibinfo{author}{\bibfnamefont{N.}~\bibnamefont{Nagaosa}},
  \bibinfo{journal}{Phys. Rev. B} \textbf{\bibinfo{volume}{79}},
  \bibinfo{pages}{060505(R)} (\bibinfo{year}{2009}{\natexlab{a}}).

\bibitem[{\citenamefont{Sato and Fujimoto}(2009)}]{sato2009topological}
\bibinfo{author}{\bibfnamefont{M.}~\bibnamefont{Sato}} \bibnamefont{and}
  \bibinfo{author}{\bibfnamefont{S.}~\bibnamefont{Fujimoto}},
  \bibinfo{journal}{Phys. Rev. B} \textbf{\bibinfo{volume}{79}},
  \bibinfo{pages}{094504} (\bibinfo{year}{2009}).

\bibitem[{\citenamefont{Sato et~al.}(2009)\citenamefont{Sato, Takahashi, and
  Fujimoto}}]{sato2009non}
\bibinfo{author}{\bibfnamefont{M.}~\bibnamefont{Sato}},
  \bibinfo{author}{\bibfnamefont{Y.}~\bibnamefont{Takahashi}},
  \bibnamefont{and} \bibinfo{author}{\bibfnamefont{S.}~\bibnamefont{Fujimoto}},
  \bibinfo{journal}{Phys. Rev. Lett.} \textbf{\bibinfo{volume}{103}},
  \bibinfo{pages}{020401} (\bibinfo{year}{2009}).

\bibitem[{\citenamefont{Tanaka et~al.}(2010)\citenamefont{Tanaka, Mizuno,
  Yokoyama, Yada, and Sato}}]{PhysRevLett.105.097002}
\bibinfo{author}{\bibfnamefont{Y.}~\bibnamefont{Tanaka}},
  \bibinfo{author}{\bibfnamefont{Y.}~\bibnamefont{Mizuno}},
  \bibinfo{author}{\bibfnamefont{T.}~\bibnamefont{Yokoyama}},
  \bibinfo{author}{\bibfnamefont{K.}~\bibnamefont{Yada}}, \bibnamefont{and}
  \bibinfo{author}{\bibfnamefont{M.}~\bibnamefont{Sato}},
  \bibinfo{journal}{Phys. Rev. Lett.} \textbf{\bibinfo{volume}{105}},
  \bibinfo{pages}{097002} (\bibinfo{year}{2010}).

\bibitem[{\citenamefont{Sau et~al.}(2010)\citenamefont{Sau, Lutchyn, Tewari,
  and Das~Sarma}}]{sau2010generic}
\bibinfo{author}{\bibfnamefont{J.~D.} \bibnamefont{Sau}},
  \bibinfo{author}{\bibfnamefont{R.~M.} \bibnamefont{Lutchyn}},
  \bibinfo{author}{\bibfnamefont{S.}~\bibnamefont{Tewari}}, \bibnamefont{and}
  \bibinfo{author}{\bibfnamefont{S.}~\bibnamefont{Das~Sarma}},
  \bibinfo{journal}{Phys. Rev. Lett.} \textbf{\bibinfo{volume}{104}},
  \bibinfo{pages}{040502} (\bibinfo{year}{2010}).

\bibitem[{\citenamefont{Alicea}(2010)}]{alicea2010majorana}
\bibinfo{author}{\bibfnamefont{J.}~\bibnamefont{Alicea}},
  \bibinfo{journal}{Phys. Rev. B} \textbf{\bibinfo{volume}{81}},
  \bibinfo{pages}{125318} (\bibinfo{year}{2010}).

\bibitem[{\citenamefont{Sato et~al.}(2010)\citenamefont{Sato, Takahashi, and
  Fujimoto}}]{sato2010non}
\bibinfo{author}{\bibfnamefont{M.}~\bibnamefont{Sato}},
  \bibinfo{author}{\bibfnamefont{Y.}~\bibnamefont{Takahashi}},
  \bibnamefont{and} \bibinfo{author}{\bibfnamefont{S.}~\bibnamefont{Fujimoto}},
  \bibinfo{journal}{Phys. Rev. B} \textbf{\bibinfo{volume}{82}},
  \bibinfo{pages}{134521} (\bibinfo{year}{2010}).

\bibitem[{\citenamefont{Sato and Fujimoto}(2010)}]{sato2010existence}
\bibinfo{author}{\bibfnamefont{M.}~\bibnamefont{Sato}} \bibnamefont{and}
  \bibinfo{author}{\bibfnamefont{S.}~\bibnamefont{Fujimoto}},
  \bibinfo{journal}{Phys. Rev. Lett.} \textbf{\bibinfo{volume}{105}},
  \bibinfo{pages}{217001} (\bibinfo{year}{2010}).

\bibitem[{\citenamefont{Lutchyn et~al.}(2010)\citenamefont{Lutchyn, Sau, and
  Das~Sarma}}]{lutchyn2010}
\bibinfo{author}{\bibfnamefont{R.~M.} \bibnamefont{Lutchyn}},
  \bibinfo{author}{\bibfnamefont{J.~D.} \bibnamefont{Sau}}, \bibnamefont{and}
  \bibinfo{author}{\bibfnamefont{S.}~\bibnamefont{Das~Sarma}},
  \bibinfo{journal}{Phys. Rev. Lett.} \textbf{\bibinfo{volume}{105}},
  \bibinfo{pages}{077001} (\bibinfo{year}{2010}).

\bibitem[{\citenamefont{Oreg et~al.}(2010)\citenamefont{Oreg, Refael, and von
  Oppen}}]{oreg2010}
\bibinfo{author}{\bibfnamefont{Y.}~\bibnamefont{Oreg}},
  \bibinfo{author}{\bibfnamefont{G.}~\bibnamefont{Refael}}, \bibnamefont{and}
  \bibinfo{author}{\bibfnamefont{F.}~\bibnamefont{von Oppen}},
  \bibinfo{journal}{Phys. Rev. Lett.} \textbf{\bibinfo{volume}{105}},
  \bibinfo{pages}{177002} (\bibinfo{year}{2010}).

\bibitem[{\citenamefont{Tanaka et~al.}(2009{\natexlab{b}})\citenamefont{Tanaka,
  Yokoyama, and Nagaosa}}]{PhysRevLett.103.107002}
\bibinfo{author}{\bibfnamefont{Y.}~\bibnamefont{Tanaka}},
  \bibinfo{author}{\bibfnamefont{T.}~\bibnamefont{Yokoyama}}, \bibnamefont{and}
  \bibinfo{author}{\bibfnamefont{N.}~\bibnamefont{Nagaosa}},
  \bibinfo{journal}{Phys. Rev. Lett.} \textbf{\bibinfo{volume}{103}},
  \bibinfo{pages}{107002} (\bibinfo{year}{2009}{\natexlab{b}}).

\bibitem[{\citenamefont{Nakosai et~al.}(2012)\citenamefont{Nakosai, Tanaka, and
  Nagaosa}}]{PhysRevLett.108.147003}
\bibinfo{author}{\bibfnamefont{S.}~\bibnamefont{Nakosai}},
  \bibinfo{author}{\bibfnamefont{Y.}~\bibnamefont{Tanaka}}, \bibnamefont{and}
  \bibinfo{author}{\bibfnamefont{N.}~\bibnamefont{Nagaosa}},
  \bibinfo{journal}{Phys. Rev. Lett.} \textbf{\bibinfo{volume}{108}},
  \bibinfo{pages}{147003} (\bibinfo{year}{2012}).

\bibitem[{\citenamefont{Qi et~al.}(2013)\citenamefont{Qi, Witten, and
  Zhang}}]{PhysRevB.87.134519}
\bibinfo{author}{\bibfnamefont{X.-L.} \bibnamefont{Qi}},
  \bibinfo{author}{\bibfnamefont{E.}~\bibnamefont{Witten}}, \bibnamefont{and}
  \bibinfo{author}{\bibfnamefont{S.-C.} \bibnamefont{Zhang}},
  \bibinfo{journal}{Phys. Rev. B} \textbf{\bibinfo{volume}{87}},
  \bibinfo{pages}{134519} (\bibinfo{year}{2013}).

\bibitem[{\citenamefont{Hosur et~al.}(2014)\citenamefont{Hosur, Dai, Fang, and
  Qi}}]{PhysRevB.90.045130}
\bibinfo{author}{\bibfnamefont{P.}~\bibnamefont{Hosur}},
  \bibinfo{author}{\bibfnamefont{X.}~\bibnamefont{Dai}},
  \bibinfo{author}{\bibfnamefont{Z.}~\bibnamefont{Fang}}, \bibnamefont{and}
  \bibinfo{author}{\bibfnamefont{X.-L.} \bibnamefont{Qi}},
  \bibinfo{journal}{Phys. Rev. B} \textbf{\bibinfo{volume}{90}},
  \bibinfo{pages}{045130} (\bibinfo{year}{2014}).

\bibitem[{\citenamefont{Goswami and Roy}(2014)}]{PhysRevB.90.041301}
\bibinfo{author}{\bibfnamefont{P.}~\bibnamefont{Goswami}} \bibnamefont{and}
  \bibinfo{author}{\bibfnamefont{B.}~\bibnamefont{Roy}},
  \bibinfo{journal}{Phys. Rev. B} \textbf{\bibinfo{volume}{90}},
  \bibinfo{pages}{041301} (\bibinfo{year}{2014}).

\bibitem[{\citenamefont{Foster et~al.}(2014)\citenamefont{Foster, Xie, and
  Chou}}]{PhysRevB.89.155140}
\bibinfo{author}{\bibfnamefont{M.~S.} \bibnamefont{Foster}},
  \bibinfo{author}{\bibfnamefont{H.-Y.} \bibnamefont{Xie}}, \bibnamefont{and}
  \bibinfo{author}{\bibfnamefont{Y.-Z.} \bibnamefont{Chou}},
  \bibinfo{journal}{Phys. Rev. B} \textbf{\bibinfo{volume}{89}},
  \bibinfo{pages}{155140} (\bibinfo{year}{2014}).

\bibitem[{\citenamefont{Wan and Savrasov}(2014)}]{wan2014turning}
\bibinfo{author}{\bibfnamefont{X.}~\bibnamefont{Wan}} \bibnamefont{and}
  \bibinfo{author}{\bibfnamefont{S.~Y.} \bibnamefont{Savrasov}},
  \bibinfo{journal}{Nat. commun.} \textbf{\bibinfo{volume}{5}},
  \bibinfo{pages}{4144} (\bibinfo{year}{2014}).

\bibitem[{\citenamefont{Kitaev}(2009)}]{kitaev2009periodic}
\bibinfo{author}{\bibfnamefont{A.}~\bibnamefont{Kitaev}}, in
  \emph{\bibinfo{booktitle}{AIP Conf. Proc.}} (\bibinfo{year}{2009}), vol.
  \bibinfo{volume}{1134}, p.~\bibinfo{pages}{22}.

\bibitem[{\citenamefont{Teo and Kane}(2010)}]{teo2010topological}
\bibinfo{author}{\bibfnamefont{J.~C.~Y.} \bibnamefont{Teo}} \bibnamefont{and}
  \bibinfo{author}{\bibfnamefont{C.~L.} \bibnamefont{Kane}},
  \bibinfo{journal}{Phys. Rev. B} \textbf{\bibinfo{volume}{82}},
  \bibinfo{pages}{115120} (\bibinfo{year}{2010}).

\bibitem[{\citenamefont{Stone et~al.}(2011)\citenamefont{Stone, Chiu, and
  Roy}}]{stone2011symmetries}
\bibinfo{author}{\bibfnamefont{M.}~\bibnamefont{Stone}},
  \bibinfo{author}{\bibfnamefont{C.-K.} \bibnamefont{Chiu}}, \bibnamefont{and}
  \bibinfo{author}{\bibfnamefont{A.}~\bibnamefont{Roy}}, \bibinfo{journal}{J.
  Phys. A} \textbf{\bibinfo{volume}{44}}, \bibinfo{pages}{045001}
  (\bibinfo{year}{2011}).

\bibitem[{\citenamefont{Abramovici and Kalugin}(2012)}]{abramovici2012clifford}
\bibinfo{author}{\bibfnamefont{G.}~\bibnamefont{Abramovici}} \bibnamefont{and}
  \bibinfo{author}{\bibfnamefont{P.}~\bibnamefont{Kalugin}},
  \bibinfo{journal}{Int. J. Geom. Methods Mod. Phys.}
  \textbf{\bibinfo{volume}{9}} (\bibinfo{year}{2012}).

\bibitem[{\citenamefont{Wen}(2012)}]{wen2012symmetry}
\bibinfo{author}{\bibfnamefont{X.-G.} \bibnamefont{Wen}},
  \bibinfo{journal}{Phys. Rev. B} \textbf{\bibinfo{volume}{85}},
  \bibinfo{pages}{085103} (\bibinfo{year}{2012}).

\bibitem[{\citenamefont{Altland and Zirnbauer}(1997)}]{altland1997nonstandard}
\bibinfo{author}{\bibfnamefont{A.}~\bibnamefont{Altland}} \bibnamefont{and}
  \bibinfo{author}{\bibfnamefont{M.~R.} \bibnamefont{Zirnbauer}},
  \bibinfo{journal}{Phys. Rev. B} \textbf{\bibinfo{volume}{55}},
  \bibinfo{pages}{1142} (\bibinfo{year}{1997}).

\bibitem[{\citenamefont{Volovik}(1987)}]{volovik1987}
\bibinfo{author}{\bibfnamefont{G.~E.} \bibnamefont{Volovik}},
  \bibinfo{journal}{JETP Lett.} \textbf{\bibinfo{volume}{46}},
  \bibinfo{pages}{98} (\bibinfo{year}{1987}).

\bibitem[{\citenamefont{Ryu and Hatsugai}(2002)}]{ryu2002topological}
\bibinfo{author}{\bibfnamefont{S.}~\bibnamefont{Ryu}} \bibnamefont{and}
  \bibinfo{author}{\bibfnamefont{Y.}~\bibnamefont{Hatsugai}},
  \bibinfo{journal}{Phys. Rev. Lett.} \textbf{\bibinfo{volume}{89}},
  \bibinfo{pages}{077002} (\bibinfo{year}{2002}).

\bibitem[{\citenamefont{Sato}(2006)}]{sato2006nodal}
\bibinfo{author}{\bibfnamefont{M.}~\bibnamefont{Sato}}, \bibinfo{journal}{Phys.
  Rev. B} \textbf{\bibinfo{volume}{73}}, \bibinfo{pages}{214502}
  (\bibinfo{year}{2006}).

\bibitem[{\citenamefont{Teo et~al.}(2008)\citenamefont{Teo, Fu, and
  Kane}}]{teo2008surface}
\bibinfo{author}{\bibfnamefont{J.~C.~Y.} \bibnamefont{Teo}},
  \bibinfo{author}{\bibfnamefont{L.}~\bibnamefont{Fu}}, \bibnamefont{and}
  \bibinfo{author}{\bibfnamefont{C.~L.} \bibnamefont{Kane}},
  \bibinfo{journal}{Phys. Rev. B} \textbf{\bibinfo{volume}{78}},
  \bibinfo{pages}{045426} (\bibinfo{year}{2008}).

\bibitem[{\citenamefont{B\'{e}ri}(2010)}]{beri2010topologically}
\bibinfo{author}{\bibfnamefont{B.}~\bibnamefont{B\'{e}ri}},
  \bibinfo{journal}{Phys. Rev. B} \textbf{\bibinfo{volume}{81}},
  \bibinfo{pages}{134515} (\bibinfo{year}{2010}).

\bibitem[{\citenamefont{Yada et~al.}(2011)\citenamefont{Yada, Sato, Tanaka, and
  Yokoyama}}]{yada2011}
\bibinfo{author}{\bibfnamefont{K.}~\bibnamefont{Yada}},
  \bibinfo{author}{\bibfnamefont{M.}~\bibnamefont{Sato}},
  \bibinfo{author}{\bibfnamefont{Y.}~\bibnamefont{Tanaka}}, \bibnamefont{and}
  \bibinfo{author}{\bibfnamefont{T.}~\bibnamefont{Yokoyama}},
  \bibinfo{journal}{Phys. Rev. B} \textbf{\bibinfo{volume}{83}}, \bibinfo{pages}{064505}
  (\bibinfo{year}{2011}).

\bibitem[{\citenamefont{Sato et~al.}(2011)\citenamefont{Sato, Tanaka, Yada, and
  Yokoyama}}]{sato2011topology}
\bibinfo{author}{\bibfnamefont{M.}~\bibnamefont{Sato}},
  \bibinfo{author}{\bibfnamefont{Y.}~\bibnamefont{Tanaka}},
  \bibinfo{author}{\bibfnamefont{K.}~\bibnamefont{Yada}}, \bibnamefont{and}
  \bibinfo{author}{\bibfnamefont{T.}~\bibnamefont{Yokoyama}},
  \bibinfo{journal}{Phys. Rev. B} \textbf{\bibinfo{volume}{83}},
  \bibinfo{pages}{224511} (\bibinfo{year}{2011}).

\bibitem[{\citenamefont{Schnyder and Ryu}(2011)}]{schnyder2011topological}
\bibinfo{author}{\bibfnamefont{A.~P.} \bibnamefont{Schnyder}} \bibnamefont{and}
  \bibinfo{author}{\bibfnamefont{S.}~\bibnamefont{Ryu}},
  \bibinfo{journal}{Phys. Rev. B} \textbf{\bibinfo{volume}{84}},
  \bibinfo{pages}{060504(R)} (\bibinfo{year}{2011}).

\bibitem[{\citenamefont{Wan et~al.}(2011)\citenamefont{Wan, Turner, Vishwanath,
  and Savrasov}}]{wan2011topological}
\bibinfo{author}{\bibfnamefont{X.}~\bibnamefont{Wan}},
  \bibinfo{author}{\bibfnamefont{A.~M.} \bibnamefont{Turner}},
  \bibinfo{author}{\bibfnamefont{A.}~\bibnamefont{Vishwanath}},
  \bibnamefont{and} \bibinfo{author}{\bibfnamefont{S.~Y.}
  \bibnamefont{Savrasov}}, \bibinfo{journal}{Phys. Rev. B}
  \textbf{\bibinfo{volume}{83}}, \bibinfo{pages}{205101}
  (\bibinfo{year}{2011}).

\bibitem[{\citenamefont{Yang et~al.}(2011)\citenamefont{Yang, Lu, and
  Ran}}]{yang2011quantum}
\bibinfo{author}{\bibfnamefont{K.-Y.} \bibnamefont{Yang}},
  \bibinfo{author}{\bibfnamefont{Y.-M.} \bibnamefont{Lu}}, \bibnamefont{and}
  \bibinfo{author}{\bibfnamefont{Y.}~\bibnamefont{Ran}},
  \bibinfo{journal}{Phys. Rev. B} \textbf{\bibinfo{volume}{84}},
  \bibinfo{pages}{075129} (\bibinfo{year}{2011}).

\bibitem[{\citenamefont{Burkov and Balents}(2011)}]{burkov2011weyl}
\bibinfo{author}{\bibfnamefont{A.~A.} \bibnamefont{Burkov}} \bibnamefont{and}
  \bibinfo{author}{\bibfnamefont{L.}~\bibnamefont{Balents}},
  \bibinfo{journal}{Phys. Rev. Lett.} \textbf{\bibinfo{volume}{107}},
  \bibinfo{pages}{127205} (\bibinfo{year}{2011}).

\bibitem[{\citenamefont{Hatsugai}(2010)}]{hatsugai2010symmetry}
\bibinfo{author}{\bibfnamefont{Y.}~\bibnamefont{Hatsugai}},
  \bibinfo{journal}{New J. Phys.} \textbf{\bibinfo{volume}{12}},
  \bibinfo{pages}{065004} (\bibinfo{year}{2010}).

\bibitem[{\citenamefont{Hughes et~al.}(2011)\citenamefont{Hughes, Prodan, and
  Bernevig}}]{hughes2011inversion}
\bibinfo{author}{\bibfnamefont{T.~L.} \bibnamefont{Hughes}},
  \bibinfo{author}{\bibfnamefont{E.}~\bibnamefont{Prodan}}, \bibnamefont{and}
  \bibinfo{author}{\bibfnamefont{B.~A.} \bibnamefont{Bernevig}},
  \bibinfo{journal}{Phys. Rev. B} \textbf{\bibinfo{volume}{83}},
  \bibinfo{pages}{245132} (\bibinfo{year}{2011}).

\bibitem[{\citenamefont{Turner et~al.}(2010)\citenamefont{Turner, Zhang, and
  Vishwanath}}]{turner2010entanglement}
\bibinfo{author}{\bibfnamefont{A.~M.} \bibnamefont{Turner}},
  \bibinfo{author}{\bibfnamefont{Y.}~\bibnamefont{Zhang}}, \bibnamefont{and}
  \bibinfo{author}{\bibfnamefont{A.}~\bibnamefont{Vishwanath}},
  \bibinfo{journal}{Phys. Rev. B} \textbf{\bibinfo{volume}{82}},
  \bibinfo{pages}{241102} (\bibinfo{year}{2010}).

\bibitem[{\citenamefont{Fu and Berg}(2010)}]{fu2010odd}
\bibinfo{author}{\bibfnamefont{L.}~\bibnamefont{Fu}} \bibnamefont{and}
  \bibinfo{author}{\bibfnamefont{E.}~\bibnamefont{Berg}},
  \bibinfo{journal}{Phys. Rev. Lett.} \textbf{\bibinfo{volume}{105}},
  \bibinfo{pages}{097001} (\bibinfo{year}{2010}).

\bibitem[{\citenamefont{Mong et~al.}(2012)\citenamefont{Mong, Bardarson, and
  Moore}}]{mong2012quantum}
\bibinfo{author}{\bibfnamefont{R.~S.~K.} \bibnamefont{Mong}},
  \bibinfo{author}{\bibfnamefont{J.~H.} \bibnamefont{Bardarson}},
  \bibnamefont{and} \bibinfo{author}{\bibfnamefont{J.~E.} \bibnamefont{Moore}},
  \bibinfo{journal}{Phys. Rev. Lett.} \textbf{\bibinfo{volume}{108}},
  \bibinfo{pages}{076804} (\bibinfo{year}{2012}).

\bibitem[{\citenamefont{Ringel et~al.}(2012)\citenamefont{Ringel, Kraus, and
  Stern}}]{ringel2012strong}
\bibinfo{author}{\bibfnamefont{Z.}~\bibnamefont{Ringel}},
  \bibinfo{author}{\bibfnamefont{Y.~E.} \bibnamefont{Kraus}}, \bibnamefont{and}
  \bibinfo{author}{\bibfnamefont{A.}~\bibnamefont{Stern}},
  \bibinfo{journal}{Phys. Rev. B} \textbf{\bibinfo{volume}{86}},
  \bibinfo{pages}{045102} (\bibinfo{year}{2012}).

\bibitem[{\citenamefont{Fu and Kane}(2012)}]{fu2012topology}
\bibinfo{author}{\bibfnamefont{L.}~\bibnamefont{Fu}} \bibnamefont{and}
  \bibinfo{author}{\bibfnamefont{C.~L.} \bibnamefont{Kane}},
  \bibinfo{journal}{Phys. Rev. Lett.} \textbf{\bibinfo{volume}{109}},
  \bibinfo{pages}{246605} (\bibinfo{year}{2012}).

\bibitem[{\citenamefont{Fulga et~al.}(2012)\citenamefont{Fulga, van Heck, Edge,
  and Akhmerov}}]{fulga2012statistical}
\bibinfo{author}{\bibfnamefont{I.~C.} \bibnamefont{Fulga}},
  \bibinfo{author}{\bibfnamefont{B.}~\bibnamefont{van Heck}},
  \bibinfo{author}{\bibfnamefont{J.~M.} \bibnamefont{Edge}}, \bibnamefont{and}
  \bibinfo{author}{\bibfnamefont{A.~R.} \bibnamefont{Akhmerov}},
  \bibinfo{journal}{arXiv:1212.6191}  (\bibinfo{year}{2012}).

\bibitem[{\citenamefont{Fu}(2011)}]{fu2011topological}
\bibinfo{author}{\bibfnamefont{L.}~\bibnamefont{Fu}}, \bibinfo{journal}{Phys.
  Rev. Lett.} \textbf{\bibinfo{volume}{106}}, \bibinfo{pages}{106802}
  (\bibinfo{year}{2011}).

\bibitem[{\citenamefont{Hsieh et~al.}(2012)\citenamefont{Hsieh, Lin, Liu, Duan,
  Bansil, and Fu}}]{hsieh2012topological}
\bibinfo{author}{\bibfnamefont{T.~H.} \bibnamefont{Hsieh}},
  \bibinfo{author}{\bibfnamefont{H.}~\bibnamefont{Lin}},
  \bibinfo{author}{\bibfnamefont{J.}~\bibnamefont{Liu}},
  \bibinfo{author}{\bibfnamefont{W.}~\bibnamefont{Duan}},
  \bibinfo{author}{\bibfnamefont{A.}~\bibnamefont{Bansil}}, \bibnamefont{and}
  \bibinfo{author}{\bibfnamefont{L.}~\bibnamefont{Fu}}, \bibinfo{journal}{Nat.
  commun.} \textbf{\bibinfo{volume}{3}}, \bibinfo{pages}{982}
  (\bibinfo{year}{2012}).

\bibitem[{\citenamefont{Tanaka et~al.}(2012{\natexlab{b}})\citenamefont{Tanaka,
  Ren, Sato, Nakayama, Souma, Takahashi, Segawa, and
  Ando}}]{tanaka2012experimental}
\bibinfo{author}{\bibfnamefont{Y.}~\bibnamefont{Tanaka}},
  \bibinfo{author}{\bibfnamefont{Z.}~\bibnamefont{Ren}},
  \bibinfo{author}{\bibfnamefont{T.}~\bibnamefont{Sato}},
  \bibinfo{author}{\bibfnamefont{K.}~\bibnamefont{Nakayama}},
  \bibinfo{author}{\bibfnamefont{S.}~\bibnamefont{Souma}},
  \bibinfo{author}{\bibfnamefont{T.}~\bibnamefont{Takahashi}},
  \bibinfo{author}{\bibfnamefont{K.}~\bibnamefont{Segawa}}, \bibnamefont{and}
  \bibinfo{author}{\bibfnamefont{Y.}~\bibnamefont{Ando}},
  \bibinfo{journal}{Nat. Phys.} \textbf{\bibinfo{volume}{8}},
  \bibinfo{pages}{800} (\bibinfo{year}{2012}{\natexlab{b}}).

\bibitem[{\citenamefont{Dziawa et~al.}(2012)\citenamefont{Dziawa, Kowalski,
  Dybko, Buczko, Szczerbakow, Szot, {\L}usakowska, Balasubramanian, Wojek,
  Berntsen et~al.}}]{dziawa2012topological}
\bibinfo{author}{\bibfnamefont{P.}~\bibnamefont{Dziawa}},
  \bibinfo{author}{\bibfnamefont{B.~J.} \bibnamefont{Kowalski}},
  \bibinfo{author}{\bibfnamefont{K.}~\bibnamefont{Dybko}},
  \bibinfo{author}{\bibfnamefont{R.}~\bibnamefont{Buczko}},
  \bibinfo{author}{\bibfnamefont{A.}~\bibnamefont{Szczerbakow}},
  \bibinfo{author}{\bibfnamefont{M.}~\bibnamefont{Szot}},
  \bibinfo{author}{\bibfnamefont{E.}~\bibnamefont{{\L}usakowska}},
  \bibinfo{author}{\bibfnamefont{T.}~\bibnamefont{Balasubramanian}},
  \bibinfo{author}{\bibfnamefont{B.~M.} \bibnamefont{Wojek}},
  \bibinfo{author}{\bibfnamefont{M.}~\bibnamefont{Berntsen}},
  \bibnamefont{et~al.}, \bibinfo{journal}{Nat. Mater.}
  \textbf{\bibinfo{volume}{11}}, \bibinfo{pages}{1023} (\bibinfo{year}{2012}).

\bibitem[{\citenamefont{Xu et~al.}(2012)\citenamefont{Xu, Liu, Alidoust,
  Neupane, Qian, Belopolski, Denlinger, Wang, Lin, Wray
  et~al.}}]{xu2012observation}
\bibinfo{author}{\bibfnamefont{S.-Y.} \bibnamefont{Xu}},
  \bibinfo{author}{\bibfnamefont{C.}~\bibnamefont{Liu}},
  \bibinfo{author}{\bibfnamefont{N.}~\bibnamefont{Alidoust}},
  \bibinfo{author}{\bibfnamefont{M.}~\bibnamefont{Neupane}},
  \bibinfo{author}{\bibfnamefont{D.}~\bibnamefont{Qian}},
  \bibinfo{author}{\bibfnamefont{I.}~\bibnamefont{Belopolski}},
  \bibinfo{author}{\bibfnamefont{J.~D.} \bibnamefont{Denlinger}},
  \bibinfo{author}{\bibfnamefont{Y.~J.} \bibnamefont{Wang}},
  \bibinfo{author}{\bibfnamefont{H.}~\bibnamefont{Lin}},
  \bibinfo{author}{\bibfnamefont{L.~A.} \bibnamefont{Wray}},
  \bibnamefont{et~al.}, \bibinfo{journal}{Nat. Commun.}
  \textbf{\bibinfo{volume}{3}}, \bibinfo{pages}{1192} (\bibinfo{year}{2012}).

\bibitem[{\citenamefont{Fang et~al.}(2012)\citenamefont{Fang, Gilbert, and
  Bernevig}}]{fang2012bulk}
\bibinfo{author}{\bibfnamefont{C.}~\bibnamefont{Fang}},
  \bibinfo{author}{\bibfnamefont{M.~J.} \bibnamefont{Gilbert}},
  \bibnamefont{and} \bibinfo{author}{\bibfnamefont{B.~A.}
  \bibnamefont{Bernevig}}, \bibinfo{journal}{Phys. Rev. B}
  \textbf{\bibinfo{volume}{86}}, \bibinfo{pages}{115112}
  (\bibinfo{year}{2012}).

\bibitem[{\citenamefont{Slager et~al.}(2012)\citenamefont{Slager, Mesaros,
  Juri{\v{c}}i{\'c}, and Zaanen}}]{slager2012space}
\bibinfo{author}{\bibfnamefont{R.-J.} \bibnamefont{Slager}},
  \bibinfo{author}{\bibfnamefont{A.}~\bibnamefont{Mesaros}},
  \bibinfo{author}{\bibfnamefont{V.}~\bibnamefont{Juri{\v{c}}i{\'c}}},
  \bibnamefont{and} \bibinfo{author}{\bibfnamefont{J.}~\bibnamefont{Zaanen}},
  \bibinfo{journal}{Nat. Phys.} \textbf{\bibinfo{volume}{9}},
  \bibinfo{pages}{98} (\bibinfo{year}{2012}).

\bibitem[{\citenamefont{Jadaun et~al.}(2013)\citenamefont{Jadaun, Xiao, Niu,
  and Banerjee}}]{jadaun2013topological}
\bibinfo{author}{\bibfnamefont{P.}~\bibnamefont{Jadaun}},
  \bibinfo{author}{\bibfnamefont{D.}~\bibnamefont{Xiao}},
  \bibinfo{author}{\bibfnamefont{Q.}~\bibnamefont{Niu}}, \bibnamefont{and}
  \bibinfo{author}{\bibfnamefont{S.~K.} \bibnamefont{Banerjee}},
  \bibinfo{journal}{Phys. Rev. B} \textbf{\bibinfo{volume}{88}},
  \bibinfo{pages}{085110} (\bibinfo{year}{2013}).

\bibitem[{\citenamefont{Fang et~al.}(2013{\natexlab{a}})\citenamefont{Fang,
  Gilbert, and Bernevig}}]{fang2013entanglement}
\bibinfo{author}{\bibfnamefont{C.}~\bibnamefont{Fang}},
  \bibinfo{author}{\bibfnamefont{M.~J.} \bibnamefont{Gilbert}},
  \bibnamefont{and} \bibinfo{author}{\bibfnamefont{B.~A.}
  \bibnamefont{Bernevig}}, \bibinfo{journal}{Phys. Rev. B}
  \textbf{\bibinfo{volume}{87}}, \bibinfo{pages}{035119}
  (\bibinfo{year}{2013}{\natexlab{a}}).

\bibitem[{\citenamefont{Chiu et~al.}(2013)\citenamefont{Chiu, Yao, and
  Ryu}}]{chiu2013classification}
\bibinfo{author}{\bibfnamefont{C.-K.} \bibnamefont{Chiu}},
  \bibinfo{author}{\bibfnamefont{H.}~\bibnamefont{Yao}}, \bibnamefont{and}
  \bibinfo{author}{\bibfnamefont{S.}~\bibnamefont{Ryu}},
  \bibinfo{journal}{Phys. Rev. B} \textbf{\bibinfo{volume}{88}},
  \bibinfo{pages}{075142} (\bibinfo{year}{2013}).

\bibitem[{\citenamefont{Liu and Zhang}(2013)}]{liu2013topological}
\bibinfo{author}{\bibfnamefont{C.-X.} \bibnamefont{Liu}} \bibnamefont{and}
  \bibinfo{author}{\bibfnamefont{R.-X.} \bibnamefont{Zhang}},
  \bibinfo{journal}{arXiv:1308.4717}  (\bibinfo{year}{2013}).

\bibitem[{\citenamefont{Alexandradinata
  et~al.}(2014)\citenamefont{Alexandradinata, Fang, Gilbert, and
  Bernevig}}]{alexandradinata2014spinless}
\bibinfo{author}{\bibfnamefont{A.}~\bibnamefont{Alexandradinata}},
  \bibinfo{author}{\bibfnamefont{C.}~\bibnamefont{Fang}},
  \bibinfo{author}{\bibfnamefont{M.~J.} \bibnamefont{Gilbert}},
  \bibnamefont{and} \bibinfo{author}{\bibfnamefont{B.~A.}
  \bibnamefont{Bernevig}}, \bibinfo{journal}{arXiv:1402.6323}
  (\bibinfo{year}{2014}).

\bibitem[{\citenamefont{Teo and Hughes}(2013)}]{teo2013existence}
\bibinfo{author}{\bibfnamefont{J.~C.~Y.} \bibnamefont{Teo}} \bibnamefont{and}
  \bibinfo{author}{\bibfnamefont{T.~L.} \bibnamefont{Hughes}},
  \bibinfo{journal}{Phys. Rev. Lett.} \textbf{\bibinfo{volume}{111}},
  \bibinfo{pages}{047006} (\bibinfo{year}{2013}).

\bibitem[{\citenamefont{Ueno et~al.}(2013)\citenamefont{Ueno, Yamakage, Tanaka,
  and Sato}}]{ueno2013symmetry}
\bibinfo{author}{\bibfnamefont{Y.}~\bibnamefont{Ueno}},
  \bibinfo{author}{\bibfnamefont{A.}~\bibnamefont{Yamakage}},
  \bibinfo{author}{\bibfnamefont{Y.}~\bibnamefont{Tanaka}}, \bibnamefont{and}
  \bibinfo{author}{\bibfnamefont{M.}~\bibnamefont{Sato}},
  \bibinfo{journal}{Phys. Rev. Lett.} \textbf{\bibinfo{volume}{111}},
  \bibinfo{pages}{087002} (\bibinfo{year}{2013}).

\bibitem[{\citenamefont{Zhang et~al.}(2013)\citenamefont{Zhang, Kane, and
  Mele}}]{zhang2013topological}
\bibinfo{author}{\bibfnamefont{F.}~\bibnamefont{Zhang}},
  \bibinfo{author}{\bibfnamefont{C.~L.} \bibnamefont{Kane}}, \bibnamefont{and}
  \bibinfo{author}{\bibfnamefont{E.~J.} \bibnamefont{Mele}},
  \bibinfo{journal}{Phys. Rev. Lett.} \textbf{\bibinfo{volume}{111}},
  \bibinfo{pages}{056403} (\bibinfo{year}{2013}).

\bibitem[{\citenamefont{Liu and Law}(2013)}]{liu2013majorana}
\bibinfo{author}{\bibfnamefont{X.-J.} \bibnamefont{Liu}} \bibnamefont{and}
  \bibinfo{author}{\bibfnamefont{K.~T.} \bibnamefont{Law}},
  \bibinfo{journal}{arXiv:1310.5685}  (\bibinfo{year}{2013}).

\bibitem[{\citenamefont{Benalcazar et~al.}(2013)\citenamefont{Benalcazar, Teo,
  and Hughes}}]{benalcazar2013classification}
\bibinfo{author}{\bibfnamefont{W.~A.} \bibnamefont{Benalcazar}},
  \bibinfo{author}{\bibfnamefont{J.~C.~Y.} \bibnamefont{Teo}},
  \bibnamefont{and} \bibinfo{author}{\bibfnamefont{T.~L.}
  \bibnamefont{Hughes}}, \bibinfo{journal}{arXiv:1311.0496}
  (\bibinfo{year}{2013}).

\bibitem[{\citenamefont{Tsutsumi et~al.}(2013)\citenamefont{Tsutsumi, Ishikawa,
  Kawakami, Mizushima, Sato, Ichioka, and Machida}}]{tsutsumi2013upt3}
\bibinfo{author}{\bibfnamefont{Y.}~\bibnamefont{Tsutsumi}},
  \bibinfo{author}{\bibfnamefont{M.}~\bibnamefont{Ishikawa}},
  \bibinfo{author}{\bibfnamefont{T.}~\bibnamefont{Kawakami}},
  \bibinfo{author}{\bibfnamefont{T.}~\bibnamefont{Mizushima}},
  \bibinfo{author}{\bibfnamefont{M.}~\bibnamefont{Sato}},
  \bibinfo{author}{\bibfnamefont{M.}~\bibnamefont{Ichioka}}, \bibnamefont{and}
  \bibinfo{author}{\bibfnamefont{K.}~\bibnamefont{Machida}},
  \bibinfo{journal}{J. Phys. Soc. Jpn.} \textbf{\bibinfo{volume}{82}},
  \bibinfo{pages}{113703} (\bibinfo{year}{2013}).

\bibitem[{\citenamefont{Sato et~al.}(2014)\citenamefont{Sato, Yamakage, and
  Mizushuma}}]{sato2014mirror}
\bibinfo{author}{\bibfnamefont{M.}~\bibnamefont{Sato}},
  \bibinfo{author}{\bibfnamefont{A.}~\bibnamefont{Yamakage}}, \bibnamefont{and}
  \bibinfo{author}{\bibfnamefont{T.}~\bibnamefont{Mizushuma}},
  \bibinfo{journal}{Physica E} \textbf{\bibinfo{volume}{55}},
  \bibinfo{pages}{20} (\bibinfo{year}{2014}).

\bibitem[{\citenamefont{Morimoto and Furusaki}(2013)}]{morimoto2013topological}
\bibinfo{author}{\bibfnamefont{T.}~\bibnamefont{Morimoto}} \bibnamefont{and}
  \bibinfo{author}{\bibfnamefont{A.}~\bibnamefont{Furusaki}},
  \bibinfo{journal}{Phys. Rev. B} \textbf{\bibinfo{volume}{88}},
  \bibinfo{pages}{125129} (\bibinfo{year}{2013}).

\bibitem[{\citenamefont{Mong et~al.}(2010)\citenamefont{Mong, Essin, and
  Moore}}]{mong2010antiferromagnetic}
\bibinfo{author}{\bibfnamefont{R.~S.~K.} \bibnamefont{Mong}},
  \bibinfo{author}{\bibfnamefont{A.~M.} \bibnamefont{Essin}}, \bibnamefont{and}
  \bibinfo{author}{\bibfnamefont{J.~E.} \bibnamefont{Moore}},
  \bibinfo{journal}{Phys. Rev. B} \textbf{\bibinfo{volume}{81}},
  \bibinfo{pages}{245209} (\bibinfo{year}{2010}).

\bibitem[{\citenamefont{Mizushima et~al.}(2012)\citenamefont{Mizushima, Sato,
  and Machida}}]{mizushima2012symmetry}
\bibinfo{author}{\bibfnamefont{T.}~\bibnamefont{Mizushima}},
  \bibinfo{author}{\bibfnamefont{M.}~\bibnamefont{Sato}}, \bibnamefont{and}
  \bibinfo{author}{\bibfnamefont{K.}~\bibnamefont{Machida}},
  \bibinfo{journal}{Phys. Rev. Lett.} \textbf{\bibinfo{volume}{109}},
  \bibinfo{pages}{165301} (\bibinfo{year}{2012}).

\bibitem[{\citenamefont{Mizushima and Sato}(2013)}]{mizushima2013topological}
\bibinfo{author}{\bibfnamefont{T.}~\bibnamefont{Mizushima}} \bibnamefont{and}
  \bibinfo{author}{\bibfnamefont{M.}~\bibnamefont{Sato}}, \bibinfo{journal}{New
  J. Phys.} \textbf{\bibinfo{volume}{15}}, \bibinfo{pages}{075010}
  (\bibinfo{year}{2013}).

\bibitem[{\citenamefont{Fang et~al.}(2013{\natexlab{b}})\citenamefont{Fang,
  Gilbert, and Bernevig}}]{fang2013topological}
\bibinfo{author}{\bibfnamefont{C.}~\bibnamefont{Fang}},
  \bibinfo{author}{\bibfnamefont{M.~J.} \bibnamefont{Gilbert}},
  \bibnamefont{and} \bibinfo{author}{\bibfnamefont{B.~A.}
  \bibnamefont{Bernevig}}, \bibinfo{journal}{Phys. Rev. B}
  \textbf{\bibinfo{volume}{88}}, \bibinfo{pages}{085406}
  (\bibinfo{year}{2013}{\natexlab{b}}).

\bibitem[{\citenamefont{Liu}(2013)}]{liu2013antiferromagnetic}
\bibinfo{author}{\bibfnamefont{C.-X.} \bibnamefont{Liu}},
  \bibinfo{journal}{arXiv:1304.6455}  (\bibinfo{year}{2013}).

\bibitem[{\citenamefont{Kotetes}(2013)}]{kotetes2013classification}
\bibinfo{author}{\bibfnamefont{P.}~\bibnamefont{Kotetes}},
  \bibinfo{journal}{New J. Phys.} \textbf{\bibinfo{volume}{15}},
  \bibinfo{pages}{105027} (\bibinfo{year}{2013}).

\bibitem[{\citenamefont{Fang et~al.}(2014)\citenamefont{Fang, Gilbert, and
  Bernevig}}]{fang2013new}
\bibinfo{author}{\bibfnamefont{C.}~\bibnamefont{Fang}},
  \bibinfo{author}{\bibfnamefont{M.~J.} \bibnamefont{Gilbert}},
  \bibnamefont{and} \bibinfo{author}{\bibfnamefont{B.~A.}
  \bibnamefont{Bernevig}}, \bibinfo{journal}{Phys. Rev. Lett.}
  \textbf{\bibinfo{volume}{112}}, \bibinfo{pages}{106401}
  (\bibinfo{year}{2014}).

\bibitem[{\citenamefont{Zhang and Liu}(2014)}]{zhang2014topological}
\bibinfo{author}{\bibfnamefont{R.-X.} \bibnamefont{Zhang}} \bibnamefont{and}
  \bibinfo{author}{\bibfnamefont{C.-X.} \bibnamefont{Liu}},
  \bibinfo{journal}{arXiv:1401.6922}  (\bibinfo{year}{2014}).

\bibitem[{\citenamefont{Atiyah et~al.}(1964)\citenamefont{Atiyah, Bott, and
  Shapiro}}]{atiyah1964clifford}
\bibinfo{author}{\bibfnamefont{M.~F.} \bibnamefont{Atiyah}},
  \bibinfo{author}{\bibfnamefont{R.}~\bibnamefont{Bott}}, \bibnamefont{and}
  \bibinfo{author}{\bibfnamefont{A.}~\bibnamefont{Shapiro}},
  \bibinfo{journal}{Topology} \textbf{\bibinfo{volume}{3}}, \bibinfo{pages}{3}
  (\bibinfo{year}{1964}).

\bibitem[{\citenamefont{Atiyah}(1966)}]{atiyah1966x}
\bibinfo{author}{\bibfnamefont{M.~F.} \bibnamefont{Atiyah}},
  \bibinfo{journal}{Quart. J. Math.} \textbf{\bibinfo{volume}{17}},
  \bibinfo{pages}{367} (\bibinfo{year}{1966}).

\bibitem[{\citenamefont{Dupont}(1969)}]{dupont1969symplectic}
\bibinfo{author}{\bibfnamefont{J.~L.} \bibnamefont{Dupont}},
  \bibinfo{journal}{Math. Scand.} \textbf{\bibinfo{volume}{24}},
  \bibinfo{pages}{27} (\bibinfo{year}{1969}).

\bibitem[{\citenamefont{Karoubi}(2008)}]{karoubi2008k}
\bibinfo{author}{\bibfnamefont{M.}~\bibnamefont{Karoubi}},
  \emph{\bibinfo{title}{K-theory: An introduction}}, vol. \bibinfo{volume}{226}
  (\bibinfo{publisher}{Springer}, \bibinfo{year}{2008}).

\bibitem[{\citenamefont{Niemi and Semenoff}(1986)}]{niemi1986fermion}
\bibinfo{author}{\bibfnamefont{A.~J.} \bibnamefont{Niemi}} \bibnamefont{and}
  \bibinfo{author}{\bibfnamefont{G.~W.} \bibnamefont{Semenoff}},
  \bibinfo{journal}{Phys. Rep.} \textbf{\bibinfo{volume}{135}},
  \bibinfo{pages}{99} (\bibinfo{year}{1986}).

\bibitem[{\citenamefont{Shiozaki et~al.}(2012)\citenamefont{Shiozaki, Fukui,
  and Fujimoto}}]{shiozaki2012index}
\bibinfo{author}{\bibfnamefont{K.}~\bibnamefont{Shiozaki}},
  \bibinfo{author}{\bibfnamefont{T.}~\bibnamefont{Fukui}}, \bibnamefont{and}
  \bibinfo{author}{\bibfnamefont{S.}~\bibnamefont{Fujimoto}},
  \bibinfo{journal}{Phys. Rev. B} \textbf{\bibinfo{volume}{86}},
  \bibinfo{pages}{125405} (\bibinfo{year}{2012}).

\bibitem[{\citenamefont{Tewari and Sau}(2012)}]{tewari2012topological}
\bibinfo{author}{\bibfnamefont{S.}~\bibnamefont{Tewari}} \bibnamefont{and}
  \bibinfo{author}{\bibfnamefont{J.~D.} \bibnamefont{Sau}},
  \bibinfo{journal}{Phys. Rev. Lett.} \textbf{\bibinfo{volume}{109}},
  \bibinfo{pages}{150408} (\bibinfo{year}{2012}).

\bibitem[{\citenamefont{Wong and Law}(2012)}]{wong2012}
\bibinfo{author}{\bibfnamefont{C.~L.~M.} \bibnamefont{Wong}} \bibnamefont{and}
  \bibinfo{author}{\bibfnamefont{K.~T.} \bibnamefont{Law}},
  \bibinfo{journal}{Phys. Rev. B} \textbf{\bibinfo{volume}{86}},
  \bibinfo{pages}{184516} (\bibinfo{year}{2012}).

\bibitem[{\citenamefont{Nakosai et~al.}(2013)\citenamefont{Nakosai, Budich,
  Tanaka, Trauzettel, and Nagaosa}}]{nakosai2013}
\bibinfo{author}{\bibfnamefont{S.}~\bibnamefont{Nakosai}},
  \bibinfo{author}{\bibfnamefont{J.~C.} \bibnamefont{Budich}},
  \bibinfo{author}{\bibfnamefont{Y.}~\bibnamefont{Tanaka}},
  \bibinfo{author}{\bibfnamefont{B.}~\bibnamefont{Trauzettel}},
  \bibnamefont{and} \bibinfo{author}{\bibfnamefont{N.}~\bibnamefont{Nagaosa}},
  \bibinfo{journal}{Phys. Rev. Lett.} \textbf{\bibinfo{volume}{110}},
  \bibinfo{pages}{117002} (\bibinfo{year}{2013}).

\bibitem[{\citenamefont{Kopnin and Salomaa}(1991)}]{kopnin1991}
\bibinfo{author}{\bibfnamefont{N.~B.} \bibnamefont{Kopnin}} \bibnamefont{and}
  \bibinfo{author}{\bibfnamefont{M.~M.} \bibnamefont{Salomaa}},
  \bibinfo{journal}{Phys. Rev. B} \textbf{\bibinfo{volume}{44}},
  \bibinfo{pages}{9667} (\bibinfo{year}{1991}).

\bibitem[{\citenamefont{Avron et~al.}(1983)\citenamefont{Avron, Seiler, and
  Simon}}]{avron1983homotopy}
\bibinfo{author}{\bibfnamefont{J.~E.} \bibnamefont{Avron}},
  \bibinfo{author}{\bibfnamefont{R.}~\bibnamefont{Seiler}}, \bibnamefont{and}
  \bibinfo{author}{\bibfnamefont{B.}~\bibnamefont{Simon}},
  \bibinfo{journal}{Phys. Rev. Lett.} \textbf{\bibinfo{volume}{51}},
  \bibinfo{pages}{51} (\bibinfo{year}{1983}).

\bibitem[{\citenamefont{Salomaa and Volovik}(1987)}]{salomaa1987quantized}
\bibinfo{author}{\bibfnamefont{M.~M.} \bibnamefont{Salomaa}} \bibnamefont{and}
  \bibinfo{author}{\bibfnamefont{G.~E.} \bibnamefont{Volovik}},
  \bibinfo{journal}{Rev. Mod. Phys.} \textbf{\bibinfo{volume}{59}},
  \bibinfo{pages}{533} (\bibinfo{year}{1987}).

\bibitem[{\citenamefont{Silaev}(2009)}]{silaev2009spectrum}
\bibinfo{author}{\bibfnamefont{M.~A.} \bibnamefont{Silaev}},
  \bibinfo{journal}{JETP Lett.} \textbf{\bibinfo{volume}{90}},
  \bibinfo{pages}{391} (\bibinfo{year}{2009}).

\bibitem[{\citenamefont{Nagai et~al.}(2012)\citenamefont{Nagai, Nakamura, and
  Machida}}]{nagai2012}
\bibinfo{author}{\bibfnamefont{Y.}~\bibnamefont{Nagai}},
  \bibinfo{author}{\bibfnamefont{H.}~\bibnamefont{Nakamura}}, \bibnamefont{and}
  \bibinfo{author}{\bibfnamefont{M.}~\bibnamefont{Machida}},
  \bibinfo{journal}{arXiv:1211.0125}  (\bibinfo{year}{2012}).

\bibitem[{\citenamefont{Stewart et~al.}(1984)\citenamefont{Stewart, Fisk,
  Willis, and Smith}}]{stewart1984}
\bibinfo{author}{\bibfnamefont{G.~R.} \bibnamefont{Stewart}},
  \bibinfo{author}{\bibfnamefont{Z.}~\bibnamefont{Fisk}},
  \bibinfo{author}{\bibfnamefont{J.~O.} \bibnamefont{Willis}},
  \bibnamefont{and} \bibinfo{author}{\bibfnamefont{J.~L.} \bibnamefont{Smith}},
  \bibinfo{journal}{Phys. Rev. Lett.} \textbf{\bibinfo{volume}{52}},
  \bibinfo{pages}{679} (\bibinfo{year}{1984}).

\bibitem[{\citenamefont{Machida et~al.}(2012)\citenamefont{Machida, Itoh, So,
  Izawa, Haga, Yamamoto, Kimura, Onuki, Tsutsumi, and Machida}}]{machida2012}
\bibinfo{author}{\bibfnamefont{Y.}~\bibnamefont{Machida}},
  \bibinfo{author}{\bibfnamefont{A.}~\bibnamefont{Itoh}},
  \bibinfo{author}{\bibfnamefont{Y.}~\bibnamefont{So}},
  \bibinfo{author}{\bibfnamefont{K.}~\bibnamefont{Izawa}},
  \bibinfo{author}{\bibfnamefont{Y.}~\bibnamefont{Haga}},
  \bibinfo{author}{\bibfnamefont{E.}~\bibnamefont{Yamamoto}},
  \bibinfo{author}{\bibfnamefont{N.}~\bibnamefont{Kimura}},
  \bibinfo{author}{\bibfnamefont{Y.}~\bibnamefont{Onuki}},
  \bibinfo{author}{\bibfnamefont{Y.}~\bibnamefont{Tsutsumi}}, \bibnamefont{and}
  \bibinfo{author}{\bibfnamefont{K.}~\bibnamefont{Machida}},
  \bibinfo{journal}{Phys. Rev. Lett.} \textbf{\bibinfo{volume}{108}},
  \bibinfo{pages}{157002} (\bibinfo{year}{2012}).

\bibitem[{\citenamefont{Tsutsumi et~al.}(2012)\citenamefont{Tsutsumi, Machida,
  Ohmi, and Ozaki}}]{tsutsumi2012}
\bibinfo{author}{\bibfnamefont{Y.}~\bibnamefont{Tsutsumi}},
  \bibinfo{author}{\bibfnamefont{K.}~\bibnamefont{Machida}},
  \bibinfo{author}{\bibfnamefont{T.}~\bibnamefont{Ohmi}}, \bibnamefont{and}
  \bibinfo{author}{\bibfnamefont{M.}~\bibnamefont{Ozaki}}, \bibinfo{journal}{J.
  Phys. Soc. Jpn.} \textbf{\bibinfo{volume}{81}}, \bibinfo{pages}{074717}
  (\bibinfo{year}{2012}).

\bibitem[{\citenamefont{Hor et~al.}(2010)\citenamefont{Hor, Williams,
  Checkelsky, Roushan, Seo, Xu, Zandbergen, Yazdani, Ong, and Cava}}]{hor2010}
\bibinfo{author}{\bibfnamefont{Y.~S.} \bibnamefont{Hor}},
  \bibinfo{author}{\bibfnamefont{A.~J.} \bibnamefont{Williams}},
  \bibinfo{author}{\bibfnamefont{J.~G.} \bibnamefont{Checkelsky}},
  \bibinfo{author}{\bibfnamefont{P.}~\bibnamefont{Roushan}},
  \bibinfo{author}{\bibfnamefont{J.}~\bibnamefont{Seo}},
  \bibinfo{author}{\bibfnamefont{Q.}~\bibnamefont{Xu}},
  \bibinfo{author}{\bibfnamefont{H.~W.} \bibnamefont{Zandbergen}},
  \bibinfo{author}{\bibfnamefont{A.}~\bibnamefont{Yazdani}},
  \bibinfo{author}{\bibfnamefont{N.~P.} \bibnamefont{Ong}}, \bibnamefont{and}
  \bibinfo{author}{\bibfnamefont{R.~J.} \bibnamefont{Cava}},
  \bibinfo{journal}{Phys. Rev. Lett.} \textbf{\bibinfo{volume}{104}},
  \bibinfo{pages}{057001} (\bibinfo{year}{2010}).

\bibitem[{\citenamefont{Sasaki et~al.}(2011)\citenamefont{Sasaki, Kriener,
  Segawa, Yada, Tanaka, Sato, and Ando}}]{sasaki2011topological}
\bibinfo{author}{\bibfnamefont{S.}~\bibnamefont{Sasaki}},
  \bibinfo{author}{\bibfnamefont{M.}~\bibnamefont{Kriener}},
  \bibinfo{author}{\bibfnamefont{K.}~\bibnamefont{Segawa}},
  \bibinfo{author}{\bibfnamefont{K.}~\bibnamefont{Yada}},
  \bibinfo{author}{\bibfnamefont{Y.}~\bibnamefont{Tanaka}},
  \bibinfo{author}{\bibfnamefont{M.}~\bibnamefont{Sato}}, \bibnamefont{and}
  \bibinfo{author}{\bibfnamefont{Y.}~\bibnamefont{Ando}},
  \bibinfo{journal}{Phys. Rev. Lett.} \textbf{\bibinfo{volume}{107}},
  \bibinfo{pages}{217001} (\bibinfo{year}{2011}).

\bibitem[{\citenamefont{Hao and Lee}(2011)}]{hao2011}
\bibinfo{author}{\bibfnamefont{L.}~\bibnamefont{Hao}} \bibnamefont{and}
  \bibinfo{author}{\bibfnamefont{T.~K.} \bibnamefont{Lee}},
  \bibinfo{journal}{Phys. Rev. B} \textbf{\bibinfo{volume}{83}},
  \bibinfo{pages}{134516} (\bibinfo{year}{2011}).

\bibitem[{\citenamefont{Hsieh and Fu}(2012)}]{hsieh2012}
\bibinfo{author}{\bibfnamefont{T.~H.} \bibnamefont{Hsieh}} \bibnamefont{and}
  \bibinfo{author}{\bibfnamefont{L.}~\bibnamefont{Fu}}, \bibinfo{journal}{Phys.
  Rev. Lett.} \textbf{\bibinfo{volume}{108}}, \bibinfo{pages}{107005}
  (\bibinfo{year}{2012}).

\bibitem[{\citenamefont{Yamakage et~al.}(2012)\citenamefont{Yamakage, Yada,
  Sato, and Tanaka}}]{yamakage2012}
\bibinfo{author}{\bibfnamefont{A.}~\bibnamefont{Yamakage}},
  \bibinfo{author}{\bibfnamefont{K.}~\bibnamefont{Yada}},
  \bibinfo{author}{\bibfnamefont{M.}~\bibnamefont{Sato}}, \bibnamefont{and}
  \bibinfo{author}{\bibfnamefont{Y.}~\bibnamefont{Tanaka}},
  \bibinfo{journal}{Phys. Rev. B} \textbf{\bibinfo{volume}{85}},
  \bibinfo{pages}{180509(R)} (\bibinfo{year}{2012}).

\bibitem[{\citenamefont{Vollhardt and W\"{o}lfle}(1990)}]{vollhardt1990}
\bibinfo{author}{\bibfnamefont{D.}~\bibnamefont{Vollhardt}} \bibnamefont{and}
  \bibinfo{author}{\bibfnamefont{P.}~\bibnamefont{W\"{o}lfle}},
  \emph{\bibinfo{title}{The Superfulid Phases of Helium-3}}
  (\bibinfo{publisher}{Taylor and Francis, London}, \bibinfo{year}{1990}).

\bibitem[{\citenamefont{Turner et~al.}(2012)\citenamefont{Turner, Zhang, Mong,
  and Vishwanath}}]{turner2012quantized}
\bibinfo{author}{\bibfnamefont{A.~M.} \bibnamefont{Turner}},
  \bibinfo{author}{\bibfnamefont{Y.}~\bibnamefont{Zhang}},
  \bibinfo{author}{\bibfnamefont{R.~S.~K.} \bibnamefont{Mong}},
  \bibnamefont{and}
  \bibinfo{author}{\bibfnamefont{A.}~\bibnamefont{Vishwanath}},
  \bibinfo{journal}{Phys. Rev. B} \textbf{\bibinfo{volume}{85}},
  \bibinfo{pages}{165120} (\bibinfo{year}{2012}).

\bibitem[{\citenamefont{Roy}(2010)}]{roy2010topological}
\bibinfo{author}{\bibfnamefont{R.}~\bibnamefont{Roy}}, \bibinfo{journal}{Phys.
  Rev. Lett.} \textbf{\bibinfo{volume}{105}}, \bibinfo{pages}{186401}
  (\bibinfo{year}{2010}).

\bibitem[{\citenamefont{Roy}(2009{\natexlab{b}})}]{roy2009topological}
\bibinfo{author}{\bibfnamefont{R.}~\bibnamefont{Roy}}, \bibinfo{journal}{Phys.
  Rev. B} \textbf{\bibinfo{volume}{79}}, \bibinfo{pages}{195322}
  (\bibinfo{year}{2009}{\natexlab{b}}).

\bibitem[{\citenamefont{Wang et~al.}(2010)\citenamefont{Wang, Qi, and
  Zhang}}]{wang2010}
\bibinfo{author}{\bibfnamefont{Z.}~\bibnamefont{Wang}},
  \bibinfo{author}{\bibfnamefont{X.-L.} \bibnamefont{Qi}}, \bibnamefont{and}
  \bibinfo{author}{\bibfnamefont{S.-C.} \bibnamefont{Zhang}},
  \bibinfo{journal}{New J. Phys.} \textbf{\bibinfo{volume}{12}},
  \bibinfo{pages}{065007} (\bibinfo{year}{2010}).

\bibitem[{\citenamefont{Ryu et~al.}(2012)\citenamefont{Ryu, Moore, and
  Ludwig}}]{ryu2012electromagnetic}
\bibinfo{author}{\bibfnamefont{S.}~\bibnamefont{Ryu}},
  \bibinfo{author}{\bibfnamefont{J.~E.} \bibnamefont{Moore}}, \bibnamefont{and}
  \bibinfo{author}{\bibfnamefont{A.~W.~W.} \bibnamefont{Ludwig}},
  \bibinfo{journal}{Phys. Rev. B} \textbf{\bibinfo{volume}{85}},
  \bibinfo{pages}{045104} (\bibinfo{year}{2012}).

\bibitem[{\citenamefont{Wang et~al.}(2011)\citenamefont{Wang, Qi, and
  Zhang}}]{wang2011topological}
\bibinfo{author}{\bibfnamefont{Z.}~\bibnamefont{Wang}},
  \bibinfo{author}{\bibfnamefont{X.-L.} \bibnamefont{Qi}}, \bibnamefont{and}
  \bibinfo{author}{\bibfnamefont{S.-C.} \bibnamefont{Zhang}},
  \bibinfo{journal}{Phys. Rev. B} \textbf{\bibinfo{volume}{84}},
  \bibinfo{pages}{014527} (\bibinfo{year}{2011}).

\bibitem[{\citenamefont{Nomura et~al.}(2012)\citenamefont{Nomura, Ryu,
  Furusaki, and Nagaosa}}]{nomura2012cross}
\bibinfo{author}{\bibfnamefont{K.}~\bibnamefont{Nomura}},
  \bibinfo{author}{\bibfnamefont{S.}~\bibnamefont{Ryu}},
  \bibinfo{author}{\bibfnamefont{A.}~\bibnamefont{Furusaki}}, \bibnamefont{and}
  \bibinfo{author}{\bibfnamefont{N.}~\bibnamefont{Nagaosa}},
  \bibinfo{journal}{Phys. Rev. Lett.} \textbf{\bibinfo{volume}{108}},
  \bibinfo{pages}{026802} (\bibinfo{year}{2012}).

\bibitem[{\citenamefont{Shiozaki and
  Fujimoto}(2013)}]{shiozaki2013electromagnetic}
\bibinfo{author}{\bibfnamefont{K.}~\bibnamefont{Shiozaki}} \bibnamefont{and}
  \bibinfo{author}{\bibfnamefont{S.}~\bibnamefont{Fujimoto}},
  \bibinfo{journal}{Phys. Rev. Lett.} \textbf{\bibinfo{volume}{110}},
  \bibinfo{pages}{076804} (\bibinfo{year}{2013}).

\bibitem[{\citenamefont{Ryu et~al.}(2010)\citenamefont{Ryu, Schnyder, Furusaki,
  and Ludwig}}]{ryu2010topological}
\bibinfo{author}{\bibfnamefont{S.}~\bibnamefont{Ryu}},
  \bibinfo{author}{\bibfnamefont{A.~P.} \bibnamefont{Schnyder}},
  \bibinfo{author}{\bibfnamefont{A.}~\bibnamefont{Furusaki}}, \bibnamefont{and}
  \bibinfo{author}{\bibfnamefont{A.~W.~W.} \bibnamefont{Ludwig}},
  \bibinfo{journal}{New J. Phys.} \textbf{\bibinfo{volume}{12}},
  \bibinfo{pages}{065010} (\bibinfo{year}{2010}).

\bibitem[{\citenamefont{Ran}(2010)}]{ran2010weak}
\bibinfo{author}{\bibfnamefont{Y.}~\bibnamefont{Ran}},
  \bibinfo{journal}{arXiv:1006.5454}  (\bibinfo{year}{2010}).

\bibitem[{\citenamefont{Ho{\v{r}}ava}(2005)}]{hovrava2005stability}
\bibinfo{author}{\bibfnamefont{P.}~\bibnamefont{Ho{\v{r}}ava}},
  \bibinfo{journal}{Phys. Rev. Lett.} \textbf{\bibinfo{volume}{95}},
  \bibinfo{pages}{016405} (\bibinfo{year}{2005}).

\bibitem[{\citenamefont{Zhao and Wang}(2013)}]{zhao2013topological}
\bibinfo{author}{\bibfnamefont{Y.~X.} \bibnamefont{Zhao}} \bibnamefont{and}
  \bibinfo{author}{\bibfnamefont{Z.~D.} \bibnamefont{Wang}},
  \bibinfo{journal}{Phys. Rev. Lett.} \textbf{\bibinfo{volume}{110}},
  \bibinfo{pages}{240404} (\bibinfo{year}{2013}).

\bibitem[{\citenamefont{Zhao and Wang}(2014)}]{zhao2014topological}
\bibinfo{author}{\bibfnamefont{Y.~X.} \bibnamefont{Zhao}} \bibnamefont{and}
  \bibinfo{author}{\bibfnamefont{Z.~D.} \bibnamefont{Wang}},
  \bibinfo{journal}{Phys. Rev. B} \textbf{\bibinfo{volume}{89}},
  \bibinfo{pages}{075111} (\bibinfo{year}{2014}).

\bibitem[{\citenamefont{Hatsugai}(1993)}]{hatsugai1993}
\bibinfo{author}{\bibfnamefont{Y.}~\bibnamefont{Hatsugai}},
  \bibinfo{journal}{Phys. Rev. Lett.} \textbf{\bibinfo{volume}{71}},
  \bibinfo{pages}{3697} (\bibinfo{year}{1993}).

\bibitem[{\citenamefont{Essin and Gurarie}(2011)}]{essin2011bulk}
\bibinfo{author}{\bibfnamefont{A.~M.} \bibnamefont{Essin}} \bibnamefont{and}
  \bibinfo{author}{\bibfnamefont{V.}~\bibnamefont{Gurarie}},
  \bibinfo{journal}{Phys. Rev. B} \textbf{\bibinfo{volume}{84}},
  \bibinfo{pages}{125132} (\bibinfo{year}{2011}).

\bibitem[{\citenamefont{Chung and Zhang}(2009)}]{chung2009detecting}
\bibinfo{author}{\bibfnamefont{S.~B.} \bibnamefont{Chung}} \bibnamefont{and}
  \bibinfo{author}{\bibfnamefont{S.-C.} \bibnamefont{Zhang}},
  \bibinfo{journal}{Phys. Rev. Lett.} \textbf{\bibinfo{volume}{103}},
  \bibinfo{pages}{235301} (\bibinfo{year}{2009}).

\bibitem[{\citenamefont{Nagato et~al.}(2009)\citenamefont{Nagato, Higashitani,
  and Nagai}}]{nagato2009}
\bibinfo{author}{\bibfnamefont{Y.}~\bibnamefont{Nagato}},
  \bibinfo{author}{\bibfnamefont{S.}~\bibnamefont{Higashitani}},
  \bibnamefont{and} \bibinfo{author}{\bibfnamefont{K.}~\bibnamefont{Nagai}},
  \bibinfo{journal}{J. Phys. Soc. Jpn.} \textbf{\bibinfo{volume}{78}},
  \bibinfo{pages}{123603} (\bibinfo{year}{2009}).

\bibitem[{\citenamefont{Shindou et~al.}(2010)\citenamefont{Shindou, Furusaki,
  and Nagaosa}}]{shindou2010}
\bibinfo{author}{\bibfnamefont{R.}~\bibnamefont{Shindou}},
  \bibinfo{author}{\bibfnamefont{A.}~\bibnamefont{Furusaki}}, \bibnamefont{and}
  \bibinfo{author}{\bibfnamefont{N.}~\bibnamefont{Nagaosa}},
  \bibinfo{journal}{Phys. Rev. B} \textbf{\bibinfo{volume}{82}},
  \bibinfo{pages}{180505(R)} (\bibinfo{year}{2010}).

\bibitem[{\citenamefont{Volovik}(2010)}]{volovik2010}
\bibinfo{author}{\bibfnamefont{G.~E.} \bibnamefont{Volovik}},
  \bibinfo{journal}{JETP Lett.} \textbf{\bibinfo{volume}{91}},
  \bibinfo{pages}{201} (\bibinfo{year}{2010}).

\bibitem[{\citenamefont{Mizushima and Machida}(2011)}]{mizushima2011}
\bibinfo{author}{\bibfnamefont{T.}~\bibnamefont{Mizushima}} \bibnamefont{and}
  \bibinfo{author}{\bibfnamefont{K.}~\bibnamefont{Machida}},
  \bibinfo{journal}{J. Low. Temp. Phys.} \textbf{\bibinfo{volume}{162}},
  \bibinfo{pages}{204} (\bibinfo{year}{2011}).

\bibitem[{\citenamefont{Silaev}(2011)}]{silaev2011}
\bibinfo{author}{\bibfnamefont{M.~A.} \bibnamefont{Silaev}},
  \bibinfo{journal}{Phys. Rev. B} \textbf{\bibinfo{volume}{84}},
  \bibinfo{pages}{144508} (\bibinfo{year}{2011}).

\bibitem[{\citenamefont{Zhang and Kane}(2013)}]{zhang2013anomalous}
\bibinfo{author}{\bibfnamefont{F.}~\bibnamefont{Zhang}} \bibnamefont{and}
  \bibinfo{author}{\bibfnamefont{C.~L.} \bibnamefont{Kane}},
  \bibinfo{journal}{arXiv:1310.5281}  (\bibinfo{year}{2013}).

\bibitem[{\citenamefont{Atiyah and Segal}(1969)}]{atiyah1969equivariant}
\bibinfo{author}{\bibfnamefont{M.~F.} \bibnamefont{Atiyah}} \bibnamefont{and}
  \bibinfo{author}{\bibfnamefont{G.~B.} \bibnamefont{Segal}},
  \bibinfo{journal}{J. Diff. Geom.} \textbf{\bibinfo{volume}{3}},
  \bibinfo{pages}{1-18} (\bibinfo{year}{1969}).

\bibitem[{\citenamefont{Freed and Moore}(2013)}]{freed2012twisted}
\bibinfo{author}{\bibfnamefont{D.~S.} \bibnamefont{Freed}} \bibnamefont{and}
  \bibinfo{author}{\bibfnamefont{G.~W.} \bibnamefont{Moore}},
  \bibinfo{journal}{Ann. H. Poincar{\'e}} {\bf 14},1927 
  (\bibinfo{year}{2013}).

\bibitem[{\citenamefont{Ishikawa and Matsuyama}(1986)}]{ishikawa1986}
\bibinfo{author}{\bibfnamefont{K.}~\bibnamefont{Ishikawa}} \bibnamefont{and}
  \bibinfo{author}{\bibfnamefont{T.}~\bibnamefont{Matsuyama}},
  \bibinfo{journal}{Z. Phys. C} \textbf{\bibinfo{volume}{33}},
  \bibinfo{pages}{41} (\bibinfo{year}{1986}).

\bibitem[{\citenamefont{Nakahara}(2003)}]{nakahara2003geometry}
\bibinfo{author}{\bibfnamefont{M.}~\bibnamefont{Nakahara}},
  \emph{\bibinfo{title}{Geometry, topology and physics}}
  (\bibinfo{publisher}{CRC Press}, \bibinfo{year}{2003}).

\bibitem[{\citenamefont{Fu and Kane}(2006)}]{fu2006time}
\bibinfo{author}{\bibfnamefont{L.}~\bibnamefont{Fu}} \bibnamefont{and}
  \bibinfo{author}{\bibfnamefont{C.~L.} \bibnamefont{Kane}},
  \bibinfo{journal}{Phys. Rev. B} \textbf{\bibinfo{volume}{74}},
  \bibinfo{pages}{195312} (\bibinfo{year}{2006}).

\bibitem[{\citenamefont{Chen et~al.}(2013)\citenamefont{Chen, Gu, Liu, and
  Wen}}]{PhysRevB.87.155114}
\bibinfo{author}{\bibfnamefont{X.}~\bibnamefont{Chen}},
  \bibinfo{author}{\bibfnamefont{Z.-C.} \bibnamefont{Gu}},
  \bibinfo{author}{\bibfnamefont{Z.-X.} \bibnamefont{Liu}}, \bibnamefont{and}
  \bibinfo{author}{\bibfnamefont{X.-G.} \bibnamefont{Wen}},
  \bibinfo{journal}{Phys. Rev. B} \textbf{\bibinfo{volume}{87}},
  \bibinfo{pages}{155114} (\bibinfo{year}{2013}).

\end{thebibliography}
\end{document}